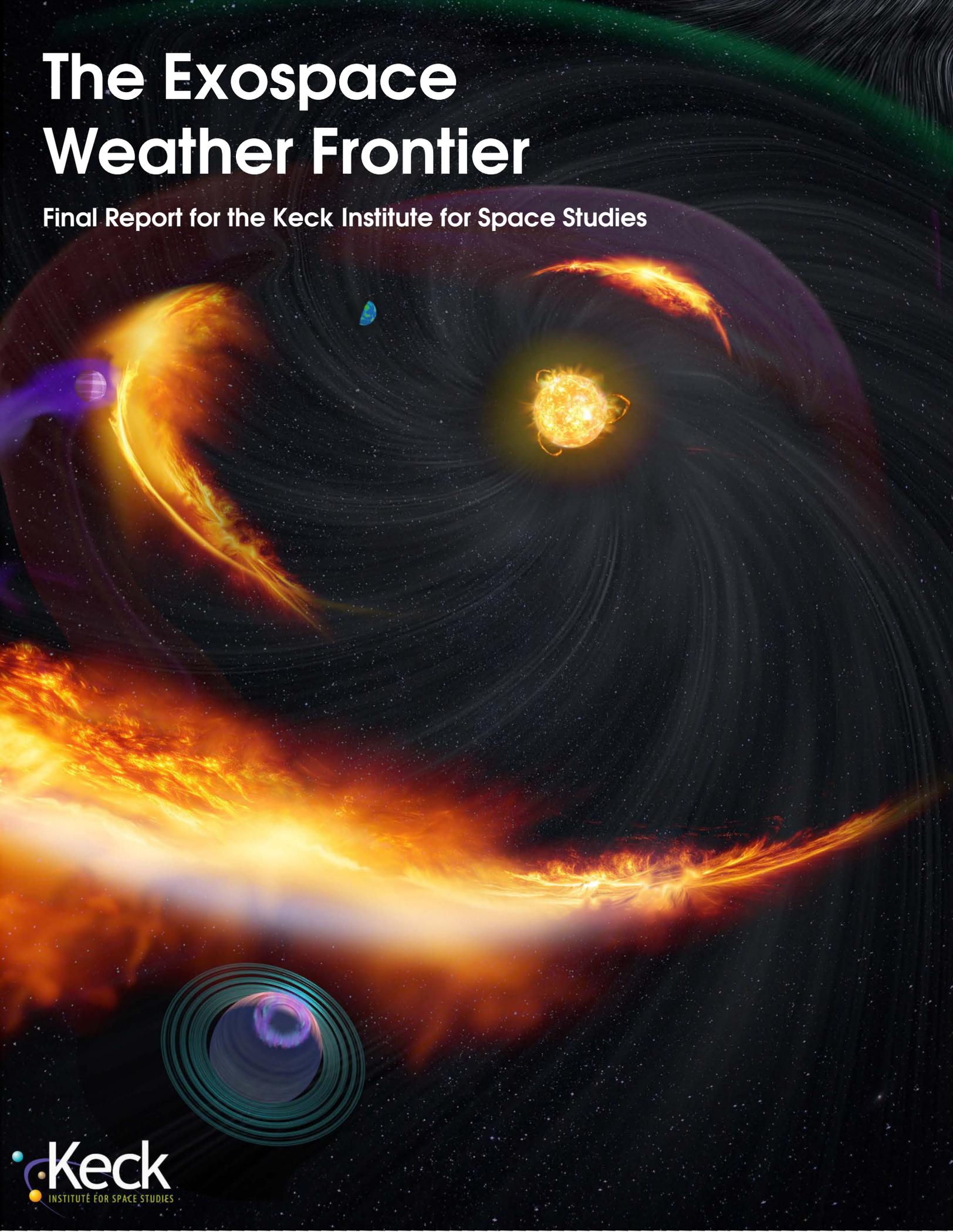

# The Exospace Weather Frontier

Final Report for the Keck Institute for Space Studies

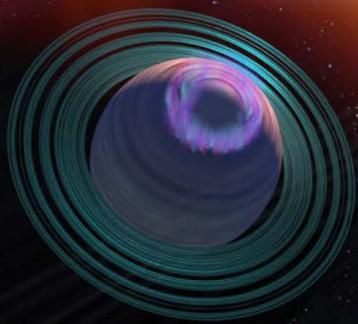

Keck
INSTITUTE FOR SPACE STUDIES

# The Exospace Weather Frontier

# Authors

**Study leads**

R. O. Parke Loyd (Eureka Scientific, Inc., astroparke@gmail.com)

Gregg Hallinan (California Institute of Technology, gh@astro.caltech.edu)

T. Joseph W. Lazio (Jet Propulsion Laboratory, California Institute of Technology, joseph.lazio@jpl.nasa.gov)

Evgenya L. Shkolnik (Arizona State University, shkolnik@asu.edu)

**Working group leads**

Magnetism: Julián Alvarado-Gómez (Leibniz Institute for Astrophysics Potsdam)

Exoplanets: Laura Amaral (Arizona State University)

Transients: Ivey Davis (California Institute of Technology)

Quasi-Steady: Alison Farrish (NASA Goddard)

Programmatics: Jim Green (NASA [Retired])

**Coauthors**

Dave Brain (University of Colorado, Boulder)

Bin Chen (New Jersey Institute of Technology)

Christina Cohen (California Institute of Technology)

Shannon Curry (University of Colorado, Boulder)

Karin Dissauer (NorthWest Research Associates)

Arika Egan (Applied Physics Laboratory, Johns Hopkins University)

Nat Gopalswamy (NASA/Goddard Space Flight Center)

Guillaume Gronoff (SSAI/NASA Langley Research Center)

Shadia Habbal (University of Hawai'i)

Renyu Hu (Jet Propulsion Laboratory, California Institute of Technology)

Meng Jin (Lockheed Martin Solar & Astrophysics Lab)

James Paul Mason (Applied Physics Laboratory, Johns Hopkins University)

Ruth Murray-Clay (University of California, Santa Cruz)

Kosuke Namekata (National Astronomical Observatory of Japan)

Rachel Osten (Space Telescope Science Institute)

Antígona Segura (Universidad Nacional Autónoma de México)

Astrid Veronig (University of Graz)

Aline Vidotto (Leiden University)

Maurice Wilson (High Altitude Observatory)

Yu Xu (Peking University)

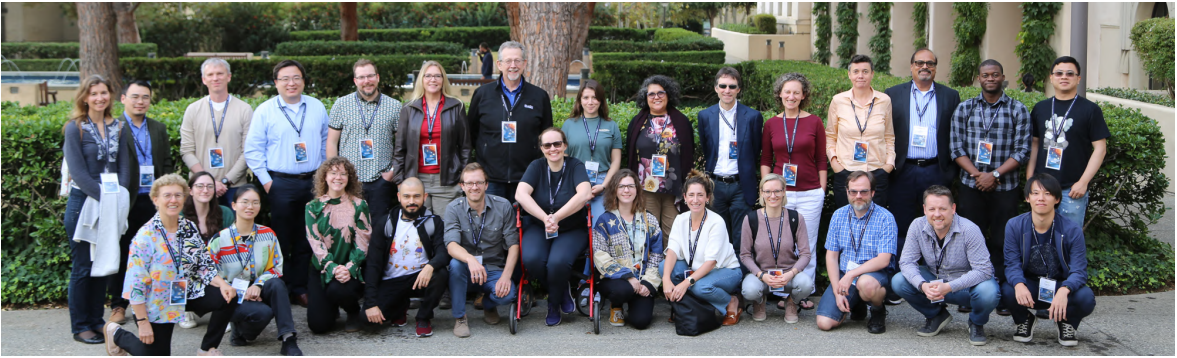

Authors of this report, the participants of the Blazing Paths held at the Keck Institute for Space Studies,

2023 November 6–10

Front Row (left to right):
Shadia Habbal (University of Hawai'i), Alison Farrish (NASA/Goddard Space Flight Center), Yu Xu (Peking University), Arika Egan (Applied Physics Laboratory, Johns Hopkins University), Julián Alvarado-Gómez (Leibniz Institute for Astrophysics Potsdam), R. O. Parke Loyd (Eureka Scientific, Inc.), Ruth Murray-Clay (University of California, Santa Cruz), Ivey Davis (California Institute of Technology), Shannon Curry (University of Colorado, Boulder), Guillaume Gronoff (SSAI/NASA Langley Research Center), Gregg Hallinan (California Institute of Technology), Kosuke Namekata (National Astronomical Observatory of Japan)

Back Row (left to right):
Aline Vidotto (Leiden University), Bin Chen (New Jersey Institute of Technology), David Brain (University of Colorado, Boulder), Renyu Hu (Jet Propulsion Laboratory, California Institute of Technology), James Paul Mason (Applied Physics Laboratory, Johns Hopkins University) Christina Cohen (California Institute of Technology), James Green (NASA [Retired]), Laura Neves Ribeiro do Amaral (Arizona State University), Antígona Segura (Universidad Nacional Autónoma de México) Joseph Lazio (Jet Propulsion Laboratory, California Institute of Technology), Evgenya Shkolnik (Arizona State University), Astrid Veronig (University of Graz), Nat Gopalswamy (NASA/Goddard Space Flight Center), Maurice Wilson (High Altitude Observatory), Meng Jin (Lockheed Martin Solar & Astrophysics Lab)

Not shown:
Heather Knutson (California Institute of Technology), Rachel Osten (Space Telescope Science Institute)

# Acknowledgments

The "Blazing Paths to Observing Stellar and Exoplanet Particle Environments" study was made possible by the W. M. Keck Institute for Space Studies, and by the Jet Propulsion Laboratory, California Institute of Technology, under contract with the National Aeronautics and Space Administration.

The study leads gratefully acknowledge the outstanding support of Harriet Brettle, Executive Director of the Keck Institute for Space Studies, as well as her dedicated staff, who made the study experience invigorating and enormously productive. Many thanks are also due to Bethany Ehlmann and the KISS Steering Committee for seeing the potential of our study concept and selecting it.

We thank all of the workshop participants for their time, enthusiasm, and contributions to the workshop and this report. The workshop was a memorable experience and set the stage for fruitful collaborations between people who would likely not have crossed paths were it not for the Keck Institute for Space Studies.

We thank study participant James Mason for creating a theme song for this workshop.[1]

We thank Yu-Chia Lin for her review of the calculations regarding stellar coronagraphy.

The content of this document is to be considered pre-decisional information and intended for planning and discussion purposes only.

---

[1] Theme song available at https://www.kiss.caltech.edu/workshops/exoplanet-particles/Blazing%20Paths%20Theme%20Song.mp4. Music is by Suno AI with lyrics and prompting by James Mason.

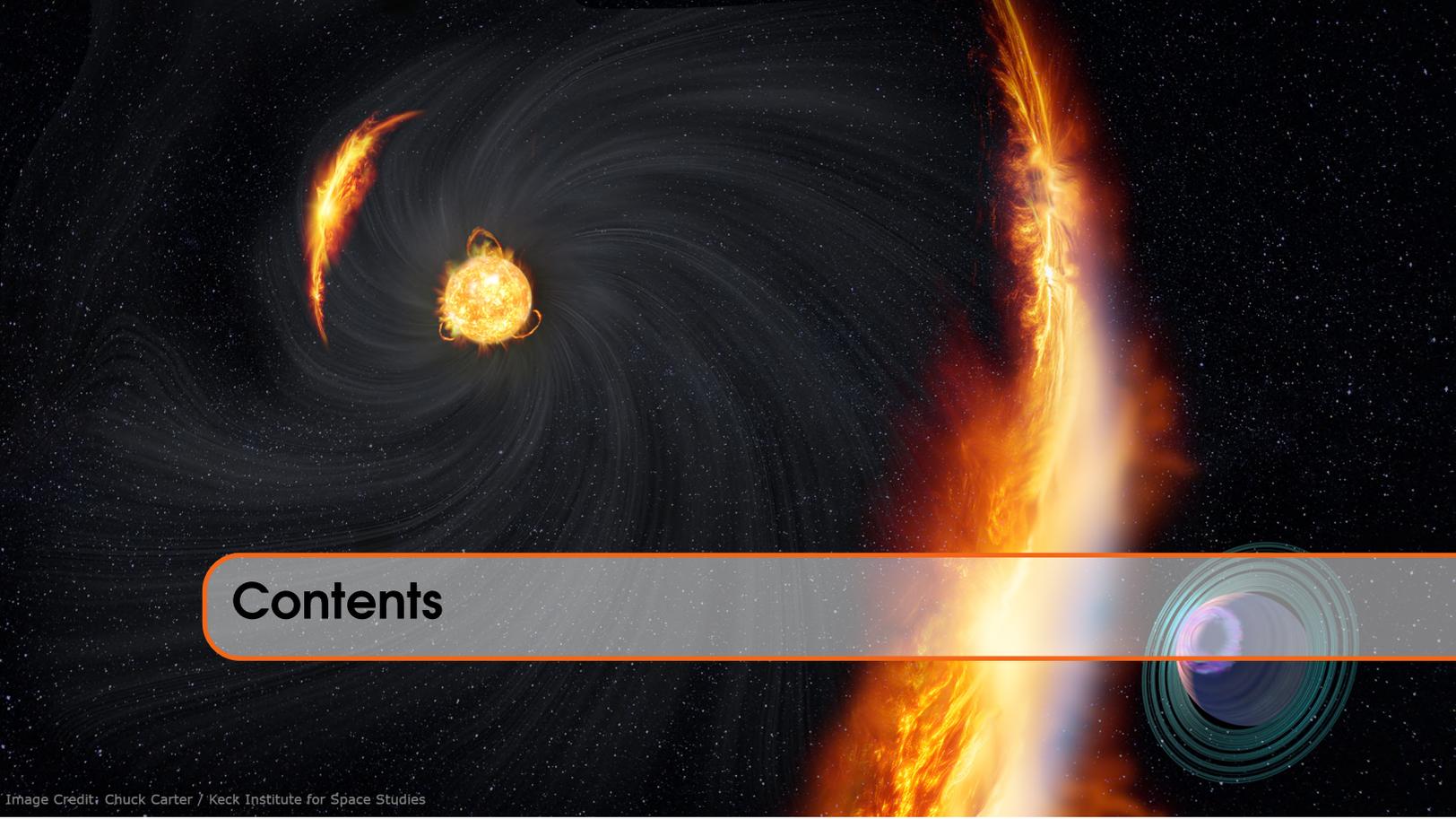

Image Credit: Chuck Carter / Keck Institute for Space Studies

# Contents

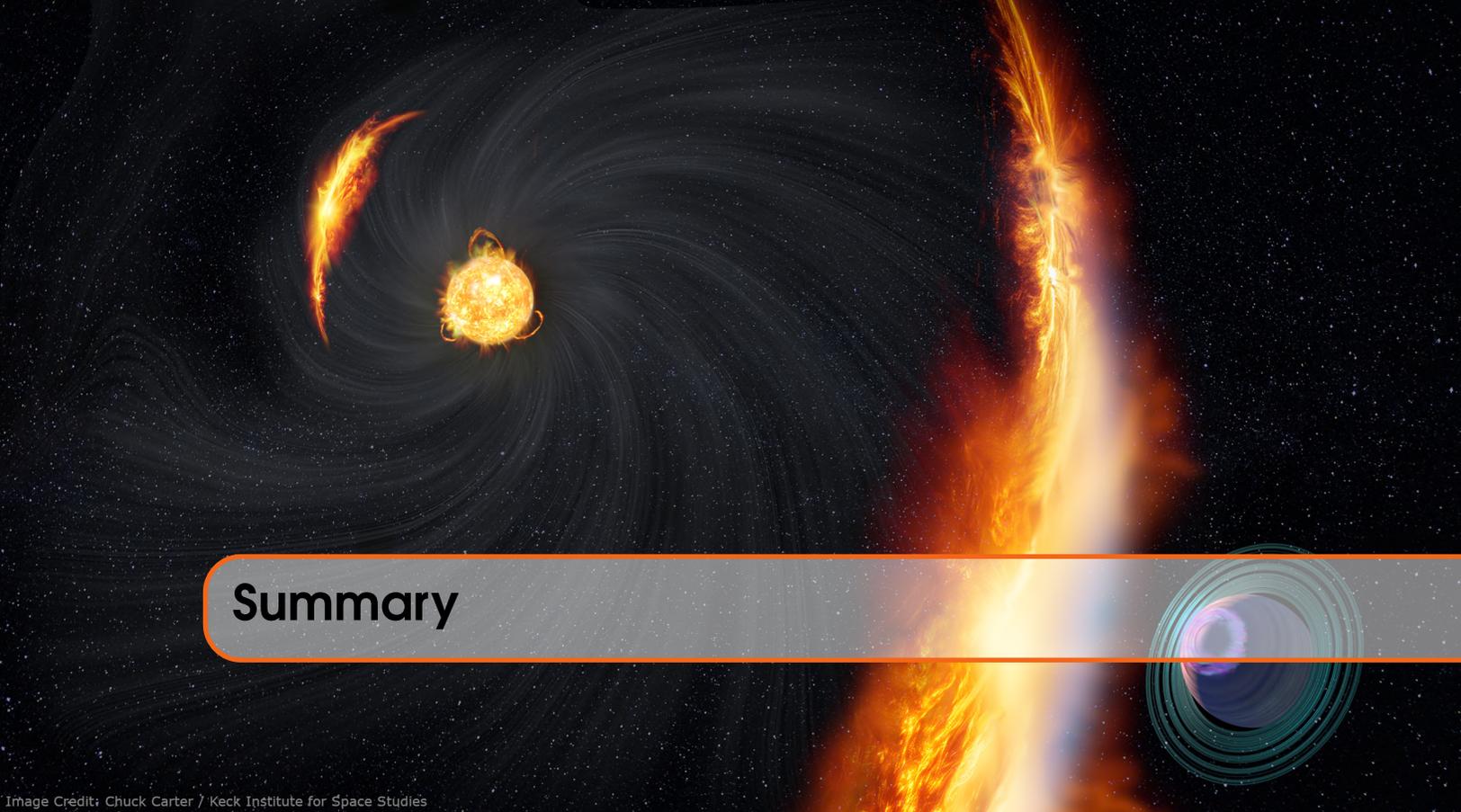

# Summary

## Space Weather Matters for Stars, Planets, and Life

Space weather is among the most powerful and least understood forces shaping planetary atmospheres. We observe its effects directly on Solar System bodies through atmospheric escape, chemical disruption, cometary tails, and auroral displays. Yet for exoplanets, we lack the tools and data to robustly assess how space weather influences their evolution, habitability, and potential biosignatures. Even the past space weather of the Solar System is shrouded in unknowns.

The Sun emits a constant outflow of charged particles embedded in magnetic fields, known as the solar wind. The fast-moving particles in the winds of the Sun and stars can erode planetary atmospheres over time, stripping away volatiles essential for climate and life. Potentially even more disruptive are explosive events, such as flares and coronal mass ejections (CMEs), which hurl vast amounts of energy and particles into space. Collectively, these phenomena define the space weather environment, and for our Sun, they are continuously tracked by a fleet of dedicated spacecraft.

Decades of observations reveal that space weather is not unique to the Sun, but is ubiquitous among stars. Many stars exhibit "exospace weather" through flaring activity, sometimes by orders of magnitude more intense than anything seen from the Sun. These stars must then also host stellar winds and many likely produce CMEs.

The study of exospace weather sits at the intersection of heliophysics, planetary science, astrophysics, and astrobiology. Observing space weather in exoplanetary systems, in combination with the beautifully resolved, in-situ context of our own system, is necessary to illuminate how stars and planets evolve, interact with their environments, and shape the conditions for life. Doing so will benefit heliophysics, planetary science, astrophysics, and astrobiology alike.

The time has come to establish exospace weather as a new pillar of exoplanet and stellar research, one that bridges stellar physics, planetary evolution, and the search for life beyond Earth. The science gaps are clear, and many of the tools to close them already exist or can be developed. What is needed now are coordinated, interdisciplinary research efforts.

## Framing the Challenge: Interdisciplinary Knowledge and Emerging Opportunities

The past decade has seen extraordinary advances in observing exoplanet atmospheres and detecting terrestrial-size planets in habitable zones. Yet the dynamic space environments of plasma and high-energy particles in which these planets reside remain largely inaccessible to direct observation. Without this context, models of planetary evolution remain incomplete.

Recognizing this need, the W. M. Keck Institute for Space Studies convened a cross-disciplinary team of scientists and instrument leaders from the planetary science, astrophysics, and heliophysics communities in 2023–2024 to explore opportunities for significant advances in exospace weather science. Organizing the subject into five core themes—magnetism and modeling, planets, quasi-steady space weather, transient space weather, and programmatics—our team synthesized foundational knowledge and identified a wide array of opportunities for progress.

This report is the product of that effort, offering new and creative scientific directions. It assembles essential background from related disciplines; outlines a range of observational campaigns, modeling innovations, methodological advances, and instrument developments poised to advance the field; and highlights outstanding theoretical challenges, proposing coordinated paths for community-wide progress. Together, these elements define a comprehensive scientific and implementation strategy to develop exospace weather into a challenging yet tractable research program and a new foundational element for the fields of heliophysics, astronomy & astrophysics, planetary science, and astrobiology.

## Understanding Exospace Weather: Priorities for Action

The following seven findings and recommendations distill the workshop's conclusions. Together, they define the current state of the field and identify specific priorities for progress.

### Finding 1

The impacts of the stellar particle environment affecting orbiting planets depend on a complex set of factors, many of which remain poorly characterized, including planetary atmospheric composition, stellar and planetary magnetic fields, and the system's evolutionary stage.

### Recommendation 1

Advance research on how exospace weather shapes planetary atmospheres and potential habitability. Focus on:
1. Modeling long-term atmospheric evolution under varied space weather conditions;
2. Observing atmospheric escape driven by stellar winds, coronal mass ejections, and stellar energetic particles;
3. Assessing how space weather may trigger or suppress conditions favorable to life.

### Finding 2

Modeling is essential for interpreting both exospace weather phenomena and their potential planetary impacts. These models must bridge sparse, indirect observations with complex underlying physics and are often highly sensitive to boundary conditions, such as stellar magnetic structure, coronal dynamics, and flare properties, as well as assumptions about planetary characteristics when relevant.

## Recommendation 2

Support the development and refinement of models that advance our understanding of stellar space weather and its potential planetary consequences. Improve existing models and develop new ones that incorporate the latest stellar and planetary physics. Focus areas include:

1. Magnetic boundary conditions across stellar types;
2. Heating mechanisms and structure in stellar coronae;
3. Emission measure distributions and flare evolution;
4. Coupled models that simulate atmospheric escape, chemistry, and long-term planetary evolution where applicable.

## Finding 3

No current observatory has been purpose-built to observe stellar particle environments. As a result, our constraints on key phenomena like stellar winds and coronal mass ejections remain speculative, often relying on instruments not designed for these detections.

## Recommendation 3

Prioritize the design and deployment of instruments explicitly optimized for detecting stellar particle events. Fund dedicated missions and instrumentation across radio, ultraviolet (UV), and X-ray regimes with capabilities such as:

1. High-contrast imaging of stellar coronae and winds;
2. Low-frequency radio arrays for stellar eruption bursts;
3. UV/X-ray spectrographs to track coronal dimming and outflows.

## Finding 4

As new techniques for detecting stellar space weather emerge, the field faces a critical credibility gap: many methods rely on indirect or proxy signals that have yet to be systematically validated, especially outside the solar context. Without rigorous testing and cross-comparison, scientific inferences risk being built on unverified assumptions.

## Recommendation 4

Establish frameworks to validate, compare, and benchmark stellar space weather diagnostics. This includes:

1. Testing new methods against solar analogs with known particle outputs;
2. Conducting intercomparison campaigns across multiple instruments and techniques;
3. Developing theory-informed metrics to evaluate the physical plausibility of claimed detections;
4. Supporting long-term monitoring to distinguish signal from astrophysical noise.

## Finding 5

The study of stellar energetic particles and coronal mass ejections is in its infancy. These phenomena are difficult to detect directly, but multiple indirect observational signatures, validated against the Sun, could open paths to studying them in other stars, especially with multiwavelength campaigns and coordinated modeling. Targeting extreme stars, such as young Suns, active M dwarfs, and planetary systems with biosignature potential, can maximize science return.

### Recommendation 5

Advance observational efforts to detect and characterize energetic particles and coronal mass ejections from stars beyond the Sun. Key approaches include:

1. Coordinated multiwavelength campaigns to capture transient space weather events;
2. Comparative analyses across stellar types to identify scaling relations and trends.

### Finding 6

Progress in exospace weather science requires sustained, deliberate investment in coordinated observational infrastructure. Existing facilities lack the sensitivity, cadence, or spectral coverage needed to characterize particle and magnetic environments or to capture key time-domain phenomena across diverse stellar types.

### Recommendation 6

Develop a strategic, multi-platform instrumentation plan to advance exospace weather research. Emphasis should be placed on:

1. Leveraging synergies across space- and ground-based observatories and wavebands;
2. Enabling long-duration time-domain monitoring of active and representative stellar populations;
3. Integrating exospace weather objectives into mission design and science traceability matrices from the earliest planning stages.

### Finding 7

Exospace weather science is inherently cross-disciplinary, yet existing programmatic structures often silo relevant expertise across astrophysics, heliophysics, planetary science, exoplanets, and astrobiology. These barriers limit conceptual integration, funding, and personnel pathways.

### Recommendation 7

Strengthen interdisciplinary integration across astrophysics, heliophysics, planetary science, exoplanets and astrobiology to advance exospace weather and star–planet interaction research. Support efforts such as:

1. Cross-domain data, model sharing, and co-analysis initiatives;
2. Development of shared terminology and conceptual frameworks;
3. Additional joint conferences, workshops, and collaborative research programs.

---

By recognizing exospace weather as a core astrophysical and exoplanetary science challenge, this report lays the foundation for a field poised to transform our understanding of planetary systems. Moving forward will require building up a sustained and diverse research program, new instruments, and a scientific culture that values and actively bridges disciplines.

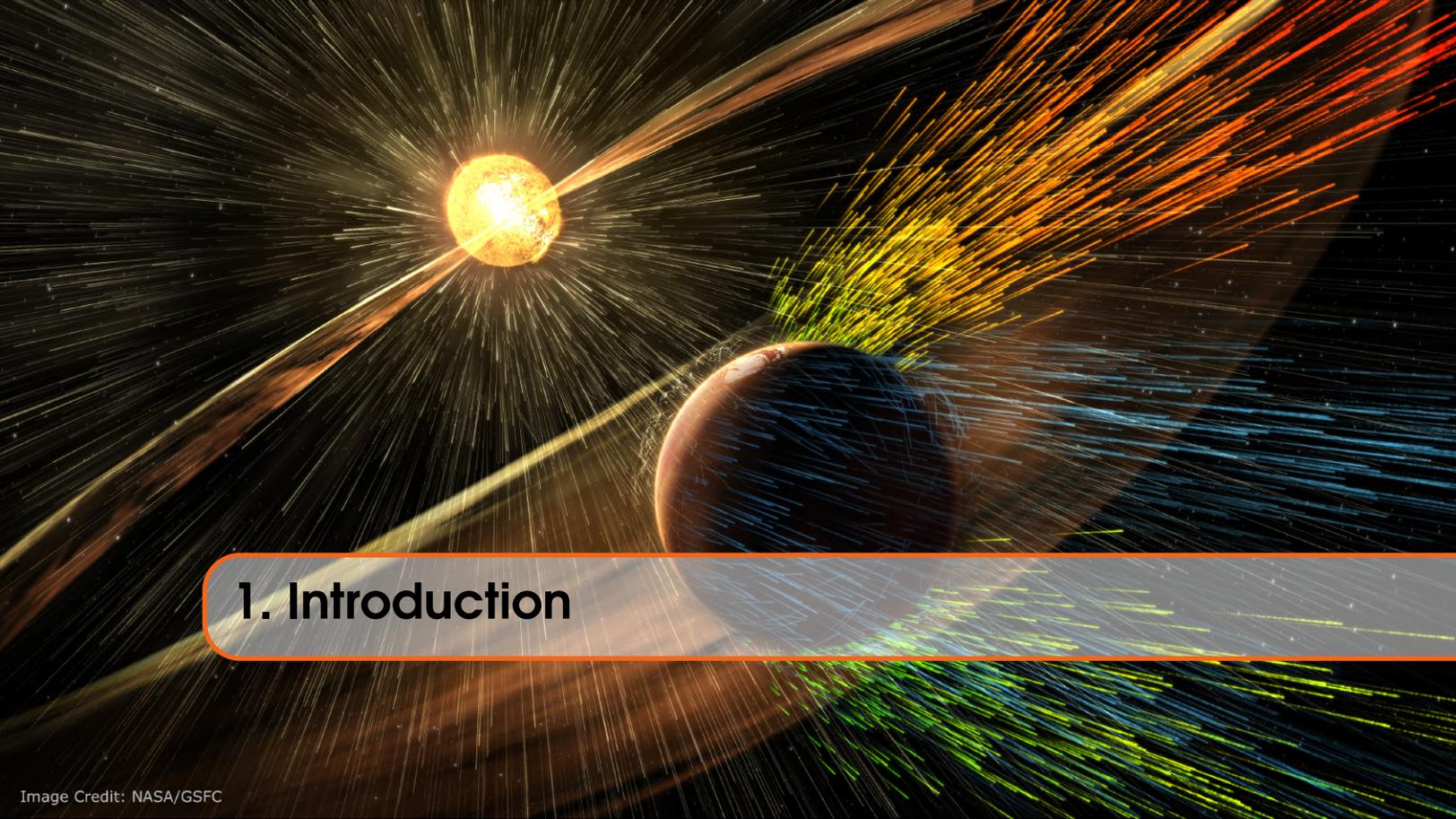

# 1. Introduction

With the confluence of new Solar System monitoring programs, renewed solar, stellar, and planet modeling efforts, and the high community priority to understand exoplanet atmospheres, tackling the problem of detecting and characterizing particles in the space environments of planets has never been more needed.

Stellar space weather produces environments that directly impact the evolution and present state of planets. It erodes and chemically modifies planetary atmospheres, affecting climate, water invento­ries, surface radiation, biogenesis, and biosignatures—all key science drivers of several current and future research endeavors, including NASA's next flagship, the Habitable Worlds Observatory (HWO) (National Academies of Sciences, Engineering, and Medicine 2021). Meanwhile, the stars themselves are impacted by the space weather they generate. Their particle winds and ejections shed angular momentum, weakening the rotationally driven magnetic activity responsible for generating their space weather. This produces a feedback of decreasing stellar activity, producing a drop in both the photon and particle emissions from stars as they age.

Now that astronomers are seeking to measure the composition and evolution of exoplanet atmo­spheres, the time has come to devote greater attention to the space environments provided by the stars they orbit. In the Solar System, the importance of the space environment is highlighted by observed impacts on planets, such as Earth and Mars, and the intensive monitoring of the Sun's emissions through a wide range of in-situ and remote sensing observatories. This suite of sensors track quasi-steady X-ray and extreme ultraviolet emission (EUV), the solar wind, and galactic cosmic rays (GCRs), along with impulsive events like flares, coronal mass ejections (CMEs), and solar energetic particle (SEP) events.

For stars other than our Sun, understanding stellar space weather requires a two-pronged approach: leveraging observations, theory, and modeling to constrain space weather properties across different stellar types, and second, by studying how these energetic processes shape planetary atmospheres. While the Solar System provides a well-monitored case study of space weather effects, extending this knowledge to exoplanetary systems requires new observational and modeling strategies.

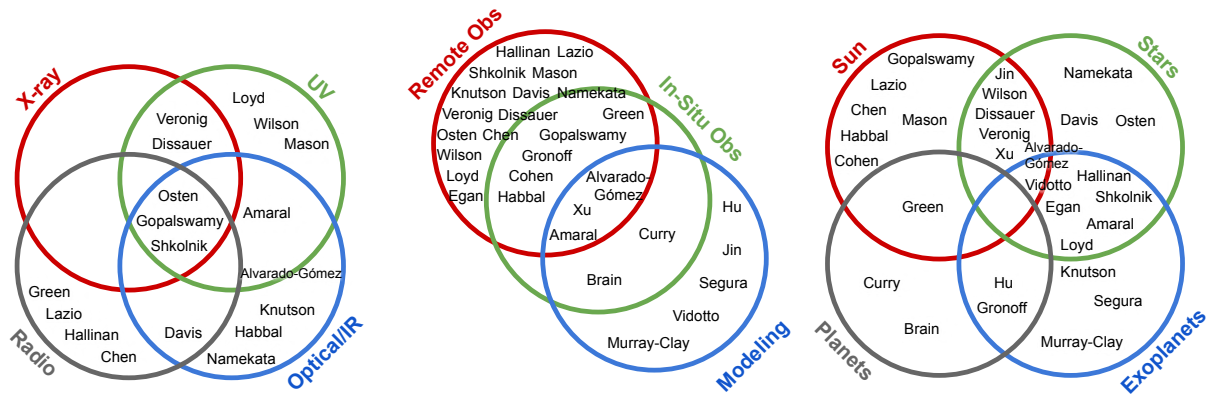

Figure 1.1: Distribution of expertise among the study team.

## 1.1   Motivation and Guiding Questions

A new era of (exo)planetary science demands a deeper grasp of the particle environments shaping planets, both in our Solar System and beyond. During 2023–2025, The Keck Institute for Space Studies supported a research study entitled "Blazing Paths to Observing Stellar and Exoplanet Particle Environments," which consisted of two in-person workshops and extensive asynchronous analysis and discussion. This effort brought together 30 scientists whose expertise spanned key aspects of the space weather topic: from observations to impacts, from the Solar System to stellar systems, from observations to theory and modeling, and from X-ray to radio wavelengths (Figure 1.1). The study's goal was to clarify and map the problems of observing the particle space weather of other stars and to identify promising paths forward using new or existing approaches.

We were guided by these three questions to identify and mature ways to observationally constrain and model space weather:

1. What is the relative importance between quasi-steady stellar winds and energetic, transient mass loss events to the local space environment of exoplanets, and how does this change with stellar type and age?

2. How do observable stellar properties——such as mass, rotation, activity metrics, metallicity, and magnetic field——shape stellar particle environments, including stellar winds and CMEs?

3. To what extent do energetic particles contribute to planet evolution and habitability, including prebiotic chemistry and the detection of biosignatures?

These questions serve as drivers for future efforts, integrating insights from astrophysics, heliophysics, and planetary science. Each of these disciplines has emphasized the need for observational constraints on planetary-system space weather in their respective decadal surveys (National Academies of Sciences, Engineering, and Medicine, 2021, 2022, 2024) and the Exoplanetary Science Strategy Report (National Academies of Sciences, Engineering, and Medicine, 2018).

Whereas a wide array of dedicated observing programs have begun building toward a comprehensive picture of the photon component of stellar space weather (Section 1.2), constraints on stellar particle emission lag far behind. Observations of winds, CMEs, and SEPs from stars other than the Sun are scarce at best. This absence of data is striking because most nearby stars are more magnetically active than the Sun, suggesting they produce more intense space weather. Identifying accurate and efficient detection techniques is foundational.

This KISS study aims to bridge the gap between stellar space weather observations and models and exoplanetary science by addressing three core objectives:

**Summarizing the state of the field** – providing an overview of what is currently known about stellar space weather and its impact on exoplanet atmospheres.

**Identifying observational and modeling gaps** – highlighting areas where new data and modeling are needed to improve our understanding of particle radiation environments.

**Proposing pathways forward** – outlining strategies for future observations, modeling efforts, and interdisciplinary collaborations that will accelerate progress in this field.

This report moves from the motivations to study exospace weather to production mechanisms, then measurement, and finally to programmatic considerations. It begins by outlining why particle space weather matters for exoplanets, then examines its stellar origins and how these can be observed and modeled. Subsequent chapters present techniques to detect and model particle environments and their planetary impacts. The final sections address programmatic strategies and opportunities for cross-disciplinary collaboration, culminating in recommendations for scientists, mission leaders, and funding agencies.

> **Not So Fast: Contested assumptions and common misconceptions.**
>
> Throughout the report, these boxes are used to identify and explain theories or assumptions often encountered in casual conversation, popular press, or even academic literature, that may not be accurate. A relevant example is the idea that magnetic fields shield planetary atmospheres from erosion by space weather. These boxes serve both as cautions and as invitations for renewed and deepened research inquiry.

## 1.2 What We Can Already Measure: The Photons of Planetary Environments

Planetary space environments consist of two key components: photon radiation and particle radiation, both of which strongly influence planetary evolution. High-energy photon radiation from stars, specifically X-ray and the near-UV, plays a crucial role in planet atmospheric heating, escape, and photochemistry. In comparison to particle radiation, photon emission at these short wavelengths is relatively well-characterized across stellar types and activity levels using space-based telescopes, including Galaxy Evolution Explorer (GALEX), NASA's Hubble and Chandra Space Telescopes (HST; CST), and ESA's X-ray Multi-Mirror (XMM-Newton) Mission, among others. However, photon-driven atmospheric processes do not act in isolation. The presence of energetic particles emitted through stellar winds, CMEs, and SEP events adds complexity to planetary environments that is still largely unquantified and remains a frontier in both observation and theory.

Photon radiation can both compete with and compound the impacts of particle radiation on planets. Far- and near-UV (FUV; NUV) emission sets the photochemical equilibrium of planetary atmospheres, establishing a baseline that particle space weather can disrupt. An example is ozone generated by UV photons. A planetary atmosphere's ozone can be destroyed by particle events (Section 2.3). Atmospheric loss is also affected. X-ray and extreme ultraviolet (together XUV) radiation heat the upper atmospheres of planets, inflating them to heights where they are more vulnerable to ion escape driven by stellar winds and CMEs (Lammer et al., 2008; Owen & Wu, 2017).

Both photon and particle radiation originate from the same underlying stellar magnetic activity. This connection in solar data reveals correlations between X-ray and UV flare flux and CME mass, speed, and kinetic energy on the Sun (Aarnio et al. 2011, Park et al. 2025). Similar relationships are expected for other Sun-like stars. These have motivated the development of a variety of scaling relationships between various high-energy wavelength ranges and particle flux in other stars, yet these should be approached with a measure of caution due to the lack of direct observational constraints. (Section 5.4). Beyond a few tentative detections, stellar particle data remain extremely limited, leaving a major gap in our understanding of exoplanet space weather and thus, exoplanetary atmosphere evolution.

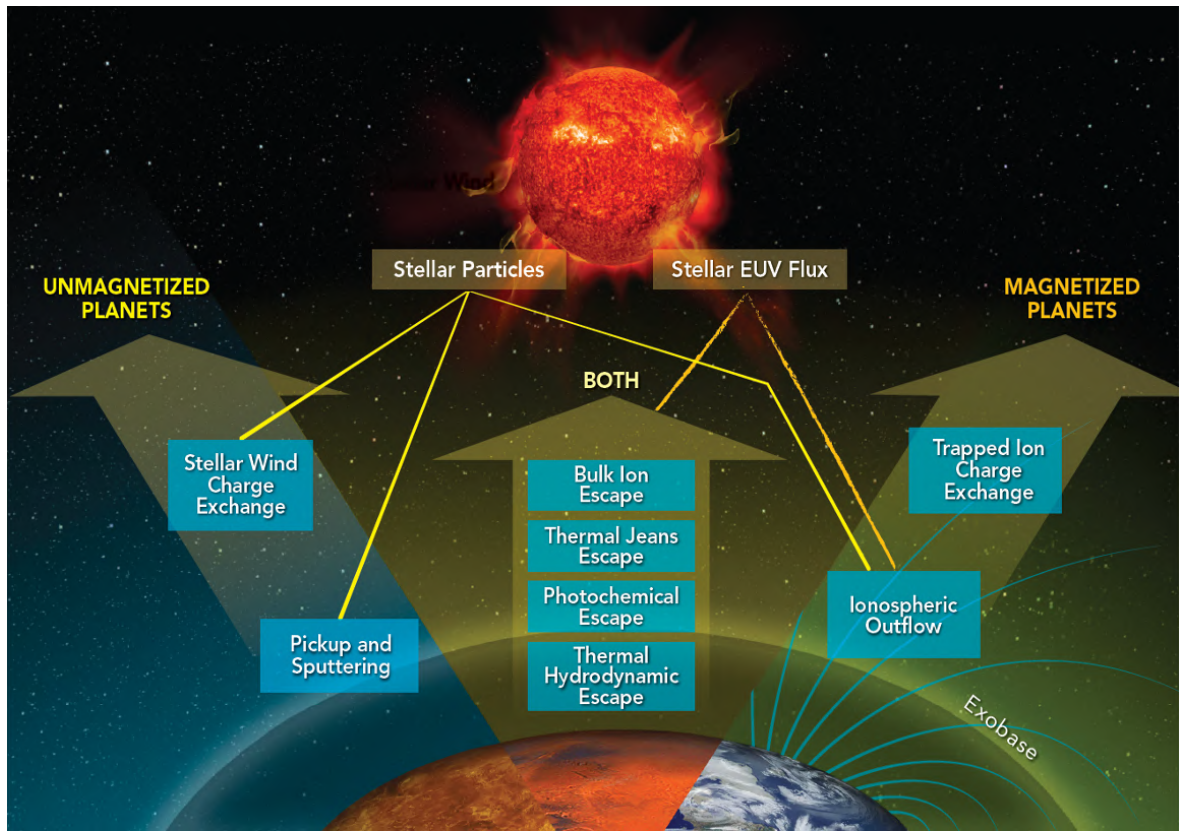

Figure 1.2: Pathways of atmospheric escape and modification driven by stellar radiation and particles. This schematic illustrates how stellar high-energy photon radiation and energetic particles affect planetary atmospheres through a range of processes. Both magnetized and unmagnetized planets are subject to escape mechanisms such as thermal hydrodynamic outflow, photochemical escape, and bulk ion loss. Unique processes arise depending on magnetic context: unmagnetized planets experience pickup and sputtering or charge exchange with the stellar wind, while magnetized planets exhibit trapped ion interactions and ionospheric outflow. Figure from Gronoff et al. (2020).

This KISS study occurred following two decades of strong and continuing growth in community effort devoted to characterizing the high-energy radiation environment of exoplanets. Earlier efforts from the *Sun-in-Time* project sought to constrain the past radiation environment of Solar System planets (Ribas et al., 2005). Following the exoplanet revolution of NASA's Kepler Space Telescope, numerous efforts have worked to characterize the X-ray and UV photometry and spectroscopy of stars across mass and age using primarily the GALEX and Hubble space telescopes, including the HAbitable Zones and M dwarf Activity across Time program (HAZMAT) (Shkolnik & Barman, 2014; Loyd et al., 2021), the Measurements of the Ultraviolet Spectral Characteristics of Low-mass Exoplanetary Systems (MUSCLES) surveys (France et al., 2014; France et al., 2016), the Far Ultraviolet M-dwarf

Evolution Survey (FUMES) (Pineda et al., 2021), mega-MUSCLES (Wilson et al., 2021), the Mors code (Johnstone et al., 2021), and several ongoing programs, not to mention efforts dedicated to characterizing individual targets such as the young, active, low-mass exoplanet host-star AU Mic (Feinstein et al., 2022). (See also the recent book on this topic by Linsky (2025).)

Despite these advances, major gaps remain, particularly in the EUV observations (10–100 nm). The existing few EUV observations of stars originate from the now-defunct Extreme Ultraviolet Explorer (EUVE), which ceased operations in 2001. Additionally, interstellar absorption makes it difficult to impossible to directly measure EUV flux from the vast majority of known exoplanet host stars. As such, most of the effort is put towards FUV and NUV observations with existing (e.g., HST, Swift) and upcoming (e.g., Star-Planet Activity Research CubeSat [SPARCS], Ultraviolet Transient Astronomy Satellite [ULTRASAT], Ultraviolet Explorer [UVEX]) space telescopes.

Another under-explored area is time variability of photon emission, components of which include rotation, activity cycles, and, especially, impulsive flares. For flares, the rate and energy of events span many orders of magnitude and cover a wide span of wavelengths that make them difficult to comprehensively characterize.

As efforts continue to characterize the variability and spectral diversity of photon emission, increased attention must be paid to the particle component of stellar space weather. While photon radiation largely governs atmospheric heating and photochemistry, energetic particles contribute additional ionization, chemical modification, and nonthermal escape that can reshape atmospheric evolution. The next chapter examines these processes in detail.

---

### Not Lost in Translation

A variety of terms are used within the communities and the literature for the same or related phenomena. Within this document, we use the term "space weather" to refer to the physical state within the Solar System that results from solar or nonsolar activity, and particularly as it affects the space environments near the planets in the Solar System. This usage is consistent with that in the Solar & Space Physics Decadal Survey *The Next Decade of Discovery in Solar and Space Physics: Exploring and Safeguarding Humanity's Home in Space*. The term "exospace weather" is used in a more general sense when discussing the physical state near stars, particularly in the context of exoplanet studies.

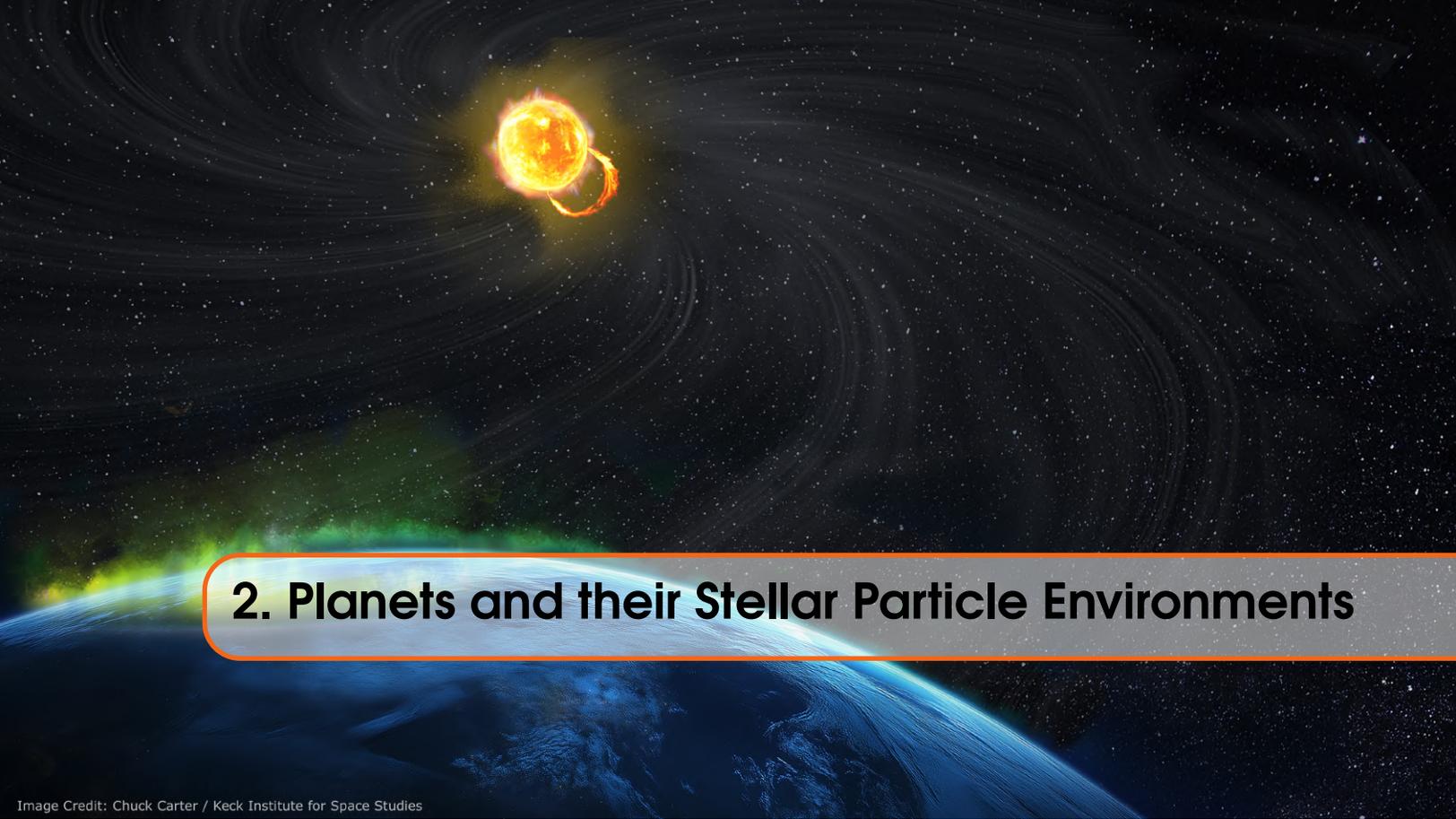

## 2. Planets and their Stellar Particle Environments

### 2.1 Introduction

Particle space weather can profoundly affect planetary atmospheres and their evolution, including atmospheric loss, different types of aurorae, photochemistry, haze formation, and more (e.g., Airapetian et al., 2016; Linsky, 2019; Airapetian et al., 2020; Gronoff et al., 2020). Depicted artistically in Figure 2.1, the intensity of these effects varies dramatically from star to star, planet to planet, and epoch to epoch. Given this breadth, developing a complete understanding of planetary evolution requires studies of the Sun and stars, planets, and exoplanets.

Exoplanetary systems provide access to a wide range of stellar types, ages, and activity levels for the study of these effects. Not only this, but entire populations of planets, such as hot rocky planets and warm "sub-Neptunes" ($1.6R_\oplus$–$3.0R_\oplus$ in radius) have no Solar System counterparts. For the foreseeable future, these classes of planets can only be studied from afar, yet Solar System planets provide an up-close view of the physics and chemistry of space weather effects that is indispensable to understanding and modeling the atmospheres and evolution of unfamiliar classes of planets. For many exoplanets, particularly those that may provide habitable conditions for life, Solar System planets provide direct natural analogs where the effects of space weather can be studied in a comparatively high degree of detail. Conversely, studying the effects of particles on exoplanets from more active stars can shed light on the history of our Solar System, providing insight into its youth when the Sun was more active and likely generated a harsher particle environment that more strongly influenced the atmospheres of Earth and its siblings.

The effects that particles have on planets depend largely on their particles' characteristic energies, as these generally determine the depth to which the particles can penetrate a planet's atmosphere. Figure 2.2 shows the energy spectra of a variety of particle radiation sources and the corresponding regions of Earth's atmosphere where they are absorbed. How this figure may translate to other planets will depend not just on differences in the incident fluxes from the host stars but also on the atmospheric

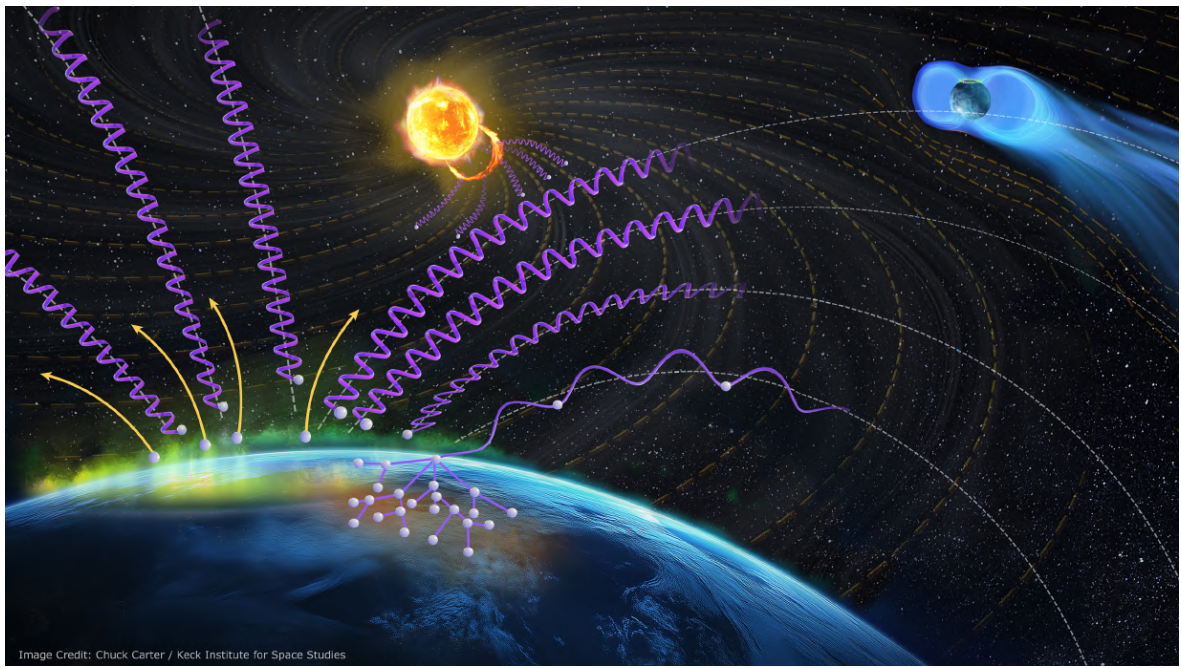

Image Credit: Chuck Carter / Keck Institute for Space Studies

Figure 2.1: Artist's illustration of a variety of planetary effects of space weather. In the background, the stellar host produces an eruption and associated energetic particles (purple spirals). In the foreground, particles from a previous event arrive at a terrestrial planet, funneled toward the visible pole by the planet's magnetic field (white dotted lines), where they generate auroral emission. A higher energy particle (loose spiral to the right) penetrates deeper into the atmosphere, where it initiates networks of nonthermal chemical reactions. In the top right, a gaseous planet closer to the star loses atmosphere in a bulk outflow that is partially shaped by the stellar wind and its spiraling magnetic field. Figure credit: Keck Institute for Space Studies / Chuck Carter. Full resolution figure available at `https://www.kiss.caltech.edu/artwork.html` and may be reused or modified on the condition of appropriate image credit.

and magnetospheric properties of the planet. Three key sources likely to impact most planets are the stellar wind, including their transient activities such as the Coronal Mass Ejections (CMEs), stellar energetic particles (SEPs; sometimes referred to as StEPs to differentiate the source as stellar rather than solar), and externally sourced galactic cosmic rays (GCRs).

In the Solar System, the solar wind (Section 4.1) is comprised of mostly ions and electrons with energies of the order of keV. Particles at these energies affect the thermospheres/ionospheres (i.e., the upper atmospheres, with densities $\lesssim 10^{13}$ species cm$^{-3}$ (about 0.03 Pa), not to be confused with the meteorologists' upper atmosphere in the upper layer of the troposphere) and exospheres (densities $\lesssim 10^6$ species cm$^{-3}$) of all planets, where they can serve as a source of ionization and dissociation, which leads to chemical reactions and a variety of nonthermal atmospheric escape processes Section 2.2). In addition, the presence of crustal magnetic fields for a planet helps the particle from the solar wind to get accelerated and enhances the energy released to the upper layers of the atmospheres. Auroral precipitation is driven by CMEs and other solar wind transients. However, CMEs do not directly impact the upper atmospheres of magnetized planets due to their magnetospheres. In contrast, unmagnetized planets receive CME particles more directly—though at lower flux, as they still traverse

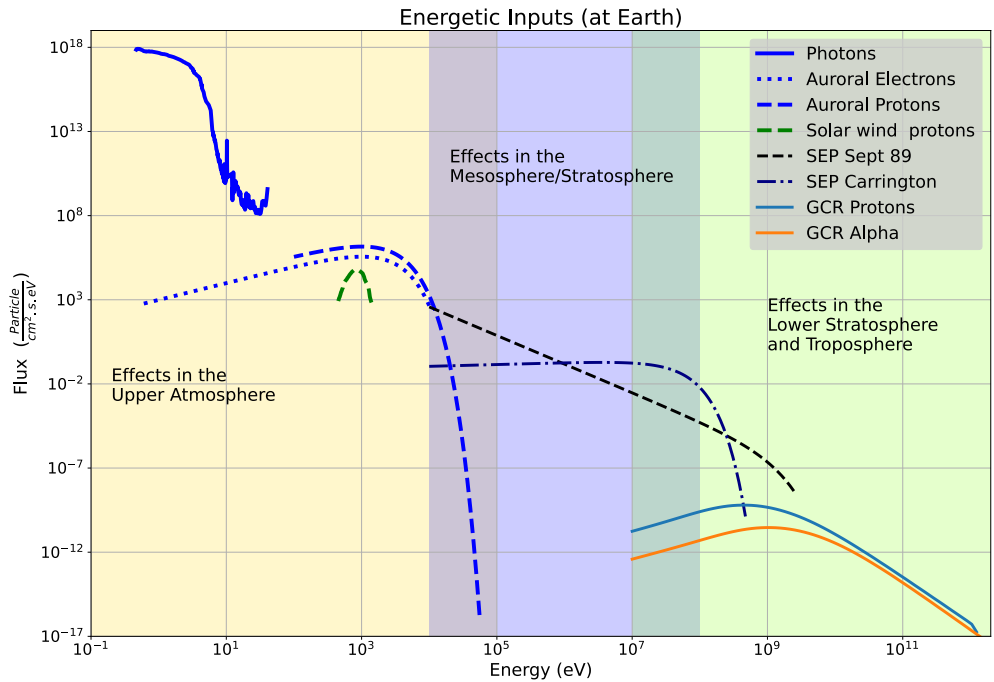

Figure 2.2: The fluxes of particle precipitation at Earth (including the photons) as a function of their energies and with the typical altitude where their effects are the most important. The CME impact on the magnetosphere will lead to particle acceleration and therefore to auroral protons/electron precipitation: those are not directly CME particles. Some of the most energetic solar energetic particle (SEP) events are included in this figure. The effects in the upper atmosphere are above $\approx 100$ km, in the mesosphere/stratosphere above $\approx 40$ km, and the other effects are down to the ground. The colored area shows the typical altitudes where the peak ionization is happening as a function of the energy: below $\approx 10^5$ eV in the upper atmosphere, then in the mesosphere/stratosphere and in the lower stratosphere/troposphere above $\approx 10^7$ eV. Credit: G. Gronoff

the bow shock. In magnetized planets, magnetic reconnection and wave-particle interactions amplify particle flux into auroral regions. For this reason, we do not treat CMEs as a distinct category here. Importantly, this applies only to the CME body; SEPs generated by CMEs behave differently.

SEPs (the solar equivalent of StEPs; Section 5.3) have energies in the megaelectron volt (MeV)—gigaelectron volt (GeV) range and typically penetrate to mesosphere levels. Here, densities are high enough to enable rapid recombination following ionization by a SEP particle, driving nonthermal chemistry. Effects include the creation of hazes and "prebiotic" molecules that might play a role in biogenesis (Sections 2.3, 2.5,(Airapetian et al., 2016; Gronoff et al., 2009a) )

Finally, the portion of GCRs that reach Earth (i.e., those not deflected by the heliosphere or solar wind; Section 4.2.1) have energies in the GeV–Tetraelectron Volt (TeV) range. Those penetrate down to the lower stratosphere ($\approx 20$ kPa), all the way to the Earth's surface, where they ionize molecules, modify chemistry, and serve as a damaging source of radiation (Airapetian et al., 2016; Gronoff et al., 2009a). However, their fluxes are too low to pose a serious hazard to life, and whether they produce any long-term effects on Earth's climate remains unclear (Gronoff et al., 2020; Airapetian et al., 2016,

Table 2.1: Observational signatures of planetary interactions with space weather, including both real-time and cumulative effects. Note that although listed, features of extrasolar observations could be tied to a space weather process; their interpretation is generally not unique, motivating further observations and modeling.

| Observational Signature | Solar System Planets | Exoplanets |
|---|---|---|
| Aurorae | Sandel & Broadfoot (1981); Mauk et al. (2002); Bertaux et al. (2005); Deighan et al. (2018) | Possible but not yet observed (though see Kao et al. 2016; Faherty et al. 2024 for brown dwarf observations) |
| Nonthermal Atmospheric Escape | Luhmann et al. (2007); Yau et al. (1985); Jakosky et al. (2018) | Possible but not yet observed |
| Interactions with Atmospheric Outflows | Not possible to use | Bourrier & Lecavelier des Etangs (2013); Schreyer et al. (2024) |
| Stripped Secondary Atmosphere | perhaps Mercury, e.g., Jasinski et al. (2020) | Zieba et al. (2023); Greene et al. (2023) |
| Chemical Perturbations | Sinnhuber et al. (2012) | Possible but not yet observed |
| Surface Space Weathering | Hapke (2001); Khurana et al. (2007); Denevi et al. (2023) | Lyu et al. (2024) |

References are examples, not exhaustive lists.

2020). However, nearby supernovae could have major effects on the atmosphere (e.g., the ozone layer), but it would be the effect of both particles and X-/gamma rays. This chapter does not further address GCRs; however, it is worth noting that GCRs could affect the chemistry of planets (Gronoff et al., 2009a,b; Loison et al., 2015; Dobrijevic et al., 2016). Other effects have been suggested, but impacts greater than those from SEPs are unlikely (Gronoff et al., 2020, and references therein).

Linking stellar particle environments to planetary evolution requires not just modeling, but observations of both the drivers and the consequences of space weather across diverse systems (Chapters 4 and 5) and may even make use of planets as real-time space-weather probes (e.g., Sections 4.3.4, 5.5.8) Space weather erosion of atmospheres could be detected by the lack of an atmosphere for a planet where one may otherwise be expected. Quantifying the role of space weather in the removal of atmospheres may require firm constraints on other potentially dominant sources and sinks, such as geological outgassing and the initial reservoir of volatile species, and how these factors vary across populations. Another avenue is to observe the impact of stellar particle fluxes on the chemical composition of planetary atmospheres, particularly a departure from thermochemical equilibrium. NASA's James Webb Space Telescope (JWST) and other powerful telescopes are key to observing atmospheric compositions through transmission and emission spectroscopy in order to infer these departures. Here again, firm constraints on all potentially dominant sources and sinks, such as photochemistry, aqueous chemistry, and geology, are essential.

In addition, space weather impacts can also provide an opportunity to better understand the fundamental properties of planetary magnetic fields. Radio auroral emission powered by space weather interactions directly probes the magnetic field strength of a planet, and is a key motivation for the development of sensitive, low-frequency observatories (Section 2.4)

Table 2.1 summarizes several promising observational avenues for detecting space weather interactions with planets, including examples from both the Solar System and exoplanetary systems. The remainder of this chapter explores in greater depth the mechanisms by which space weather shapes planetary environments: atmospheric escape, atmospheric chemistry, and the potential genesis and detection of life.

## 2.2 Atmospheric Escape

The space weather environment of Solar System planets has dramatic impacts on the retention of their atmospheres. The lessons learned from studying these bodies provide guidance for contemplating the likely impact of space weather on exoplanets, which live in a much wider array of stellar environments.

Atmospheres are critical to the character and habitability of terrestrial planets. While the retention of primordial gas envelopes depends primarily on the (thermal) hydrodynamic escape driven primarily by stellar extreme-ultraviolet (EUV), and to a lesser extent X-ray radiation (together called XUV, e.g., Owen, 2019), the long-term evolution of outgassed, secondary atmospheres can be substantially affected by the strength of nonthermal removal processes. For example, it has been suggested that atmospheric loss driven by stellar winds could completely strip away outgassed atmospheres from the TRAPPIST-1 planets, the only known planetary system hosting habitable-zone and rocky exoplanets that are potentially within the reach of JWST for detailed atmospheric characterization (Dong et al., 2018). The initial JWST observations of the rocky planets closest to the host star in that system suggested a lack of any atmosphere, notionally consistent with a strong cumulative impact by the stellar winds (Greene et al., 2023; Zieba et al., 2023). The likelihood of an atmosphere on the habitable-zone planets in that system, and thus their potential habitability and the scientific value to be pursued with large allocations of JWST observation time, critically depends on the relative strength between the particle-driven atmospheric loss and the geological processes that could replenish the atmosphere (e.g., Krissansen-Totton, 2023). More broadly speaking, whether rocky exoplanets across a wide range of size, temperature, and stellar type could retain their volatile endowment and host an atmosphere is becoming a central endeavor of today's exoplanet characterization, exemplified by the numerous JWST programs dedicated to test the "cosmic shoreline" hypothesis (Zahnle & Catling, 2017). As nonthermal escape driven by the solar wind has shaped the volatile retention of Solar System rocky planets, we consider it essential to understand how space weather in planetary systems orbiting other stars impacts the atmospheres of the planets they host.

> **Not So Fast: Magnetic fields do not necessarily protect planetary atmospheres from escape.**
>
> Statements that Earth's magnetic field has shielded it from stellar wind, thus protecting its atmosphere and surface water, can be found widely. In reality, it is far from certain whether Earth's or any planet's magnetic field plays a significant role in protecting the atmosphere (e.g., Gunell et al., 2018; Brain et al., 2021), and the Solar System itself provides evidence against the idea in the form of the thick atmosphere retained by unmagnetized Venus.

In the Solar System, we can use Venus, Earth, and Mars as probes to understand how dynamic space weather impacts terrestrial planets at different distances, with and without the presence of a magnetic field. For example, isotopic observations of the atmospheres on Mars and Venus have been used to infer extensive atmospheric escape that has occurred on these planets (e.g., Avice et al., 2022; Thomas et al., 2023). Atmospheric retention and the evolution of atmospheric composition on terrestrial planets in the Solar System are determined, primarily, by thermal processes at the earliest stage and nonthermal processes throughout the evolutionary history. These processes are driven by high-energy radiation or by particles and often mediated by magnetic fields. Figure 2.3 depicts a variety of processes producing atmospheric escape. In general, heavy ion escape is typically driven by the solar wind and its transient enhancement, while UV radiation still plays an essential role in setting up the thermosphere (Section 1.2). The radiation-driven escape of light species (Jeans and hydrodynamic escape) is also subject to the influence of the stellar wind environment. For an overview of the physics of these processes, we refer the reader to Gronoff et al. (2020).

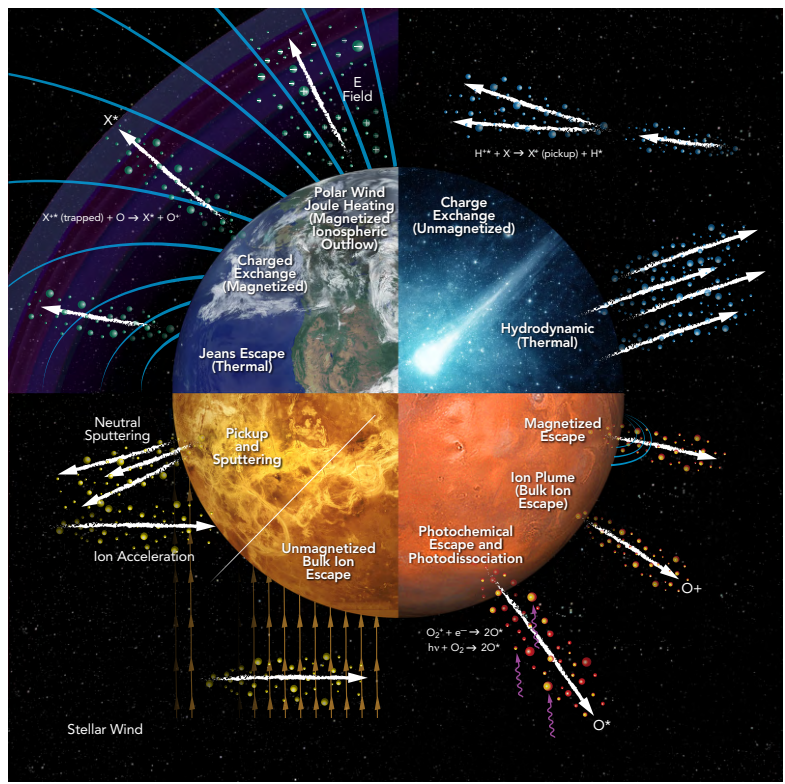

Figure 2.3: The variety of atmospheric escape processes. Shown are the major escape processes in magnetized planets (upper-left quarter), comets and hydrodynamically escaping planets (upper right), small unmagnetized planets (lower right), and larger unmagnetized planets under higher solar wind (lower left). (From Gronoff et al., 2020)

Figure 2.4, from the review by Ramstad & Barabash (2021), shows how the escape rates of heavy ions in these three planets, overall, change during the solar cycle. Along the solar cycle, the EUV flux, the frequency of transient events, such as flares and CMEs, and the solar wind conditions change. In the Earth's case, the escape rate depends also on the geomagnetic conditions, and ranges between about $10^{24}\,\mathrm{s}^{-1}$ during low EUV conditions (i.e., solar minimum) to about $10^{25}\,\mathrm{s}^{-1}$ in the average state (Slapak et al., 2017; Schillings et al., 2019) to as much as $10^{26}\,\mathrm{s}^{-1}$ in the solar maximum (Yau et al., 1985; Nilsson et al., 2012). When we compare the escape rates in different EUV levels, Earth's ion escape increases with the EUV flux, but also with the dynamic pressure of the solar wind by a factor of $\sim 10$ (Figure 2.5).

Even though Mars and Venus do not have an intrinsic dipole planetary magnetic fields like the Earth, many of the same physics of atmospheric escape can be applied. Their present-day escape rates range between $10^{24}\,\mathrm{s}^{-1}$ and $10^{26}\,\mathrm{s}^{-1}$; however, the leading mechanisms and limiting factors for each planet vary significantly (Lammer & Kasting, J.F., 2008; Curry et al., 2015a; Gronoff et al., 2020; Curry et al., 2025).

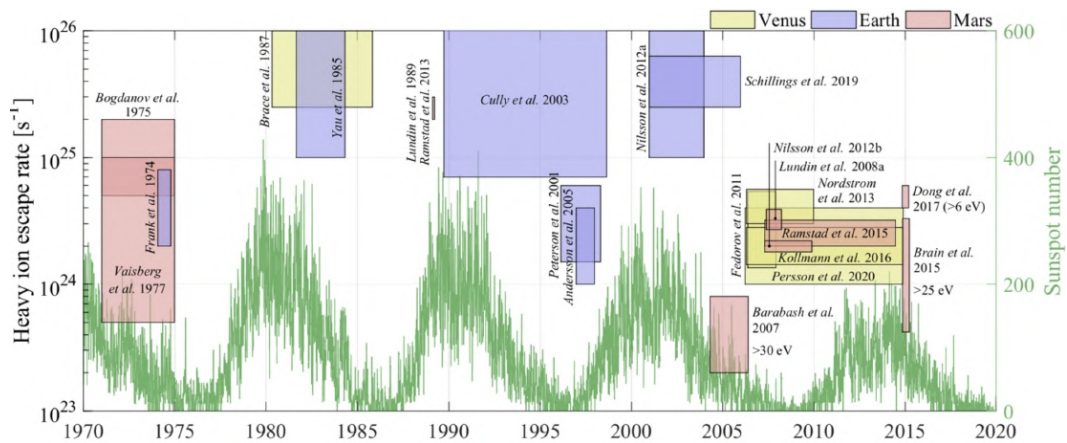

Figure 2.4: In-situ measurements of heavy ions ($O^+$ and heavier) escape rates from Venus, Earth, and Mars throughout several solar cycles. Figure from Ramstad & Barabash (2021), Springer Nature, CC BY license.

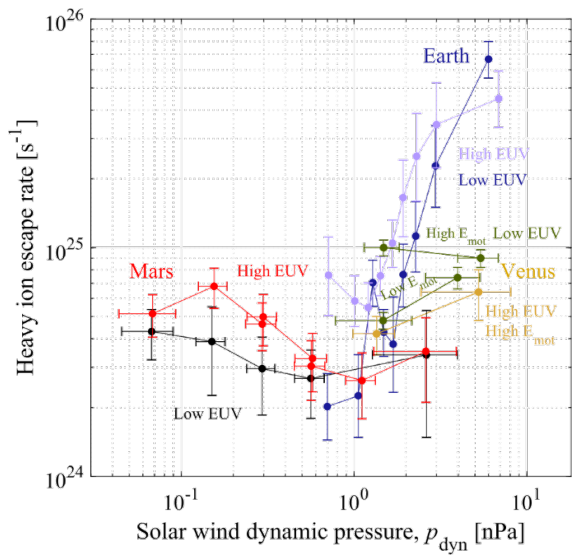

Figure 2.5: In-situ measurements of heavy ions ($O^+$ and heavier) escape rate from Venus (Masunaga et al., 2019), Earth (Schillings et al., 2019), and Mars (Dong et al., 2017b; Ramstad et al., 2018) as a function of solar wind dynamic pressure for low and high EUV levels. Figure from Ramstad & Barabash (2021), Springer Nature, CC BY license.

Transient solar events exert additional effects on the terrestrial planets of our Solar System. Interplanetary CMEs have been repeatedly observed to significantly enhance ion outflow and atmospheric escape (Curry et al., 2015b; Luhmann et al., 2017). Additionally, for Mars, a stream interaction region (SIR) during the solar minimum can perturb the plasma environment of the planet more than during the maximum, enhancing the ion outflow by one order of magnitude. This may happen due to the ionization levels of the upper atmosphere during the maximum being high, which can hold the plasma environment in place (Kajdič et al., 2021).

The atmospheres of many known terrestrial exoplanets are also likely battered by space weather orders of magnitude more intense due to more active hosts or closer orbital distances. For example, Dong et al. (2017a) found through application of magnetohydrodynamic (MHD) models simulating both the stellar wind and planetary ion escape, that Proxima Centauri's terrestrial, habitable-zone planet could experience oxygen escape two orders of magnitude greater than the Solar System terrestrials for a $CO_2$-rich atmosphere. Recent observations of TRAPPIST-1 b & c, two terrestrial planets slightly larger than Earth in close-in orbits of the parent star, indicate the absence of a thick atmosphere (Zieba et al.,

> **Not So Fast: Stellar winds may not only subtract from a planetary atmosphere.**
>
> Recent kinetic simulations (i.e., simulations properly accounting for situations in which the fluid approximation breaks down), indicate the interaction with the solar wind may result in the deposition of more hydrogen than is removed from the Martian atmosphere (Hinton et al., 2024).

2023; Greene et al., 2023). A $CO_2$-rich outgassed atmosphere on these planets could be quickly removed by intense ion escape powered by the stellar winds (Dong et al., 2018).

The gaseous exoplanets that have ongoing hydrodynamic atmospheric escape can produce observable outflows of hydrogen and helium (Dos Santos, 2022; Orell-Miquel et al., 2024). Studies indicate that stellar XUV radiation can provide sufficient energy to power these outflows (Murray-Clay et al., 2009). The interaction of these outflows with the surrounding space environment strongly influences their observability, as well as what we can ascertain about the processes and physics that drive them (e.g., Bisikalo et al., 2013; Matsakos et al., 2015; McCann et al., 2019; Carolan et al., 2021). This, in turn, lets us extrapolate these observations to dynamics, such as the atmospheric mass loss rate.

Simulations of both terrestrial nonthermal escape process and gaseous planet thermal escape have, thus far, generally relied on assumed stellar wind conditions or winds based on MHD models constrained primarily from magnetic observations of the host star (Section 3). For most exoplanet hosts, these models lack observational constraints on key outputs needed for planet models, e.g., stellar wind ram pressure and energetic particle flux. This represents a critical uncertainty in downstream simulations and hampers any effort to evaluate the impact of atmospheric escape processes on the long-term evolution of the planets. In the Solar System, observational constraints on the particle space weather of young solar analogs are critical to reconstructing the atmospheric evolution of Venus, Earth, and Mars. Improving these observational constraints is essential for advancing our understanding of planetary evolution.

## 2.3  Atmospheric Chemistry

High-energy particles impact the atmospheres and surfaces of terrestrial bodies in the Solar System, driving both chemical and physical changes. Particles accelerated by stellar activity are accompanied by ionizing radiation (XUV), which enhances their effects—for example, by contributing to atmospheric escape processes (Section 2.2).

On Earth, the $N_2$-rich atmosphere absorbs XUV radiation above 90 km, while energetic particles can penetrate all the way to the surface (Mironova et al., 2015). After XUV, the most likely agents for breaking the nitrogen molecule are lightning and high-energy particles, which act through indirect processes (Sinnhuber et al., 2012; Mironova et al., 2015). Once produced, nitrogen atoms combine with oxygen to form $NO_x$ species (N, NO, $NO_2$, $NO_3$, $N_2O_5$) that destroy ozone through catalytic reactions. Particles also dissociate water, forming $HO_x$ species (H, OH, $HO_2$, $H_2O_2$), which are highly reactive and deplete both $O_3$ and $CH_4$.

High-energy particles typically enter Earth's atmosphere near the poles, producing $HO_x$ and $NO_x$ in the mesosphere and stratosphere. $HO_x$ compounds are short-lived and act locally, while $NO_x$ can persist for days and spread to lower altitudes and latitudes, where they drive most ozone depletion. During SEP events, $NO_x$ can reduce ozone for months (Sinnhuber et al., 2012; Mironova et al., 2015). Large energies are required to break the $N_2$ triple bond (9.8 keV), and photodissociation occurs between 60 and 100 nm——corresponding to EUV wavelengths. These EUV photons are absorbed at high altitudes and contribute to the formation of thermospheres in dense atmospheres (Yung & DeMore, 1999).

The role of particles is more evident during energetic transient events that can accelerate more particles and compress Earth's magnetic field. For example, during the Carrington Event (Section 5), a powerful solar storm in 1859 and the most energetic flare recorded in recent history, aurorae were observed at very low latitudes (Hayakawa et al., 2020) and the whole telegraph network was damaged (Giegenback, 2015) as a result of the shower of high-energy particles that reached Earth's atmosphere. Numerical models calculated the $NO_x$ created during the event depleted the stratospheric ozone by 20% weeks after the event and warmed part of Russia and Europe by 7 K (Calisto et al., 2013).

Despite several missions dedicated to the study of Venus and Mars, there have been few investigations into the bulk atmospheric chemistry changes initiated by the impact of high-energy particles. Most studies of space weather effects have focused on atmospheric escape (Section 2.2), aurorae (Section 2.4), the interaction of the solar wind with the magnetic field it induces at Venus (Luhmann, 1986; Futaana et al., 2017), and with Mars' fossil magnetic field and atmospheric ionization (Luhmann, 1990; Crider, 2004).

The effect of space weather on the neutral atmospheric chemistry of Venus and Mars is relevant for the future characterization of $CO_2$-rich exoplanetary atmospheres. Nakamura et al. (2023) provided the first calculations of neutral chemistry for Mars initiated by high-energy particles. Their numerical 1D simulations showed the atmospheric chemistry is dominated by $CO_2$ ionization and the catalytic cycle recombines CO back to $CO_2$. The main difference of the neutral chemistry initiated by high-energy particles on Mars compared to Earth would be the dominance of $HO_x$ on the ozone chemistry. These numerical results could potentially be tested with observations from European Space Agency's (ESA) ExoMars Trace Gas Orbiter (Nakamura et al., 2023). The studies of space weather effects on Venus have been mostly focused on the atmospheric escape and ionization (e.g., Airapetian et al., 2020), but some models predict the ion clusters formed by high-energy particles are relevant as nucleation sites for clouds and indicate the ion chemistry may have an impact in the neutral chemistry (Aplin, 2013).

For all terrestrial planets and moons, the atmospheric composition depends on the initial budget, atmospheric escape, atmosphere-surface interactions, exchange with the planetary interior, temperature profile, and atmospheric chemical reactions. Atmospheric compositions for such rocky bodies can go from noble gases for very cold planets (< 50 K, like Pluto) to variable combinations of $N_2$, $CO_2$/CO/$CH_4$, $O_2$, and $H_2O$ for atmospheres with $T_{eq} < 300$ K and contributions of $SO_2$, Na, and SiO for hottest atmospheres (Wordsworth & Kreidberg, 2022). Warm, potentially habitable planets (surface temperatures around 300 K) are usually considered to have atmospheres composed of $CO_2$-$N_2$-$H_2O$ (Kopparapu et al., 2013). An interesting proposal is the existence of exo-Titans (Mandt et al., 2022), exoplanets similar to the largest Saturn satellite, Titan. This satellite has a dense atmosphere composed of $N_2$ and $CH_4$, where particles accelerated by the magnetosphere of Saturn break $N_2$, while $CH_4$ is photolyzed by solar UV, generating a rich hydrocarbon chemistry that creates hazes that give the satellite its characteristic orange color (Waite et al., 2007). The effects of space weather on this type of exo-atmosphere has not yet been studied.

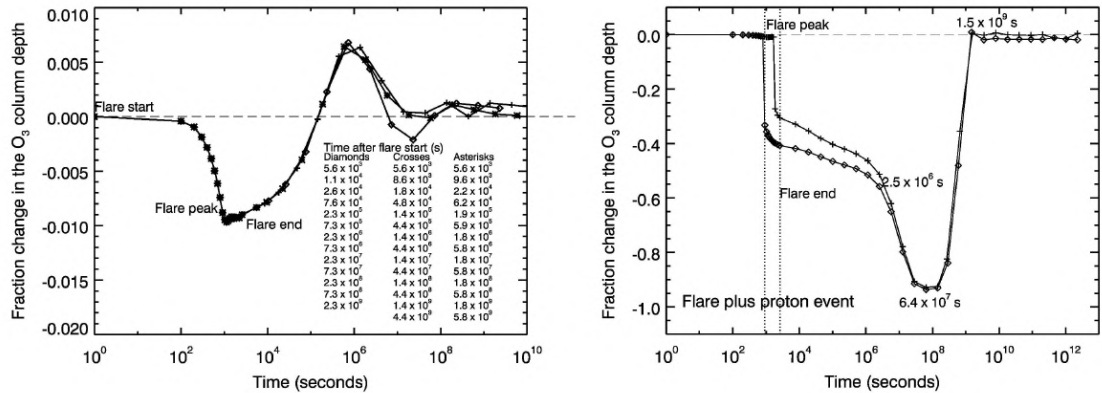

Figure 2.6: Fraction of ozone depleted by an energetic flare considering only the emitted UV flux *(left)* and the UV plus a proton event associated with the flare. Figure from Segura et al. (2010).

### 2.3.1 Exospace Weather impacts on Planets Orbiting M dwarfs

Due to their favorability to observations, the study of the effect of space weather in terrestrial exoplanets has been mostly focused on understanding its effect on habitability and biosignatures on planets in the habitable zone of M dwarfs. M dwarfs emit energetic flares more frequently than the Sun (e.g., Ramsay & Doyle (2015)) and, based on our knowledge of the Sun, we expect the emission of high-energy particles during these events. The first work regarding the effect of high-energy particles during a transient event in the atmosphere of a habitable-zone Earth-like planet was performed by Segura et al. (2010). Using a 1D photochemical model coupled to a 1D radiative convective model, they calculated the effect of an energetic flare from AD Leonis (AD Leo) on a planetary atmosphere with the same bulk composition present in Earth's atmosphere (0.21 $O_2$, 0.78 $N_2$). The simulations showed that while the flare photons did not significantly perturb atmospheric ozone, the $NO_x$ produced by high-energy protons (>10 eV; $\lambda < 124$ nm) depleted almost all the ozone two years after the flare, and the atmosphere recovered after 50 years, during which the planet lost ozone as a detectable biosignature (Fig. 2.6). Tilley et al. (2019) calculated the effect of a series of flares, finding the combination of flares and particles would deplete the ozone by 94% in 10 years. These initial works did not include a planetary magnetic field or the effect of a resonant planetary orbit; both would affect the ozone abundance and its spatial distribution, which requires 3D general circulation models.

The effect of UV flares and particles on atmospheres like present-day Earth's has been explored by Chen et al. (2021) and Ridgway et al. (2023) with 3D atmospheric models using the solar proton flux from the SOLARIS-HEPPA collaboration (e.g., Funke et al. (2024)). They found long-term ozone depletion and $N_2O$ production, but their impact on planetary transmission spectra is not detectable with current telescopes. The depletion of $O_3$ because of $NO_x$ produced by particles means more UV at the planetary surface in all the simulations, but the duration and intensity of the UV depend on the total energy of the flare, the density, and the atmospheric composition (Yamashiki et al., 2019).

Solar and stellar energetic particle events are not the only source of relativistic particles. Galactic cosmic rays (GCRs) are another source of energy with impacts on planets. Initially, the term "cosmic rays" referred to both GCRs and SEPs, since they have some overlapping range in energy (typical SEP range from below the MeV to a couple of GeV; GCR range from the MeV to $\sim 10^{23}$ eV, though for the applications that concern us, we can stop at around 1 TeV). For the Solar System, the GCR flux outside the heliosphere can be considered effectively constant on human timescales. Within the heliosphere, changes in the interplanetary magnetic field modulate the flux of GCRs (Section 4.2).

The GCR environment for terrestrial exoplanets was initially studied by calculating the magnetic moment of a tidally locked planet around an M dwarf, which resulted in 15% less magnetic moment than that of present-day Earth. This translates into an order-of-magnitude larger flux of cosmic rays than those received on our planet (Grießmeier et al., 2005). Models that considered cosmic ray transport in the astrosphere of M dwarfs indicated variable responses to cosmic rays, from suppression below present-day Earth levels (Mesquita et al., 2022b) to 10 times more intensity than Earth (Herbst et al., 2020). Recent modeling of the conditions present over a range of stellar ages from 0.6–2.9 Gyr found that stellar particle fluxes always dominate over galactic cosmic rays (Rodgers-Lee et al., 2021a), and that the impact of stellar particles increases with more active and/or younger stars.

For specific exoplanet cases, simulations of TRAPPIST-1 e with $CO_2$-dominated atmospheres showed that GCRs have minimal chemical impact. In contrast, a single SEP event equivalent to the historical Carrington Event could significantly alter abundances of $NO_2$, $CH_4$, $H_2O$, and $HNO_3$, producing spectral features detectable by JWST only after at least 30 transit observations (Herbst et al., 2024). Similar SEP-driven effects were predicted for Proxima Cen b (Scheucher et al., 2020).

## 2.4    Aurorae: Illuminating Planets with Space Weather

In the Solar System, energetic particles precipitating into the magnetospheres and upper atmospheres of planets generate emission from UV to radio wavelengths. This process yields *aurorae*, spectacular light displays visible to fortunate observers on (or orbiting) Earth as the "Northern Lights" (*aurorae borealis*) and "Southern Lights" (*aurorae australis*). Aurorae and associated emission having to do with particle precipitation provide a unique opportunity in the study of exoplanetary evolution, simultaneously providing information on the planet's particle environment, atmospheric composition, thermal structure, and its magnetic field. This section provides an introduction to aurorae and their utility in understanding planets; we save the discussion of aurorae's utility as space weather probes for Section 5.5.8.

Well before the advent of spaceflight, auroral displays were linked to charged particles precipitating into the Earth's atmosphere from the Sun (Birkeland, 1896). Dynamics Explorer-1 observations eventually confirmed the close linkage between energetic particle precipitation, radio wavelength emissions generated as they propagate along magnetic field lines, and resulting UV emission from an aurora that was obtained (Huff et al., 1988).

Aurorae, in the strict sense of the word, originate from the impact of particles on atmospheric species, leading to the creation of an excited state that then radiatively decays, emitting light. Both precipitating electrons and protons can produce this effect.

> **Not So Fast: Planetary aurorae do not imply the presence of a magnetic field.**
>
> Even in the absence of a magnetic field, emission resembling that of magnetically controlled aurora is possible. Thus, for an exoplanet, UV, optical, or infrared (IR) emission characteristic of aurorae in the Solar System cannot be taken to mean that the planet has an intrinsic magnetic field. Electron cyclotron maser instability (ECMI) emission is a different matter.

For example, in Earth's atmosphere, excitation of oxygen into the $O(^1D)$ and $O(^1S)$ states leads to the familiar red (630 nm) and green (557.7 nm) emission as these states radiatively decay. When protons are involved, they can undergo charge exchange, capturing electrons and producing emission as those electrons cascade through energy levels. These excited states last for a few hundred seconds and a few seconds for the O($^1$D) and O($^1$S) states, respectively, providing time for collisions to de-excite the atom and "quench" the emission in regions of sufficiently high density and causing aurorae to preferentially appear at high altitudes with lower densities.

The generation of aurorae through particle precipitation can involve complex sequences of interactions, such as the creation blue aurorae at Earth that requires the ionization of an $N_2$ molecule at low altitudes by a precipitating particle, the lofting of the generated $N_2^+$ to altitudes of 500 km–600 km by ambipolar diffusion, and then the resonant scattering (excitation and immediate radiative decay) of 427.8 nm photons from the Sun (Beaudoin et al., 2025). Gronoff et al. (2012); Gronoff et al. (2025) provide more details regarding the transport of electrons and other particles in the upper atmospheres of planets and the creation of airglow.

For gas giants or other hydrogen-rich atmospheres, another source of auroral emission is triatomic hydrogen $H_3^+$. Here, precipitation particles ionize (diatomic) molecular hydrogen, which then reacts with a neutral hydrogen molecule to produce $H_3^+$. The dense rotational-vibrational (ro-vibrational) energy states of this molecule are responsible for strong thermal emission at 2 $\mu$m and 4 $\mu$m, and can act as a major coolant in planetary thermospheres (Maillard & Miller, 2011). Jupiter (Drossart et al., 1989), Saturn (Geballe et al., 1993), and Uranus (Trafton et al., 1993) all exhibit $H_3^+$ emission.

Because auroral emission comes from specific electronic (e.g., O[$^1$D]) or ro-vibrational (e.g., $H_3^+$) transitions, it holds promise as a tool for positively identifying chemical species within a planetary atmosphere. Moreover, it can probe the physical conditions (temperature, density) of the atmosphere, due to the fact that many forms of auroral emission are sensitive to the rate of collisions, which might de-excite, excite, or destroy the emitter before it can emit.

Magnetic fields play an important role in the generation of aurorae. The electrons precipitating at Earth and leading to the aurorae are more energetic than the particles of the solar wind. This increase in energy is due to the interaction of the solar wind plasma with the magnetosphere, and the definition of aurorae is sometimes restricted to only apply if such accelerations have taken place. Mechanisms for this acceleration can include magnetic reconnection, electromagnetic waves, and field-aligned currents. Planets with intrinsic magnetic fields, versus those lacking them, will dissipate more energy through these mechanisms due to the larger area of magnetic interactions with the impinging wind (e.g. Maggiolo et al., 2022).

Magnetic fields also play a key role in the generation of radio emission as particles ultimately bound to generate aurorae travel along those fields. This emission arises through the electron cyclotron maser instability (ECMI), and it is present among all of the magnetized planets in the Solar System. This emission's dependence on the magnetic field means that it is highly circularly polarized—as much as 100% at the fundamental frequency. This polarization property can be indispensable for distinguishing ECMI emission from other emission processes. ECMI's dependence on the magnetic field also means that the frequency at which this emission is observed

> **Not So Fast: A planet need not have an intrinsic magnetic field to have a magnetosphere or magnetic interaction with the stellar wind.**
>
> The currents generated by the passing of the magnetized stellar wind around a planet produce induced magnetic fields. Mars, for example, which lacks an intrinsic magnetic dynamo, still imposes an obstacle to the magnetized solar wind, resulting in a draped field and "induced magnetosphere," including a magnetic bow shock (e.g., Brain et al. 2010).

is directly dependent on the magnetic field strength, scaling like $\nu = 2.8 s B$ MHz, where $B$ is the field strength in Gauss and $s$ is the harmonic. Hence, observations of ECMI emission provide a direct measurement of the magnetic field strength of a planet.

Despite the importance of magnetic fields, a planet need not have an intrinsic magnetic field to produce auroral (or auroral-like) emission. The observation of aurora-like processes at Venus shows the solar wind particles can be diverted towards the nightside and lead to auroral phenomena. This is due to the induced magnetic field generated in Venus's ionosphere, and any planet with an ionosphere can be expected to experience a similar effect. However, the acceleration of particles by an induced magnetosphere is generally far weaker than what is possible with an intrinsic magnetic field, such as that of Earth, Jupiter, and even the localized, remnant crustal fields of Mars.

Although magnetic fields of several hot Jupiters have been detected (Shkolnik et al., 2005) and measured (Cauley et al., 2019), observers have searched extensively for auroral and related radio emission from exoplanets in the past few decades with no success yet (Shkolnik et al., 2006; Richey-Yowell et al., 2025). Lenz et al. (2016) and Gibbs & Fitzgerald (2022) also found no evidence of $H_3^+$ emission from the archetypal hot Jupiter HD 209458 b. France et al. (2010) and France et al. (2011) detected emission at 1582 Å from the HD 209458 system, but ultimately attributed it to photoexcitation of molecular hydrogen $H_2$ by O,I emission from the host star. This emission would correlate with stellar activity despite the fact that it does not involve particle precipitation. Current telescopes likely will not obtain significantly stronger constraints, but future telescopes, such as the 30-m class ground-based telescopes or the Habitable Worlds Observatory (HWO), might be able to detect such emission.

There is an even longer history of searches for radio emission from exoplanets, pre-dating the discovery of any exoplanets, thus far with no success (Lazio, 2024, provides a summary of all such efforts). Although radio emission has been detected from exoplanetary systems that could be consistent with that expected from an exoplanet, it is more likely coming directly from the host star and/or due to a star-planet interaction (e.g., Vedantham et al., 2020).

While the detection of radio emission directly from an exoplanet would provide a powerful indication of not only energetic particle effects, but also potential means of probing the interior structure of exoplanets through the detection of an intrinsic magnetic field, previous efforts likely have been conducted with insufficient sensitivity and almost certainly at wavelengths that are either too short or at frequencies that are too high.

Jupiter's large magnetic field strength results in ECMI emission at a fundamental frequency $\nu \lesssim 40\,\text{MHz}$ that is observable from ground-based observatories. All other Solar System planets produce ECMI emission that is below the Earth's ionospheric cutoff of $\approx 10\,\text{MHz}$, and hence require space-based observations.

The nature of this emission also means there are limits on what the plasma density local to the planet can be if ECMI emission is to escape; it is generally required that the cyclotron frequency $\nu_B \gtrsim 2\nu_e$ for escape, where $\nu_e$ is the local (electron) plasma frequency. While this may not be as much of a limiting factor for Solar System planets where the plasma densities of the ambient solar wind and planetary ionospheres are relatively diffuse, this may be an important factor in exoplanetary systems, particularly for planets orbiting close to their host stars (e.g., Daley-Yates & Stevens, 2018).

As an intermediate step on the road to detecting auroral and related radio emission from exoplanets, observations of ultra-cool (M, L, T, and Y) dwarfs have yielded intriguing results. Radio observations have detected ECMI emission from a number of these objects (Kao et al., 2016; Kao & Pineda, 2024). Many of these objects are not known to have a companion, indicating an intrinsic source of particle precipitation, though the presence of a binary companion enhances the likelihood of emission (Kao & Pineda, 2024). $H_3^+$ observations of these same and other brown dwarfs have resulted only in non-detections, possibly indicating that the precipitating particles have energies sufficiently high to penetrate to dense levels of the atmospheres where significant populations of $H_3^+$ cannot be sustained (Pineda et al., 2024).

While the efforts to date have not succeeded in detecting auroral or auroral-adjacent emission, at any wavelength, they have revealed potential confounding effects that may complicate any future efforts to associate auroral emission directly with particle effects.

For Earth, Venus, and Mars, airglow or "chemiluminescence," driven by a series of reactions involving the photolysis and recombination of oxygen molecules, emits at the same wavelengths as aurorae.

A comprehensive understanding of the several confounding processes will be essential to any effort to unambiguously associate planetary emission with an auroral process.

## 2.5  Impact of Space Weather on the Origins of Life

Particle-driven chemistry (Section 2.3) may play a key role in prebiotic processes relevant to the origins of life. Experiments and models suggest high-energy particles can initiate hydrogen cyanide (HCN) chemistry, leading to the production of molecules such as amino acids, both in comets and on early Earth (Kobayashi et al., 1998, 2023; Airapetian et al., 2016). This chemistry also contributes to the formation of hazes that influence climate and UV shielding, as seen on Saturn's moon Titan (Waite et al., 2007), and possibly on early Earth (Arney et al., 2016).

Unlike UV light or lightning, high-energy particles can generate prebiotic molecules even in the absence of methane, a key reduced (H-bearing) compound (Kobayashi et al., 2023). This is relevant because the $CH_4$ abundance in Earth's prebiotic atmosphere is poorly constrained (Olson et al., 2018), and methane is likely to be scarce in terrestrial exoplanets (Guzmán-Marmolejo et al., 2013; Thompson et al., 2022).

Reduced atmospheres, rich in hydrogen and other easily oxidized species, generally yield more prebiotic-relevant molecules (Trainer, 2013). However, potentially habitable terrestrial planets are expected to have been only weakly reduced, transitioning toward neutral atmospheres. In the habitable zones of M dwarfs, planets with $CO_2$-dominated atmospheres may contain more CO than their Sun-like counterparts. This is due to higher fluxes of photons below 200 nm, a range less abundant around M dwarfs. As a result, CO may accumulate (Hu et al., 2020). In such atmospheres, solar protons are a more efficient energy source for producing prebiotic molecules than GCRs or lightning (Nava-Sedeño et al., 2016).

The role of high-energy particles is not limited to atmospheres. On icy bodies like Europa and Enceladus, their interaction with the frozen surface can produce reactive species such as $O_2$ and $HO_x$, which may contribute to life's origin or serve as metabolic energy sources. While Enceladus receives a lower particle flux than Europa, its active plume activity may help deliver these oxidants into its subsurface ocean (Chyba, 2000; Hand et al., 2009, 2020; Pavlov et al., 2024).

Efforts to detect life on exoplanets often focus on gaseous biosignatures——molecular byproducts of metabolism detectable via remote sensing (Schwieterman et al., 2018). For Earth-like atmospheres (0.21 $O_2$, 0.78 $N_2$), GCRs (0.1–8 GeV) generally do not hinder detection of key biosignatures such as $O_3$ or $N_2O$ (Grenfell et al., 2007; Grießmeier et al., 2016), though they may increase $HNO_3$ levels (Scheucher et al., 2018). Numerical simulations indicate that flares combining particles and photons can enhance nitrogen-bearing biosignatures in $O_2$-rich atmospheres (Chen et al., 2021; Herbst et al., 2019).

Numerical simulations indicate the combined effect of particles and photons during a flare may enhance nitrogen-bearing biosignatures in $O_2$ rich atmospheres like present Earth (Chen et al., 2021;

Herbst et al., 2019). Molecular oxygen and its photochemical product, ozone ($O_3$), are considered one of the most robust biosignatures (Meadows et al., 2018) that can be seriously affected by highly energetic particles. For example, a planet with an atmosphere similar to present-day Earth's ($O_2$ and $N_2$ dominated) could lose its ozone because of the particles accelerated during a flare, leaving the planet without protection from UV photons that destroy organics, therefore harming surface life (Segura et al., 2010; Tilley et al., 2019).

While much of this section focuses on how particle-driven chemistry may support the origin of life on a planet or moon, space weather may also play a role in an alternate theory: panspermia. This hypothesis proposes that microorganisms may be transported through space via dust, comets, or planetesimals and delivered to planets, where they could seed life (Kawaguchi, 2019; Mcnichol & Gordon, 2012). Panspermia does not explain how life began, only how it might spread. The viability of panspermia depends on the properties of the interstellar medium—composed of dust and gas structures with varied densities and temperatures—and on how stellar space weather shapes planetary system boundaries. Stellar winds and the astrospheres they generate determine what external material can enter planetary systems (§4.2, Figure 4.2). The size of interstellar dust grains is critical: small grains are charged and easily deflected by stellar winds, while larger grains can penetrate more deeply into astrospheres and potentially deliver biological material to terrestrial planets (Czechowski & Mann, 2003; Slavin et al., 2012; Sterken et al., 2013).

Radiation exposure is another limiting factor. Microbes riding in small grains receive higher radiation doses and have shorter survival times, while those embedded deeper in larger bodies may be shielded for much longer journeys.

> **Not So Fast: The flashiest forms of space weather should not be the sole consideration in exoplanetary impacts.**
>
> The relative impact of various forms of space weather depend on the nature of the impact in question, the properties of the planet(s) of interest, and the timescale of concern. In particular, forms of space weather that often garner attention due to their short-lived intensity, like flares and CMEs, do not necessarily dominate long-term cumulative effects. Atmospheric escape from Venus provides an example. Meanwhile, stream interaction regions, a less commonly discussed transient, also play a prominent role (§5.1; Edberg et al. 2011).

## 2.6  Paths Forward to Linking Stellar Activity and Planetary Evolution

The many ways particle environments shape planetary atmospheres, through both composition and escape, make observational constraints essential. Planetary impacts help us prioritize which environmental properties matter most. In planetary science, the most useful progress will come from improved constraints on key observables.

Stellar wind density, velocity, and magnetic field are core inputs in current models, and also govern CME interactions with planets. We group these as "intensity." Since CMEs vary in intensity, their statistical distribution and frequency may be the most relevant model inputs. The role of SEPs is less certain. Though SEPs are often represented by time-averaged spectra in the literature (e.g., Desai & Giacalone, 2016), they are in fact discrete events. Whether this simplification is valid remains an open question.

In the realm of atmospheric chemistry, a key need is to move beyond the scaling relationships developed from solar data used to estimate hypothetical SEP fluxes based on the photon radiation of flares. In place of these, real observational constraints are needed for exoplanetary systems.

There also remains much progress to be made, not just in observing the inputs but in modeling the impacts. Increased modeling of particle chemistry within a range atmospheric compositions, such as Titan-like $N_2$-$CH_4$ atmosphere, would be an example of this. Laboratory experiments provide important inputs and validation to chemical models. Because the origin and detection of life depend on chemistry, these paths forward apply to the subject of life as well.

Observations of exoplanetary aurorae could be a powerful step forward in that they simultaneously provide information valuable to multiple exoplanet science cases. Atmospheric chemistry, the genesis and detection of life, and, to a lesser extent, atmospheric escape could all benefit from constraints on particle radiation that auroral observations, with sufficient development, may yield, all while these same aurorae probe properties of the planets. The development of space-based radio arrays to avoid Earth's ionospheric cutoff, along with a drive to greater sensitivities, could yield breakthroughs in this area. Meanwhile, work to better understand and disentangle potential confounding signals will be essential for interpreting auroral signals, whenever they become available.

Solar System planetary science will continue to have a key role in informing efforts to observe the inputs and impacts of exoplanetary particle environments. Our system will, for the foreseeable future, remain the only one where both inputs and impacts can be observed in-situ, and there remains much to be learned that can inform exoplanetary science. An example highlighted in this chapter are measurements of chemical changes to SEPs at Venus and Mars that, if available, could test models and provide insight into the impact of StEPs on $CO_2$-rich exoplanetary atmospheres.

Planetary science (solar and extrasolar) and efforts to observe the stellar particle radiation influencing planets must feed back on each other in an inherently iterative process. Efforts exploring and assessing the impact of extreme environments relative to our own through simulations have provided some initial guidance. The next step is to better observationally constrain those environments, so that planetary simulations can be improved, key inputs in need of tighter constraints identified, and iterative progress made. Efforts like NASA's interdisciplinary program "Retention of Habitable Atmospheres in Planetary Systems" (PI, D. Brain; 2023–2028) are intended to explore potential impacts that can then guide observational efforts.

Only through sustained exchange where planetary science informs observational priorities, and new observations refine models, will meaningful progress in understanding the evolution and present state of planets—and the life they may support—be possible.

---

**Zooming Out: Related Findings and Recommendations (FRs)**

Understanding the full impact of dynamic plasma, energetic particles, and magnetic field environments on planets and life requires continued advancement in observing and modeling interactions between planets and space weather (FR1). This, in turn, requires an accurate picture of space weather itself, developed through observations, models, and supporting instrumentation (FR2, FR3, FR4, FR5, FR6). Answers to questions involving the interplay between space weather, planets, and life require input from distinct disciplines, and so especially benefit from programmatic structures that promote interdisciplinary collaboration (FR7).

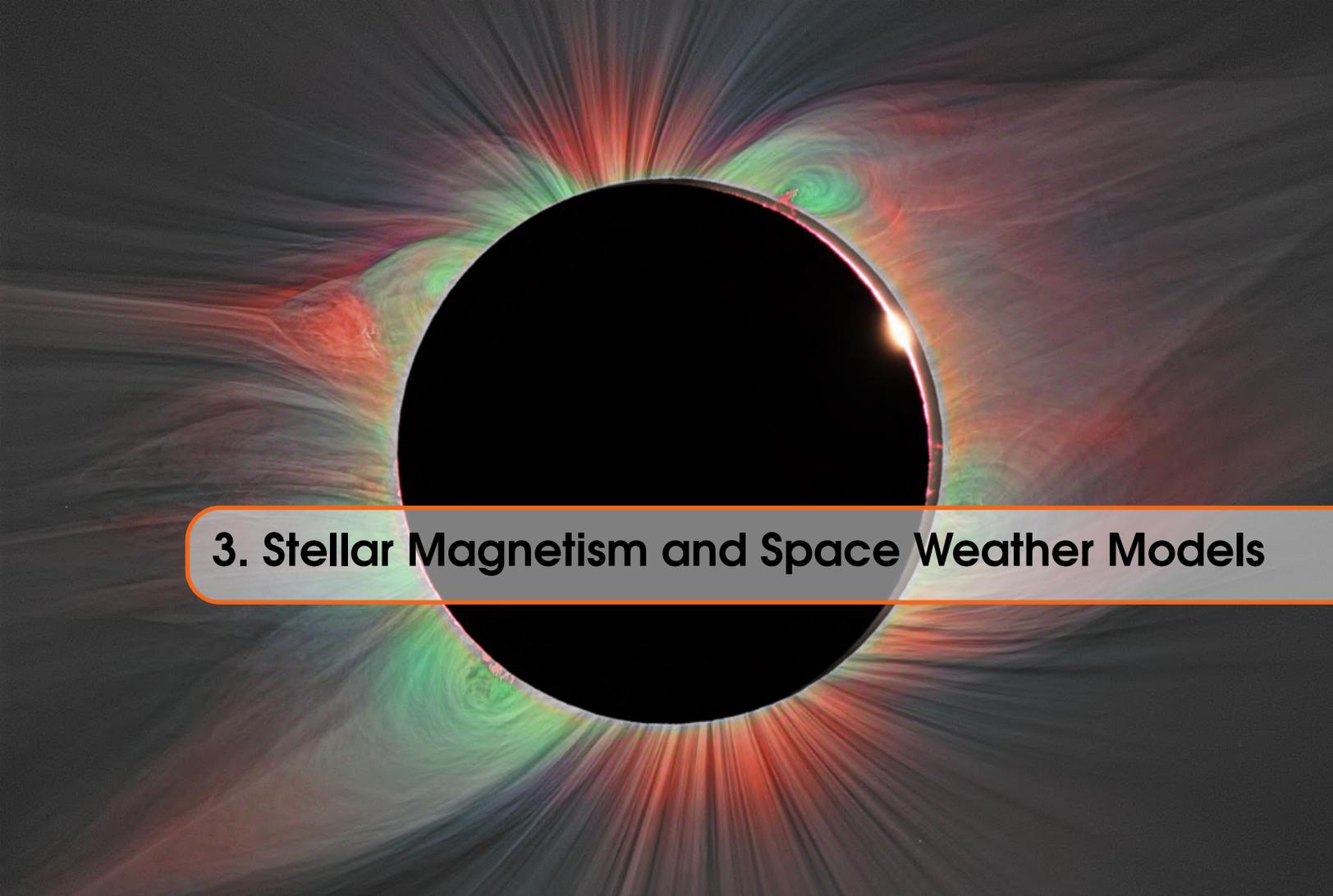

# 3. Stellar Magnetism and Space Weather Models

## 3.1 Solar Magnetism and Space Weather

The Sun's magnetism and its variability create the space weather in the Solar System. Knowledge of the Sun's magnetism has roots in the early 20[th] century when it became feasible to measure both the polarity and the approximate strength of the magnetic field in sunspots (Hale et al., 1919). By the 1950s, the first full-disk magnetic field maps (magnetograms) were being created Babcock 1953), and continuous daily full-disk observations began in the 1970s (Howard, 1974; Livingston et al., 1976). Stanford University has been providing daily full-disk maps since 1976, with ongoing observations at the Wilcox Solar Observatory (Scherrer et al., 1977). The National Solar Observatory at Kitt Peak and the Mount Wilson Observatory have also produced a long series of photospheric magnetic field maps (Pevtsov et al., 2021). Space-based instruments like the Solar and Heliospheric Observatory/Michelson Doppler Imager (SOHO/MDI) and, more recently, the Solar Dynamics Observatory/Helioseismic Magnetism Imager (SDO/HMI), provide observations free from atmospheric interference with high spatial and temporal resolution.

Despite this extensive coverage, these continuous full-disk observations span only four activity cycles or two magnetic cycles. The solar cycle, approximately 11 years in duration, is characterized by the periodic fluctuation in solar activity, such as the number of sunspots, faculae, prominences, flares, coronal mass ejections (CMEs), and energetic particle events (EPs), among others. The solar magnetic cycle, spanning roughly 22 years, encompasses two solar cycles. During each 11-year solar cycle, the polarity of the small- and large-scale solar magnetic field reverses. Thus, after two solar cycles, the Sun's magnetic field returns to its original orientation. At the beginning of the activity cycle, sunspots appear at higher latitudes and gradually move towards the equator as the cycle progresses. This migration pattern is often referred to as the butterfly diagram. Hathaway (2015) provides a more extensive review of the solar cycle.

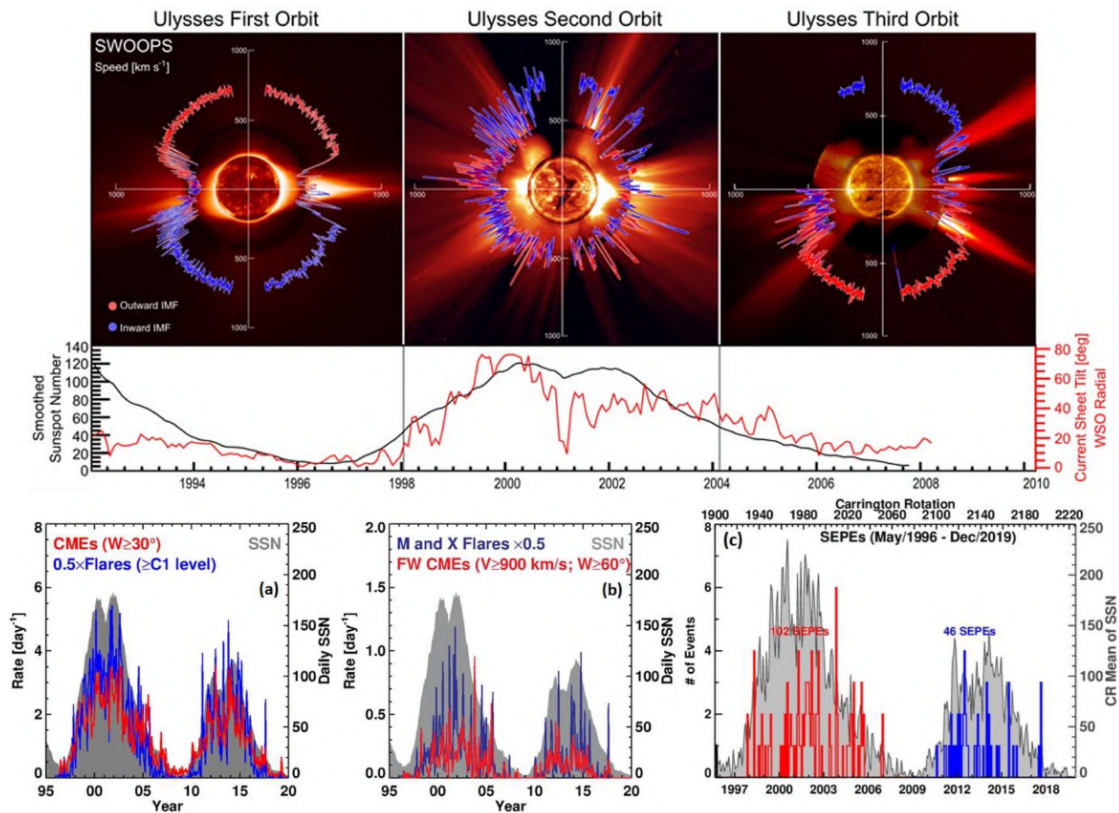

Figure 3.1: Modulation of the particle space weather around the Sun due to its magnetic cycle. (*top*) Polar plots of the solar wind speed (in $km\,s^{-1}$), colored by the interplanetary magnetic field polarity for Ulysses' three polar orbits. (*middle*) Contemporaneous values for the smoothed sunspot number (black) and heliospheric current sheet tilt (red), alined to match the top figures. (*bottom*) Solar cycle variation of transient phenomena (flares, CMEs, EPs): a) Daily rate for the general CME population (width $\geq 30°$; red) and the number of soft X-ray flares (GOES class $\geq$ C1.0; blue). b) Fast and wide CMEs (speed $\geq 900\,km\,s^{-1}$; width $\geq 60°$; red) and major soft X-ray flares (M- and X-class; blue). c) Number of large solar energetic particle events (EPs) detected in the $> 10$ MeV GOES channel with intensity exceeding 10 pfu (blue: Cycle 23, red: Cycle 24. The sunspot number (SSN) is shown in the background (gray). Adapted from McComas et al. (2008a, AGU) and Zhang et al. (2021, Springer Nature, CC BY license).

The observations to date have made a close link between the Sun's magnetism and its space weather indisputable (Figure 3.1). The coronal high-energy radiation (EUV/X-rays), the quasi-steady solar wind, as well as transient phenomena such as flares, CMEs, and SEPs, are all connected directly or indirectly to different properties of the solar magnetic field. Evidence of this connection can be seen at all temporal and spatial scales, from extremely small and short reconnection events (micro, nano-flares) to the cycle-induced size modulation of the heliosphere. However, due to the Earth's limited heliolatitude range, accurately measuring polar fields remains challenging (Petrie 2015). Furthermore, as only one side of the Sun is visible from Earth or near-Earth orbits, our monitoring of the entire Sun's magnetic field at any given time is typically ∼ 50% incomplete. Consequently, even for the Sun, our capacity to study the evolution of the photospheric magnetic field and trace the physical driver of solar space weather is still limited.

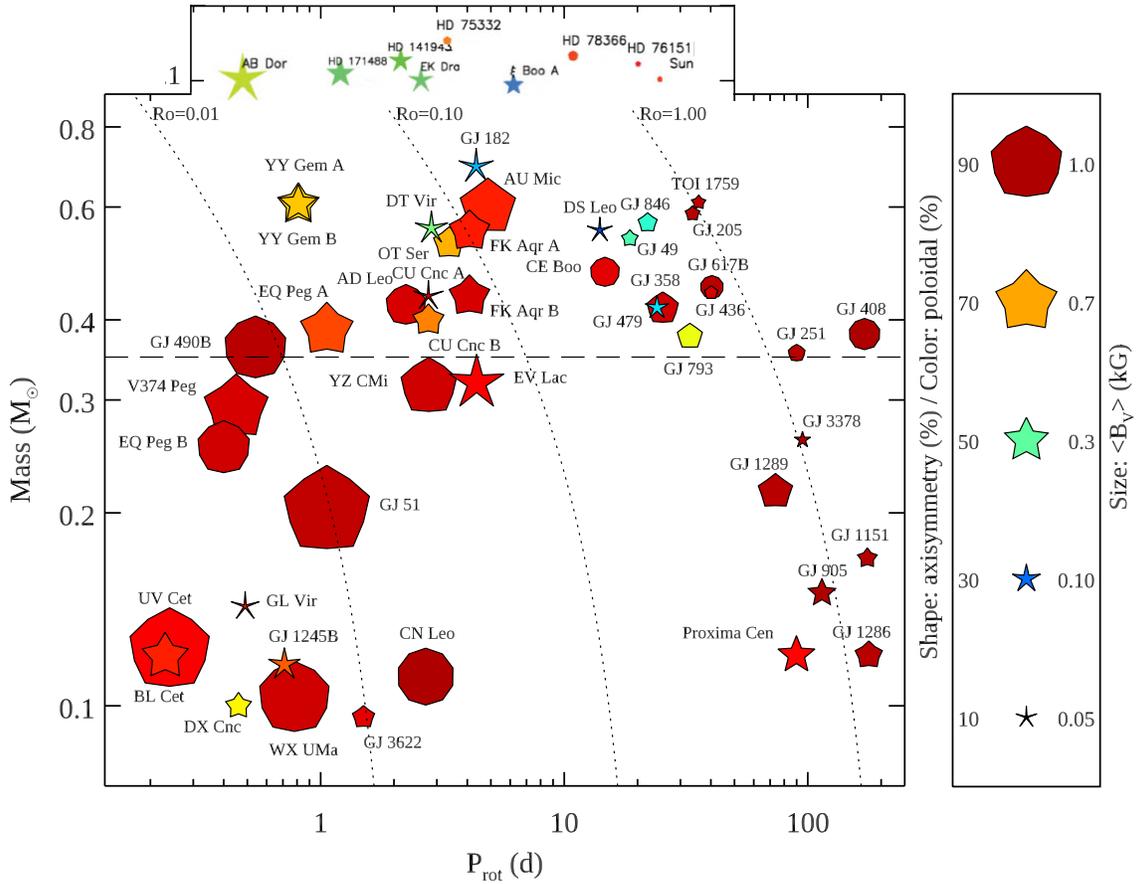

Figure 3.2: Observational constraints on the large-scale magnetic field structure of cool main-sequence stars obtained through spectropolarimetry and ZDI. These measurements are presented on the stellar mass and rotation period plane. Symbol sizes represent the relative strength of the large-scale magnetic field, spanning three orders of magnitude (from 3 G to 1.0 kG, with the Sun being the smallest data point). The color scale indicates the main geometry of the surface field, where red and blue correspond to purely poloidal and toroidal fields, respectively, and other colors represent a mix of these two fundamental topologies. Symbol shapes indicate the symmetry level of the large-scale field, with decagons for purely axisymmetric configurations and star-shaped symbols for purely non-axisymmetric fields. Curved dashed lines indicate regions of constant Rossby number ($R_o$) while the horizontal segmented line corresponds to the fully convective limit. Adapted from Donati (2011) and Kochukhov (2021), Springer Nature, CC BY license.

It is natural to think that if other stars display similar signs of magnetism as the Sun, they will also host similar space weather in their systems. Conversely, if the properties of the stellar magnetic field deviate largely from what has been observed on the Sun, it is possible that the space weather conditions of those systems would be more diverse, with some phenomena or processes never observed or perhaps not even possible in our planetary environment. In this context, it is crucial to obtain knowledge on the properties of magnetic fields on other stars and exploit their connection as a driver of space weather to obtain a proper understanding of the particle environments of other planetary systems.

## 3.2 Observations of Stellar Magnetic Fields: The State of the Art

In the stellar regime, high-resolution spectroscopy and spectropolarimetry have unveiled the widespread prevalence of magnetic fields in cool main-sequence stars. Techniques such as Zeeman broadening/intensification (ZB/ZI) along with Zeeman-Doppler imaging (ZDI) have shown that active stars exhibit surface field strengths notably surpassing those of the Sun (Donati & Landstreet, 2009; Reiners, 2012; Kochukhov, 2021). Both techniques are based on the Zeeman effect, but the former measures the additional broadening/depth in sensitive spectral lines due to the presence of magnetic fields, while the latter takes advantage of the induced polarization in the resulting Zeeman components (see Kochukhov 2018 for further details). These intensified magnetic fields are not confined solely to small-scale formations, such as sunspots and active regions, but extend to the magnetization of the entire star. Moreover, as ZDI provides insights into the spatial distribution of the surface's large-scale field, it enables the examination of the photospheric field's complexity across diverse stellar parameters.

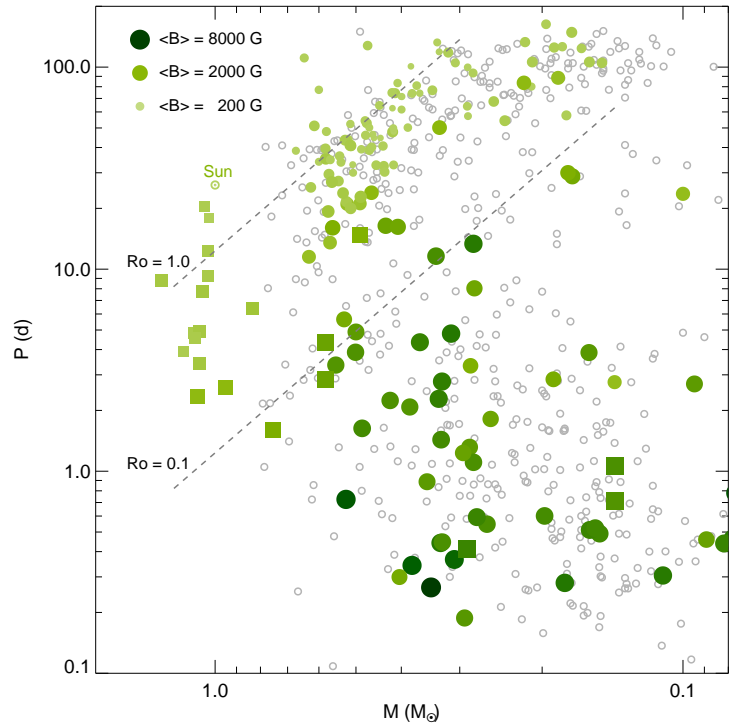

Figure 3.3: Period-mass diagram displaying measurements of the total (unsigned) magnetic field via ZB/ZI in cool main-sequence stars. Colors and symbol sizes denote different magnitudes of the average field strength as indicated. The squares correspond to similar measurements from previous investigations (Shulyak et al., 2017; Kochukhov et al., 2020a). The gray points in the background correspond to a sample of stars with known rotation (Newton et al., 2017). Regions of constant $R_o$ are highlighted by dashed lines. Credit: Reiners et al. (2022a).

In the context of Sun-like stars ($M_\star \simeq 1\,M_\odot$), complex and relatively weak ($< 1\,\mathrm{kG}$) large-scale surface field configurations have been documented. Slow rotators ($P_{rot} > 15\,\mathrm{d}$) exhibit the simplest and weakest large-scale fields. Field complexity escalates in more rapidly rotating stars (as depicted by star shapes in Figure 3.2), intensifying up to kG levels for the most active objects. Many maps of fast-rotating stars exhibit robust surface toroidal (or azimuthal) fields; a phenomenon lacking a clear counterpart on the Sun, likely indicating a shift in the underlying stellar dynamo (Käpylä et al., 2023).

As illustrated in Figure 3.2, the strongest magnetic fields (up to a few $10^3$ G) have been detected primarily in the low-mass segment of the main-sequence ($M_\star < 0.6 M_\odot$), predominantly populated by fully (or nearly fully) convective M dwarfs. ZDI reconstructions indicate that as rotation period diminishes, very low-mass objects with similar stellar attributes seem to exhibit two distinct classes of magnetic topologies: one tending towards robust and straightforward magnetic geometries (mostly poloidal and axisymmetric configurations), and another with much weaker fields, featuring a notable presence of non-axisymmetric configurations, including toroidal components. These characteristics are prevalent in early, mid, and late M dwarfs, and have recently been observed in less active stars of this spectral type (e.g., Bellotti et al., 2023, Figure 7). The peculiarities observed in the magnetic field have been construed as indicative of dynamo bi-stability occurring in these low-mass objects (Gastine et al., 2013), which have also been recently reported in ZB observations (Shulyak et al., 2017). However, recent results seem to indicate an even more complex picture for these objects, with the detection of strong and variable fields for very slowly rotating M dwarfs (Lehmann et al., 2024).

Complementing the ZDI reconstructions, ZB and ZI techniques have revealed larger magnetic field strength values on Sun-like stars (Kochukhov et al., 2020b) and M dwarfs (Reiners et al. 2022b; see Figure 3.3). These are typically interpreted as a measure of the mean strength of the small-scale field, as they are sensitive to the unsigned magnetic field on the visible stellar surface (typically missed by ZDI reconstructions due to cancellation effects; see Kochukhov 2016). In a recent development, Kochukhov et al. (2023) report the first map of the small-scale magnetic field in a cool star, opening new possibilities for the characterization of the surface magnetism in the stellar regime. At present, approximately 200 cool main-sequence stars have had large-scale magnetic field measurements conducted via spectropolarimetry, and roughly half of those have ZDI reconstructions. For a small set of targets, multiple ZDI observations have been retrieved, revealing the presence of polarity reversals and magnetic cycles (Jeffers et al., 2023; Bellotti et al., 2024). In terms of the mean surface field strength, close to 400 stars have measurements via ZB or ZI techniques.

### 3.2.1 Instrumentation for Mapping Stellar Magnetic Fields

The large majority of the results discussed in the previous section have been obtained with ground-based instrumentation mounted on relatively small telescopes (2–4 m apertures). A brief overview of the instruments currently available for stellar magnetic field studies is provided in Table 3.1. We will restrict ourselves to high-resolution spectropolarimeters, as those capabilities are required for performing large-scale magnetic field reconstructions via ZDI. In addition, as explained by Kochukhov (2018), ZB or ZI techniques can be performed using regular high-resolution optical or near-IR spectroscopy, provided that other broadening agents in the spectral lines (e.g., thermal and rotational Doppler broadening, instrumental smearing) are well understood and the data are of sufficient signal-to-noise.

In terms of upcoming instrumentation, none of the high-resolution first-light spectrographs planned for the next generation of optical/NIR ground-based observatories (30–40 m class) contemplate polarimetric capabilities in their design (Fanson et al., 2022; Padovani & Cirasuolo, 2023; Heptonstall et al., 2024). While detailed instrumental polarimetric analyses have been performed for all facilities (Extremely Large Telescope [ELT], de Juan Ovelar et al. 2014; Thirty Meter Telescope [TMT], Atwood et al. 2014; Anche et al. 2018; Giant Magellan Telescope [GMT], Anche et al. 2023b), a recent investigation suggests an advantage for the GMT over the other two ELTs for sensitive polarimetry (Anche et al., 2023a). This leaves the door open for possible second-generation instruments targeting high-resolution spectropolarimetry. Similarly, there are various instrument concepts planned for space-borne spectropolarimetry (see Neiner et al. 2025), with the notable case of Pollux: a UV-Visible-IR high-resolution spectropolarimeter for NASA's Habitable Worlds Observatory (Muslimov et al., 2024). All of these developments are crucial, as having access to this observing mode in the era of ground-based ELTs and the next flagship space missions could provide an edge in the fields of stellar magnetism and particle environments to US-based researchers in an otherwise Europe-dominated field.

Table 3.1: Currently Available High-resolution Spectropolarimeters

| Instrument | Telescope Diameter | Spectral Resolution | Wavelength Coverage [nm] | Main Scientific Objective |
|---|---|---|---|---|
| Narval@TBL (*Pic du Midi, France*) | 2.0 m | 65,000 | 375–1050 | Magnetic field measurements in stars |
| ESPaDOnS@CFHT (*Mauna Kea, Hawai'i, USA*) | 3.6 m | 65,000 | 370–1050 | Stellar magnetism, exoplanetary atmospheres |
| SPIRou@CFHT (*Mauna Kea, Hawai'i, USA*) | 3.6 m | 70,000 | 980–2350 | Exoplanet detection, stellar magnetism |
| HARPSpol@ESO3.6m (*La Silla, Chile*) | 3.6 m | 115,000 | 378–691 | High-precision stellar radial velocities, magnetic fields |
| PEPSI@LBT (*Mt. Graham, Arizona, USA*) | $2 \times 8.4$ m | 220,000 | 383–907 | Stellar magnetic fields, exoplanet studies |
| CRIRES+@VLT (*Cerro Paranal, Chile*) | 8.2 m | 100,000 | 950–5500 | High-resolution spectroscopy in the infrared, exoplanets, stellar physics |

## 3.3 Modeling Particle Environments via Stellar Magnetism

Models based on observations of solar and stellar magnetic fields can be used to simulate the winds, CMEs, and EPs of the Sun and stars. Models of this sort are a key component of studying particle environments and their impacts on planets. No star's space weather, even the Sun's, is ever likely to be observed with complete coverage in both space and time. Models, guided by the observations that are available, can both fill these gaps and explore novel, unobserved regimes.

The Sun, being the closest star to Earth, holds a unique position for constructing and validating such models, as it does for all astronomical studies. It is the only star whose magnetic field (Section 3.1) and particle environment can be observed with high spatial resolution and directly measured in-situ. The foundational knowledge of the Sun and heliosphere is crucial for studying other star-planet systems across the universe. Consequently, the modeling of stellar space weather is primarily based on our understanding of solar space weather and its three key elements: the quiescent solar wind, a continuous stream of charged particles originating from the solar atmosphere that propagates throughout the Solar System; coronal mass ejections (CMEs), which are large expulsions of plasma and magnetic fields from the Sun's corona into space; and solar energetic particles (SEPs), high-energy ions and electrons accelerated by solar flares and CMEs. This section will begin with an overview of the modeling of solar particle environments, including the quiescent solar wind (Section 3.3.1), solar CME modeling (Section 3.3.2), and SEP modeling (Section 3.3.3), before progressing to models applicable to stellar contexts (Section 3.3.4). More detailed descriptions of the solar wind, CMEs, and SEPs are available in Sections 4.1, 5.2, and 5.3.

### 3.3.1 Quiescent Solar Wind

A natural starting place for constructing models of circumstellar particle environments is the quiescent solar wind. Such a model must handle the wide range of densities, temperatures, field strengths, velocities, and other physical parameters, spanning several orders of magnitude, that the solar wind exhibits between the Sun, the planets, and beyond (see Figure 3.1). This large dynamical range presents challenges in physically modeling the solar wind within a single comprehensive model. As a result, different modeling approaches are adopted to handle the various physical conditions encountered. Kinetic solar wind models focus on small-scale interactions within the plasma, such as those between waves, ions, and the background magnetic field, offering insights into the physics of local heating and dissipation phenomena at small spatiotemporal scales (refer to Marsch 2006; Echim et al. 2011 for a detailed review). In contrast, global MHD solar wind models offer a fluid description on a global scale, simplifying solar wind heating and acceleration processes. There are a variety of global MHD solar wind models, differing primarily in their treatment of coronal heating terms (refer to Gombosi et al. 2018 for a review). State-of-the-art global MHD models typically incorporate coronal heating via Alfvén wave dissipation, integrating additional wave kinetic equations to model the exchange of momentum and energy between the plasma and wave fields (e.g., Sokolov et al. 2013; van der Holst et al. 2014b; Lionello et al. 2014; Réville et al. 2020). These models are often driven by observations at the inner boundary to capture the large-scale structures of the solar wind plasma and are extensively used for space weather forecasting and data-model comparison studies.

Here, we use the Alfvén Wave Solar Model (AWSoM) as an example of global MHD solar wind models. Developed within the Space Weather Modeling Framework (SWMF) at the University of Michigan, AWSoM employs a phenomenological approach to Alfvén wave dissipation, where the wave spectrum is not resolved, and instead, the total energy densities of the counter-propagating waves are calculated. The energy dissipated from waves heats the corona and the wave pressure gradient accelerates the solar wind. Additionally, the model separates electron and proton temperatures, and the two species are coupled by collision. Its simulation domain starts in the upper chromosphere and extends into the heliosphere, employing adaptive mesh refinement to resolve structures like current sheets and shocks. The inner boundary condition is specified by observed magnetic maps to reproduce background solar wind conditions. Figure 3.4 illustrates the modeled global solar wind in 2D (as a cut in the 3D model), 3D, and through in-situ comparisons with data from the Parker Solar Probe (PSP).

While most global solar wind models use fixed inner boundary conditions, rendering them incapable of replicating the dynamics of the solar wind (only the large-scale structures are modeled), future developments aim to integrate time-dependent inner boundary conditions driven by time series of magnetic maps (e.g., Lionello et al. 2023; Mason et al. 2023; Wang et al. 2025). Additionally, alternative coronal and solar wind heating mechanisms (e.g., by magnetic reconnection; Raouafi et al. 2023) should be considered in the future.

### 3.3.2 Solar CME Modeling

Modeling solar CMEs presents additional challenges relative to the solar wind due to their complex structure in space and time. Their key component, and driving force, is thought to be the magnetic flux rope (MFR). However, due to the challenges in measuring the coronal magnetic field, most observations of MFRs rely on EUV and/or soft X-ray imagers, leaving the structure of erupting magnetic flux ropes not fully understood (see Section 5.2 for CME observations). As a result, most CME models focus on simulating the MFR configuration, initiation, and evolution. Broadly, these models can be categorized into three groups: data-inspired, data-constrained, and data-driven.

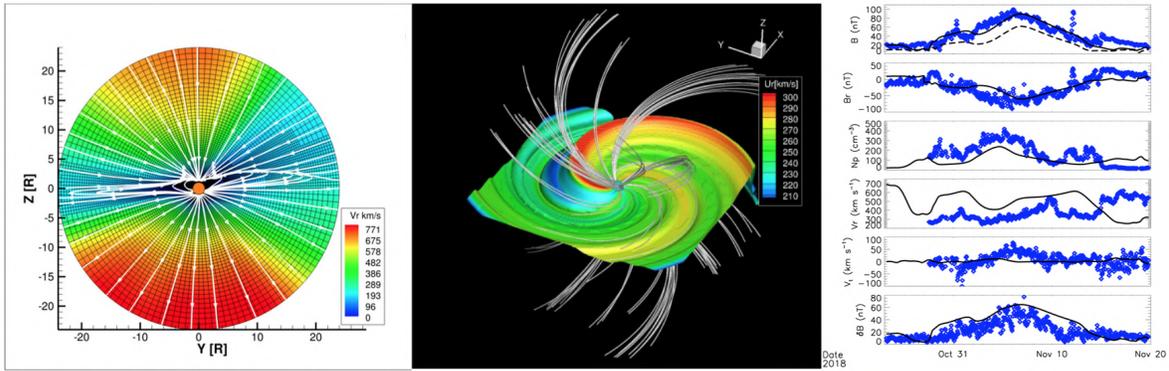

Figure 3.4: *(left)* Steady-state solar wind radial velocity for Carrington Rotation (CR) 2107 (2011 February 16 to 2011 March 16) of the meridional slice at X = 0 with magnetic field lines in white. The axes are in units of solar radii. The black grid shows the simulation cells (figure adapted from Jin et al. 2017a); *(middle)* 3D solar wind for CR 2063 (2007 November 4 to 2007 December 1) to a heliocentric distance of 2 AU. The surface shows the location of the current sheet (where the radial magnetic field strength is zero) colored by the radial speed (figure adapted from Oran et al. 2013); *(right)* Comparison of the AWSoM model output (black lines) with hourly PSP data (blue diamonds) including (*from top to bottom*) total magnetic field strength, radial magnetic field component, proton number density, radial velocity, azimuthal velocity, and Alfvén wave magnetic field perturbation amplitude. The dashed line indicates the nominal AWSoM background field strength without Alfvén magnetic field fluctuations (implemented based on data from the first PSP perihelion passage). Credit: van der Holst et al. (2022).

Data-inspired CME models aim to establish simplified scenarios or analytical models that replicate common features observed across many observational studies. Notable among these are the Gibson & Low (GL; Gibson & Low 1998) and Titov Demoulni (TD; Titov & Démoulin 1999) analytical flux rope models, along with their subsequent modifications (e.g., Titov et al. 2014; Sokolov & Gombosi 2023). The GL model is based on the three-part density structure of CMEs observed in white light (Figure 3.5a), whereas the TD model can reproduce the filament channel seen along the polarity inversion line in $H_\alpha$ images. To model an eruption, these analytical CME flux rope models need to be integrated into a background solar atmosphere and wind model (see Section 3.3). They are also used in zero-beta MHD simulations to better understand fundamental CME physics (see white paper by Török et al. 2023). Note that the existence of a flux rope before an eruption remains debatable, leading to other data-inspired models such as the breakout model (Antiochos et al., 1999) and the tether-cutting model (Moore et al., 2001), which propose that the flux rope forms during the eruption.

Data-constrained flux rope models depend on measurements of the photospheric magnetic field, specifically the vector field, at a given time. Techniques like magnetofriction (e.g., van Ballegooijen 2004; Bobra et al. 2008; Su et al. 2009; Savcheva & van Ballegooijen 2009a) and nonlinear force-free field relaxation (e.g., Canou et al. 2009; Guo et al. 2010; Malanushenko et al. 2014) allow for the reconstruction of flux rope structures that align with observations in EUV and X-ray imagery. Similar to data-inspired models, these reconstructions could serve as initial conditions in global MHD simulations for modeling CME eruptions. See Figure 3.5b for an example.

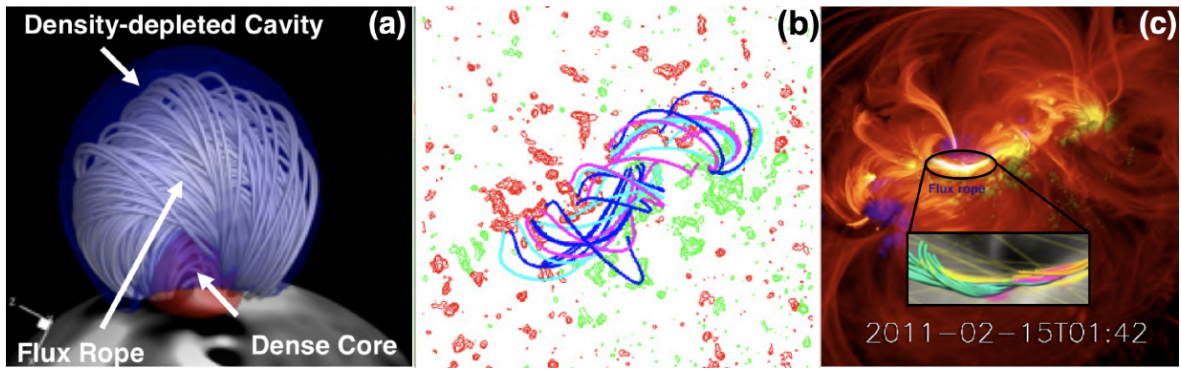

Figure 3.5: (a) Gibson-Low flux rope configuration inserted into an active region, adapted from Jin et al. (2017b); (b) Flux rope configuration from nonlinear force-free modeling, adapted from (Savcheva & van Ballegooijen, 2009b); (c) Data-driven flux rope by the Canadian Galactic Emission Mapper (CGEM), adapted from (Fisher et al., 2015). The figure shows a synthesized mock coronal image based on a proxy for emissivity calculated from the current density distribution in the model. A current-carrying flux rope is just starting to lift off from AR 11158 (Chintzoglou et al., 2019).

Data-driven models represent the cutting edge in CME simulation, where the flux rope is formed and erupted self-consistently in response to time-dependent photospheric drivers, either through flux transport models (e.g., Yeates et al. 2008) or sequential measurements of the vector magnetic field (e.g., Cheung & DeRosa 2012; Jiang et al. 2016). See Figure 3.5c for an example.

All these types of models have demonstrated reasonable accuracy in comparison with observations to varying degrees. For instance, data-inspired flux rope models have been widely utilized in global MHD models aimed at providing space weather forecasts. Moreover, data-driven models have delivered promising outcomes, capable of generating M- and X-class flares with time-dependent boundary driving, albeit without the particle beams thought to be a central component of such flares. Future enhancements in CME modeling will stem not only from the development of models (e.g., coupling models to encompass a broader range of scales) but will also depend heavily on comprehensive data coverage of the Sun, particularly the magnetic field, which is essential for driving global MHD models. In addition, more reliable coronal magnetic field measurements are needed to assess the magnetic field of the flux rope.

### 3.3.3 Solar SEP Modeling

Solar energetic particles (SEPs) are generated through particle acceleration during solar eruptions, which mainly occurs via two mechanisms: magnetic reconnection during solar flares and shock waves driven by CMEs through the diffusive shock acceleration process (e.g., Axford et al. 1977, Section 5.3). Consequently, SEP models could be generally classified into two types based on these processes. Section 5.3 discusses SEPs and their observed properties in more detail.

For particle acceleration associated with solar flares, the prevailing model utilizes MHD simulations of magnetic reconnection combined with a "test particle" approach. In this method, individual particle trajectories are traced through evolving electromagnetic fields derived from MHD solutions (Pontin & Priest, 2022). However, a significant limitation of the test particle approach is the lack of interaction between the accelerated particles and the background electromagnetic field. To address this, recent advancements include the particle-in-cell (PIC) kinetic approach, which is integrated with MHD simulations to create hybrid fluid-kinetic models, albeit at a higher computational cost (Guo et al., 2023). In the context of CME-driven shocks, particles repeatedly traverse the shock front due to interaction with magnetic turbulence near the shock, gaining energy through the diffusive shock acceleration mechanism. After escaping the acceleration regions, their movement through the solar wind is described by the focused transport equation.

Current SEP models fall into two categories: empirical models like *protons* (the operational SEP proton prediction model at the National Oceanic and Atmosphere Administration Space Weather Prediction Center [NOAA SWPC] forecast office) and Proton Prediction System (PPS; model developed by the Air Force Research Laboratory), which are primarily used for forecasting and focus on providing probabilities of an SEP occurrence based on observations of solar flares/CMEs. Physics-based SEP models, such as the improved Particle Acceleration and Transport in the Heliosphere (iPATH) Hu et al. 2017) or Multiple Field Line Advection Model for Particle Acceleration (M-FLAMPA) (Borovikov et al. 2018), on the other hand, aim to simulate particle acceleration and transport within the heliosphere by solving the focused transport equation and incorporating multiple physical processes such as magnetic focusing, adiabatic cooling, convection, pitch-angle scattering, and cross-field diffusion. These processes rely heavily on the background magnetic field, necessitating coupling with MHD models of the solar wind. Notably, since high-energy particles are primarily accelerated close to the Sun (within 10 solar radii) where CME-driven shocks are most intense, accurately modeling large SEP events requires detailed representations of early-stage CME-driven shock evolution in MHD models (Li et al., 2021). For a comprehensive review of current SEP models, see Whitman et al. (2023).

As illustrated in Figure 3.6, current SEP modeling yields promising results in reproducing observed SEP characteristics during certain solar SEP events. By integrating the Global MHD model AWSoM with the particle model iPATH, the combined model has successfully reproduced SEP intensity profiles at both Earth and Solar TErrestrial RElations Observatory (STEREO-A) locations. However, there are instances where the model does not align with observations. A significant challenge in SEP modeling is limited information about shock properties, particularly during the early stage of shock evolution. The structure of the ambient solar wind further complicates the 3D shock structure. Moreover, some critical parameters for particle acceleration (e.g., seed population) and particle transport (e.g., cross-field diffusion) are difficult to get from observations and remain largely as free parameters in the SEP models. These limitations exacerbate in their application to stellar environments, introducing significant uncertainties. Future improvements in solar SEP modeling will depend on a better understanding of particle acceleration and transport processes, informed by more in-situ SEP measurements at various distances and heliospheric longitudes. Enhanced models of the solar wind and CMEs will also be crucial in advancing SEP modeling.

### 3.3.4 Modeling in the Stellar Regime

Given the success of global MHD models of the solar wind and particular CME/SEP models in reasonably reproducing solar and heliospheric observations, various groups have extended their usage to the stellar domain. An example is shown in Figure 3.7). These extensions incorporate information on the modeled star's magnetic field, mainly in the form of ZDI reconstructions. The approach has been used to investigate a variety of topics, including

- Coronal and stellar wind structures across spectral types (e.g., Alvarado-Gómez et al., 2016a; Cohen et al., 2017; Marsden et al., 2023);

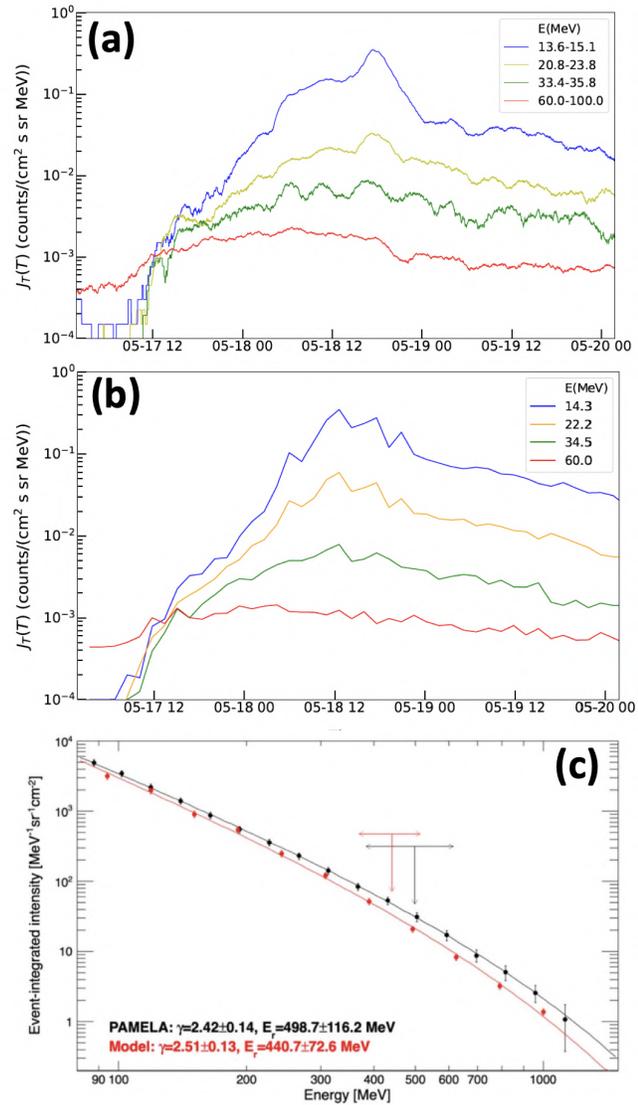

Figure 3.6: (a) Proton time-intensity profiles for four energy bins observed at STEREO-A during the 2012 May 17 SEP event. (b) Modeled proton time-intensity profiles from AWSoM+iPATH. (c) Spectrum comparison between Payload for Antimatter Matter Exploration and Light-nuclei Astrophysics (PAMELA) observation and model simulation. The black and red arrows indicate the location of the roll-over energies. Figures adapted from Li et al. 2021.

- Associated mass and angular momentum losses (e.g., Garraffo et al., 2016; Evensberget et al., 2023);
- The mass loss-activity relationship (Alvarado-Gómez et al., 2016b; Chebly et al., 2023);
- The rotational evolution of cool stars (Garraffo et al., 2018; Evensberget & Vidotto, 2024);
- Various types of star-planet interactions mediated by the stellar wind (e.g., Harbach et al., 2021; Kavanagh et al., 2021; Vidotto et al., 2023);
- Magnetic confinement of CMEs in active stars (Alvarado-Gómez et al., 2018, 2019, 2020a);
- The impact on CMEs on exoplanetary atmospheres/outflows (Hazra et al., 2022; Alvarado-Gómez et al., 2022b; Cohen et al., 2022); and
- The propagation of SEPs in highly magnetized stellar environments (Fraschetti et al., 2019, 2022; Hu et al., 2022).

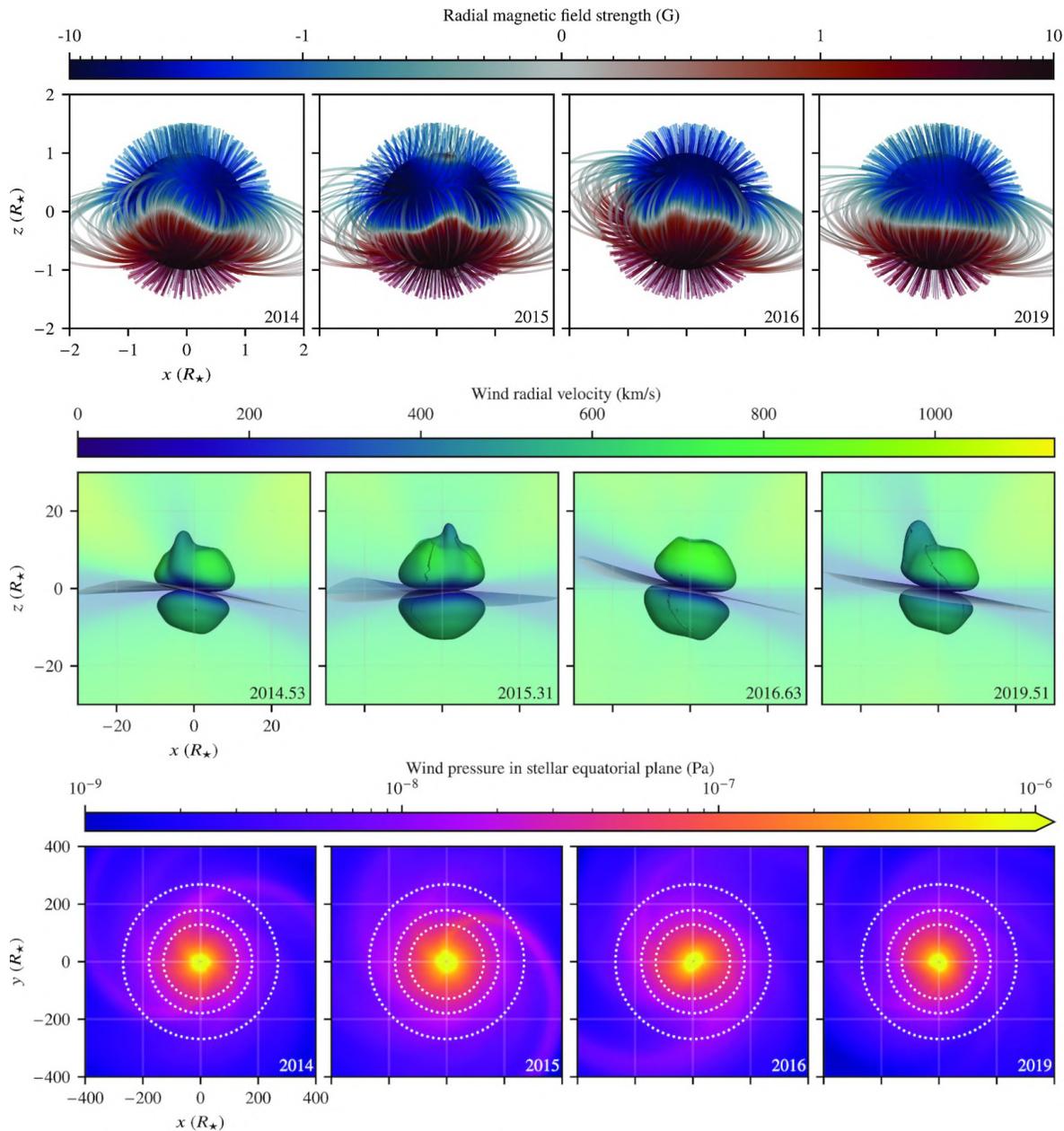

Figure 3.7: Example stellar magnetic field and wind reconstruction, assembled from Marsden et al. (2023) Figures 11-13. The star in this example is the late F star χ Draconis A. Dotted white lines in the bottom row represent the hypothetical orbits of Venus, Earth, and Mars for reference. Note the different scales of each row and that the bottom row is a top-down versus edge-on view.

Despite this progress, there are still several open questions and issues that need to be addressed in order to improve the applicability and realism of these models for the stellar regime. A discussion is provided by Lynch et al. (2023). The most pressing aspects are briefly summarized below:

**Magnetic Boundary Conditions:** Stellar modeling studies usually employ ZDI maps as an inner boundary condition for the photospheric magnetic field. This approach presents various challenges that range from limitations of the technique itself to practical aspects around the data. ZDI is only sensitive to the largest scales on the stellar surface and only measures the residual magnetic field (i.e., the field that survives after cancellation effects). Because of this, it is known that the derived maps only recover a very small fraction of the photospheric magnetic flux (with estimates ranging between $10\% - 0.1\%$ depending on the overall activity level of the star; see Kochukhov et al. 2020a).

Additionally, the spatial resolution achievable in the final map not only depends on the properties of the object of interest (namely, its rotational velocity) but also on the instrumental resolution, phase coverage, and S/N of the observations (Kochukhov, 2016). The latter will also influence the iterative goodness-of-fit procedure performed by ZDI (with no consensus in the community on a stopping criterion; see the discussion in Alvarado-Gómez et al. 2015), which sets the maximum field strength retrieved in the final map.

Another complication is the possible arbitrary inclination of the stellar rotation axis, which limits the fraction of the visible surface that can be mapped. This angle is typically assumed to be $60°$ even for objects for which the actual inclination is known (e.g., the case of AU Mic; see Klein et al. 2021).

On a more practical front, despite their publication, current ZDI maps are not available as publicly accessible data products. This limits the capacity of different groups to perform comparative studies with identical surface magnetic field input.

All in all, while the robustness of ZDI has been established in multiple different investigations, several improvements are still necessary to transform these magnetic reconstructions to stellar counterparts of solar magnetograms (typical inputs for steady-state corona and wind simulations).

**Thermodynamic Boundary Conditions:** Most corona/solar wind models require the specification of base values of plasma temperature and density. In the case of AWSoM, these thermodynamic properties are tailored to match conditions at the top of the solar chromosphere (see van der Holst et al. 2014a) and fine-tuned for validation purposes based on EUV imagery and solar wind conditions at 1 au (Jivani et al., 2023). However, these validation procedures are not possible in the stellar regime so that these base properties effectively transform into free parameters. This issue is exacerbated by the lack of clarity in how such parameters should be adjusted for different types of stars, as well as different activity levels (age). This situation, combined with the limited spatial and magnetic resolution of the input photospheric magnetic field (i.e., ZDI maps), dramatically increases the uncertainty of model outcomes in the stellar case.

These drawbacks are clearly evidenced in the emission measure distributions (EMDs) resulting from stellar models. The EMD, a key parameter in optically thin plasma emission, can be determined from X-ray and UV observations and compared to models. Coffaro et al. 2020 analyzed EMDs of the Sun and the young solar-like star $\varepsilon$ Eridani, noting that while the solar corona peaks at about 1 MK, $\varepsilon$ Eridani's corona shows a higher peak temperature of over 3 MK, with a significant portion of the plasma exceeding 10 MK, as indicated by EUVE data. However, MHD models based on ZDI maps often underestimate the emission measure (EM) of higher temperature plasmas, typically generated in active regions (Airapetian et al., 2021). This highlights the importance of EMDs in refining global MHD models, particularly for stellar environments where spatially resolved observations are not possible. Similarly, this calls for future Sun-as-a-star validation studies based on disk-integrated EMDs (instead of spatially resolved data) to better understand the capabilities and limitations of the model in the stellar regime.

**Assumptions behind the quasi-steady coronal heating/wind acceleration:** Apart from the base thermodynamic properties, MHD models typically require some additional parameters responsible for heating the corona and accelerating the stellar wind (see 3.3.1). For the specific case of AWSoM, Alfvén wave turbulence and its dissipation are responsible for coronal heating and wind acceleration, employing limited free parameters (Alfvén wave correlation length and Poynting flux) to phenomenologically model Alfvén wave dissipation. The latest implementation of the model, motivated by PSP measurements (Figure 3.4), requires four additional input parameters (van der Holst et al., 2022). As with the base density and temperature, information on these Alfvén wave-related properties is not available for other stars (although it might be possible to infer some of these characteristics based on far-ultraviolet (FUV) spectral lines; see Boro Saikia et al. 2023).

To complicate things further, the processes powering the corona and winds of other stars could be very different from the ones assumed dominating the solar case (Drake & Stelzer, 2023). As an example, the coronae of active stars (particularly M dwarfs) appear to be continuously flaring, and their light curves can often be described using a superposition of flares with almost no quiescent intervals (e.g., Caramazza et al. 2007; Huenemoerder et al. 2010). This challenges the assumption that Alfvén waves are the primary source of coronal heating and stellar wind acceleration for these stars, possibly implying the need to include additional effects due to transient events such as flares.

**Observational constraints for the stellar wind:** As can be seen in Figure 3.4, model validations for the solar wind are performed on individual physical properties such as plasma density, velocity, and magnetic field strength. This approach can disentangle the various physical processes at play in the simulation, enabling adjustments if needed. In the stellar regime, constraints on the stellar wind properties are scarce at best (Chapter 4). These are limited only to estimates of the mass loss rate (i.e., product of the density and flow velocity) without any information on the actual density or wind velocity or its magnetic field.

Furthermore, while multiple vantage points are used for solar wind model-observations comparisons (at different heliocentric distances), the large majority of stellar wind constraints are derived at the astrosphere-interstellar medium (ISM) boundary (i.e., the hydrogen wall; Section 4). Apart from the intrinsic uncertainties and degeneracies associated with these methods, stellar wind constraints at such large distances (i.e., several astronomical units) do not contain any information pertaining to the processes powering and accelerating the stellar wind. As such, in their current form, observations of the stellar wind provide only limited constraints for modeling studies of outflows in cool stars.

**Limitations in the CME/EPs modeling for stars:** The issues described previously not only affect simulations of quasi-steady stellar environments, but also drastically limit the modeling of transient phenomena such as CMEs and SEPs. Because of this, studies in the stellar regime so far have used the most simplified type of solar eruption models available (i.e., data-inspired models; see Section 3.3.2). As discussed by Lynch et al. (2023), various strategies have been employed to compensate for the lack of small-scale structures and magnetic flux in nominal stellar data (i.e., ZDI). This also includes the use of state-of-the-art dynamo solar/stellar models to inform the photospheric magnetic field boundary condition hosting the MFR responsible for the eruptions (Alvarado-Gómez et al., 2022a; Xu et al., 2024). Still, without improved information on the surface magnetic field (particularly its variability on relevant time scales) and actual constraints on CME parameters (i.e., mass, velocity, kinetic energy), very little progress can be expected in this area. The situation is even more complicated for SEP modeling on other stars, where the lack of any useful observational constraints limits their applicability to exploratory studies completely bound to the phenomenology observed in the Solar System.

## 3.4 Paths Forward in Observing Stellar Magnetism and Modeling Exoplanetary Particle Environments

Models are the only way to produce pictures of space weather that are comprehensive in space and time. Such comprehensive pictures are essential to evaluating long-term impacts on planets. Stellar magnetism drives space weather, and accurate modeling, therefore, requires accurate observations of stellar magnetic fields. A direct assessment of accuracy is possible within the heliosphere, where model outputs can be benchmarked against in-situ measurements. This chapter has outlined the state of the art and laid out opportunities for advancement in a number of areas.

One such area is magnetic field measurements of stars. Both the breadth and depth of these observations can be advanced by installing polarimeters on more sensitive observatories and with a greater number of resources to more densely sample in time the rotation of a star. On the analysis side, the use of accurate inclinations (supported by greater efforts to measure stellar inclinations) could improve accuracy. Meanwhile, inter-comparison and cross-validation of ZDI maps could be enabled by making ZDI maps publicly available. Establishing a standardized stopping criterion would help ensure this subjective choice is not the source of differences.

Another area is observations of the Sun, which provides the best validation case for comprehensive modeling of circumstellar particle environments. In comparison to stars, the Sun and solar space weather are well observed; in an absolute sense, they remain very sparsely sampled. Improving the spatial and temporal coverage of particle and magnetic field measurements will have a commensurate impact on model validation. Measurements of magnetic fields in the corona, particularly in developing flux ropes, would be especially valuable, as would further observational insights into processes accelerating SEPs.

Modeling stellar space weather remains a challenging area of active development with a variety of paths forward. An area deserving of focus is model boundary conditions. Time-dependent boundary conditions and the exploration of a variety of potential heating mechanisms in stellar cases, particularly for highly flare-active stars, will be important to explore. Finding methods for matching the high-temperature component of stellar EMDs will be helpful, and here the use of Sun-as-a-star analyses could help. Coupling models across scales will be essential for tracking stellar space weather from the stellar surface to the astropause, and some efforts here have already begun.

Finally, a key path forward is tighter observational constraints on actual stellar winds, CMEs, and similar products of stellar magnetism. This, of course, was the original motivation for the KISS workshop and is the topic of a large fraction of this report, with many paths forward identified. In order to serve as good constraints on models, however, these observations will need to target the most fundamental properties possible. For example, rather than a globally averaged total mass loss of the stellar wind, the wind velocity and density at a particular location could be more constraining to a global 3D MHD model. The magnetic field of the stellar wind is particularly valuable, yet also particularly hard to constrain, with no foreseeable solutions identified as part of our workshop. Although observing these properties is challenging, and some may seem impossible at present, their value justifies continued effort to identify, develop, and advance new and existing techniques.

**Zooming Out: Related Findings and Recommendations (FRs)**

Only models can provide comprehensive pictures of space weather environments and their underlying physics, making sustained model development an essential piece of the exospace weather field (FR2). This development can be aided by new observational constraints, enabled through advances in methods (FR4, FR5) and hardware (FR3, FR6). Collaboration between modeling, observational, solar, and stellar communities is key to this work and should be incentivized at a programmatic level (FR7).

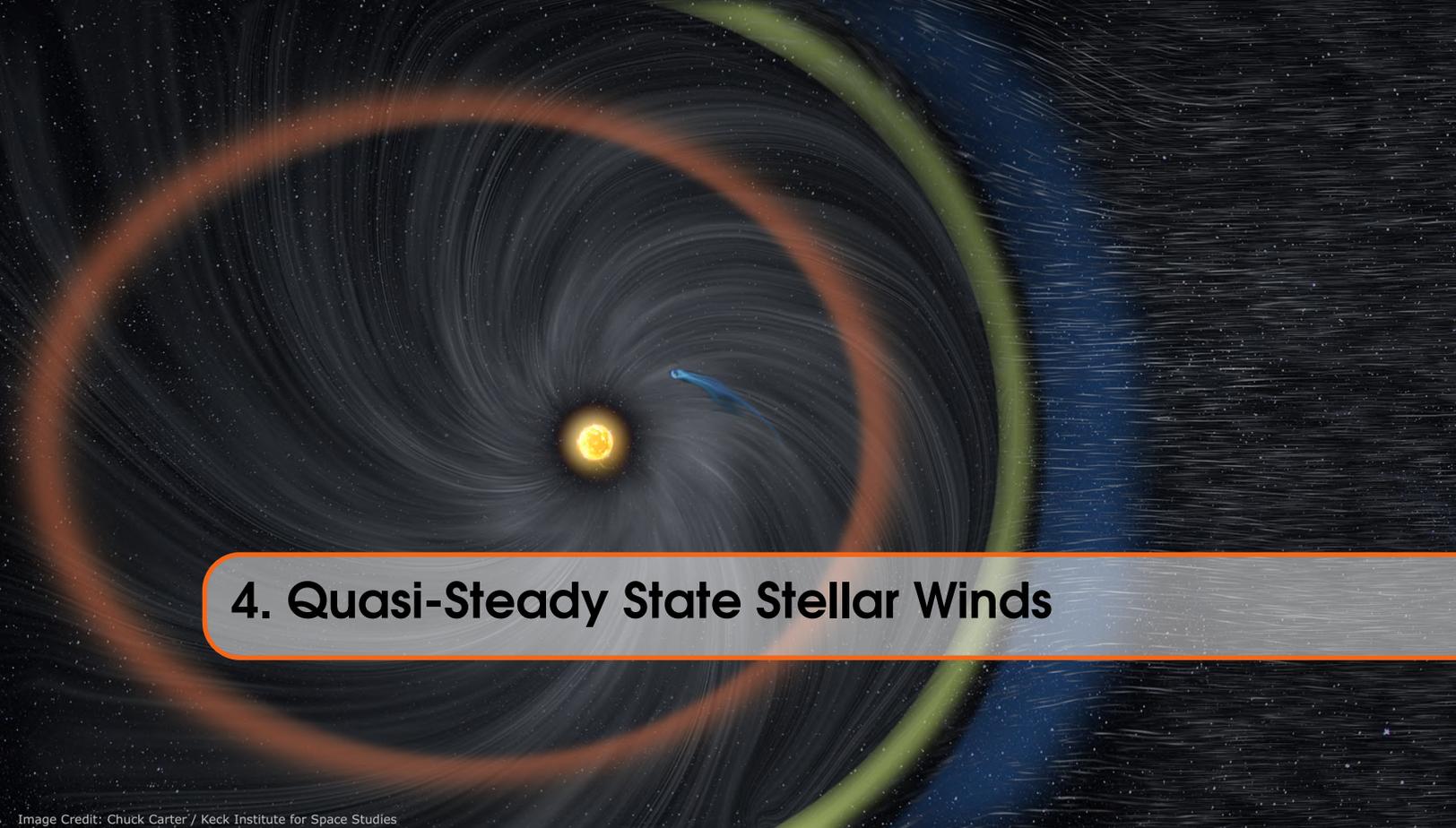

## 4.1  Overview

In our Solar System, the Sun's magnetic activity gives rise to a solar wind that carries plasma, magnetic fields, and angular momentum away from the Sun and out into interplanetary space (Figure 4.1). The solar wind continuously sweeps past the Earth and other Solar System bodies, providing energy, particles, and magnetic interactions as inputs to the coupled magnetosphere, ionosphere, and atmosphere systems of the planets (Temmer, 2021). The region of space dominated by the Sun's magnetic field is known as the heliosphere. The heliosphere interacts with and carves a path through the interstellar medium (ISM), and shields the Solar System from the influence of galactic cosmic rays (GCRs). (The analogous structure for other stars, the astrosphere, is discussed in Section 4.2.) The evolution of the Sun's magnetic field and all of the dynamic space weather phenomena it produces are integral components of the history of Solar System planets and our understanding of the habitable conditions that have allowed life to evolve and persist on Earth (Güdel, 2007).

The solar wind varies on transient, short-term, and long-term timescales as the Sun's magnetic field erupts, cycles, and evolves. Transient space weather phenomena such as flaring and coronal mass ejections will be reviewed in Chapter 5. This chapter deals with the quasi-steady particle environments of the Sun and other stars. The term *quasi-steady* is used here to reflect the more slowly-varying aspects of space weather in the Solar System and in exoplanet systems, chiefly the solar/stellar wind and the modulation of GCRs by the heliosphere/astrosphere. The solar wind is not truly steady-state but is known to exhibit day-to-day variations in structure, density, and speed; additionally, the solar wind varies on solar cycle and stellar evolutionary timescales as the solar magnetic field changes in time (see, e.g., McComas et al. 2008b; Vidotto 2021).

Single cool stars are known to spin down and become less magnetically active across the gigayear (Gyr) timescales of stellar evolution over their main-sequence lifetimes (Skumanich, 1972; Tu et al., 2015; Johnstone et al., 2021). Given the relative dearth of observations of stellar winds for stars other than the Sun, the exact dependence of stellar wind parameters such as velocity, density, and interplanetary magnetic field (IMF) strength and orientation on host star parameters such as age and spectral type is largely unconstrained. There is, therefore, significant scientific interest in investigating the interrelated phenomena of stellar magnetic evolution and stellar wind variability. Additionally, the Sun exhibits a solar activity cycle of a roughly 11-year period. While it is known that the overall mass-loss rate of the Sun varies only weakly with the solar cycle (Cohen, 2011), the prevalence, periods, and amplitudes of activity cycles for other cool stars create a challenging parameter space to probe observationally and to interpret physically (See et al., 2016). The variability of stellar wind conditions in response to changing stellar magnetic activity is an important area of active research in understanding the solar-stellar connection (Brun & Browning, 2017).

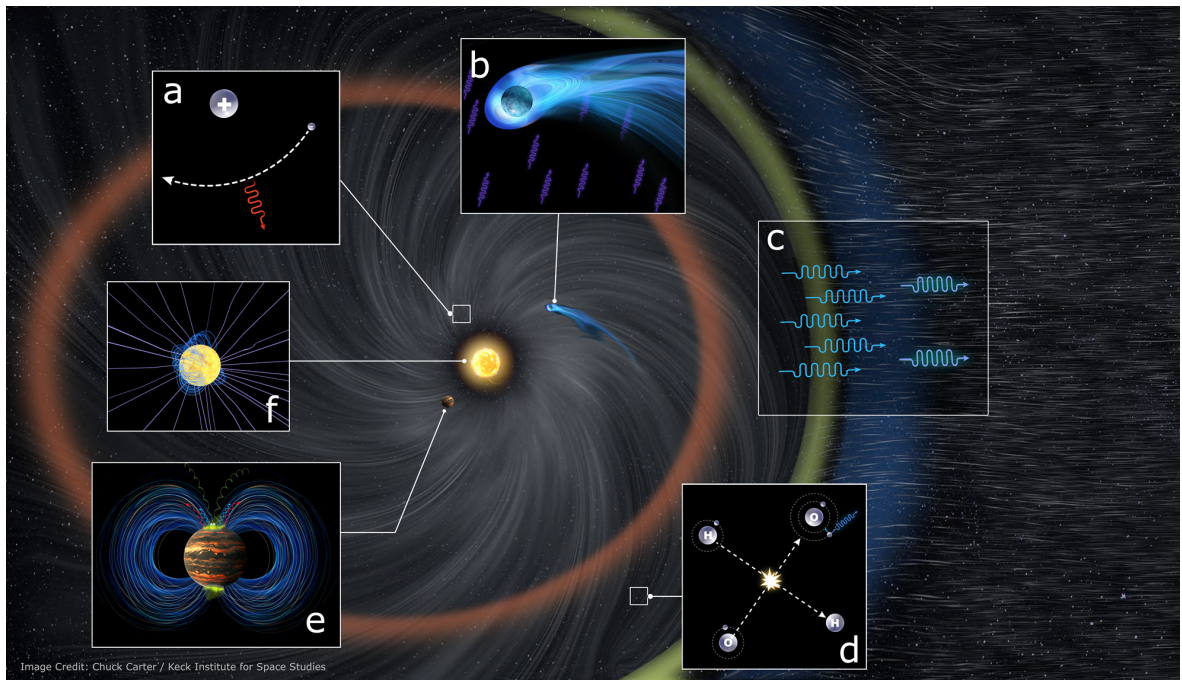

Figure 4.1: Artist's concept of a stellar wind, depicted as the spiraling white wisps[*], showing key features and a selection of potential observables. The colored boundaries depict, from inner to outer, the wind termination shock (orange), the astropause (green), and the region of compressed ISM material, or "hydrogen wall," (blue) behind the bow shock or bow wave. Insets symbolize (a) free-free emission (Section 4.4.3), (b) absorption of light by a wind-deflected planetary outflow (Section 4.3.4) (c) absorption of Lyman-$\alpha$ photons by compressed ISM material (Section 4.3.1), (d) emission from ISM-wind charge exchange (Section 4.3.3), and (e) planetary auroral emission (applicable to winds but more typically associated with transients, Section 5.5.8). Inset (f) represents the magnetism of the star that is the ultimate engine of space weather (Chapter 3). Figure credit: Keck Institute for Space Studies / Chuck Carter. Full resolution figure available at `https://www.kiss.caltech.edu/artwork.html` and may be reused or modified on the condition of appropriate image credit.

[*] The spirals shown in the figure contrast with the radially emanating lines used in many other schematic representations of solar or stellar winds. The spirals may be thought of as streaklines (the paths traced if a dye were injected into a fluid flow), and are meant to capture the effect of stellar rotation. Lines strictly tracing flow velocity vectors, known as streamlines, would emanate radially outward.

It is important to constrain stellar winds not only to understand their relationship to stellar magnetism, but also in order to determine their impact on exoplanet environments and aspects of habitability such as planetary atmospheric retention. The influence of stellar winds on the exoplanets residing in the environments of other stars has been discussed in depth in Chapter 2. Placing constraints on the quasi-steady space weather of a stellar system would help to better inform stellar wind models that are primarily constrained at their lower boundary by observations of stellar magnetic fields, as discussed in Chapter 3. Additionally, new constraints would allow for a more detailed understanding of the transient events that perturb the quasi-steady states of stellar space weather, as we will explore in Chapter 5.

This chapter will discuss the state of the art of several stellar wind observational methods, with a discussion of the feasibility and science yield of each method. A selection of these is illustrated in Figure 4.1. We begin in Section 4.2 with a review of the concept of astrospheres, including a discussion of the role of galactic cosmic rays in the quasi-steady state of a stellar system and the environments of associated                                                                           exoplanets.

We then discuss current capabilities and promising avenues of research into the quasi-steady behavior of exoplanet host stars and young Sun analogs of the nearby solar neighborhood, chiefly in the form of the stellar wind long-term (non-eruptive) variability on timescales of years to gigayears. In Section 4.3 we discuss current and near-term capabilities for observing and constraining stellar winds. Section 4.4 details several additional stellar wind observational techniques, with longer-term prospects such as needed instrumentation and speculative scientific methods. The structure of this chapter is to review and suggest observational methods by which stellar winds can be characterized and place them in context. We close with some discussion of paths forward in this research area in Section 4.5.

> **Not So Fast: Age is not always predictive of stellar activity.**
>
> The initial spin state of a star when it forms can make a very large difference in how its activity evolves with age. This is particularly true for G-type stars, where at ages of a few hundred megayear (Myr) the difference in X-ray luminosity can be over an order of magnitude (Tu et al., 2015; Johnstone et al., 2021). To first order, rotation rate is likely most predictive of activity for stars of a given mass. Tidal interactions with companions and metallicity are some additional complicating factors (See et al., 2024).

## 4.2    Recurring Concept: Astrospheres

The outward pressure of a star's wind opens a cavity within the surrounding interstellar medium in which the star resides (Figure 4.1). This cavity is known as the astrosphere or, for the Sun, the heliosphere. Astrospheres are shaped by the relative motion between the star and the ISM. The boundary area between these two flows includes a range of interactions between the particles (Section 4.2.1), dust grains (with a possible connection to an exogenic origin of life, Section 2.5), and magnetic fields of each region, as shown schematically in Figure 4.2. The astrosphere and its interactions with the interstellar medium make good fodder for observable signatures of stellar winds, as discussed later in this chapter in Section 4.3.1. For further reading, Wood (2004) reviews stellar astrospheres as observed through their Lyman-$\alpha$ absorption and Richardson et al. (2022) reviews the heliosphere observed in-situ by the Voyager and New Horizons spacecraft.

Three distinct boundaries are defined in the region where the stellar wind and ISM encounter each other, as shown in Figures 4.1 & 4.2. The surface defining where the pressure of the stellar wind and the ISM balance—effectively, where the two flows physically collide—is the astropause. Inside the astropause, the stellar wind must decelerate before coming to a halt, resulting in a termination shock as the flow transitions from supersonic to subsonic. Beyond the astropause in the direction of the stellar wind's upwind motion relative to the ISM, the ISM must experience the same deceleration. This can form a bow shock or a bow wave, depending on whether the motion of the star does or does not exceed the sound speed of the ISM, respectively. In this region between the astropause and the bow shock/wave, neutral hydrogen from the ISM piles up, becoming heated and compressed to form a "hydrogen wall."

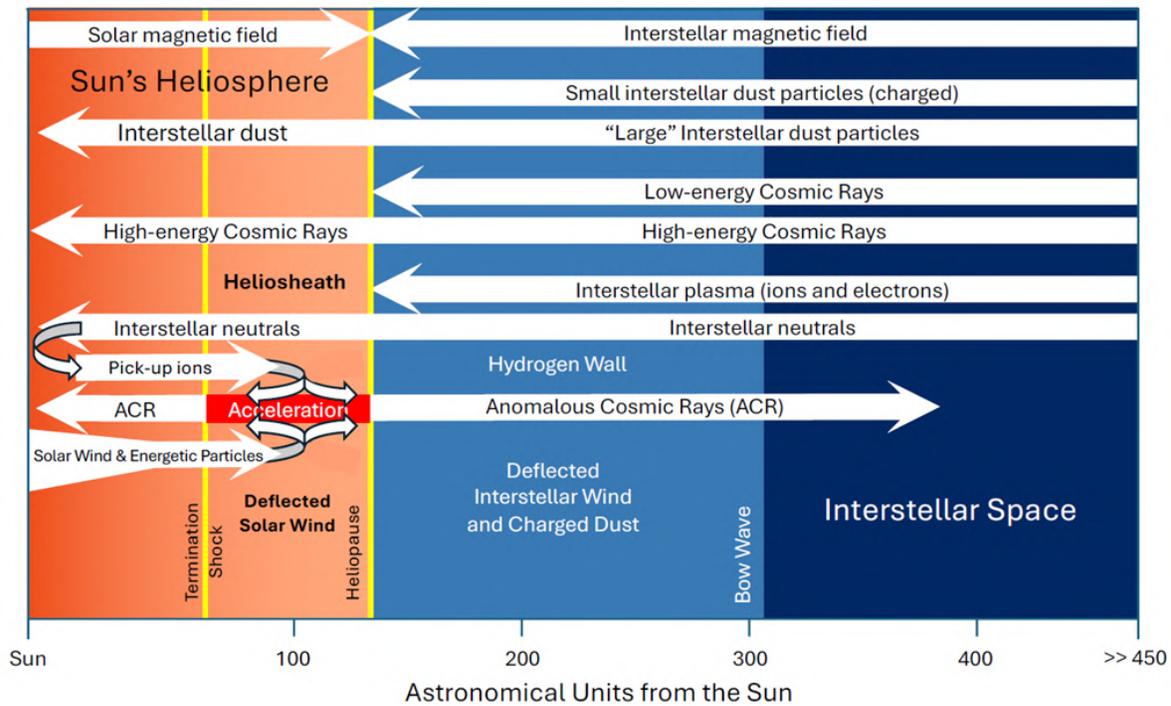

Figure 4.2: Schematic of the role the heliosphere boundary (with analogy to astrosphere) plays in controlling the transport of matter and magnetic field between the domain of the Sun/star and the ambient ISM. The heliopause presents a barrier to some elements of the interstellar medium, while allowing interstellar dust, high-energy cosmic rays, and the interstellar neutrals to flow into the heliosphere. Distances are based on Voyager 1 and Voyager 2 data (Stone et al., 2005; Li et al., 2008; Burlaga & Ness, 2014; Dialynas et al., 2019), the bow wave determination is based on McComas et al. (2012), and the representation of anomalous cosmic rays is based on Pesses et al. (1981); Stone et al. (2019); McComas et al. (2019); Giacalone et al. (2022). Credit: Updated and adapted by James Green from Figure 11 of (Frisch, 2000), originally credited to Linda Huff.

The size of the astrosphere is determined by a balance of stellar wind pressure against the pressure of the ISM. Both of these can vary considerably, yielding orders-of-magnitude variations in the size of astrospheres. Figure 4.3 provides a comparison between the heliosphere and a much smaller astrosphere. The Sun's astrosphere, the heliosphere, extends about 90 AU to the termination shock in the upwind direction (McComas et al., 2021). For some stars, observations indicate astrospheres as large as ∼1000 AU in the upwind direction, and (Wood et al., 2005) notes that from Earth the angular extent of some astrospheres is similar to the Moon. A star's astrosphere will vary in size over time as the strength of its wind evolves and as it transits through ISM "clouds" of differing density.

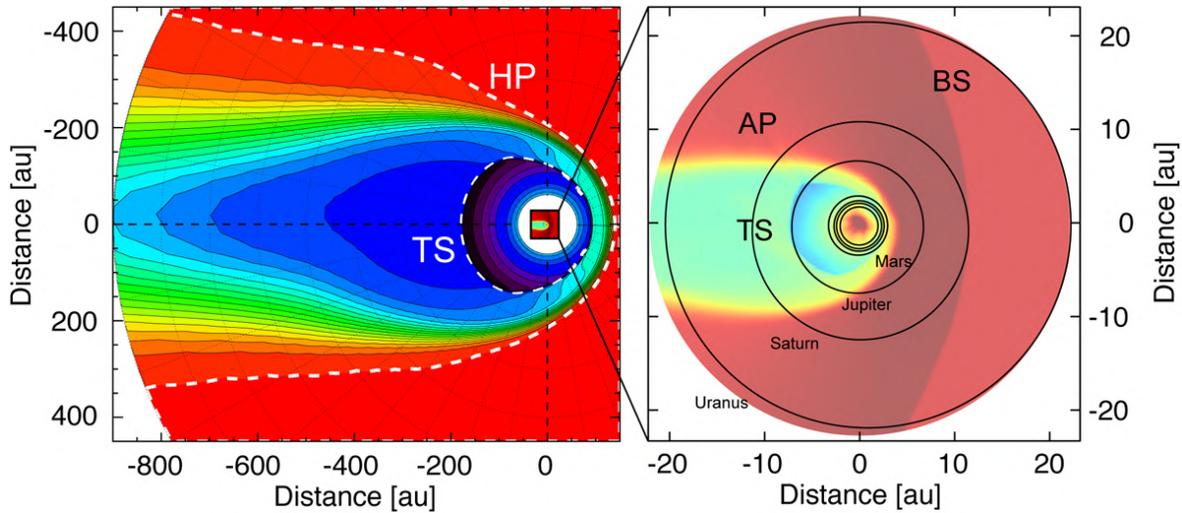

Figure 4.3: Comparison of the size and geometry of the heliosphere *(left)* with a zoom-in panel *(right)* showing a modeled astrosphere for the M dwarf star LHS 1140, with the orbits of Solar System planets included for scale (Scherer et al., 2025). Abbreviations are heliosphere (HP), termination shock (TS), bow shock (BS), and astropause (AP).

### 4.2.1 Galactic Cosmic Rays and their Transport

Beyond their utility in the observation of stellar winds, astrospheres also play a significant role in exoplanet space weather environments by influencing the flux of galactic cosmic rays (GCRs) into the interior of a star-planet system. GCRs are energetic ionized particles that do not originate from the host star, but instead come from external sources in the galaxy, hence their designation as "galactic" cosmic rays. Their origins likely include supernovae and other very energetic events across the galaxy, although challenges to this paradigm of GCR production have recently been discussed in the literature (Gabici et al., 2019).

GCRs span a large range of energies; low-energy GCRs are scattered at the outer boundaries of astrospheres, shielding the inner astrosphere from their effects. Higher-energy particles can penetrate into the astrosphere and are then subject to propagation through the magnetic field of the stellar wind, resulting in GCR fluxes that vary with activity cycles and transient wind structures. GCR transport within the astrospheres of different stars vary according to their different stellar magnetic field and stellar wind properties (Rodgers-Lee et al., 2021b). GCR flux also varies with distance within the astrosphere, so exoplanets in different orbits around the same host star will receive different GCR fluxes (Mesquita et al., 2022a). Finally, since GCR transport is influenced by changes in stellar magnetism, GCR flux received at an exoplanet will also change with time as the host star's magnetic field and winds evolve with age (Rodgers-Lee et al., 2020).

Additionally, some particles sourced from the central host star can be accelerated and deflected at the interior of the astrosphere boundary, generating a population of "anomalous" galactic cosmic rays of local origin (these are the ACRs displayed in Figure 4.2). Recent reviews of GCR propagation in the heliosphere are provided by Engelbrecht et al. (2022) and Rankin et al. (2022).

Variations in stellar astrosphere sizes and geometries based on the wind/ISM balances are expected to play a role in long-term space weather, or "space climate," by affecting the ability of GCRs to reach the planets within the astrosphere. As stars pass through regions of dense ISM, it is possible for astrospheres to shrink to within the orbits of planets, directly exposing those planets to the flow of the interstellar medium and the impact of GCRs. Recent modeling studies, utilizing constraints from iron isotopes in the Earth's geological record, have suggested that the Sun may have experienced one or more such events in the past several Myr, exposing early Earth to greater fluxes of GCRs (Opher et al., 2024b,a).

## 4.3 Observational Methods - Near Term

### 4.3.1 Ly-alpha Astrosphere Method

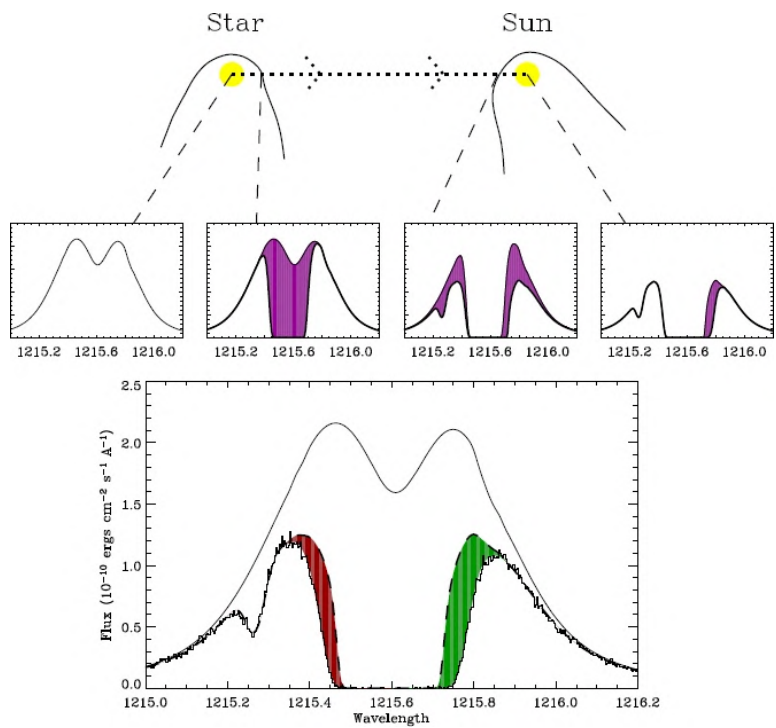

Figure 4.4: Schematic depicting, from left to right, how a star's intrinsic Ly-$\alpha$ line profile is modified by the stellar astrosphere, ISM absorption, and then the heliosphere (Wood, 2004). The lower panel demonstrates the contribution from heliospheric hydrogen (green shading) and from astrospheric hydrogen (red shading).

To date, the most successful method (in terms of number of measurements) used to constrain the winds of other stars is the observation of absorption of Ly-$\alpha$ photons through complex charge exchange reactions by neutral hydrogen that is compressed and heated into a "hydrogen wall" upwind of the astrosphere (Wood et al., 2001; Wood, 2004; Wood et al., 2021). Because the ISM itself strongly absorbs Ly-$\alpha$ photons, this absorption cannot be seen in isolation. Instead, it appears as an "excess absorption" relative to that expected from the unperturbed ISM (Figure 4.4). Geometry requires that this excess absorption appear on the blueward side of the absorption profile: for the sight line from Earth to pass through the hydrogen wall, the velocity of the star relative to the ISM (the upwind direction) must be pointed roughly in the direction of Earth, resulting in a blueward offset of the hydrogen wall (moving with the star) with respect to the ISM from the Earth's perspective. The same excess absorption appears due to the interaction of the heliosphere with the ISM when looking at stars near the velocity vector of the Sun, but on the red edge of the absorption core, given the reversed perspective.

This technique provided constraints on the winds of an impressive 31 stars (Wood et al., 2021). However, it is subject to a number of limitations (Wood, 2004). Targets must be relatively nearby, generally within 10 pc. With growing distances, the width of the ambient ISM's optically thick absorption increases until it eventually subsumes the thermally broadened absorption of hydrogen near the astrosphere. Further, outside of neutral ISM clouds, particularly in the mostly ionized local bubble, a lack of neutral hydrogen to produce a neutral hydrogen wall can prevent a detection. The method is also subject to a key geometry constraint. The line of sight must pass near enough to the "upwind" direction, the direction of the star's motion through the ISM, to pass through a sufficient population of heated neutral hydrogen. In practice, the angle between the upwind vector and the line of sight to the star must be under about 120° (about a quarter of the unit sphere) to yield detections. Slow, relative velocities between the star and the ISM are also prohibitive. Finally, weak stellar winds also carve out smaller astrospheres, piling up less dense hydrogen walls, and are therefore also less likely to produce detectable signals than stronger stellar winds. It can be difficult to translate non-detections from such stars into mass loss limits, as they often can also be attributed to the ionization state of the ISM near the star.

These observational constraints mean this method cannot be applied to just any star, particularly less magnetically active stars with weaker winds. Improved instrumentation, such as a high-resolution spectrograph with an effective area at least as large as NASA Hubble's Space Telescope Imaging Spectrograph (STIS), could yield further detections in the future. However, detections will diminish as additional targets become farther away, where ISM absorption is greater, particularly beyond the mostly ionized Local Bubble in the ISM.

Improvements to the astrosphere method itself could improve the science yield of this type of observation. One key assumption adopted to infer the stellar mass-loss rate from Ly-$\alpha$ absorption is a constant wind speed of 400 km/s. Although there is reason to believe that winds should generally have velocities near the escape velocity of the star, as in the case of the 400 km/s slow solar wind, observational validation of this assumption is necessary to verify these results; indeed, modeling studies based on ZDI stellar magnetic field observations predict larger stellar wind speeds (Chebly et al., 2023).

The application of more than one stellar wind observational method described in this chapter to the same target star could help to break the degeneracies in wind speed and density that make this simplifying assumption necessary. There is also scientific interest in the concept of revisiting the same target stars with the Ly-$\alpha$ astrosphere method years or decades after initial observations, in order to investigate the possibility of secular changes in stellar mass-loss rate due to stellar cycle activity, or other sources of astrosphere variability such as stellar transient activity and changes in the local ISM flow.

### 4.3.2  Direct Imaging of Astrospheres

Section 4.2 set the stage for the detection of astrospheres as the dominant approach astronomers use to constrain quasi-steady stellar winds observationally. Because astrospheres and their interactions with the ISM can have large angular scales (Wood et al., 2002, note that $\varepsilon$ Eri's astrosphere extends 42', which would be larger than the full Moon), emissions from this region may be directly observable. Although the densities are low in the interaction region, volumes are large, meaning the total number of emitted photons may be detectable with sufficient sensitivity. Meanwhile, the region is not only large in an absolute sense, but appears large enough on the plane of the sky for some nearby targets that even modest spatial resolution would be sufficient to resolve the physical structure. Section 4.3.3 describes the possibility of detecting charge exchange-induced X-ray emission from astrospheres. This section addresses the possibility of direct detection in optical light.

A rough estimate of emission in the hydrogen Balmer line ($H\alpha$), a line with a long heritage of observations using ground-based telescopes, is possible by assuming optically thin emission. As detailed in Section A.1, a flux density of $10^{-23}$ erg cm$^{-2}$ s$^{-1}$ Å$^{-1}$ for an R magnitude brightness of 35.6 is possible in a 1" square pixel. This emission is currently too faint to detect with the Dragonfly Telephoto Array, which can perform R and G band imaging and get down to 30 mag/arcsec$^2$ in a 1" square pixel in an exposure of 10 hours (Abraham & van Dokkum, 2014). However, there is a possibility that implementing a narrowband filter centered on the $H\alpha$ line with a width of a few Ångstroms, as well as summing emission from adjacent regions of the sky, could bring this idea closer to feasibility.

### 4.3.3 Charge Exchange-Induced X-ray Emission from the Astrospheres

As described in Section 4.2, the interaction of the stellar wind with the neutral ISM is a potential mechanism to detect quasi-steady stellar mass loss. The idea to look for X-ray emission from charge exchange scattering as a signature of a stellar wind arises from a confluence of two events. First is the recognition of the large angular extent of nearby astrospheres. There is a parallel with X-ray cometary emission, originating from charge exchange from solar wind heavy ions with neutral cometary material (Cravens, 1997). Stellar winds are potentially detectable via extended X-ray emission through their interaction with the neutral ISM.

Wargelin & Drake (2001) first put forward this idea, noting that for the closest stars, it would be possible to spatially and spectrally resolve this emission from the bright coronal emission if the charge-exchange emission is sufficiently extended and the X-ray detector is sensitive enough. The charge exchange emission is expected to be thousands of times weaker than stellar coronal emission, requiring such a search to be performed far out in the wings of the point spread function, where the expected charge exchange wind signal would dominate over background contribution. Sensitive spectroscopy can resolve the charge exchange from coronal emissions due to their different spectral signatures. Carbon, nitrogen, and oxygen are the most abundant metals expected in stellar winds, and there is a characteristic cluster of transitions diagnostic of charge-exchange processes near O VII at 0.56 keV, where current large X-ray telescopes have adequate sensitivity, spectral resolution, and spatial resolution to probe this emission. This technique was first attempted on the nearest star outside our Solar System—Wargelin & Drake (2002) applied the method to Proxima Centauri using extant data from NASA's Chandra X-ray Observatory. They derived a one sigma upper limit of 4 $\dot{M}_\odot$ with the assumption of a factor of 3 in model uncertainties. This is still larger than the 0.2 $\dot{M}_\odot$ upper limit reported by Wood et al. (2001) from an astrospheric hydrogen wall measurement.

Recent reinvigoration of this technique has produced interesting results. Kislyakova et al. (2024) reported spectral detections of soft X-ray excesses in spatial regions ranging from 1-11 arcminutes away from the central star, in three cases of nearby stars with previously detected astrospheres using the Lyman-$\alpha$ technique. This excess emission occurred in the region near the 0.56 keV oxygen K$\alpha$ triplet line above an estimate of the scaled-down stellar coronal flux. This discovery does not spatially resolve the astrospheric emission. Instead, a rather large annulus is used to collect photons far from the central star and generate a spectrum that can be referenced against the stellar emission in search of an additional emission component. The results compare reasonably well, within a factor of two in most cases, for stars with results from previous investigations using the Lyman $\alpha$ astrosphere method.

More recently, Lisse et al. (ApJ, under revision) report on spatially resolved astrospheric charge exchange emission from the relatively nearby zero age main sequence (ZAMS) star HD 61005 (Figure 4.5). While this star is further away than the samples mentioned above (distance of 36 pc), it is a young X-ray bright star that has accompanying evidence for a strong interaction between the local interstellar medium (LISM) and the dust disk. The LISM around this star is roughly 500 times denser than solar neighborhood conditions, making this a favorable case to observe stellar wind charge exchange-induced X-ray emission. The astrosphere is spatially resolved with a diameter, $d \sim 220$ AU in extent, and the X-ray emission morphology is roughly spherical, as expected for an astrosphere dominated by a strong stellar wind outflow. The spectral energy distribution of the halo photons shows an excess at low energies, being dominated by line emission from O VIII (650–850 eV) and Ne IX (900–910 eV).

There are several advantages of this technique; the first is that it is roughly independent of stellar activity level. The detectability does scale with stellar emission, and highly active stars have enhanced levels of coronal emission, so there is an indirect dependence. There is a geometrical advantage as well, since the line of sight does not need to be near the nose of the astrosphere. Another advantage stems from not needing to contend with discrete ISM absorption (as in the case of the Lyman-$\alpha$ astrospheres; Section 4.3.1). Like the Lyman-$\alpha$ astrospheres method, however, the local ISM does need to be mostly neutral for sufficient charge exchange to occur. Additionally, as for the Ly-alpha astrospheric method, the charge exchange detections can only be translated to quantities of interest related to the stellar wind (i.e., $\dot{M}$) via model-dependent estimates of both the geometry of the system, as well as the density of astrospheric neutral hydrogen. Another disadvantage is that it requires a large effective area X-ray telescope with relatively high spatial resolution and low X-ray background. While the Chandra X-ray Observatory's superb spatial resolution is unmatched, over the course of its 25 years in space, the low-energy sensitivity has degraded. The XMM-Newton Observatory also satisfies the effective area requirement, but has significantly lower angular resolution.

This method is potentially much more general than the highly successful Lyman-$\alpha$ astrosphere method (Section 4.3.1), and future instrumentation will almost certainly lead to advances. At the time of this writing, a probe-class mission concept selected for phase A study called the Advanced X-Ray Imaging Satellite (AXIS) [1] promises a field-of-view-averaged spatial resolution that is six times better than Chandra, with an effective area four times that of XMM-Newton below 2 keV. The New Advanced Telescope for High-ENergy Astrophysics (NewAthena) mission concept in Europe (Cruise et al., 2024), recently endorsed by the European Space Agency and under study for official adoption, would provide similar spatial resolution as XMM-Newton currently provides, but with a greatly enhanced effective area.

### 4.3.4   Exoplanet Outflows as Probes for Stellar Winds

Close-in exoplanets have been used with some preliminary success to probe the wind environments near their host stars. Hydrodynamic outflows of escaping planetary atmospheric material can be shaped by the stellar wind as the planet traverses the near-star environment. Extreme close-in exoplanets experience enhanced stellar wind densities, strong stellar magnetic fields, and high Keplerian velocities relative to the local stellar wind flow (Cohen et al., 2011; Kislyakova et al., 2014; Vidotto et al., 2015).

---

[1] https://blog.umd.edu/axis/

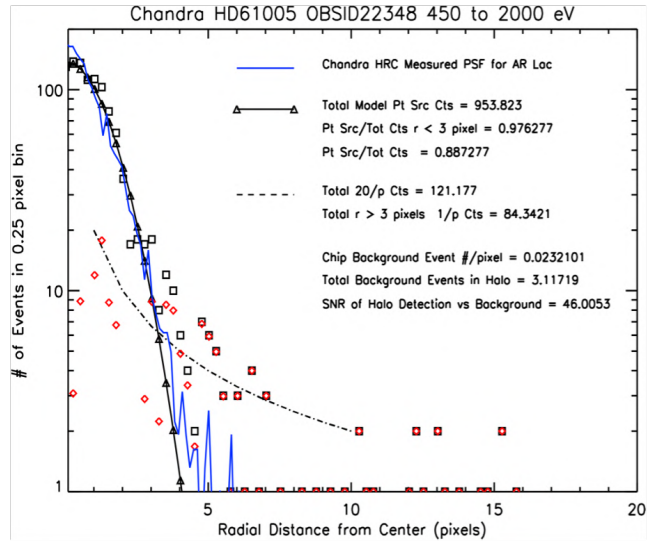

Figure 4.5: Figure 5 of Lisse et al. (submitted) showing the radial distribution of X-ray photons in the region around HD 61005. The data are indicated by boxes; triangles and the solid line show the best-fit point source model, while red diamonds show the residuals after subtraction. The residuals follow a trend decreasing with the inverse of projected distance from the central star.

While these strong stellar wind conditions could be detrimental to the potential for habitability, they can also provide conditions for observable stellar wind-exoplanet interactions.

To date, models have appealed to stellar wind-dependent interactions between a planetary outflow and the local stellar wind to explain Hubble Space Telescope observations of three systems. These planets include the hot Jupiters HD 209458 b (orbiting a host star of spectral type F8, Kislyakova et al. (2014); Villarreal D'Angelo et al. (2018)) and HD 189733 b (host star type K2, Bourrier & Lecavelier des Etangs (2013)), and the warm Neptune GJ 436 b (host star type M3, Bourrier et al. (2016); Schreyer et al. (2024)). The atmospheres of these gaseous worlds are dominated by hydrogen, and their outflowing atmospheric material can be detected in Lyman-$\alpha$ as the planet transits the host star, a technique known as transmission spectroscopy. The atmospheric outflows are shaped by the ambient stellar wind as the planet moves through the near-star environment, producing an asymmetry in the shape of the planetary transit light curve (see also a discussion of transient variability of these outflows in Section 5.5.10). This interaction may occur at a bow shock between a pressure-driven Parker-wind-like fluid outflow from the planet and the stellar wind, or the interface may be determined by the planet's magnetosphere.

Additionally, charge exchange can occur between stellar wind ions and neutral hydrogen escaping the planetary atmosphere, producing a population of high-velocity neutral hydrogen that manifests as a fast blue-shifted component in the Lyman-$\alpha$ profile (Holmström et al., 2008). Combining these two observational signatures with models of the ambient stellar wind properties, transmission spectroscopy of these exoplanets has produced model-dependent constraints on the stellar wind environment at the orbits of these planets. Modeling the transmission spectroscopy profile in Lyman-$\alpha$ for the F8 host star HD 209458 in the context of a magnetospheric wind interface yielded local stellar wind speeds of 400 km s$^{-1}$, similar to that of the Sun's present-day slow solar wind, but with a relatively large wind density at the extreme close-in orbit of HD 209458 b (Kislyakova et al., 2014). Additionally, a recent modeling study suggests planetary magnetic field strength and orientation can also play a role in shaping outflows and could be probed along with stellar wind strength using high precision Lyman-$\alpha$ transit measurements (Presa et al., 2024).

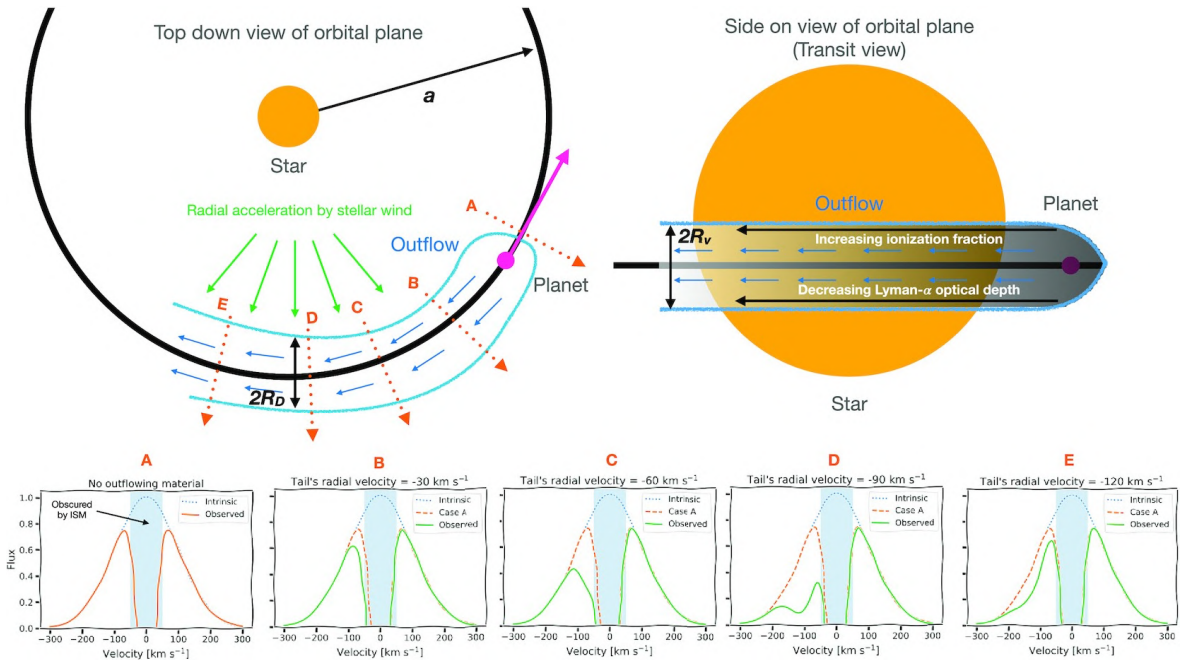

Figure 4.6: Illustration of bulk stellar wind acceleration of material outflowing from a planetary atmosphere. In models of strong pressure-driven planetary winds, the outflowing planetary gas remains collisional far from the planet. Radial acceleration of the resulting "tail" of material by the stellar wind (top left; green arrows) causes the observed line-of-sight velocity of the transiting gas to evolve, modifying the Lyman-$\alpha$ transit profile (bottom panels). Figure from (Owen et al., 2023), Figure 1; see (Schreyer et al., 2024) for a more detailed version of the model geometry.

It is, however, worth noting that the wind speeds and densities derived from modeling the outflow profiles of transiting exoplanets have substantial degeneracy, i.e., a wide range of stellar wind speeds and densities coupled with a range of planetary outflow properties could produce the neutral hydrogen column density constraints from Lyman-$\alpha$ transmission spectroscopy (Schreyer et al., 2024; Vidotto, 2021). Additionally, ISM absorption causes Lyman-$\alpha$ transits to be observed typically only in the blue-shifted wing, limiting the ability of this method to probe the origin and acceleration of the outflows (Owen et al., 2023). Some theoretical work in this area has probed the morphology of planetary outflows as a function of stellar wind properties, such as magnetic field strength, density, velocity, and temperature (Harbach et al., 2021), with reasonable agreement with observations of the outflow from the hot Neptune AU Mic b, which displays signatures of outflowing material ahead of the planet rather than in its tail (Rockcliffe et al., 2023).

Going forward, there are many opportunities for progress in using exoplanetary outflows to constrain stellar winds, some of which are already underway. The Hubble Space Telescope has just begun a treasury-scale program to observe exoplanets expected to be undergoing hydrodynamic escape. The Survey of Transiting Exoplanets in Lyman-$\alpha$ (STEL$\alpha$; Loyd et al. 2024) will broaden the parameter space of observed outflows. Increased yields and firmer constraints on transit duration would be possible with a dedicated UV mission, such as the recent Ultraviolet Spectroscopic Characterization Of Planets and their Environments (UV-SCOPE) concept (Ardila et al., 2022). An observatory with a greater effective area at Lyman-$\alpha$ such as NASA's Habitable Worlds Observatory (HWO), or an approach utilizing stacking transits, could yield breakthroughs if it were able to resolve changes in the velocity structure of the absorption as an outflow transits the stellar disk (Schreyer et al., 2024). Other wavelengths could also yield new results, most notably the metastable helium triplet near 10830 Å that can be observed with ground-based telescopes (e.g., Nail et al., 2024). It could be worthwhile to explore the utility of other UV lines and perhaps even the helium ionization continuum short of 504 Å for probing outflow-wind interactions. As a final note, the direct imaging of outflows may be worth exploring as well, as their spatial extent can be very large (e.g., see again Nail et al. 2024), and for systems within 10 pc, planets with outflows may reside outside the inner working angle of future high-contrast imaging systems (Mennesson et al., 2024). Variability in exoplanetary outflows may provide a means of detecting CMEs, a topic addressed further in Section 5.5.10.

### 4.3.5  Radio Scintillation

The idea of radio scintillation as a potential probe of the spatial extent of nearby stellar astrospheres stems from the confluence of a couple of topics. The first is the long-standing use of background quasars to probe the plasma conditions in the Solar System. Termed "interplanetary scintillation" (Hewish et al., 1964), and exactly analogous to the reason stars twinkle at visible wavelengths, density irregularities in the solar wind or in coronal mass ejections (Tokumaru et al., 2003) produce an intensity modulation from distant, compact sources of radio wavelength radiation. In these cases, the object with the density variations is known (solar wind or CME), and some properties of this plasma screen or scattering screen can be derived from the observational properties of the scintillation.

The second related topic is the discovery from several different analyses of scintillating quasars that the location of the inferred scattering screens is in the solar neighborhood, generally less than 10 pc (Bignall et al., 2006; Macquart & de Bruyn, 2007). This led Linsky et al. (2008) to speculation about three quasars exhibiting large amplitude intraday variability and annual scintillation variability due to the Earth's motion relative to turbulent plasma screens within a few parsecs of the Sun. They found these lines of sight pass through the edges of partially ionized, warm interstellar clouds, where two or more clouds may interact. The relative radial and transverse velocities of these clouds are known, and large enough to generate the turbulence responsible for the observed scintillation. In addition, the authors concluded that the flow velocities of the warm interstellar clouds were consistent with fits to the transverse flows of the radio scintillation signals. More recently, Wang et al. (2021) identified new scintillating objects, five of which exhibit a linear arrangement on the sky, demonstrating that multiple scintillators can lie behind the same phase screen.

A stellar wind dominated by CMEs could plausibly be subject to substantial density variations (See Section 3.3.2). Although the spatial scales and timescales would grow as the events traverse the distance from the star, these density perturbations might be detectable as scintillations from a field of background quasars, schematically shown in Figure 4.7. Calculations in Section A.3 determine the approximate distance from a star where the density from a CME-dominated wind would exceed the local ISM electron density, and find these minimum $d_{crit}$ values to be 2.4-6.4 AU. The $d_{crit}$ values are estimated based on measurements from a single solar CME. At present, this measurement seems restricted to only the nearest stars, as the angular diameter extent is only a few arcseconds for stars out to 5 pc distance. This makes it unlikely that sufficient background quasars would be found. While this is potentially an independent method to measure the extent of stellar astrospheres, and attractive from the standpoint of being an analog of a well-developed Solar System technique, it does not appear promising as an approach to provide considerably more information about stellar winds.

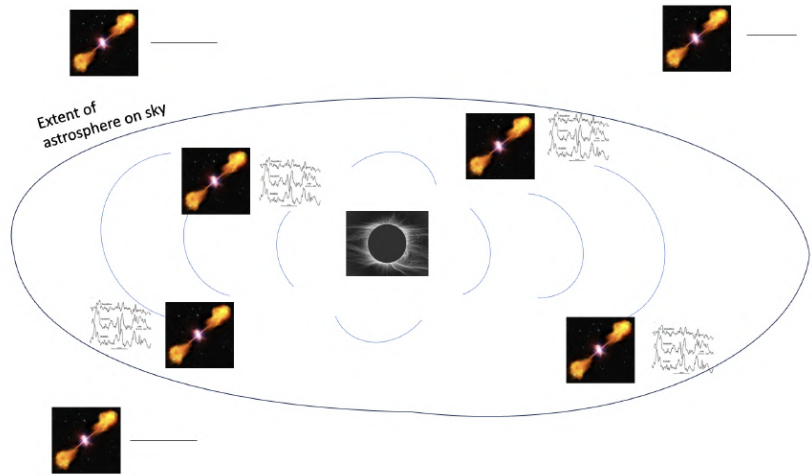

Figure 4.7: Schematic example of how the physical extent of astrospheres around nearby stars could be detected with a constellation of compact quasars in distant galaxies. Those quasars with sightlines passing through the extent of the astrosphere would experience some amount of scintillation, while those outside of the astrosphere would not. The lines next to each quasar inside the astrosphere are examples of interplanetary scintillation noted in Vitkevich & Vlasov (1970). Artist's conception of quasar radio image by ESA/C. Carreau[a], and solar corona image from NCAR HAO[b].

[a] `https://sci.esa.int/s/8ZkyzBA`

[b] `https://scied.ucar.edu/image/solar-corona-eclipse-and-uv-images`

## 4.4 Observational Methods - Long Term

### 4.4.1 Circumstellar Dust as a Wind Probe

Efforts to develop direct imaging technologies for exoplanets have recognized the role of dust in the interplanetary volume surrounding the star, so-called zodiacal dust, as an important background source when attempting to measure the reflected light of a planet (Figure 4.8a, e.g., Cash, 2006). In fact, for starshades, exozodiacal dust likely is the single limiting factor in the ability to extract planetary signals (Damiano et al., 2024). This may be an example of "one astronomer's noise is another astronomer's treasure." Strong stellar winds could be capable of clearing out exozodiacal dust, meaning that the possible imaging of the exozodiacal dust background around stars by starlight-suppressing instruments could be an opportunity to glean information on the presence and strength of a stellar wind.

Direct imaging might not even be required to glean such information. Dust in debris disks, such as the Solar System's Edgeworth-Kuiper Belt, can also produce detectable excess infrared emission in the stellar spectrum (Figure 4.8b Eiroa et al., 2013). There are, however, considerable, perhaps insurmountable, challenges to obtaining quantitative information on stellar winds based on the presence of zodiacal dust in a system. Gustafson (1994); Jewitt (2024) provide reviews of the physics of zodiacal dust (and other "debris") on which we base much of the remainder of this discussion. Another review with a focus on extrasolar observations is provided by Mann et al. (2006).

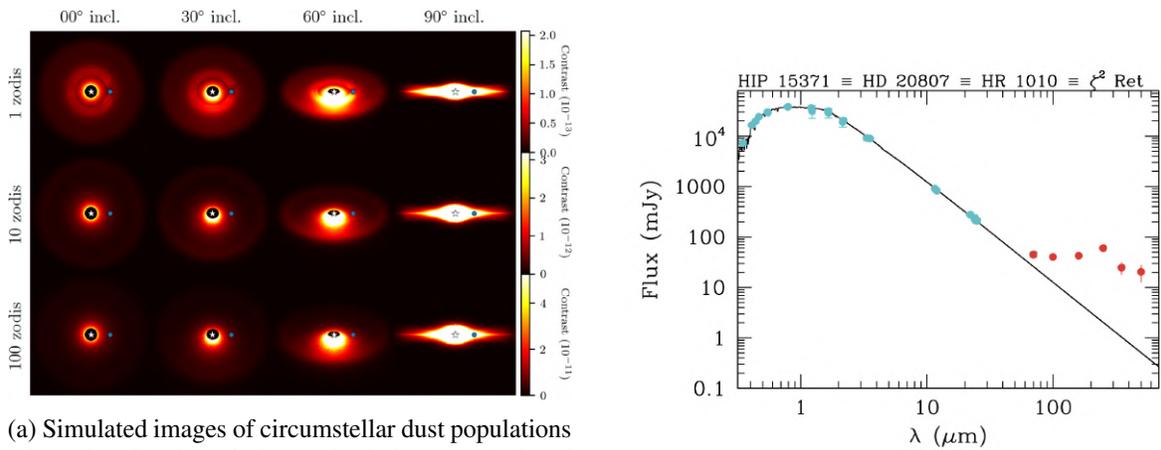

(a) Simulated images of circumstellar dust populations of varying density relative to the Solar case (units of "zodis") and varying viewing angle from Currie et al. (2023). Simulations include a range of dynamical effects on the dust population, such as collisions and Poynting–Robertson drag. The blue dot indicates the location of a planet at 1 AU. Note the varying contrast scales.

(b) Photometry of a system showing a clear infrared excess (red points) in comparison to a model of the stellar photospheric spectrum (black line fit observations shown as blue points). The infrared excess indicates the presence of a dust population. Figure from Eiroa et al. (2013).

Figure 4.8: Examples of simulated and real data capable of yielding constraints on dust in stellar systems.

A number of forces act on dust particles that can remove them from their host system (Figure 4.9a), and the balance of these forces depends both on the size of the dust grains as well as the properties of the stellar wind and radiation field. These forces are both radial and transverse (drag). In the Solar System, radial pressure on dust particles from light is roughly four orders of magnitude greater than that of protons in the solar wind. However, except for particles near micrometer sizes in which radial forces can overcome gravity to "blow out" the dust, radial pressures only act to modify the orbital equilibrium of dust grains, without actually removing them. Instead, for larger particles, it is Poynting-Robertson drag that acts as the primary dust sink. A drag force is also exerted by collisions with protons in the solar wind, but it is lower by roughly a factor of three. Hence, if observing our own system from afar, dust content would be a poor probe of the solar wind.

The same might not be true for other stars, where the balance of radiation- and wind-based drag could tip in favor of the wind. Plavchan et al. (2006) noted that excess infrared emission (at 11.7 μm) is conspicuously absent among four M dwarfs they observed. The rate of dust removal due to radiative forces should be lower in these systems due to their lower luminosities, raising the question of why they detect no dust emission. They suggest the resolution to this question is that drag from stellar winds with mass flow rates greater than the Sun's is removing dust in these systems.

The overarching challenge of this method, however, has not to do with how dust is removed from a system, but how it is added. Knowledge of the dust source rate is needed to infer what the removal rate must be from an observed steady-state population of dust (Figure 4.9b). Dust removal timescales are generally much shorter than the ages of stars, so an observable population of dust can only be sustained if resupplied by an ongoing source (Hughes et al., 2018). In the Solar System, the population of zodiacal dust is thought to be sustained primarily through collisions of minor bodies and the sublimation of comets as they approach the Sun, with some contribution from interstellar dust and the collision of that

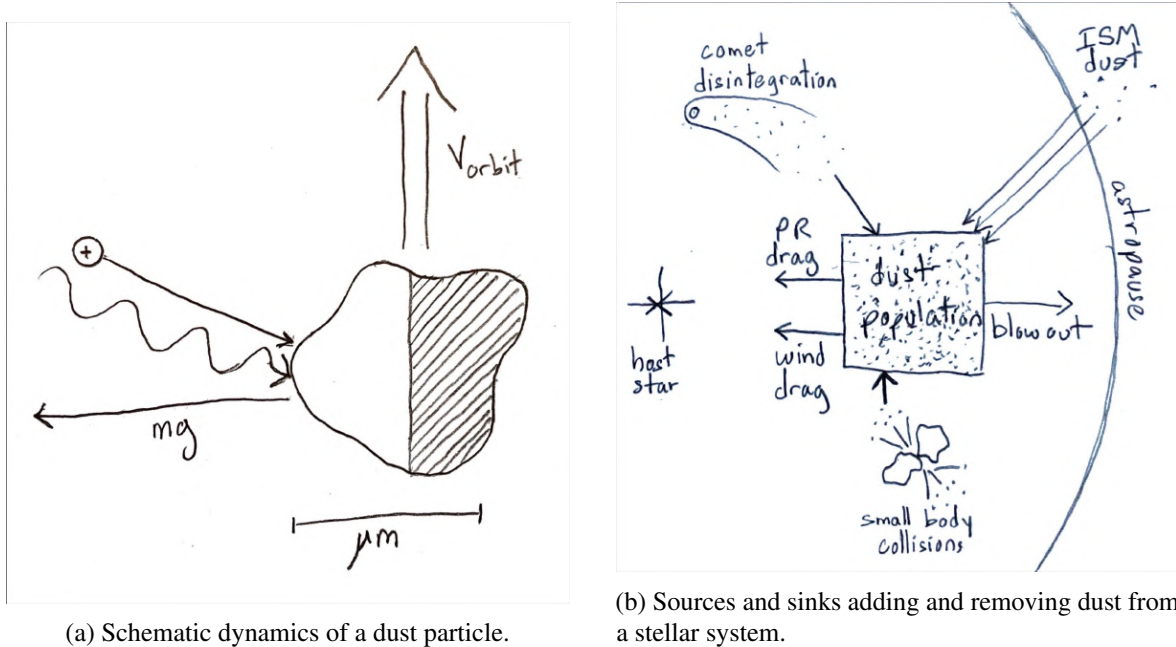

(a) Schematic dynamics of a dust particle.

(b) Sources and sinks adding and removing dust from a stellar system.

Figure 4.9: Relevant factors influencing the population of zodiacal dust in a stellar system. Schematics by R. O. Parke Loyd.

dust with Kuiper belt objects (Mann et al., 2006). However, even the Solar System production rates are highly uncertain. For other systems, these rates are likely to be dependent on the system architecture, particularly the presence of minor bodies, hence they could vary widely. Without knowledge of the rate at which dust is being added to the system, no constraint is possible on the rate of dust removal that could then be tied back to the stellar wind. This topic could be revisited in future decades when direct imaging capabilities have expanded and improved, to explore whether knowledge of system architectures, coupled with modeling validated on the Solar System, could yield estimates of dust creation rates adequate to constrain the properties of the stellar wind in systems where that wind is the dominant cause of dust removal.

### 4.4.2 Stellar Coronagraphy

For roughly half a century, the Sun's wind (and its transient structures) has been observed with human-made coronagraphs, and long before that, through eclipses of the Moon. With the glare of the Sun blocked, light scattered by the wind itself becomes visible. Coronagraphic instruments like the Large Angle and Spectrometric COronagraph (LASCO) on the Solar and Heliospheric Observatory (SOHO) spacecraft are now a cornerstone of the study of the solar wind, especially for detecting and estimating the mass and speed of CMEs (Sections 5.2 & 5.5.1). An example coronagraphic image, showing the wind along with an embedded CME, is shown in Figure 4.10.

Given the ubiquitous use of coronagraphy to observe the Sun, a natural question for the study of stellar winds is whether coronagraphs or other "starlight suppression" systems can be used. There is no immediate answer to this question, as the vastly larger distance of stars means that the scales involved differ by many orders of magnitude (Figure 4.11). A detailed comparison of the differences in the parameters and design of solar coronagraphs and "starshades" (effectively, external stellar

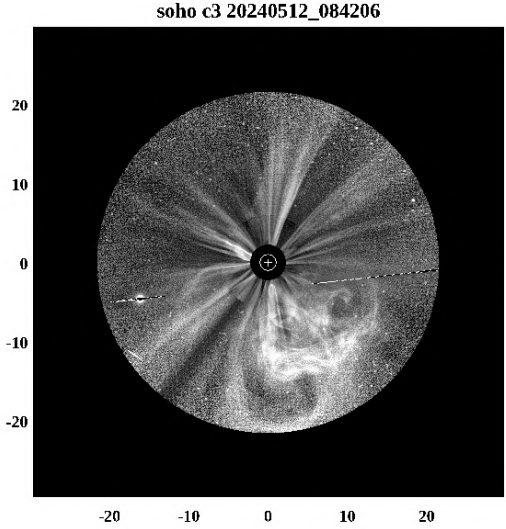
soho c3 20240512_084206

Figure 4.10: An example coronagraphic image of the solar wind, including a CME (bottom right), obtained by the LASCO instrument aboard the SOHO spacecraft. The figure combines the wide field of view of the C3 coronagraph with the narrower field of the C2 coronagraph. The (occulted) Sun is represented by the white circle, and the axes are in units of $R_\odot$. Image processed by Huw Morgan.[a]

[a] https://solarphysics.aber.ac.uk/people/huw_morgan.php

coronagraphs) is available in Aime et al. (2024). A key design parameter of starlight suppression systems, both starshades and internal coronagraphs, is the achievable contrast as a function of angular distance from the star. This is the amount of starlight present at that distance due to diffraction (and, in practice, scattering), relative to the brightness the star would have at its center in an image without suppression. For a stellar wind to be detectable, its brightness must exceed that of diffracted (or scattered) starlight within the contrast capability of the suppression system.

An estimate of stellar wind brightness due to the mechanism responsible for the light in most coronagraphic images of the Sun, Thomson scattering by electrons, is possible. Following the framework first established in Section 5.5.1, we derive a first-order estimate of this contrast for stellar winds

$$C = \frac{F_{\text{wind}}}{F_\star} = \frac{\dot{M}\lambda^2 d^2\sigma}{8\pi^2 D^2 r_\perp^3 v m_p}. \tag{4.1}$$

where $C$ is the contrast, $\dot{M}$ is the mass loss rate of the wind, $\lambda$ is the wavelength of observation, $d$ is the distance of the system, $\sigma$ is the scattering cross section, $D$ is the diameter of the telescope aperture, $r_\perp$ is the closest approach of the line of sight to the star, $v$ is the wind velocity, and $m_p$ is the proton mass. The details of this derivation are available in Appendix A.5.

The star $\varepsilon$ Eridani can serve as a case for the potential detectability of a stellar wind. It is an active star at 3.22 pc with a mass loss rate estimated at 30 $\dot{M}_\odot$ where $\dot{M}_\odot = 2 \times 10^{-14}$ $M_\odot$ yr$^{-1}$ (Wood et al., 2021). Given its close proximity and high mass loss rate, $\varepsilon$ Eridani likely represents one of the most promising targets for high-contrast imaging of a stellar corona. Figure 4.12 plots the contrast curve for $\varepsilon$ Eridani's wind, assuming a wind speed of 400 km s$^{-1}$ (following Wood et al. 2021), a wavelength of 500 nm, and a 6 m telescope aperture such as is currently planned for HWO. For comparison, we plot the approximate bounds from laboratory measurements of a number of internal starlight suppression systems under consideration for HWO, as well as an external occulter ("starshade") (Mennesson et al., 2024). The result indicates it is unlikely that internal suppression systems will be able to image a stellar wind in the foreseeable future. A starshade, in contrast, could succeed, although the formation flying such a system requires would likely limit the number of targets it could visit.

Future consideration of this possibility should address signal-to-noise concerns and include an estimate of where in such a system Thomson scattering from the wind will begin to exceed scattering from dust ("exozodiacal light"), though strong winds might remove this dust (Section 4.4.1). It might be worthwhile to explore scattering by ions in the wind, which could provide larger scattering cross sections at the cost of fewer photons, lower abundances, and lower instrument effective areas (for likely operation at X-ray wavelengths). Systems that combine interferometry with starlight suppression might also be a fruitful area of investigation. A simple next step could be to incorporate a treatment of winds along with zodiacal dust in an existing starlight suppression simulation package such as Toy Coronagraph Lin (2024).

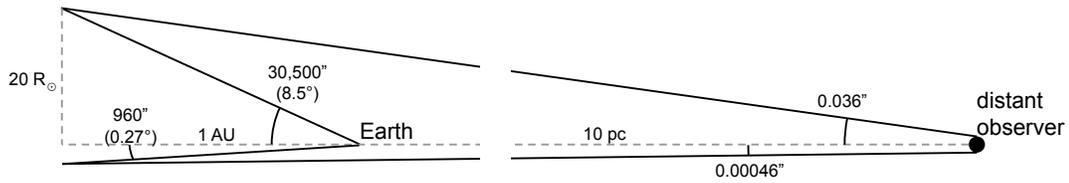

Figure 4.11: Comparison of the wildly different scales of a coronagraphic image of the solar wind and a hypothetical coronagraphic image of the same extent of a stellar wind. Figure by R. O. Parke Loyd, inspired by Figure 1 of Aime et al. (2024).

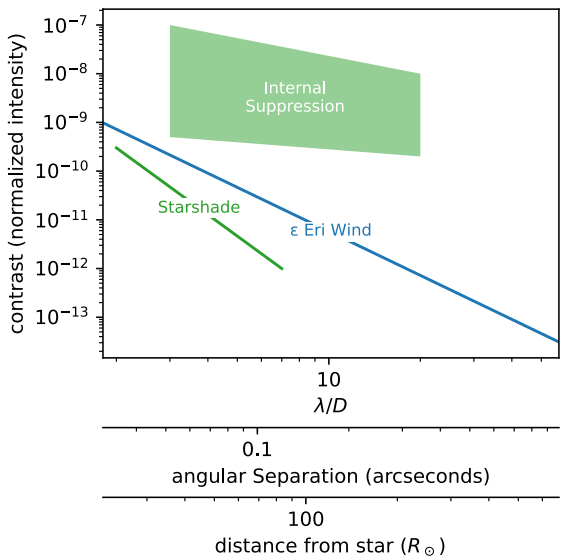

Figure 4.12: Estimate of the contrast of the stellar wind of $\varepsilon$ Eridani relative to the intensity of the star at the center of the Airy disk when imaged by a 6 m telescope at optical wavelengths (blue lines), compared to the range observable with various internal suppression setups (green shaded region) and a starshade (green line) based on laboratory prototype tests (Mennesson et al., 2024). The contrast is plotted as a function of angular separation in units of $\lambda/D$ for easy comparison to other contrast curves from starlight suppression systems, as well as in physical units of angular separation and distance from the star. This illustrates that a starshade might enable detection of the wind of $\varepsilon$ Eridani if the mass loss rate inferred from observations of Ly$\alpha$ astrospheric absorption is accurate. Figure by R. O. Parke Loyd.

As transient enhancements of the wind density, CMEs may prove more detectable than stellar winds through coronography (Section 5.5.1).

### 4.4.3 Radio Free-Free Emission

An ionized stellar wind naturally produces continuum bremsstrahlung (free-free) radiation across the electromagnetic spectrum, with its innermost, densest region emitting free-free radiation at radio frequencies (Güdel, 2002). The idea of detecting stellar winds in radio originated from studies of winds of early-type (hot) stars (Panagia & Felli, 1975; Wright & Barlow, 1975; Olnon, 1975). These seminal studies showed that in the case of an optically thick wind ($\tau_\nu \gg 1$ at low frequencies $\nu$), a stellar wind expanding at constant speed has a flux density that depends on the frequency as $S_\nu \propto \nu^{0.6}$. In the optically thin regime ($\tau_\nu \ll 1$ at high $\nu$), this dependence becomes $S_\nu \propto \nu^{-0.1}$. For winds of low-mass stars, the assumption of constant speed in the innermost region of the stellar wind is not valid, and thus the spectral indices change, with simulations showing indices ranging from 1.2 to 1.6 in the low-frequency, optically thick regime (Ó Fionnagáin et al., 2019). At 1 GHz, Ó Fionnagáin et al. (2019) estimate radio flux densities for solar analogs of

$$S_\nu \simeq 0.24 \ \mu\text{Jy} \left[ \frac{\Omega}{\Omega_\odot} \right]^{0.39} \left[ \frac{10 \ \text{pc}}{d} \right]^2, \tag{4.2}$$

here $\Omega$ is the angular rotation rate of the star and $d$ is the distance. Based on the spin-down rate of solar analogs, this equates to

$$S_\nu \simeq 0.24 \ \mu\text{Jy} \left[ \frac{4.6 \ \text{Gyr}}{t} \right]^{0.39} \left[ \frac{10 \ \text{pc}}{d} \right]^2, \tag{4.3}$$

where $t$ is stellar age. This implies that if we were to place the Sun at 10 pc, the radio flux of its wind at 1 GHz would be less than 1 $\mu$Jy. Younger Suns, which are expected to have denser winds, would have emissions with higher flux densities. However, the distance-squared decay still plays a dominant role in the detectability of winds using radio observations. [2]

Due mostly to the limiting sensitivity of current radio telescopes, radio free-free emission has not been a viable technique to date to detect and fully characterize low-mass stellar winds. However, one important advantage of this method is that even in the case of non-detections, the method can still provide upper limits on stellar mass loss (subject to wind model assumptions), which are meaningful and direct constraints of winds of low-mass stars. From radio observations of solar analogs, Fichtinger et al. (2017) were able to place upper limits on the mass-loss rate of the stars in their sample. The limits they derived are nonetheless several orders of magnitude higher than constraints based on rotational evolution models (see Figure 4.13).

Unfortunately, the expected weak emission of a stellar wind competes against several emission mechanisms potentially in existence in cool stellar microwave emission (e.g., gyrosynchrotron, gyroresonance, bremsstrahlung from the chromosphere), even outside of flaring events. Villadsen et al. (2014), for example, were able to detect thermal radio emission from three solar-like stars at 34.5 GHz—however, they explain this emission as optically thick, chromospheric thermal free-free emission. The distinction between a weak wind signature and the thermal component from close to the stellar surface requires extremely accurate radio spectra (Drake et al., 1993; Villadsen et al., 2014).

We note that Section 4.3.5 details a seemingly related topic: the possibility of observing radio scintillation in the light of background quasars from density inhomogeneities in the astrosphere encountered as that light traverses the astrosphere of the star on its way to be detected at Earth by a radio telescope. That is a propagation effect of external radio waves through the plasma in the astrosphere. The topic of discussion in this section is the innate thermal emission by the stellar wind itself.

---

[2]The optically thick part of the wind creates a "radio photosphere" that surrounds the star in the form of a paraboloidal shape (Kavanagh et al., 2019). A source of radio emission, such as a close-in exoplanet (Farrell et al., 1999), would have its emission attenuated or fully absorbed when it enters the radio photosphere of its host star. Although this might be problematic for planet detection in radio, close-in exoplanets can be useful to probe the inner regions of stellar winds (Vidotto & Donati, 2017; Kavanagh & Vidotto, 2020).

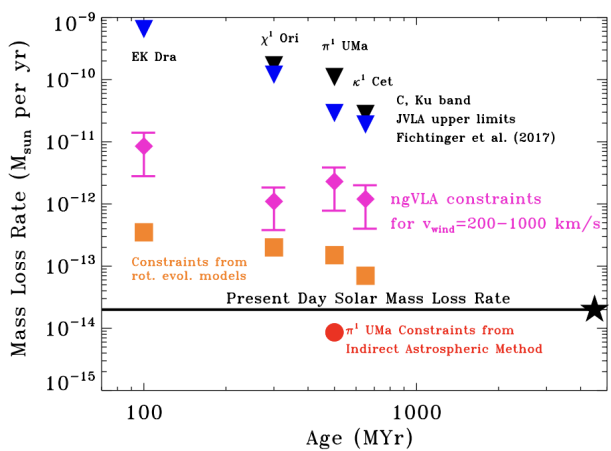

Figure 4.13:  Figure 1 of Osten & Crosley (2017), showing observational constraints on winds from solar-analog stars from recent Karl G. Jansky Very Large Array (JVLA) observations. Expected mass-loss rates due to rotational evolution models for these stars are indicated in orange, while the present-day solar mass-loss rate is marked as a horizontal line. Expectations from optically thick bremsstrahlung emission for 12-hour observations with the Next Generation Very Large Array (ngVLA), across a range of assumed wind speeds, are shown in magenta. The mass-loss rate inferred from the indirect Ly$\alpha$ astrosphere method (see Section 4.3.1) is significantly lower than any of these.

With the advances in radio technologies and the construction of more sensitive radio telescopes, detecting radio emission from stellar winds might be possible. It would require higher-frequency observations ($v > 10$ GHz) and significant amounts of integration time per star. Osten & Crosley (2017) provided some sample scenarios for solar-analog stars as well as nearby M dwarfs, under the more optimistic assumption that the wind emission is optically thick, and for ~12 hours on source for each target.

### 4.4.4  Gazing Far Ahead: In-Situ Measurements of Exoplanet Space Weather

While current strategies rely entirely on remote sensing and modeling, it is worth considering the potential for future in-situ measurements of exoplanetary space weather environments. This capability is not feasible today, but exploring it as a long-term goal can guide technology development and inform mission architecture. In-situ data could offer direct measurements of stellar wind properties, particle fluxes, and magnetic interactions, particularly valuable for stars that are not of solar type or solar age, providing critical benchmarks for validating remote observations and improving model fidelity.

Although speculative, such measurements would complement the approaches outlined in this chapter by resolving key ambiguities and testing the limits of our current indirect techniques.

By obtaining direct data from these environments, we can gain a more accurate and comprehensive understanding of space weather phenomena and their implications for exoplanet atmospheres and surface life beyond Earth. Of course, this will not be easy, and at this time is not technologically feasible. Such an effort to send probes to other stellar systems will require intense investment in new technological capabilities and enormous patience, and of course, a nearby target.

Proxima Centauri (Proxima Cen), also known as Alpha Centauri C, is the third star, by mass, in the Alpha Centauri system and is the closest stellar neighbor to the Sun at "just" 4.2 light-years (1.3 parsecs). As such, it offers the best opportunities for in-situ exploration. Additionally, Proxima Cen hosts a known terrestrial planet within its habitable zone, Proxima Centauri b (Proxima Cen b) (Anglada-Escudé et al., 2016). Proxima Cen is an M dwarf star and is known to have frequent flares, which likely indicate a powerful particle environment that will influence the habitability of Proxima Cen b.

On the other hand, Alpha Cen A is a Sun-like star (G type), potentially offering a more familiar environment for life, but it does not yet have any confirmed planets. The relative stability of the stellar environment of Alpha Cen A and its wider habitable zone, which is more accessible to direct imaging or astrometric searches for planets, makes it a valuable target for several existing mission concepts and from the ground using extremely large telescopes (Wagner et al., 2021).

Alpha Cen B is a K-type star, a spectral type that has been heralded as possibly creating a "super-habitable"[3] environment for its planets, compared to G- and M-type stars (Heller & Armstrong, 2014). No planet has yet been discovered around this star either. Statistically, though, there is probably at least one rocky planet orbiting Alpha Cen A or B in the habitable zone and searches are ongoing (Belikov et al., 2019).

These three stars of the Alpha Cen system provide contrasting environments (G, K, and M), all of which are critical for understanding the diversity of planetary systems and their potential for hosting life, making the system a rich place for exoplanet discovery and characterization. Their proximity makes them the most plausible targets for a first attempt at in-situ exospace weather measurements.

However, the distances between stars are vast by any human scale. The Voyager 1 and 2 missions, while groundbreaking in that they are the first human-made spacecraft to leave the Solar System, are traveling at speeds of 17 km/s, only 0.006% the speed of light. They will pass other stars eventually, in about 40,000 years, with a closest approach of about 1.7 light-years. Voyager's observations revealed the structure and dynamics of the outer heliosphere but were limited by the technology of the 1970s. A revisit of this region with modern instrumentation is essential, not only to deepen our understanding of heliospheric boundaries, but also to lay the foundation for future exospace weather missions (Stone et al., 2015).

Project Longshot was a theoretical proposal developed in the late 1980s by NASA and the US Naval Academy. This ambitious concept aimed to send a robotic probe to the Alpha Centauri system using nuclear fusion propulsion. The journey would take about 100 years. Although no active development is currently underway, Project Longshot laid important groundwork for interstellar travel concepts and highlighted the significant technological and financial challenges involved (Matloff et al., 1988; Gilster, 2004).

Longshot itself drew inspiration from the earlier British Interplanetary Society's Project Daedalus, a landmark 1970s study that first outlined the feasibility of reaching another star system using fusion propulsion within a human lifetime. Designed as a two-stage, helium-3–driven probe targeting Barnard's Star, Daedalus helped define the conceptual framework for interstellar mission design (Bond & Martin, 1978).

---

[3]Super-habitable planets are hypothesized to be even more conducive to life than Earth, potentially due to factors like longer-lived stellar hosts, thicker atmospheres, more stable climates, or higher surface water fractions.

More recently, the Breakthrough Starshot project is being developed privately by the Breakthrough Initiative. It represents a pioneering effort to send laser-accelerated tiny cameras to the Proxima Cen exoplanetary system, aiming for speeds of 25% the speed of light. The journey would take approximately 20 years, plus the 4.2 light-years to return images.

Although the Breakthrough Starshot program has many technical challenges to overcome, one key element that must be tracked is the harsh radiation that the tiny cameras will encounter on their way, including the radiation environment within the exoplanet system. This radiation will produce noise that limits the ability of the cameras to return pristine images. However, the radiation detected by these cameras would provide valuable data on the space weather in the system, including the environment around the habitable zone planet, Proxima Cen b. Coordinating measurement requirements to ensure useful scientific data can be extracted from the radiation impacting the cameras is crucial. This innovative approach highlights the potential for new technologies to expand our understanding of space weather in distant stellar systems (Lubin, 2016; Loeb & Lingam, 2018).

## 4.5  Paths Forward in Quasi-Steady Exospace Weather

In this chapter, we have discussed a variety of observational methods for probing the stellar wind environments of other stars. A number of these methods hold promise for future development, and within this arena are opportunities for synergies and links between methods and with other topics of study. They are each subject to important caveats about the errors and degeneracies inherent to interpreting these observational constraints.

Model dependency is one such caveat. Currently, the most successful techniques, based on Lyman-$\alpha$ astrospheric absorption (Section 4.3.1), X-ray emission resulting from charge exchange (Section 4.3.3), and planetary outflow transits (Section 4.3.4), are all based on interactions of the stellar wind with other material, and each requires models and particular assumptions to interpret the observations and arrive at inferred properties of stellar winds.

To test the consistency of model-based inferences, a particularly valuable avenue of future work would be to leverage more than one observational technique for a set of common target stars. (Kislyakova et al., 2024) enabled this for three targets: 61 Cyg, 70 Oph, and $\varepsilon$ Eri, though the constraints compared in that work each rely on the same model for interpretation and applied to differing observables, both resulting from the wind-ISM interaction. Comparisons based on differing interactions could be even more valuable, such as wind-planet and wind-ISM interactions. Perhaps most valuable would be the comparison of model-based inferences with future measurements with less model dependence, such as free-free radio emission or coronagraphic imagery. Where possible, focusing new efforts on bright and/or nearby stars like $\varepsilon$ Eri and $\alpha$ Cen with existing constraints provides added value through technique comparisons.

A number of techniques identified in this chapter approach the limit of what is possible with current instrumentation, and only await observatories with next-generation sensitivity or other advanced capabilities to open or widen their detection space. These include emission from wind-ISM charge exchange from metal ions at X-ray wavelengths or H$\alpha$ at optical wavelengths, radio free-free emission, and wind coronagraphy. Consequently, these techniques have a high probability of yielding future results.

In many cases, measurements may come "for free" in the form of "one astronomer's noise is another astronomer's signal." Coronagraphs intended for imaging planets may image stellar winds as well as a background source. Lyman-$\alpha$ observations of planetary transits could capture astrospheric absorption. Far in the future, images from spacecraft sent to another system may contain information on EPs as a source of image noise. Nonetheless, additional work to more explicitly define the instrument requirements that best enable the use of the methods we have explored, or others, would be valuable. If communicated to the teams conceptualizing and designing future observatories, this work could encourage simple choices that ensure observations of quasi-steady stellar space weather are possible.

Among next-generation observatories, NASA's Habitable Worlds Observatory holds promise in multiple areas. Its enhanced UV sensitivity will be helpful for cases involving the measurement of Lyman-$\alpha$ lines, potentially including astrospheric absorption and especially planetary outflow-wind interactions. Meanwhile, if paired with a starshade, it may enable detections of scattered light from stellar winds.

It is also important to note that the quasi-steady behavior of a stellar system does not act in isolation. Transients will regularly perturb this system, potentially influencing observations intended to be confined to quasi-steady phenomena. For example, a planetary outflow observed under the influence of a CME will likely yield a much different inferred wind pressure (Section 5.5.10). One possibly underexplored area is the amplitudes and timescales of variability in stellar winds and how this could potentially bias measurements. The Sun's mass-loss rate varies little across activity cycles, but the same might not be true for other stars. Meanwhile, the quasi-steady background influences the propagation of transients like CMEs and EPs, and could also influence observations intended to be made of these events. This overlap also offers opportunities for added value. For example, coronagraphic imagery could yield constraints on both winds and CMEs (Sections 4.4.2 & 5.5.1).

---

**Zooming Out: Related Findings and Recommendations (FRs)**

Our observational understanding of quasi-steady space weather could be rapidly accelerated by purpose-built observatories (FR3). As the field grows, efforts to validate new and existing techniques through strategies such as inter-comparisons, Sun-as-a-star analysis, and deep monitoring are critical to establishing confidence and quantifying the accuracy of new and existing techniques (FR4), while instrument development is essential to reveal previously inaccessible signals and to expand parameter space (FR6). These efforts are greatly assisted by promoting cross-disciplinary collaboration, particularly in the case of exospace weather observations, between solar and stellar communities (FR7).

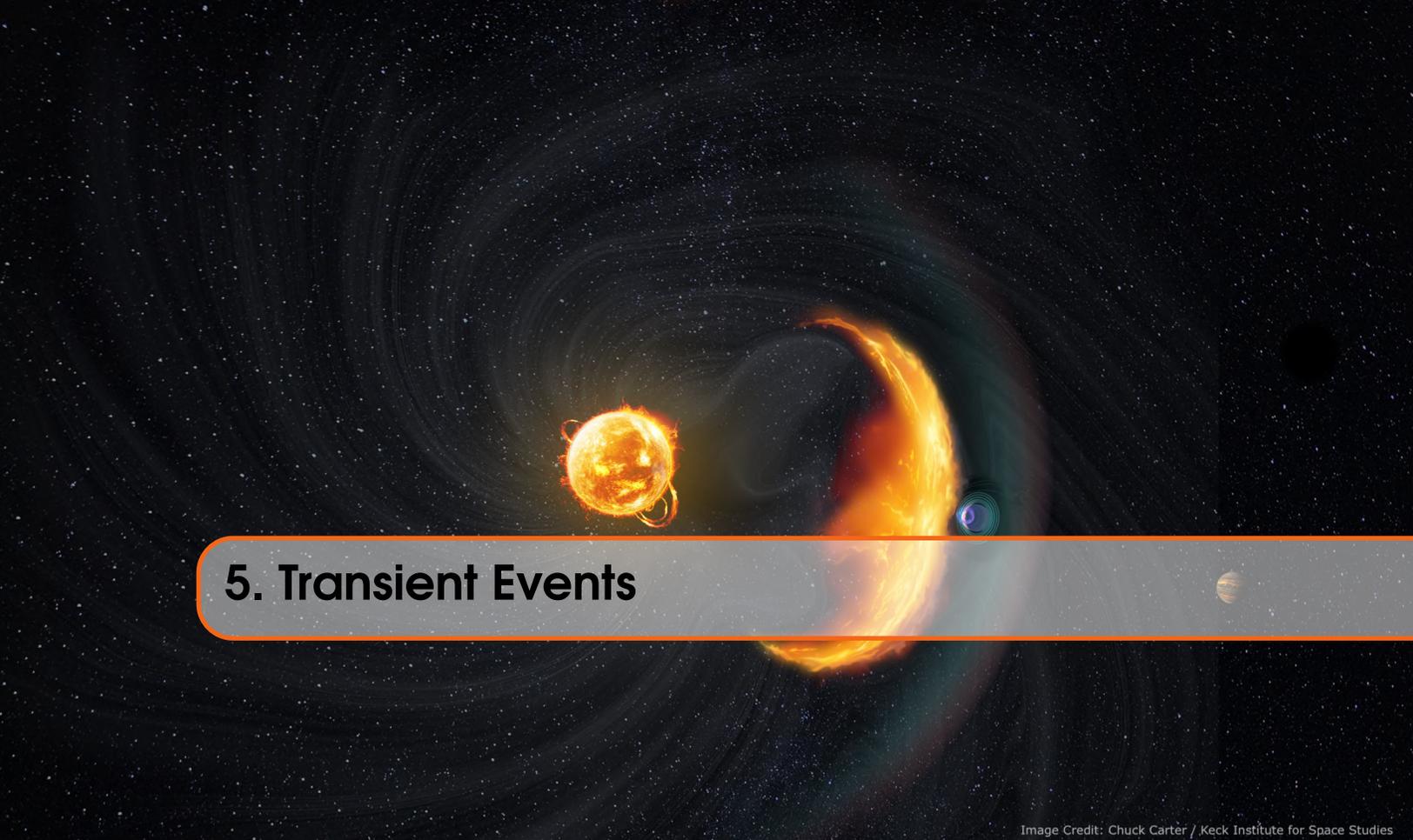

## 5.1 Overview

The Sun regularly produces transient events on timescales of seconds to hours. These are exemplified by jets, prominence eruptions, solar flares, and coronal mass ejections (CMEs). The latter two can then generate solar energetic particles (SEPs). One of the foci of this KISS study was to identify promising avenues of research in the transient particle behavior of main-sequence FGKM stars in the nearby solar neighborhood, with particular emphasis on the detection of CMEs and SEPs from exoplanet host stars and their impacts on planet(s) (see Chapter 2).

Studying transient events from the Sun offers the advantage of both in-situ measurements of their properties and spatially resolved remote-sensing observations of the magnetic structures that initiate them. With these tools, heliophysics continue to develop forecasting tools for CME and SEP events from the electromagnetic signatures that both precede and accompany these events. Unfortunately, extrasolar observations of space weather are not afforded the advantages of in-situ measurements or high levels of spatial detail in remote sensing. As such, extrasolar observations require unresolved or coarsely resolved electromagnetic signatures to serve as proxies of CME or SEP occurrence, the most successful of which have been Hα spectral line asymmetries and X-ray brightness dimming. Moschou et al. (2019a) provides a summary of candidate stellar CME events observed in Hα and X-ray and their use in developing a stellar flare-CME relation. Leitzinger & Odert (2022) review these methods and other observing techniques that have been attempted to identify stellar CMEs.

Without direct measurements or observations of particle flux, it has been difficult—if not impossible—to unambiguously conclude whether a CME rather than, for example, an extraordinary flare has occurred on another star. This difficulty is exacerbated by our limited understanding of stellar magnetic field strengths, field topologies, stellar wind properties, and how any of these parameters might impact CME and SEP generation and propagation. Similarly, it is difficult to constrain how we might distinguish between properties of the ambient wind and a transient event (see Chapter 3 for more on efforts to model space weather generated by stars and Chapter 4 for more on observing stellar winds). For instance, the strong magnetic fields of late-type M dwarfs that are popular CME-producing candidates may suppress both the escape of CMEs (Alvarado-Gómez et al., 2018) as well as the development of shocks at coronal heights that would otherwise be conducive to ground-based radio detections (Villadsen & Hallinan, 2019). Additionally, the estimated low-helicity density of these stars may not be conducive to the generation of CMEs in the first place (Jardine, 2024). The limited number of stellar candidate CME detections has inhibited the investigation of how CME occurrences and properties might vary as a function of stellar parameters, such as spectral type, age, rotation period, and metallicity. By extension, it has inhibited our ability to understand the Sun's production of transient events throughout its life and their impact on solar and planetary evolution.

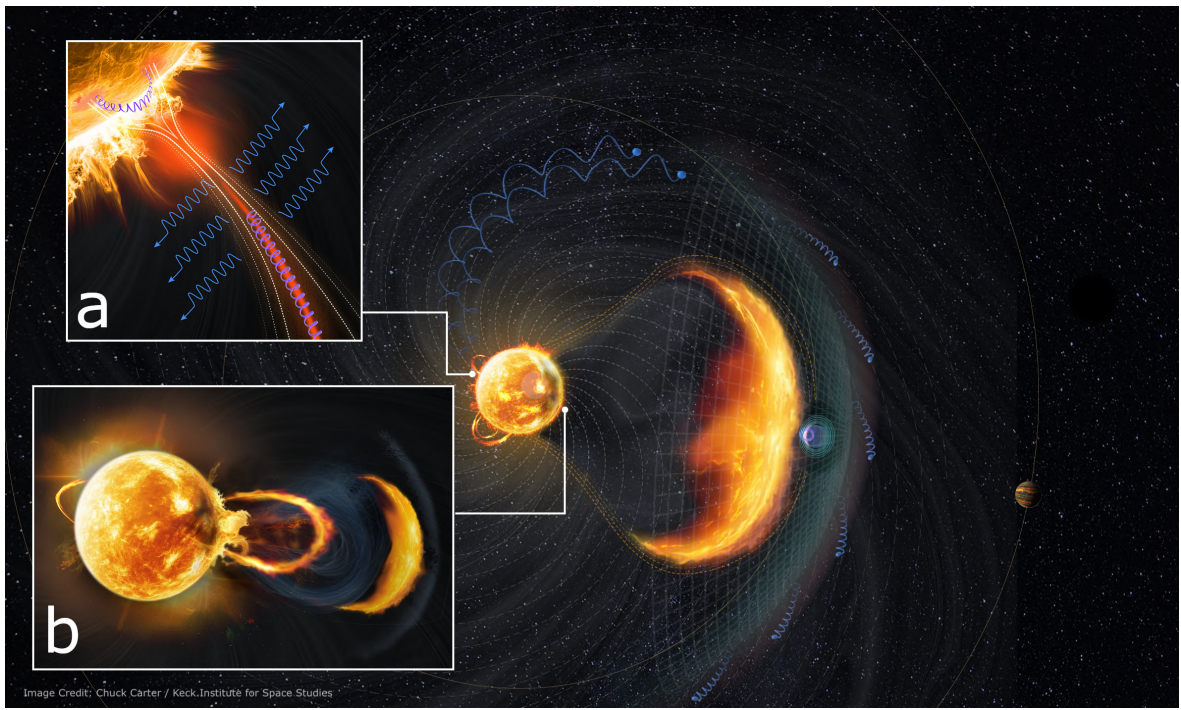

Figure 5.1: Artist's illustration of a CME and EP release in a stellar system. EP release from a reconnection site is shown schematically in inset (a). Particles trapped on closed loops near the reconnection site and heating at loop footpoints will produce emission at a variety of wavelengths (Section 5.6). Outward-beamed particles can produce type III radio bursts (Section 5.5.7). These particles may escape into interplanetary space, spiraling along the magnetic field of the stellar wind (main figure). The early evolution of a CME is shown schematically in inset (b). In the main figure, a lenticular shock ahead of this CME produces a broad swath of EPs spiraling outward along the magnetic field of the stellar wind, and the planet behind the shock experiences a potentially observable aurora (Section 5.5.8) as a result of its passage. When lower in the corona, this shock will produce radio emission as well (Section 5.5.6). An observer viewing the CME face-on (i.e., from the right side of the figure) would be optimally positioned to detect Doppler-shifted absorption or emission from the erupting filament (loop in (b), Section 5.5.2), mass-loss dimming (darkened area of the corona in (b), Section 5.5.3), or dimming due to obscuration by the filament or CME material (Section 5.5.4). A side-on observer, viewing from, e.g., above the figure, would be optimally positioned to detect the CME in coronagraphy. Figure credit: Keck Institute for Space Studies / Chuck Carter. Full resolution figure available at `https://www.kiss.caltech.edu/artwork.html` and may be reused or modified on the condition of appropriate image credit.

In addition to the events mentioned thus far, the Sun also produces temporary structures in its wind arising from the interaction of fast streams with slower wind in their path. The resulting pileups are termed stream or co-rotating interaction regions (SIRs and CIRs, Richardson, 2018). Although technically transient, CIRs can persist for several months and interact with planets at predictable intervals. While CIRs and SIRs are a component of any complete description of particle space weather, they were beyond the scope of the workshop effort.

In this chapter, we consolidate and re-evaluate previous CME detection methods, investigate the plausibility of as-of-yet unemployed (or unsuccessful) solar proxies for stellar CMEs and SEPs, and consider novel approaches that are either not utilized or not applicable in the Solar System. We provide brief descriptions of CMEs and SEPs in Sections 5.2 and 5.3, respectively, before going into the details of various detection methods in Section 5.5.

## 5.2 CME Description

CMEs are magnetic structures entraining coronal plasma that is abruptly launched from the surface of the Sun into interplanetary space. Beginning with typical scales of $\sim 10^4$ km near the Sun, they can expand to scales approaching an astronomical unit (AU) as they propagate beyond the planets, making them the largest eruptive phenomena in the Solar System (Webb & Howard, 2012; Lugaz et al., 2024).

The discovery of solar CMEs has its origins in geomagnetic storms (Gopalswamy, 2016). In the mid-19th century, it became clear these storms were linked to sunspots and their corresponding 11-year cycle (Sabine, 1852), a link that is now subject to no doubt whatsoever. In 1859, R. C. Carrington and R. Hodgson separately observed a sudden white-light flare on the solar surface within a complex sunspot group, now referred to as the "Carrington Event," that preceded an extreme geomagnetic storm (Carrington, 1859; Hodgson, 1859). It took more than a century to discover that this flare, and ones like it, were only one of the manifestations of magnetically dynamic events at the Sun that lead to geomagnetic storms.

As early as the 1940s, ground-based detectors began detecting gigaelectron volt (GeV) particles and, later, secondary neutrons thought to be associated with the Sun and the sudden brightening of solar flares (Forbush, 1946; Meyer et al., 1956). With the onset of increased satellite measurements, these events were further associated with the detection of large-scale interplanetary shocks. This ensemble of measurements underscored the undeniable role of the expansion of solar magnetic fields into space, in particular those fields associated with dynamic events such as prominence eruptions and flares.

In the early 1970s, the advent of space-based coronagraphs revealed sporadic ejecta of about $10^{16}$ g of material from the Sun—CMEs. Yet, a one-to-one correspondence between CMEs and flares was never unequivocally established. Gosling (1993) dismissed flares as playing a central role in initiating geomagnetic events, instead identifying CMEs as the likely culprit. Remote sensing observations, in particular from the Atmospheric Imaging Assembly (AIA) instrument on the Solar Dynamic Observatory, have now established a firm relationship between prominence eruptions, CMEs, and flares, with magnetic reconnection initiating and often, though not always, linking these dynamic events. Further support for this conclusion comes from multiwavelength observations of total solar eclipses (Habbal et al., 2014, 2021) that show solar prominences invariably positioned at the base of large-scale loops of active regions and the bulges of streamers (Figure 5.2). Figure 5.3 conceptually associates prominence eruptions, CMEs, and geomagnetic storms with a series of images.

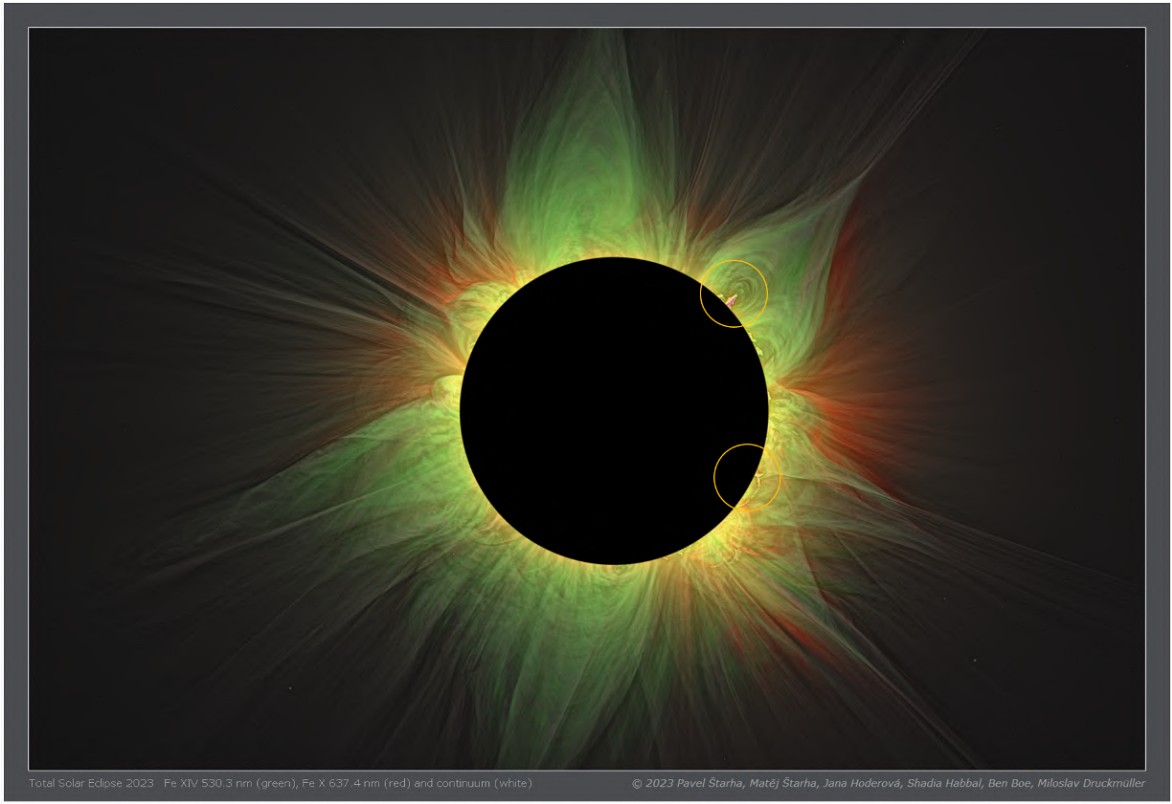

Figure 5.2: Composite processed image of the solar corona taken during the 2023 total solar eclipse in white light, Fe XIV 530.3 nm (green) at 1.8 MK, and Fe X 637.4 nm (red) at 1.0 MK. Prominences (reddish complex structures) are invariably found to be enshrouded by the hottest material in the corona at the base of streamers, as indicated by the yellow circles. Credit: Shadia Habbal, see Habbal et al. (2014).

The structure of a solar CME is, canonically, comprised of three parts: 1) a loop-shaped "plasma pileup" at its front surrounding 2) a volume of low density plasma and 3) a core of dense, cooler plasma (Figure 5.1, inset (b) Illing & Hundhausen, 1985). In common, visible-light coronagraphic images of the Sun, Thomson scattering by free electrons in the CME plasma exposes its projected structure, with dense regions appearing bright and low-density regions appearing dark (Figure 5.3; CME seen in the upper right of panel B). In reality, only around one-third of solar CMEs exhibit the three-part morphology. Others span narrow, jet-like shapes with opening angles as small as 2° to broad events with opening angles >120° that can appear global when viewed face-on, with narrower events occurring more frequently (Yashiro et al., 2004). As CMEs expand, their shape does not often change greatly, leading to the common assumption of "self-similar expansion" (Cremades & Bothmer, 2004; Kumar et al., 2024).

Typical masses for a solar CME is $10^{14}$–$10^{16}$ g, though they span masses as low as $10^{12}$ g to above $> 10^{17}$ g (Gopalswamy, 2022). Their occurrence rate is inversely related to their mass— lower mass events are more common than higher mass ones (Figure 5.4)—and varies from $\sim 0.5$ to $\sim 6 \, \mathrm{d}^{-1}$ (excluding undetected events) over the solar activity cycle, with a phase delay of 6-12 months (Figure 3.1; Gopalswamy et al., 2003; Yashiro et al., 2004; Raychaudhuri, 2005; Robbrecht et al., 2009).

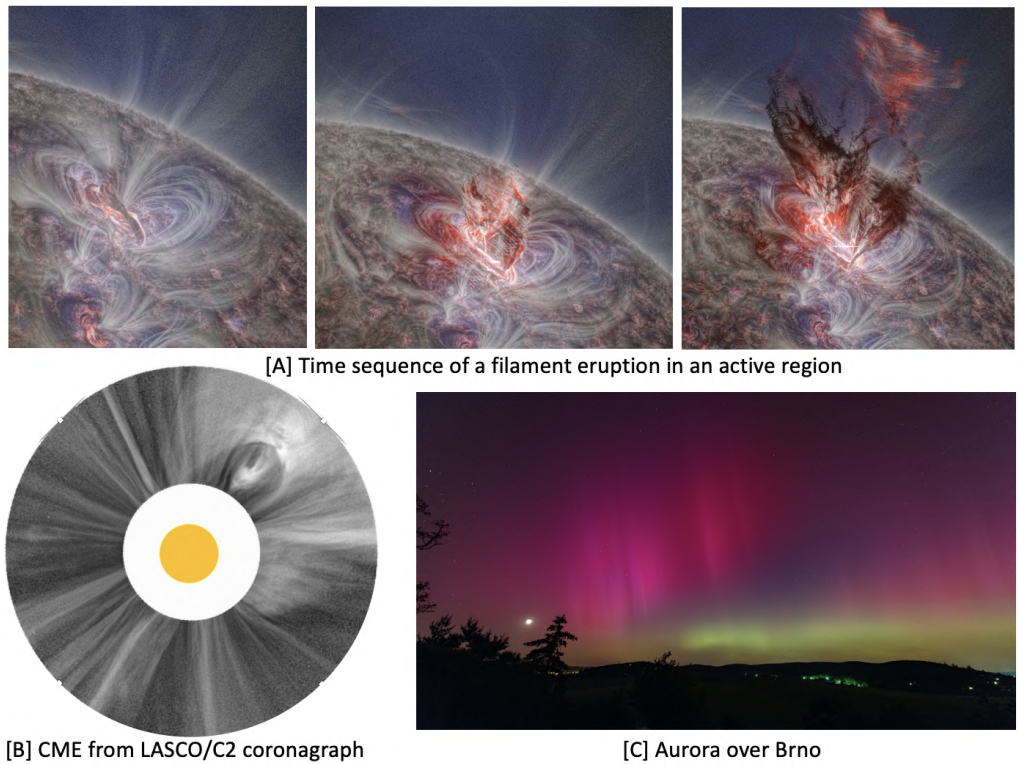

Figure 5.3: Example of a sequence of transient events leading to an aurora: (A) Prominence eruption at the Sun observed in the EUV from SDO/AIA (compiled by M. Druckmüller)]. (B) CME captured in the corona by the LASCO/C2 coronagraphs. (C) Appearance of an aurora (photo courtesy M. Druckmüller). Credit: Shadia Habbal.

The mass of a CME can increase as it propagates outward, both from the collection of mass it plows into ahead or from mass streamed into it from behind (DeForest et al., 2013; Veronig et al., 2019).

The velocities of solar CMEs, or, more precisely, the velocities of the fronts seen in coronagraphic images, range from 20 to > 2500 km s$^{-1}$. Most of their acceleration occurs in the low corona, below 2 $R_\odot$. Beyond this point, their speed varies little, though fast CMEs tend to decelerate and slow ones tend to accelerate (Webb & Howard, 2012). As they travel outward, they can experience deflections as large as ~40°, generally due to interactions with regions of higher magnetic field (Vial & Engvold, 2015; Kumar et al., 2024). Stellar CMEs could exhibit larger deflections and substantial changes in speed due to interactions with strong ambient fields (Kay et al., 2016; Alvarado-Gómez et al., 2018). The speeds and masses of solar CMEs correspond to kinetic energies spanning 10$^{29}$ to 10$^{32}$ erg, again with weaker (lower energy) events occurring more frequently (Figure 5.4). These energies generally far exceed those of associated flares (Emslie et al., 2004).

In contrast with CMEs, solar prominences are cool, dense ropes of material spanning ~ 10$^4$ km. They are suspended in the corona by a complex magnetic structure that provides gravitational support and thermal isolation. Prominence eruptions are often associated with CMEs, though the two are physically distinct (Parenti, 2014; Vial & Engvold, 2015; Gibson, 2018).

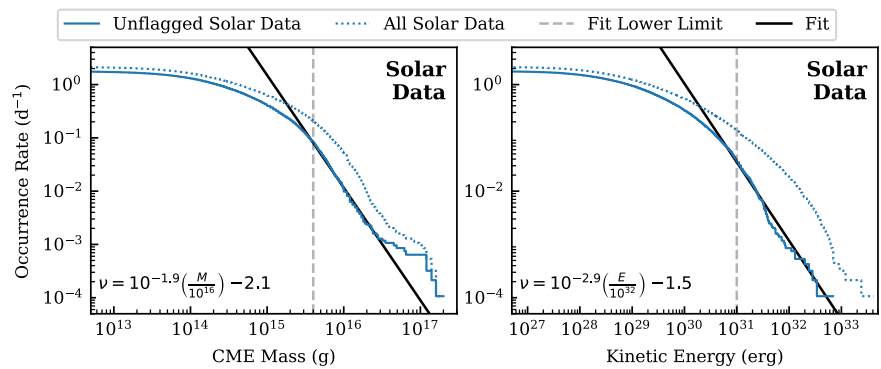

Figure 5.4: The occurrence rate of CMEs with measured mass and energy, based on events in the Coordinated Data Analysis Workshop (CDAW) catalog. The plots are cumulative, meaning they give the rate of events with mass or energy exceeding the value on the x-axis. The dotted lines include events with quality flags, and an additional 41% of events had no mass or energy in the catalog at the time this plot was created and are therefore not included in the occurrence rates. Power-law fits (black lines, based on events with values greater than the dashed vertical lines) are shown as well, with their associated formula in the lower left of each panel. From Loyd et al. (2022).

Prominences are the same as filaments—the two names correspond to whether the structure is seen standing out bright above the limb or silhouetted against the solar disk. Erupting prominences are responsible for the bright core seen in many CMEs (House et al., 1981). In a number of CME models, the presence or formation of a prominence is key to producing a CME (Section 3.3.2). Interactions between solar prominences and CMEs are complex: in some events, multiple prominence eruptions are observed with the same CME, and in others, an associated prominence is observed to collapse back onto the Sun (Webb et al., 1997; Simnett, 2000; Wang & Sheeley, 2002).

Reviews by Chen (2011); Webb & Howard (2012); Temmer (2021); Zhang et al. (2021) and Gopalswamy (2022), among others, provide a starting point into the solar CME literature.

> **Not So Fast: There is more to "halo" CMEs than just perspective.**
>
> A CME directed toward the observer will appear in coronagraphic images as an expanding ring, or a "halo." However, many CMEs can be directed well away from the observer and still appear as a halo. Kwon et al. (2015) observed that roughly two-thirds of halo events in solar cycle 24 appeared as halos in observations made $90°$ apart, likely explained by broad, $> 180°$, shocks. A broader shock increases the chances a CME will be viewed as a halo, biasing the population of CMEs observed as halos towards events producing broad shocks. Borader shocks result from CMEs that expand more widely, which is in turn promoted by greater energies or lower pressure in the interplanetary medium into which the CMEs are launched.

## 5.3 SEP Description

SEPs are one piece of the broader ion distribution originating from the Sun. This distribution spans energies ranging from keV to GeV, evolving from the solar wind to suprathermal particles to SEPs (see Figure 5.5 for a summary of SEP fluences and their phenomenological associations, and see also reviews by Desai & Giacalone 2016; Cohen et al. 2021).

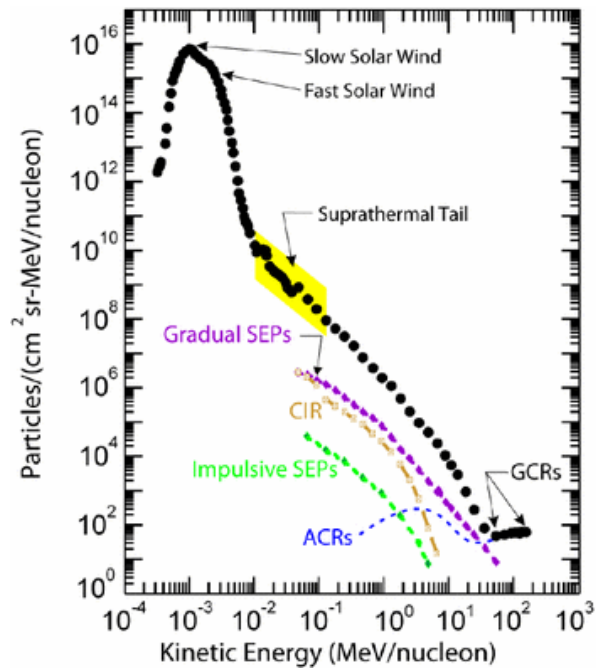

Figure 5.5: Oxygen fluences measured by several instruments on board the Advanced Composition Explorer (ACE) during three years (black data points) showing the fast and slow solar wind components (near 1 keV/nucleon) along with a suprathermal tail (highlighted in yellow) that extends to higher energies to merge into the energetic particle population. Also shown are representative particle spectra obtained for individual gradual and impulsive SEP events, corotating interaction regions (CIRs), anomalous cosmic rays (ACRs), and galactic cosmic rays (GCRs). These spectra (particularly for the SEP event and CIRs) vary significantly with time and from event to event, but when accumulated over three years, they produce a relatively smooth distribution as seen in the black data points above $\sim 0.1$ MeV/nucleon. Image from Desai & Giacalone (2016), adapted from Mewaldt et al. (2001).

The transition energy from each regime is not a set value, but generally SEPs are considered to be ions at energies greater than $\sim 0.5$ MeV/nucleon. For electrons, those above $\sim 30$ keV are considered "energetic." SEP events are primarily electrons and protons, but there are small contributions from ions of helium and heavier ions (through iron and above). Two primary acceleration mechanisms lead to energetic particle populations. The first is reconnection, generally associated with solar flares (Figure 5.1 inset (a)); the second is shock acceleration, typically at shocks in the corona or interplanetary space driven by fast CMEs (Figure 5.1 main panel; see, e.g., Desai & Giacalone (2016)). While it is expected that there are compositional signatures in the created SEP population specific to each of the acceleration mechanisms, in practice, it is difficult to evaluate the relative contributions, as large flares and fast CMEs often occur together. Additionally, the characteristics of the SEP population can be altered due to transport processes as the particles travel from the Sun to a distant observer (e.g., near Earth; see, e.g., Mason et al. 2012).

In its simplest form, the observed time profile of SEPs is generally characterized by a rapid increase in intensity to a peak, followed by an exponential decay. Events can last from minutes to days (Desai & Giacalone, 2016). The onset of an event can exhibit "velocity dispersion," where higher-energy particles arrive at the observer before the lower-energy ones. If the first arriving particles 1) experience minimal scattering, 2) are released from the acceleration site roughly at the same time, and 3) travel along the same path length, the observed velocity dispersion will be "sharp" and distinct. In such cases, the release time and path length can be derived from the arrival times and energies of the first arriving particles. The peak intensity of the event is governed by the efficiency and duration of the acceleration process, while the rate of the decay (and therefore the duration of the SEP event) is generally a function of the three-dimensional diffusion of particles through the interplanetary medium.

However, many conditions can alter this general time profile. An observer's magnetic connection to the source region strongly affects what is observed. In the case of flare-related reconnection, the acceleration is expected to be short and spatially confined. Thus, the resulting SEP population can only be detected by an observer magnetically connected to regions close to the acceleration site. In contrast, shock acceleration can occur over a much wider range of solar longitudes (and latitudes), resulting in more distantly connected observers registering the SEP event. Although shocks generally weaken as they propagate away from the Sun, they can be strong enough to accelerate significant numbers of particles continually out to 1 AU and beyond. This results in a much longer-lived SEP event, often with a broader peak.

> **Not So Fast: SEP events often cannot be cleanly classified into the two categories of gradual or impulsive.**
>
> There is a substantial heritage of classifying SEP events as either impulsive or gradual, with associated inferences about their acceleration mechanisms (Reames, 2013). However, numerous SEP events exhibit mixed properties between these two classifications (Cohen et al., 1999), and the details of SEP event generation remain an area of open inquiry. For this reason, Section 5.3 avoids the terms gradual and impulsive.

The severity of any planetary impact from an SEP event depends on some aspects. Naturally, the duration and peak intensity of an SEP event are key. As the longitudinal distribution of SEPs about the acceleration site is not uniform, the location of the planet relative to either the flaring region or the nose of the CME has a significant impact. While shock acceleration generally spreads SEPs over a wider longitudinal range than reconnection, the shock is typically strongest at the nose of the CME, resulting in the highest SEP intensities and energies at that location. The overall strength of the shock affects not only the highest energies to which it can accelerate particles, but it also generally governs how long it can accelerate significant numbers of SEPs. Reconnection can be complex and of differing durations, but generally produces significantly smaller SEP events by itself.

The planetary impact also depends on the characteristics of the SEP population. An event with large numbers (hundreds of MeVs) of high-energy particles—one with a high overall flux and a "hard" energy spectrum, indicating that the spectrum is weighted toward higher energies compared with other events—will be more impactful than one with a lower flux or softer energy spectrum. For planets with magnetospheres, higher rigidity (momentum per charge) particles penetrate further into the system than those of lower rigidities, so not only are high-energy particles a larger concern, but also ions with higher mass-to-charge ratios (e.g., iron compared to protons, see e.g., Mertens et al. (2024)). The hardness of an SEP spectrum is related to the details of its acceleration, such as where a CME begins to form a shock (Gopalswamy et al., 2016).

## 5.4 Flare-CME-SEP Scaling Relationships

Not unexpectedly, there are correlations between the three main varieties of transient events that involve magnetic reconnection—flares, CMEs, and SEPs—produced by the Sun.

For example, more energetic flares are more likely to be accompanied by a CME, and that CME is likely to have greater mass and energy (Yashiro & Gopalswamy, 2009).

For stars, the difficulty of directly observing CMEs and SEPs, combined with the relative ease of observing flares, invites the application of solar flare-CME-SEP correlations. This approach provides a baseline hypothesis of what the rates and properties of CMEs and SEPs would be if the star in question exhibited the same flare-CME-SEP correlation as the Sun. Studies of planetary impacts from SEPs, such as the ozone depletion discussed in Section 2.3, have generally relied on

> **Not So Fast: Shock-accelerated SEP events may not necessarily produce the greatest exoplanetary impacts.**
>
> By definition, the longer duration, wider spatial coverage, and CME association of shock-accelerated SEP events generally indicate a larger potential for planetary impacts (Desai & Giacalone, 2016). However, differences in stellar systems, such as high Alfvén speeds that inhibit the formation of CME-driven shocks, could alter the balance of shock- and flare-accelerated events in cumulative planetary impacts.

this approach, employing flare-SEP relationships like the one reported in Jian et al. (2013). Flare-CME scalings, in comparison to flare-SEP scalings, have perhaps seen less use in planetary applications, yet similar scaling laws have been developed based on solar events (e.g., Aarnio et al., 2011).

Applications of solar flare-CME-SEP relationships to stars, or, for that matter, even the Sun, are subject to some limitations.

1. The scatter between individual events making up these relationships is large. For example, for a given CME speed, there is a four-order-of-magnitude spread in the measured SEP peak intensities at 2 MeV (Kahler, 2001). Similarly, for a selected soft X-ray intensity, the spread of $> 30$ MeV SEP intensities is three orders of magnitude (Figure 5.6; Papaioannou et al., 2023). Herbst et al. (2019) used GOES and SphinX measurements of solar flares and proton fluxes to provide lower and upper limits for the SEPs associated with flares in G, K, and M main-sequence stars, finding a range of six orders of magnitude on the predicted proton fluxes for AD Leo, based on the X-ray intensity reported by Segura et al. (2010).

2. Flare-CME-SEP relationships can hide additional dimensions of variability. For example, two SEP events with the same peak intensity may have durations, and hence, fluences (flux integrated over time) that differ by orders of magnitude.

3. Applying these relationships to stars often requires the insertion of additional scalings between observational bandpasses for flares (e.g., Youngblood et al., 2017), accompanied by their own large degrees of scatter, before applying solar flare-SEP-CME relationships.

Solar flare-CME-SEP relationships are unlikely to be accurate in non-solar contexts, as evidenced by a wide array of modeling results (Section 3.3.4). For SEPs, the dramatically differing magnetic configurations of stars Section 3.2) could alter the possibility or efficiency of particle acceleration through reconnection and the ability of these particles to escape (e.g., Fraschetti et al., 2019). The characteristics of the background stellar wind into which a CME is launched will determine the strength of a shock and even whether a shock is formed at all, influencing SEP generation from shocks (e.g., Kahler & Reames, 2003). The stellar wind and stellar magnetic field may also affect the transport of SEPs, potentially changing their probability of reaching a planet (e.g., Cheng et al., 2024). Even the assumption that SEPs are associated with stellar flares has yet to be proven, and Kahler & Ling (2023) caution that even for the Sun, most flares are not accompanied by SEP events, resulting in the overestimation of cumulative SEP fluxes through blind application of flare-SEP relationships.

For CMEs, numerical models indicate that CMEs can be suppressed in active stars by strong overlying fields (Alvarado-Gómez et al., 2018) although not for all cases (Fraschetti et al., 2019). Meanwhile, low rates of helicity generation by stars with mild differential rotation may inhibit CME formation in the first place (Jardine, 2024). Observations already rule out the validity of flare-CME relationships for several stars with high flare rates. For these stars, applying solar flare-CME relationships yields energy loss rates that are physically implausible (Drake et al., 2013) and mass loss rates that exceed constraints on cumulative mass loss available from the Lyman-$\alpha$ astrospheric absorption method (Section 4.3.1; Odert et al., 2017).

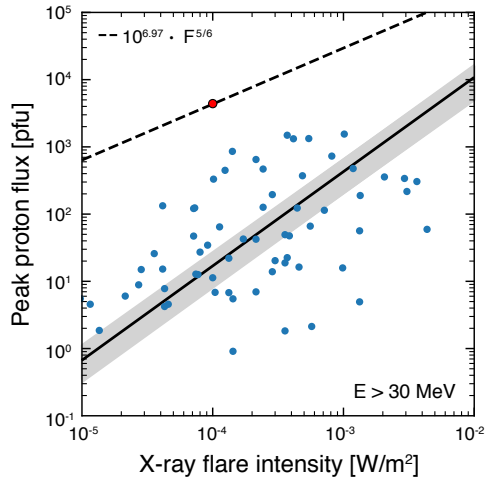

Figure 5.6: Relationship between solar X-ray flare peak flux and the peak flux protons with energies > 30 MeV in associated SEP events. The dashed line represents an estimate of the upper limit of possible SEP fluxes, anchored to the highest observed flux (red point). Modified from (Papaioannou et al., 2023).

Applying solar flare-CME-SEP scalings to stellar systems can be an insightful "what if?" investigation to explore potential impacts. However, in the ideal future scenario, the rate and intensity of CME and SEP events for systems of interest will be determined directly from observation—the very motivation for this report and the workshop behind it. These will almost certainly differ from what solar scalings predict for stars unlike the Sun in mass and activity level. Meanwhile, continued and expanded modeling efforts can explore the regimes in which such relationships are likely to break down (Section 3.3.4).

That said, flare-CME-SEP relationships will remain useful, even as observations and modeling become more comprehensive. Advances in observations and modeling may enable the calibration of these relationships to specific stars or mass-activity categories. Meanwhile, physical arguments could be used to adapt solar flare-CME-SEP scalings to specific stars, such as accounting for a threshold energy at which CMEs are able to "break out" of an overlying magnetic field. Well-calibrated flare-CME-SEP relationships will likely prove particularly helpful for assessing the cumulative effects of large numbers of transient events, where scatter averages out.

## 5.5 Techniques

There are a multitude of ways to characterize CMEs and SEPs and predict their occurrence on the Sun. Our ability to directly measure solar particle fluxes has enabled the identification of many remotely observable electromagnetic signatures that are useful for spotting the occurrence and predicting the properties of transient solar events. However, any remote EM technique used for the Sun taken on its own—or even with an additional method—is generally insufficient to conclude whether a CME or SEP has occurred on another star. Challenges include low signal strengths, confounding signatures (especially when observations are unavoidably disk-integrated), and sensitivities to different event orientations and properties. In this section, we discuss a number of existing and potential CME and SEP detection methods, considering their applicability to detecting and characterizing stellar events, with a focus primarily on CMEs. For CMEs, a summary of the methods—plus some additional methods not reviewed in depth here—are provided in Table 5.1.

| Observational Signature | Properties Probed | Sun | Stars | References for Stellar Work | Section |
|---|---|---|---|---|---|
| Coronagraph | Mass, speed | ✓ | ? | | Sec. 5.5.1 |
| Doppler shifts in coronal & chromospheric emission lines during a flare | Prominence speed | ✓ | HR 9024, EV Lac, EK Draconis | A19, C22, O20, L20, M20, N22, I23, N24 | Sec. 5.5.2 |
| Dimming from evacuated coronal material (EUV/X-ray) | Mass, speed | ✓ | Proxima Centauri, AB Dor, $\varepsilon$ Eridani, AU Mic | V21, L22 | Sec. 5.5.3 |
| Absorption dimming: increase in $N_H$ during a flare | Mass | ? | ✓ | M17 | Sec. 5.5.4 |
| Negative radio bursts | Mass | ✓ | ? | | Sec. 5.5.4 |
| Type II radio burst | CME speed | ✓ | ? | C16, C18a, C18b | Sec. 5.5.6 |
| Planetary aurorae | Occurrence? | ✓ | ? | | Sec. 5.5.8 |
| SPI radio emission variation | Occurrence? | X | ? | | Sec. 5.5.9 |
| Exoplanet outflow variability | Pressure | X | ? | L12, H22, H24 | Sec. 5.5.10 |
| Spectroscopy gamma-ray/nuclear lines | CME shock parameters? | ✓ | ? | | Sec. 5.6.1 |
| Type IV radio burst | Total magnetic flux? | ✓ | Proxima Centauri | Z20, M24 | Sec. 5.6.3 |
| Gyrosynchrotron | Magnetic field strength | rare | ? | | Sec. 5.6.3 |

✓ = has been used and demonstrated as a technique to probe CMEs
X = not possible to use
? = Possible technique but not used to date or no positive results

Table 5.1: Observational methods to detect coronal mass ejections on the Sun and Stars, adapted from Osten & Wolk (2016); Veronig et al. (2025). Reference key: A19: Argiroffi et al. (2019), C16: Crosley et al. (2016), C18a: Crosley & Osten (2018a), C18b: Crosley & Osten (2018b), C22: Chen et al. (2022), H22: Hazra et al. (2022), H24: Hazra et al. (2024), I23: Inoue et al. (2023), L12: Lecavelier des Etangs et al. (2012), L20: Leitzinger et al. (2020), L22: Loyd et al. (2022), M17: Moschou et al. (2017), M20: Muheki et al. (2020), M24: Mohan et al. (2024), N22: Namekata et al. (2022a), N24: Namekata et al. (2024), O20: Odert et al. (2020), V21: Veronig et al. (2021), Z20: Zic et al. (2020)

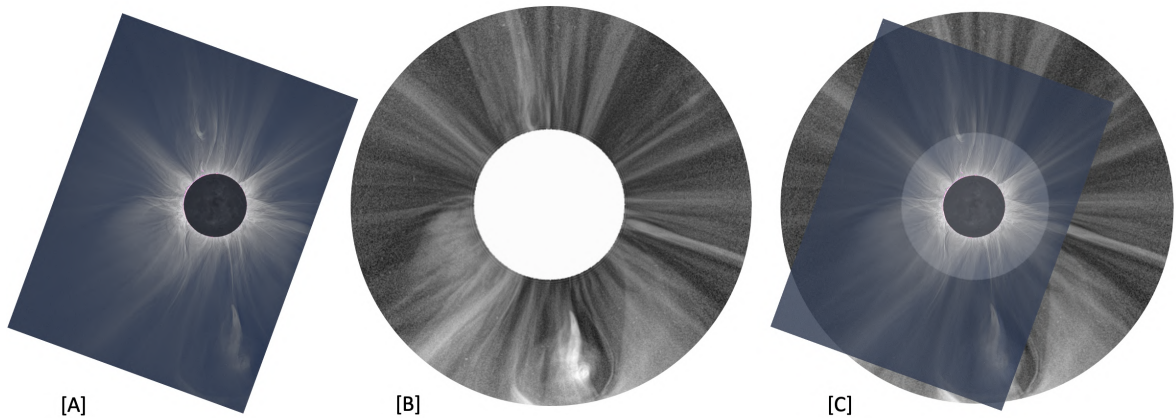

Figure 5.7: Comparison of white light images taken during a total solar eclipse [A] and with the Large Angle and Spectrometric COronagraph (LASCO/C2) [B]. The overlay between the two in [C] demonstrates how much of the inner corona is missing, especially from space-based coronagraphs. Credit: Shadia Habbal.

### 5.5.1 Coronagraphy

Coronagraphs are instruments used to block the light of the Sun or another bright central source to suppress the glare that might otherwise obscure faint objects nearby in the field of view. More generally, they are one example of a high-contrast imaging system, where contrast here refers to the ratio of light from the faint nearby source to that of the bright central object. Coronagraphs can be internal to a telescope, such as a vortex coronagraph, or external, such as the Moon during an eclipse or the starshade concept for imaging exoplanets. Aime et al. (2024) provides a review and comparison of coronagraph principles in both solar and stellar contexts.

Coronagraphs provide the possibility of measuring photospheric light that is Thomson-scattered on free electrons in solar and stellar coronae. Figure 5.7 shows example images of the solar coronae, both with a coronagraphic instrument and during an eclipse (where the Moon serves as a natural coronagraph). The scattered light is optically thin, polarized, and its brightness is linearly proportional to the electron number density along the line-of-sight (independent of the temperature of the plasma). Therefore, coronagraphs can be used to derive the (column) density and mass of structures in a corona. In addition, when observing moving features (like CMEs) with high-enough time cadence and duty cycle, coronagraphs also enable estimates of the projected plane-of-sky speed of these features. From such time-series measurements, the occurrence rate and kinetic energies of solar CMEs can be inferred.

#### 5.5.1.1 On the Sun

The serendipitous spacing and orbital alignment of the Moon has made coronagraphic observations of the Sun possible from before the advent of modern heliophysics as a result of total solar eclipses. Even today, solar eclipses provide advantages in observing the solar corona over existing man-made coronagraphs, such as a smaller inner working angle (Figure 5.7). Yet eclipses occur only once every 12 to 18 months and last for seven minutes at most. Thus, when Lyot (1939) created an artificial total solar eclipse with an optical design called a coronagraph, it amounted to a breakthrough in heliophysics.

Earth-bound coronagraphs suffer from scattered light from the sky, so their yield is significantly enhanced by placing them in space on board satellites. The first space-based coronagraphs on board the Orbiting Solar Observatory 7 (OSO-7), Skylab, Solwind (a.k.a. P78-1), and Solar Maximum Mission (SMM) in the 1970s and 1980s, led to the detection and understanding of the importance of CMEs for space weather at Earth (Koutchmy, 1988). They also provided insight into their intricate three-part magnetic structure. At present, continuous observations of CMEs by the Large Angle and Spectrometric Coronagraph (LASCO) on board the Solar and Heliospheric Observatory (SOHO) and the coronagraphs (COR1 and COR2) on board the two Solar TErrestrial RElations Observatory (STEREO) spacecraft, image the solar corona out to about $30\,R_\odot$ and $15\,R_\odot$, respectively, albeit missing the first $0.5\,R_\odot$ above the surface due to the inevitable diffraction at the edges of the occulter. Observations with LASCO have been used to create catalogs of solar CMEs spanning decades (Yashiro et al., 2004; Robbrecht & Berghmans, 2004; Gopalswamy et al., 2024).

In addition to white light images acquired with coronagraphs, the Ultraviolet Coronagraph Spectrometer (UVCS) on SOHO achieved a notable scientific breakthrough by providing the first inferences of the solar wind speed in the inner corona (Kohl et al., 1997; Habbal et al., 1997). The legacy of coronagraphs continues with their more recent incorporation on the Parker Solar Probe and the Solar Orbiter, which have captured a range of transient events beyond $10\,R_s$ and pursue the scientific legacy of UVCS in further yielding the chemical composition of CMEs.

### 5.5.1.2  On other stars

A direct image of a CME via coronagraphy would perhaps be the most unambiguous indication that a stellar CME has occurred. However, using solar CMEs as a guide for what might be observable for stars is challenging due to a mismatch in the viewing limitations of solar and stellar coronagraphs. For example, solar observations with the Large-Angle and Spectrometric Coronagraph (LASCO) extend only to about $\sim 30\,R_\odot$ due to the field of view of the instrument (Brueckner et al., 1995). Yet even for the most nearby star, this distance corresponds to a required inner working angle $< 0.1$", currently achievable only by some bands of the Keck Planet Imager and Characterizer (Mawet et al., 2016, see also Figure 4.11 for a schematic comparison of scales). Should we attempt to observe a CME when it is at a distance from its star that would be feasible to spatially resolve, the flux from the CME due to both scattered light from the star and stimulated emission would be much lower than when it is within the $30\,R_\odot$ regions within which most solar CME remote observations are made.

To understand the feasibility of using coronagraphy to directly image the Thomson-scattered emission from a stellar CME, we derived an order-of-magnitude formula for the contrast of the CME to the star, $C$, assuming the ideal case that the CME is launched off the limb (see Appendix A.4).

$$C = \frac{\sigma_{\mathrm{T}}}{4\pi r^2}\frac{M_{\mathrm{CME}}}{m_X} \tag{5.1}$$

We have used this formula to investigate contrast as a function of CME mass (which sets the total number of scattering electrons) and distance (Figure 5.8, left panel). We have also investigated contrast as a function of CME volume and electron density (Figure 5.8, right panel), accounting in this case for angular scales of CMEs that subtend multiple resolution elements of an instrument, assuming the star is at 10 pc (see Appendix A.4). The reader may note that Equation 5.1 does not explicitly depend on the distance of the star, since contrast is defined relative to the star, and both star and CME will dim equal amounts with distance from the observer. However, in practice, distance will be a key factor, as the inner working angle of an instrument will determine the minimum physical distance a CME must be from its host star to become visible. The further a star is, the greater this distance, and the lower the contrast will be of a CME when it first emerges beyond the inner working angle.

From these estimations, the CME properties required to directly image a stellar CME with modern exo-solar coronagraphy (e.g., the Gemini Planet Imager [GPI] with a contrast of $\sim 10^{-6}$ for a G0 star and apparent I-band magnitude of 8 (Macintosh et al., 2014)) are extreme—at the upper limit of estimated stellar CME masses ($\approx 10^{22}$ g, Moschou et al., 2019a) and at densities several orders of magnitude higher than we observe for solar CMEs at distances of $\sim 1$ AU (reported in e.g., Temmer et al., 2021). While current instrumentation is likely incapable of directly imaging stellar CMEs, the intended resolution and contrast of the Habitable Worlds Observatory (HWO) puts directly imaging ideal-case CMEs within the realm of possibility. At an intended contrast $\approx 10^{-10}$ (Allan et al., 2023), HWO perhaps would be able to directly image a CME originating from the $10^{19}$ g prominence reported by Namekata et al. (2024) for EK Dra. Understanding CME properties and how they may impact surrounding exoplanets (and how these properties may vary with stellar properties) is directly aligned with HWO's mission to contextualize potentially habitable systems in terms of the Solar System. As such, systems with candidate CME detections (irrespective of whether there is a confirmed exoplanet around the star) should be considered prime targets for HWO's coronagraphic observations.

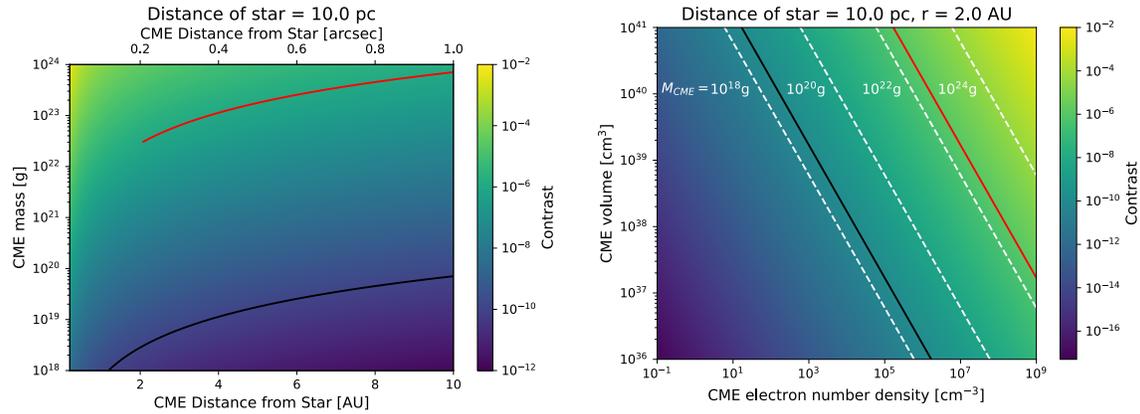

Figure 5.8: Contrast of scattered light from a CME relative to the central star as a function of a CME's mass and distance from its star *(left)* and the CME's density and volume for a CME at a distance of 2 AU *(right)* as estimated from equation A.6 and assuming a stellar distance of 10 pc. The red line indicates a contrast of $10^{-6}$ (roughly the contrast of Gemini Planet Imager/GPI) and the black line indicates a contrast of $10^{-10}$ (the intended contrast for Habitable Worlds Observatory [HWO]. The cutoff in the GPI line on the left plot indicates the lower limit of the CME distance that GPI would be able to resolve for a star at 10 pc. The white dashed lines on the right plot indicate volume and density values of constant mass. The mass range is informed by the estimated stellar CME masses reported by Moschou et al. (2019a); the range of recorded solar CME masses is not represented in these plots. Figure by Ivey Davis.

It is worth noting that while our estimates are based on total-intensity scattered light, there are additional axes of the problem space that can be leveraged. For instance, the scattering of light by CMEs is known to polarize it (Mierla et al. (2011) and references therein). It has also been shown that light that is linearly polarized by scattering can be used to identify, for example, debris disks even when there is not a detection in total intensity (Lewis et al., 2024). Linear-polarization measurements may be a way to detect CMEs even if they are not detected in total intensity. CMEs also produce intrinsic emission that may be possible to leverage. For such emission, the intensity scales like $n_{e^-}^2$ rather than $n_{e^-}$, but also depends on the temperature and ionization state of the wind and CME. Each of these techniques may also benefit from trying to detect the void left behind by the CME rather than the CME

itself—in this case, understanding properties of the wind will be crucial (see Chapter 4 and Appendix A).

Our calculations are also based on a highly simplified representation of CMEs that neglect details like the chromaticity of scattering and the inhomogeneity of electron distributions in CMEs. It also assumes that the entire CME is contained in a single pixel, although Appendix A.4 includes a description for estimating the contrast in the case that the CME spans multiple pixels. Overall, our estimates for CME direct imaging are promising but warrant a more robust investigation of the relevant physics and ways to utilize various facets of direct imaging.

### 5.5.2 Doppler-Shifted Spectral Lines from Erupting Prominences and CMEs

Doppler spectroscopy is a potentially powerful means of detecting CMEs by directly probing mass motions. However, interpretations can be challenging in light of many alternate sources of Doppler shifts in stellar emission, including bulk stellar motion (e.g., rotation or orbit around a companion), convective motions emerging at the surface (e.g., granulation), waves and oscillations (e.g., acoustic p-mode oscillations), and more (Dravins et al., 1981; Stief et al., 2019; Saar & Donahue, 1997; Meunier et al., 2010; Haywood et al., 2016; Schrijver & Zwaan, 2000; Broomhall et al., 2009; Dumusque et al., 2011). When multiple components contribute to the same line of sight, they must be disentangled to enable an accurate interpretation, a challenge for solar observations and even more so for inherently disk-integrated stellar observations. Even when only one plasma structure contributes, complications from, for example, simultaneous blueshifts and redshifts resulting either from expansion or helical motions can arise (e.g., Ciaravella et al., 2000). Most of the aforementioned blueshifts and redshifts act as sources of "astrophysical noise," with amplitudes ranging from cm s$^{-1}$ to km s$^{-1}$, whereas CME velocities may reach thousands of km s$^{-1}$. However, particularly pernicious confounding signals are blueshifts up to several hundred km/s due to chromospheric evaporation from flares (Pevtsov et al., 2007; Milligan & Dennis, 2009).

Projection effects add an extra element to the interpretation of Doppler spectroscopy. CMEs, if launched near the solar or stellar limb or behind the disk, can produce redshifts rather than blueshifts more typically associated with motion away from the star (Moschou et al., 2019b). Figure 5.9 schematically illustrates some of the key motions that could produce Doppler signals during a CME.

### 5.5.2.1 On the Sun

Spectroscopic measurements can reveal the internal physical mechanisms of solar CMEs through observed Doppler shifts (Hansen et al., 1971; Tousey et al., 1973; Gosling et al., 1974). For example, as a CME propagates, its expansion rate can be approximated from the broadening of each spectral line and its bulk motion velocity and acceleration along the observer's line of sight (LOS) from centroid shifts, provided the instrument used has sufficient wavelength resolution (Antonucci et al., 1997; Ciaravella et al., 1997; Landi et al., 2010). For solar observations, these measurements are often made with the aid of a coronagraph, thereby isolating emission from the optically thin corona (see Section 5.5.1). Figure 5.10 provides example solar CME observations by the SOHO/UVCS slit spectrometer that reveal fast (±800 km/s) blueshifts and redshifts of emission lines from O VI plasma. This example also illustrates the challenge of confounding signals, as streamers disrupted by the propagating and expanding CME share the line of sight.

When a solar CME's direction of motion is nearly aligned with the plane of the sky (POS), interpretation becomes more difficult. Expansion can still be inferred from Doppler broadening, but bulk velocity is difficult to determine since the CME's velocity component along the line of sight is small. However, this can be overcome by using emission lines that resonantly scatter coronal emission as probes of the motion of overlying coronal plasma. As a CME accelerates outward, a given line subject to radiative pumping will successively brighten and dim as the CME's motion Doppler shifts it across emission lines from the Sun. For example, O VI 1037.61 Å emission from a CME will brighten as the CME passes through velocities near 200 and 400 km s$^{-1}$ as it successively scatters Doppler-shifted emission from the Sun's 1037.02 Å and then 1036.34 Å C II lines (Li et al., 1998; Wilson et al., 2022). This is demonstrated in Figure 5.11, where the intensity ratio of the O VI 1031.91 Å to the 1037.61 Å drops in response to brightening of the 1037.61 Å component as Doppler-shifted emission is resonantly scattered.

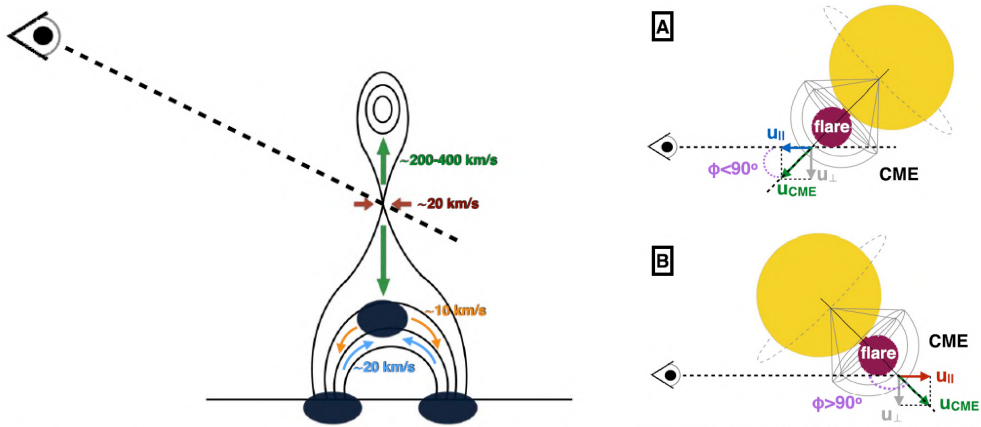

Figure 5.9: *(left)* Schematic demonstrating the standard CME-flare scenario in the Sun and the multiple flows triggered by reconnection and the subsequent atmospheric response according to Hara et al. (2011). The blue blobs indicate the X-ray emission sources. *(right)* Schematic demonstrating the scenario of a CME expanding and propagating away from the star and giving rise to either blueshift (A) or redshift (B) signatures as a result. Credit: Figure adapted from Moschou et al. (2019b).

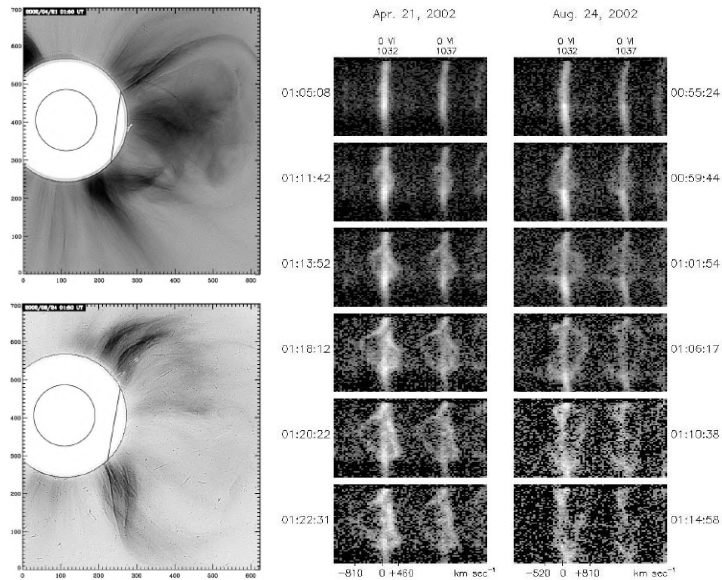

Figure 5.10: *(left)* LASCO C2 images of the 2002 April 21 and August 24 CMEs resulting from the combination of different UVCS slit positions. *(right)* Six long-slit spectra for the April and August events showing the O VI 1032 Å and 1038 Å lines, with H I Lyman $\beta$ and the second-order Si XII 521 Å lines faintly visible. The spectra illustrate how the CME disturbs the formerly stable (coronal and streamer) plasma while it expands along the LOS with thin sheets of O VI plasma. Credit: Figure adapted from Raymond et al. (2003).

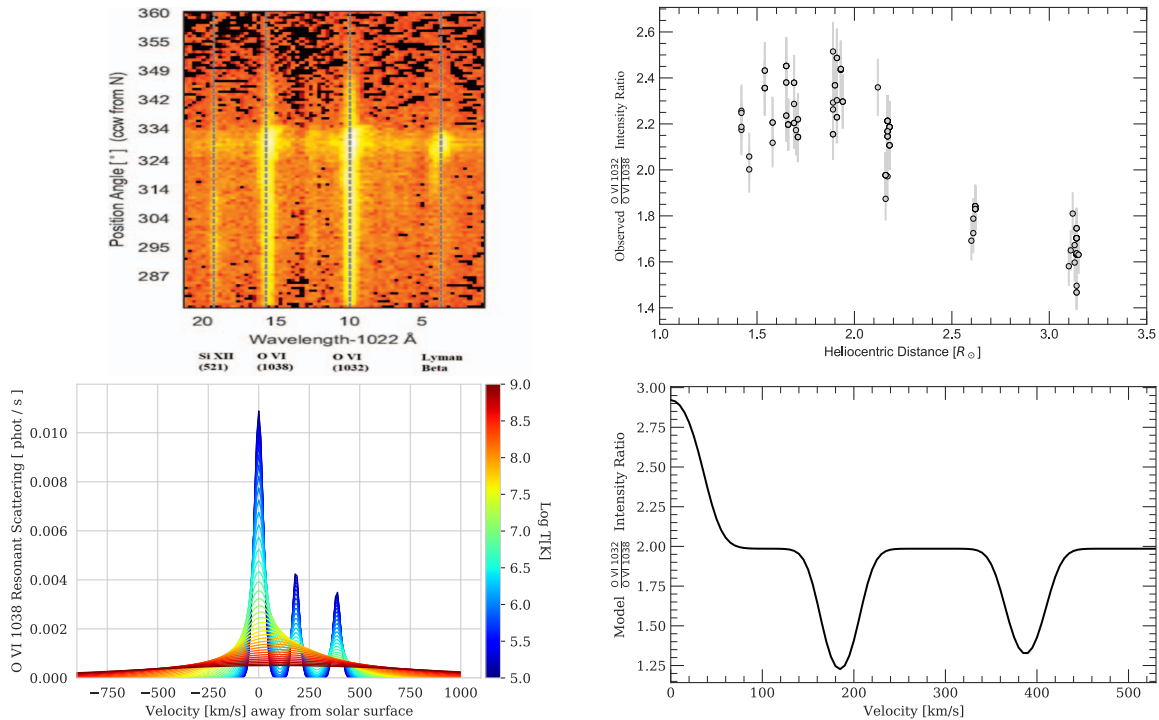

Figure 5.11: Top: Slit image spectrum at 1.4 $R_\odot$ *(left)* and intensity ratios *(right)* of individual clumps of CME prominence core plasma observed with SOHO/UVCS in the O VI emission lines. *(bottom left)* Theoretical resonant scattering component of radiation emitted by O VI ions under conditions resembling the core of a solar CME. *(bottom right)* Theoretical ratio of intensities between the spectral lines of 1031.91 Å and 1037.61 Å radiating from the collisional excitation and resonant scattering of O VI ions assuming CME-like conditions. Atomic spectroscopy models imply that the speed of the observed coronal plasma can be inferred without a Doppler shift measurement when the intensity of the 1031.91 Å line is no longer twice that of the 1037.61 Å line. Credit: Top and bottom-right figures from Wilson et al. (2022). Bottom left by Maurice Wilson.

In full-disk (i.e., Sun-as-a-star) spectra of the Sun, Doppler shifts related to CMEs are sometimes visible, even without a coronagraph. Assuming the CME is moving towards the observer, a secondary component or blue-wing enhancement/absorption caused by the moving plasma can appear in the blue wing of the lines. Previous studies have reported blue shifts of EUV line centroids during flares, interpreted as plasma motions during the flares or as possible signatures of associated CMEs (Hudson et al., 2011; Brown et al., 2016; Chamberlin, 2016; Cheng et al., 2019). The signal can manifest either as a shift in the line centroid, in which case it represents a lower bound on the CME LOS velocity, or as a secondary emission/absorption component, in which case the bulk motion of the CME can be measured more accurately (see, e.g., Xu et al. 2022; Lu et al. 2023; Xu et al. 2025).

Doppler-shifted emission/absorption also arises from prominence eruptions associated with CMEs in full-disk H$\alpha$ observations of the Sun (e.g., Namekata et al. 2022c,a; Otsu et al. 2022; Pietrow et al. 2024). H$\alpha$ spectra during filament eruptions are complicated by the optical thickness of the H$\alpha$ line. For the Sun, prominence eruptions and CMEs have an association rate of 70–80%, meaning Doppler-shifted H$\alpha$ emission/absorption due to a flare likely indicates a CME occurred (e.g., Gilbert et al., 2000; Gopalswamy et al., 2003; Raymond et al., 2003).

There are a number of paths by which future solar observations analysis could aid in the interpretation of Doppler shifts in stellar spectroscopy. Expanding the body of Sun-as-a-star spectra of CMEs, along with confounding phenomena like chromospheric evaporation, could help establish rates of true and false positives and enable searches for telltale signals that can distinguish confounding sources. Some telltales might already be present in the same spectra used for Doppler shift measurements, such as line intensity ratios that yield plasma densities (Raymond & Ciaravella, 2004; Murphy et al., 2011) or polarization signals that provide information on magnetic fields (e.g., Lin et al., 2000; Tomczyk et al., 2007; Tian et al., 2013; Yang et al., 2020; Sheoran et al., 2023, see also Chapter 3 ). Combining Doppler diagnostics with other diagnostics might yield false positive estimates that are lower than Doppler diagnostics alone. Some studies (e.g., Houdebine et al., 1990; Leitzinger et al., 2011; Argiroffi et al., 2019; Namekata et al., 2022c; Wilson & Raymond, 2022) have already provided similar recommendations or candidate detections involving one or more of these techniques. Using additional information already present in spectra used for Doppler diagnosis could also include the plasma temperature and density via atomic spectroscopy intensity ratios or the magnetic field structure via polarized light.

### 5.5.2.2   On other stars

Reports of blueshifted emission from chromospheric lines suggestive of stellar prominence eruptions have steadily increased since the 1990s, as spectroscopic observations of stellar flares have become increasingly common. These observations have mostly been made in optical wavelengths (mostly in Balmer lines; see Namekata et al. 2022b; Vida et al. 2024 for a review), but some in UV wavelengths (Hawley et al., 2007; Leitzinger et al., 2014). They have mostly targeted M dwarf stars (e.g., Houdebine et al. 1990; Vida et al. 2016b; Maehara et al. 2021; Notsu et al. 2024, see Figure 5.12a), but recent reports show signatures in G dwarfs (Namekata et al. 2022c, 2024, see Figure 5.12b) and RS CVn-type binary systems (e.g., Ayres et al. 2001; Inoue et al. 2023) as well.

> **Not So Fast: CMEs can be launched with sub-escape velocities.**
>
> Material erupting at sub-escape velocities on the Sun is often ultimately associated with an escaped CME (Gopalswamy et al., 2003). Non-gravitational forces, such as magnetic fields, can play a large role in the kinematics of a solar or stellar CME (e.g., Alvarado-Gómez et al., 2018), and must be considered when interpreting velocity measurements of erupting prominences or coronal material.

Determining whether blueshifts (or redshifts) are a result of prominence motions rather than other dynamic phenomena, such as chromospheric evaporation, remains challenging, particularly in spatially unresolved stellar observations. This challenge is elevated for M dwarf observations, where, unlike the Sun, prominence eruptions viewed against the disk may appear in emission rather than absorption due to the weaker backlight (Leitzinger et al., 2022).

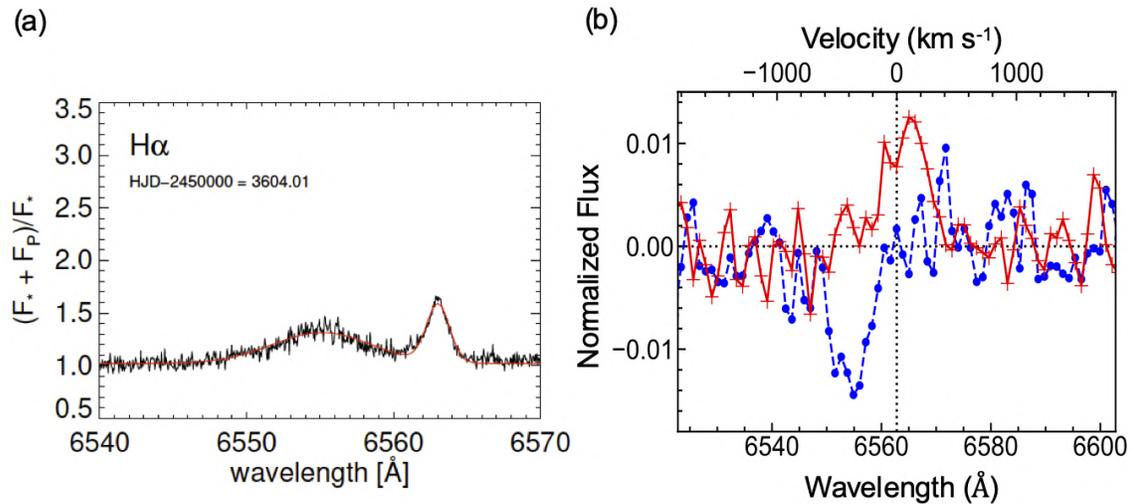

Figure 5.12: Three examples for Doppler shift observations of stellar flares. (a) Blueshifted Hα (λ = 6562.8 Å) emission from a flare on an M dwarf, V374 Pegasi. Figure from (Vida et al. 2016a; see also Vida et al. 2016a). (b) Blueshifted Hα "absorption" from a superflare on a solar-type star, EK Draconis (Namekata et al., 2022c). Red represents the flare phase, while the blue line corresponds to the post-flare phase. Figure produced from data published in (Namekata et al., 2022c).

During M dwarf flares, chromospheric lines almost exclusively exhibit blueshifted emission profiles (Figure 5.12a), making them a significant confounding factor. In contrast, both absorption and emission blueshifts are observed in G dwarfs (Figure 5.12b; Namekata et al. 2022c, 2024), providing strong evidence for the occurrence of solar-like filament/prominence eruptions.

In many cases, observed blueshift velocities do not exceed the surface escape velocity of stars, making it challenging to conclude that stellar CMEs occurred. Consequently, these observations are often not considered definitive evidence of CMEs. However, on the Sun, prominence and filament eruptions frequently show sub-escape velocities, yet are still associated with escaped CMEs (Gopalswamy et al., 2003). Solar observations also reveal that some of these slower eruptions eventually accelerate to reach escape speeds (Gopalswamy et al., 2000), highlighting that initial velocity alone does not determine whether plasma will escape.

Furthermore, although observations of stellar eruptions have detected velocities approaching or exceeding the escape velocity in some cases (e.g., Houdebine et al. 1990; Vida et al. 2016b; Inoue et al. 2023; Namekata et al. 2024), the connection between prominence eruption candidates and stellar CMEs remains uncertain due to the short lifetime of the signals, which hinders tracking the propagation of a prominence into interplanetary space. Moreover, it is conceivable that large-scale overlying magnetic fields, as suggested by Alvarado-Gómez et al. (2018), could confine plasma and prevent its escape even if it initially achieves escape velocity. Ultimately, since many forces other than gravity could be involved in the dynamics of plasma eruptions, velocity alone cannot clearly establish its eventual fate. Despite these challenges, the numerous reports of such events indicate that this line of investigation holds promise for identifying stellar CMEs.

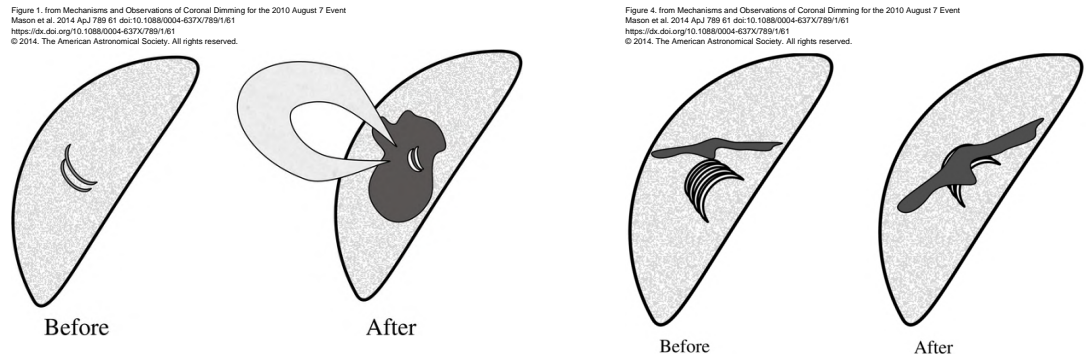

Figure 5.13: Schematic illustration of mass loss dimming *(left)* and obscuration dimming *(right)*, from Mason et al. (2014).

In recent years, Doppler shifts during flares have also been reported in X-rays (Argiroffi et al. 2019; Chen et al. 2022; Inoue et al. 2024), although distinguishing these shifts from chromospheric evaporation is challenging. Argiroffi et al. (2019) detected a Doppler shift of 90 km s$^{-1}$ (below the escape velocity) only in the lines corresponding to the steady corona temperature and interpreted it as a CME origin by combining it with 1D loop calculations. In the Sun, no X-ray Doppler shifts from CMEs have been reported to the best of our knowledge, making it impossible to use solar CMEs as a guide for interpretation. To deepen the understanding of stellar X-ray Doppler shifts, it is necessary to increase the sample size. Future missions should focus on filling these gaps and thoroughly investigating similar phenomena on the Sun.

### 5.5.3 Mass-Loss Coronal Dimming

Mass-loss coronal dimmings are observed as temporary (order of hours) reductions of the extreme ultraviolet (EUV) and soft X-ray (SXR) emission caused by the expansion and mass loss due to a CME. They were first detected in solar EUV and SXR imagery in the late 1990s (e.g., Sterling & Hudson, 1997; Thompson et al., 1998). Coronal dimmings are an attractive diagnostic of solar and stellar CMEs because of their direct relationship to the coronal mass ejection itself (see Figure 5.13). The mass that is "ejected" from the corona soon stops emitting because of its expansion and corresponding reduction in density, and it no longer contributes to the star's spatially integrated coronal emission, regardless of whether the CME remains in the field of view (Figure 5.14). An advantage of coronal dimming is that it is relatively insensitive to viewing angle and needs—in principle—only simple measurement devices (photometry). Solar and stellar coronae are by definition optically thin, and their typical temperatures (of order 1 MK) result in peak emission in the EUV and SXR regimes. As long as material is ejected from the observable hemisphere of the coronae, coronal dimming can be observed, and the same amount of emission will be lost as if the ejection had occurred anywhere else in the visible hemisphere.

#### 5.5.3.1 On the Sun

Coronal dimmings represent the early footprints of CMEs in the lower corona. They probe numerous elements of the physics of CMEs related to their initiation, magnetic connectivity, and reconfiguration, as well as mass loss and replenishment (see recent review by Veronig et al., 2025). A particular strength is that they are observed in both spatially resolved image data as extended regions of reduced emission (e.g, Thompson et al., 1998; Dissauer et al., 2018), as well as in spatially unresolved Sun-as-a-star light curves as significant dips compared to the pre-event background level (Mason et al., 2014, 2016, Figure 5.15), which makes them a promising method to detect CMEs on other stars (Harra et al., 2016; Veronig et al., 2021). Importantly, characteristic dimming parameters identified from image data (e.g., area, brightness, area growth rate) can be used to constrain the mass and speed of CMEs during their early evolution (Dissauer et al., 2019; Chikunova et al., 2020).

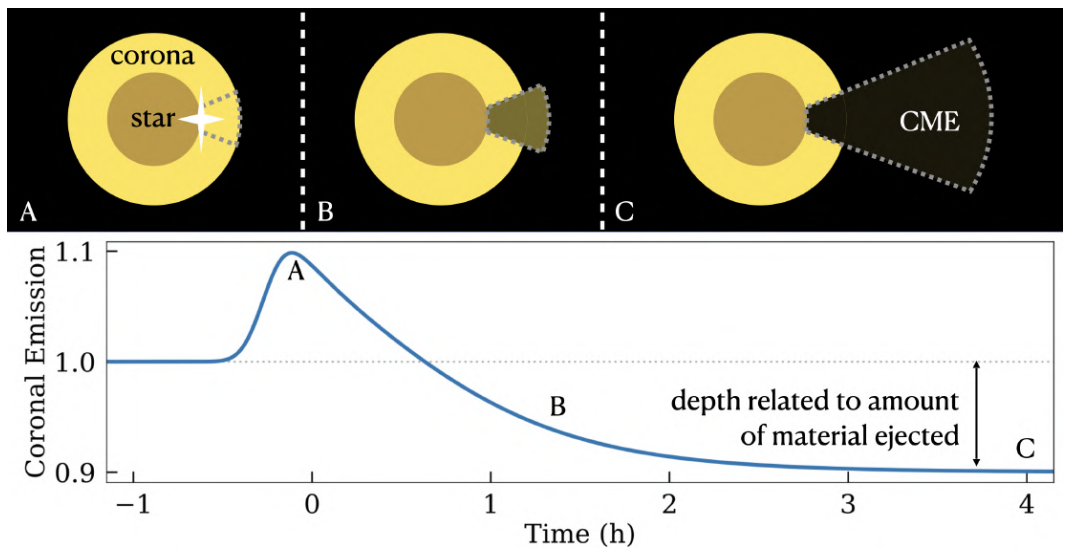

Figure 5.14: Schematic of the physical basis of coronal dimming. Figure by R. O. Parke Loyd.

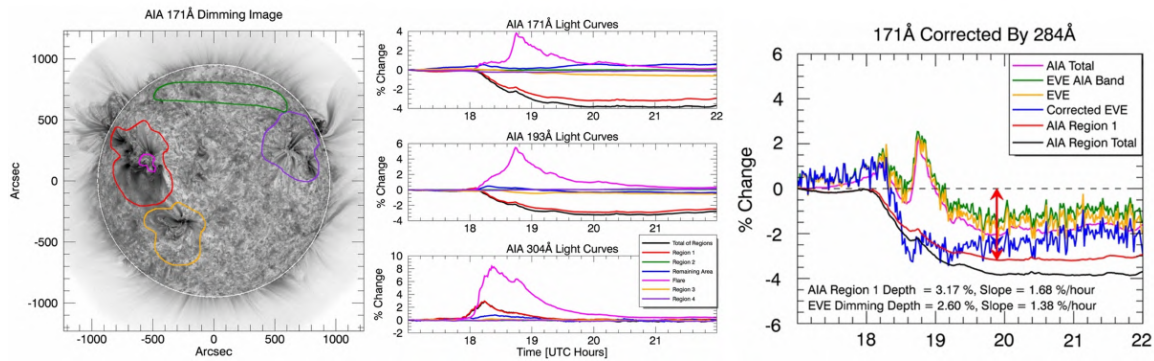

Figure 5.15: Illustrating the unique observational potential of solar coronal dimmings for detection of stellar CMEs. (*left and middle panels*) Dimmings identified in regions-of-interest (colored contours and corresponding time profiles) in full-disk integrated SDO/AIA extreme-ultraviolet 171 Å images. (*right panel*) Dimmings observed in spatially integrated irradiance measurements of SDO/EVE at the same wavelength. Adapted from Mason et al. (2014).

In resolved images of the Sun, a fully developed coronal dimming appears as a "transient coronal hole." Longer-lasting coronal holes (persisting for several solar rotations) that develop without a CME are common—it is the transience that differentiates the coronal dimming of a CME, which develops in around an hour (Dissauer et al., 2018) and dissipates on average timescales of 8–10 hr but may last >1 day (Attrill et al., 2008; Reinard & Biesecker, 2008; Veronig et al., 2021).

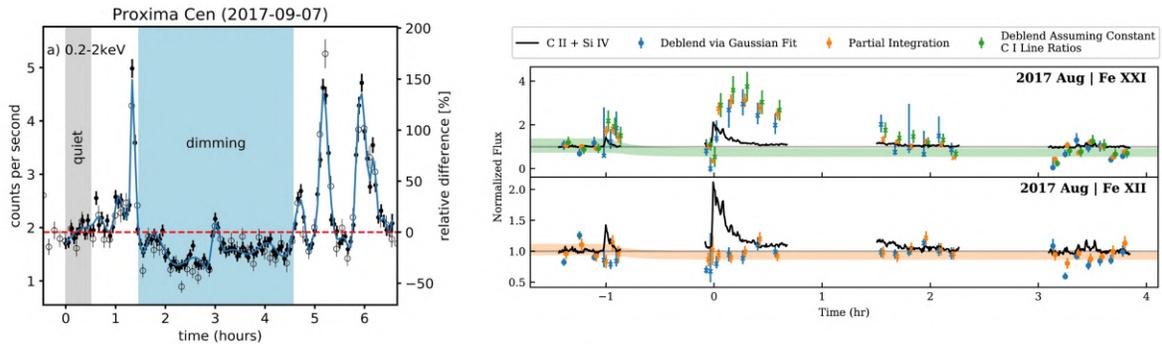

Figure 5.16: Two examples for potential post-flare stellar dimmings. *(left)* XMM-Newton X-ray (0.2–2 keV) light curve on Proxima Centauri, the possible dimming signature is marked in light blue. *(right)* Fe XXI and Fe XII HST observations on ε Eridani, potential dimming regions are indicated by translucent regions. Credit: Adapted from Veronig et al. (2021) and Loyd et al. (2022).

Resolved images of a coronal dimming event developing on the Sun show that the reality of coronal dimming is more complex than the simplified picture of Figure 5.14. The "background" environment of the corona across the Sun is constantly changing. At the eruption site, the magnetic reconnection associated with the CME will not only remove plasma, but it will also heat and compress plasma that remains in the corona. Some of these effects manifest as "noise" in the light curve, but others can generate spurious dimming signals. Fortunately, at least for the Sun, the random variations from active region evolution and the quiet Sun tend to be much smaller in amplitude than mass-loss dimming. The main source of interference in the spatially unresolved observations is the emission from the associated flare.

On the Sun, we can spatially resolve the coronal emission, and therefore distinguish between the flaring and the dimming components that are typically both present in strong flare/CME events. Veronig et al. (2021) made a systematic study of all flares ≥M5 in Sun-as-a-star observations by the Extreme ultraviolet Variability Experiment (EVE) onboard SDO, along with simultaneous EUV imaging by SDO/AIA. They found that the emission enhancement caused by the flare and the emission decrease due to the dimming start roughly simultaneously, with the flare first dominating the overall Sun-as-a-star signal. However, the dimming is longer lasting than the flare and eventually (after about 1–2 hours) becomes the dominant component (see Figure 5.14). The competing time scales here are energy release and the cooling times of the flare, versus the recovery time of the corona after a CME. Based on their solar analysis, Veronig et al. (2021) found that if a dimming is observed in full-Sun light curves in the aftermath of a large flare, the probability that a CME occurred is ~90%, suggesting that coronal dimmings are good proxies for CMEs for Sun-like stars.

Mason et al. (2019) searched four years of the SDO/EVE Sun-as-a-star data for coronal dimmings corresponding to flares ≥C1 and characterized the dimming light curves in terms of depth, slope, and duration. They found that many dimmings occur at the ambient coronal temperature. However, they also found dimmings in emission from plasma at temperatures across the full measured range ($3.84 \leq \log T \leq 7.0$) and with magnitudes as deep as 10% (though the majority occur around 1% depth).

### 5.5.3.2 On other stars

Stars more magnetically active than the Sun might be capable of producing more extreme CMEs and deeper associated dimmings. If detectable, they should be found in wavelengths sensitive to the temperature of the stellar corona of each star, which can differ from that of the Sun (Jin et al., 2019). Indeed, Veronig et al. (2021) found evidence of many events in archival X-ray and EUV data with depths greater than 10% (see Figure 5.16). In total, they found 21 candidates for CME-induced dimmings in about 200 data sets of late-type flaring stars. About half of the detections were from three stars: AB Dor, AU Mic, and Proxima Centauri. However, even non-detections of dimming provide constraints on the size (mass) and rate of CMEs. For example, Loyd et al. (2022) studied archival Hubble Space Telescope (HST) observations of $\varepsilon$ Eridani in the FUV to determine that CMEs $\gtrsim 10^{15}$ g of 1 MK plasma must occur less than a few times per day. Recently, Namekata et al. (2024) reported a post-flare dimming in X-ray emission from EK Draconis with an amplitude of about 10% that accompanied strong evidence of an erupting prominence based on H$\alpha$ Doppler shifts.

A wide range of additional differences between stars and the Sun will influence the characteristics of coronal dimming. The total mass of plasma in the corona, its spatial distribution, and its density structure will influence the depth of a coronal dimming produced by a CME of a certain mass and vice versa (Loyd et al., 2022), even in a highly simplified model like Figure 5.14. The magnetic structure of the star is another key difference. In particular, stronger large-scale fields on more active stars (especially M dwarfs) might be capable of magnetically confining an eruption, preventing the escape of a CME while still producing a dimming signature (Alvarado-Gómez et al., 2019). More work is needed to determine whether magnetically confined "failed" eruptions also produce coronal dimming signatures. Sun-as-a-star data could empirically address this question for Sun-like stars.

For observations of stellar mass-loss dimmings, the major issue is the availability of data sets, in terms of continuous high-cadence time series at EUV or SXR wavelengths with high sensitivity. The proposed ESCAPE (Extreme-ultraviolet Stellar Characterization for Atmospheric Physics and Evolution; France et al., 2022) mission would fill this gap, using EUV and FUV spectroscopy and a planned deep monitoring program for selected targets (spectral type F to M) for stellar CME occurrence via their dimming signature. Also, the next advancement in X-ray observatories is likely to enable better measurements, such as line-resolved dimming.

### 5.5.4 Obscuration Dimming

Obscuration dimming occurs when material ejected by a CME or filament eruption temporarily absorbs background emission in the line of sight. This transient effect creates a distinct observational dimming signature, as illustrated in Figure 5.13 (right panels). Obscuration dimmings are observed on the Sun, and their investigation in the stellar context is of great interest as a potential indicator of CMEs.

#### 5.5.4.1 On the Sun

Solar filament eruptions often produce dimming signatures in imaging observations, such as in He II 304 Å. Unlike mass-loss dimming, which arises from coronal plasma depletion associated with CMEs and is observed across multiple coronal wavelengths (e.g., Fe IX 171 Å, Fe XII 195 Å) sensitive to the background corona—obscuration dimming occurs when cooler, optically thick filament material in the corona absorbs background emission, producing the dimming signature (Mason et al., 2014, see Figure 5.13, right panel). Despite its distinct characteristics, a direct comparison of the properties of mass-loss and obscuration dimming in relation to CMEs remains an open research question.

The first detection of solar obscuration dimming was made in microwave emission (Gopalswamy & Yashiro, 2013), resulting from a prominence eruption. More recently, Xu et al. (2024) identified six cases of obscuration dimming within more than a thousand filament eruptions in full-disk He II 304 Å observations (Figure 5.17). The dimming regions covered approximately 3% to 6% of the solar disk, with dimming depths ranging from 1% to 6% and durations spanning 0.4 to 7.0 hours. Notably, a positive correlation was found between dimming depth and area, which may help constrain filament sizes in stellar observations.

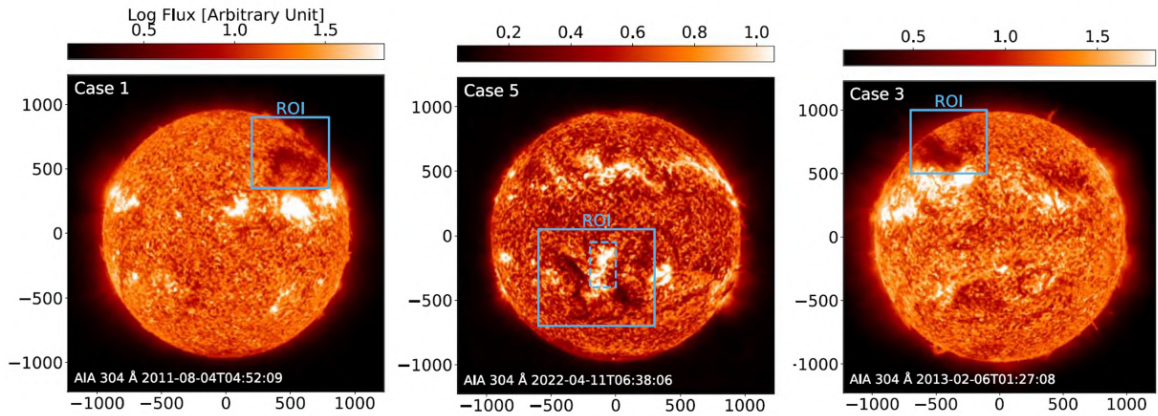

Figure 5.17: Examples of filament eruption cases on the Sun showing filament material that is obscuring part of the solar disk leading to detectable dimming signatures in Sun-as-a star flux curves of He II 304 Å. Adapted from Xu et al. (2024).

Beyond optical, EUV, and X-ray observations, "negative radio bursts" or "radio dimmings" present an intriguing possibility for detecting obscuration dimming in stellar contexts (see Figure 5.18). Grechnev et al. (2008) provided a detailed analysis of a filament eruption that covered one-fourth of the visible solar surface. The event showed numerous indications for a CME, including ejected material appearing in coronagraph images, Moreton and coronal waves, a type II radio burst, and long-lived dimmings observed with SOHO Extreme ultraviolet Imaging Telescope (EIT) at 304 Åand 195 Å, as well as in Hα. Furthermore, radio observations at $1 - -9.4$ GHz revealed a pronounced dip in flux during the decay phase of the associated flare, lasting approximately an hour. The depth of this dip varied with frequency, increasing towards lower frequencies, reaching ∼10% at 1 GHz and ∼5% at 5 GHz, with no apparent decrease at the highest frequency. Grechnev et al. (2008) modeled this effect as free-free absorption caused by overlying material and successfully matched the observed frequency-dependent dimming depths. Their mass estimates, in the range of $3 - 4 \times 10^{15}$ g, were consistent with other absorption-based measurements within methodological uncertainties. Their results suggest that the erupting material split into two components, one successfully escaping as a CME and the other draining back to the solar surface. Hou et al. (2025) published an analysis of two more recent negative radio bursts at lower frequencies (100–450 MHz, meter and decimeter wavelengths) that are also consistent with free-free absorption (Figure 5.18).

While not a primary focus in solar studies, obscuration dimming holds significant potential in the Sun-as-a-star framework, especially for advancing the detection of stellar CMEs. For example, it remains to be explored whether obscuration dimming occurs in solar X-rays, particularly during energetic events with dense, cool overlying material. Future research could also focus on modeling how filament-related absorption, clearly observed in solar EUV and UV data, would appear in unresolved stellar light curves and spectra. Moreover, exploring tell-tale signatures to distinguish between obscuration and mass-loss dimming—such as differences in timing, duration, or morphology—may help disentangle filament motion from overall CME dynamics in unresolved solar and stellar observations.

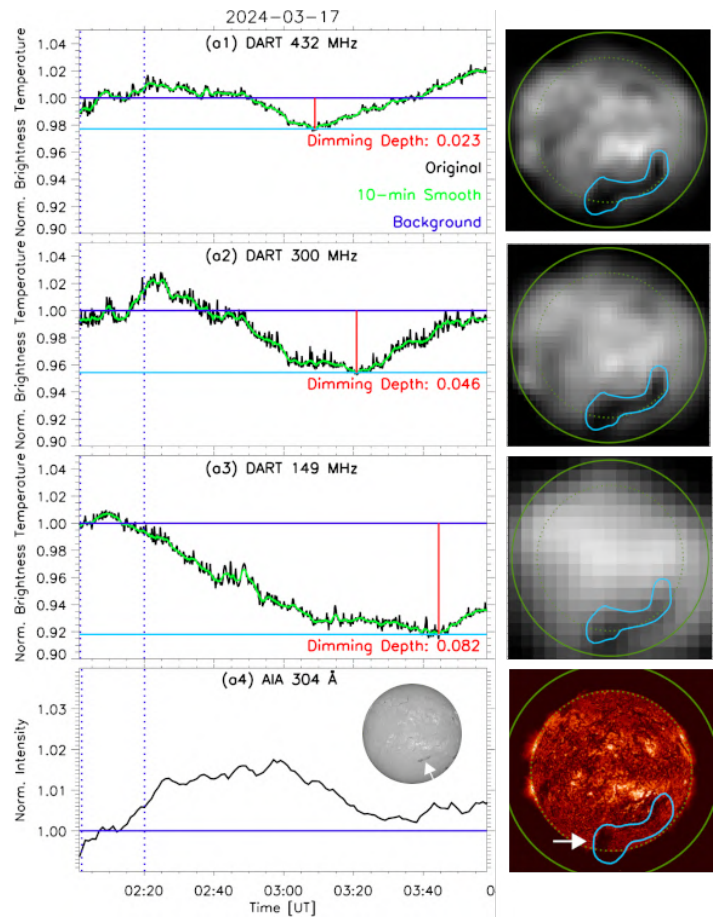

Figure 5.18: Example of "negative radio burst" from a filament eruption occurring on 2024 March 17. *(left)* Light curves at different radio frequencies and the AIA 304 Å (helium emission) band, normalized to the pre-flare intensity, showing the appearance of a frequency-dependent "dip" during the decay phase. *(right)* Images of emission in the same bands during the eruption. Within the bottom-left panel is an image in Hα made with the Chinese Hα Solar Explorer (CHASE) observatory, clearly identifying the filament structure. Credit: Figure adapted from Hou et al. (2025).

Obscuration dimmings associated with failed solar eruptions could provide valuable insights into magnetic confinement on other stars.

### 5.5.4.2 On other stars

A promising method for detecting stellar CMEs involves measuring transient increases in hydrogen column density ($N_H$) in X-ray spectra, commonly referred to as obscuration dimming, absorption dimming, or X-ray continuum absorption (Moschou et al., 2017, 2019a; Osten, 2023). If an increase in $N_H$ follows a stellar flare, it may serve as evidence that a CME is obscuring the line of sight.

Early studies suggested transient X-ray absorption signatures linked to stellar CMEs and prominence eruptions (Haisch et al., 1983; Ottmann & Schmitt, 1996; Tsuboi et al., 1998; Franciosini et al., 2001; Pandey & Singh, 2012). Favata & Schmitt (1999) reported a significant increase in the absorbing hydrogen column during a large flare in Algol. Although deficiencies in plasma codes were considered as a possible explanation, the authors deemed local absorbing material to be the more probable cause.

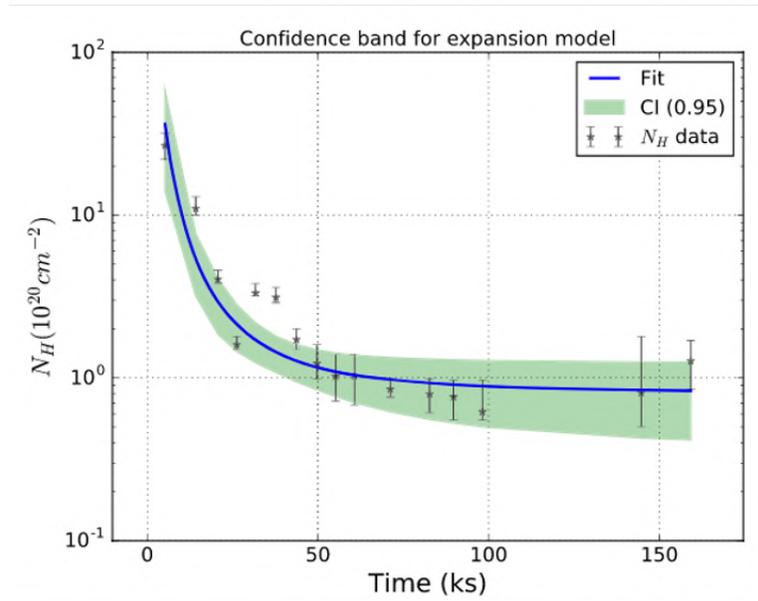

Figure 5.19: Temporal variation of hydrogen column density through the decay phase of a large X-ray flare seen on the active binary system Algol reported in Favata & Schmitt (1999) and analyzed in Moschou et al. (2017) as the passage of a CME over the X-ray flaring loops.

Moschou et al. (2017) expanded on this analysis and found that the hydrogen column decreased over time, approximately following a $t^{-2}$ dependence (see Figure 5.19). By modeling NH variations as a self-similar expansion of a CME front with uniform velocity, they estimated CME masses in the range of $2-20 \times 10^{21}$ g, several orders of magnitude larger than even the most extreme solar CMEs. A follow-up study by Moschou et al. (2019a) analyzed seven stellar CME candidates using X-ray absorption, deriving CME mass estimates between $1.2 \times 10^{16}$ g and $1.0 \times 10^{23}$ g.

Further investigation of possible obscuration dimming from stellar CMEs would greatly benefit from multiwavelength observations, particularly in the (E)UV and radio bands. Radio observations might offer a promising but underutilized alternative for detecting stellar obscuration dimming, given the "negative radio bursts" observed on the Sun. Next-generation radio arrays such as the Square Kilometer Array and upgrades to existing solar-dedicated facilities may improve prospects for detecting radio absorption events associated with stellar eruptions. Analyzing X-ray dimmings both in line and continuum emission could help establish correlations between signatures due to mass loss and obscuration. Such studies would enhance the credibility of stellar CME detections and provide further constraints on their physical parameters. Current X-ray instruments capable of detecting absorption dimming from stars include the High Throughput X-ray Spectroscopy Mission and the X-ray Multi-Mirror Mission (XMM-Newton), NASA's Chandra X-ray Observatory, NASA's Neutron Star Interior Composition Explorer (NICER), as well as the newly launched Einstein Probe and XRISM. Future missions such as ESA's New Advanced Telescope for High ENergy Astrophysics (NewAthena) (Cruise et al., 2025) could significantly advance the search for stellar CMEs using this method by providing improved sensitivity and spectral resolution.

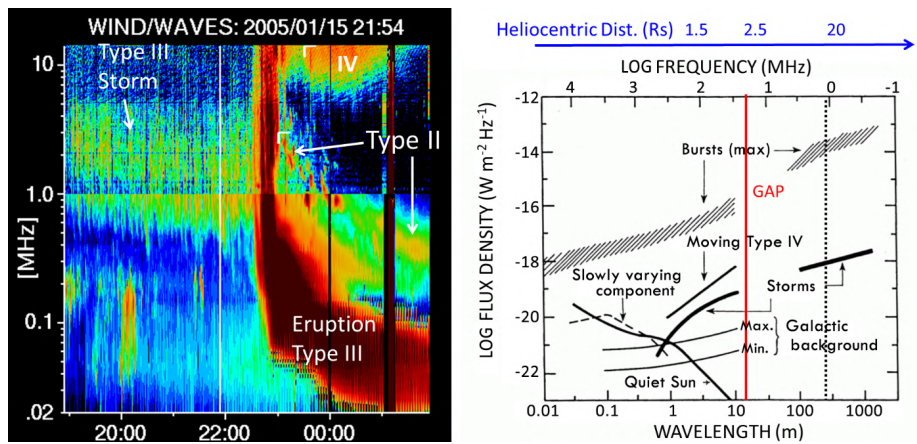

Figure 5.20: *(left)* The four types of low-frequency radio bursts (14 MHz to 20 kHz) from Wind/WAVES. Figure from Gopalswamy et al. (2019). *(right)* Summary of solar radio flux densities at all wavelengths. Wind/WAVES data fill the gap between 2 and 14 MHz that existed for decades before the launch of Wind in 1994. The vertical red and black (dashed) lines denote 14 MHz and 1 MHz, respectively. As noted at the top of the flux density plot, ground-based radio telescopes observe radio emission originating below $\sim$2 $R_\odot$ from the Sun, while the space-borne telescopes observe radio bursts originating beyond 2 $R_\odot$, extending to Earth orbit and beyond. Figure from McLean & Labrum (1985).

### 5.5.5 Radio Emission from CMEs: A Primer

CMEs and associated flares can produce a variety of nonthermal radio emission, both coherent and incoherent in nature and spanning frequencies from kHz to GHz (rarely reaching THz). Of particular note are the various types of bright radio bursts that were among the first phenomena discovered in the early days of radio astronomy in the 1940s. They are classified as type I, II, II, IV, and V, based on their appearance in a radio dynamic spectrum, with type II, III, and IV bursts relevant in the discussion of CMEs. Type II radio bursts are due to electrons accelerated in shocks driven by CMEs. Type III radio bursts are due to electrons propagating along open magnetic field lines from the electron acceleration site. Type IV bursts are caused by nonthermal electrons trapped in magnetic structures, which can be stationary, such as post-flare loops, or moving, such as CMEs. Figure 5.20 shows an event in which type II, III, and IV bursts all occurred. We discuss type II and type III bursts in Section 5.5.6 & Section 5.5.7.

Most burst types are due to coherent emission mechanisms, either due to plasma radiation at the fundamental or second harmonic of the plasma frequency ($\approx 9000 n^{1/2}$ [Hz], where $n$ is the electron number density [cm$^{-3}$]), or due to electron cyclotron maser emission at the electron gyrofrequency ($\approx 2.8 \times 10^6 B$ [Hz], where $B$ is magnetic field strength). These bursts are therefore exceptional diagnostics of the local plasma density or magnetic field strength at the location of origin of the emission. In addition to coherent bursts, there can also be incoherent thermal and nonthermal (Section 5.6.3) components of CMEs; the latter are often classified as type IV bursts.

### 5.5.6 Type II Bursts

#### 5.5.6.1 On the Sun

The most common radio signatures associated with solar CMEs are type II radio bursts, which are coherent plasma radiation produced by super-Alfvénic shocks propagating in the corona and sometimes well out into the interplanetary medium. The observed behavior of a type II burst in a dynamic spectrum, typically due to plasma radiation, traces the density in the solar corona as the associated shock propagates through the corona and IPM. Both the velocity of the shock and the density profile of the Sun are therefore encoded in the burst morphology in a dynamic spectrum, revealing shock speeds of ∼100s–1000s of km/s.

Type II bursts have been observed from decimetric to hectometric wavelengths by radio spectrographs and have been imaged at discrete frequencies (e.g., Bain et al. (2012)). However, their exact relation to the underlying CME driver has remained murky. In recent years, instruments like the general-purpose Low-Frequency Array (LOFAR) and the Murchison Widefield Array (MWA) have produced multi-frequency imaging of type IIs (Morosan et al., 2019b). Determining the relationship between type II radio bursts and the shock driver is a key science goal of arrays being commissioned at long wavelengths dedicated to solar observing, or with dedicated solar observing modes. These include the Owens Valley Radio Observatory Long Wavelength Array (OVRO-LWA; 15–88 MHz) and the Sun Radio Interferometer Space Experiment (SunRISE) Kasper et al. (2022), which will use a space-based interferometric array covering 0.1–25 MHz.

Because of their relationship with super-Alfvénic shocks, type II radio bursts with starting wavelengths in the metric to decametric regime (i.e., forming at ∼ $1.5 - 10 \, R_\odot$) are excellent indicators of the presence of fast solar CMEs. Gopalswamy et al. (2001) found that nearly *every one* of the 103 type II radio bursts observed by Wind/WAVES in the decametric-hectometric (DH) wavelengths for ≳5 years is associated with a white light CME recorded by the SOHO/LASCO coronagraph. The radio spectrograph–coronagraph combination showed that CMEs producing type II bursts are more energetic and likely travel far from the Sun, increasing the probability of impacting planets. Type II bursts extending to kilometric wavelengths are associated with CMEs responsible for producing SEPs, resulting in ground-level enhancements (GLEs, rises in the cosmic ray intensity registered by ground-based detectors; Gopalswamy et al., 2005b). This CME population has an average sky-plane speed of ∼2000 km/s, about five times faster than an average CME. The hierarchical relationship between CME speed (or kinetic energy) and the wavelength range of type II bursts indicates a close connection between shock strength and duration of particle acceleration (Gopalswamy et al., 2005a; Gopalswamy, 2006).

The fact that type II bursts extending beyond the metric domain indicate stronger shocks has been illustrated in terms of SEP association. CMEs accelerating impulsively and attaining high speeds close to the Sun (within a couple of $R_\odot$) result in high-energy SEPs. The shock formation coincides with the onset of a type II radio burst (Gopalswamy et al., 2012), indicating electron acceleration to tens of keV. Cliver et al. (2004) found that only 25% of the western hemispheric type II bursts occurring only in the metric domain were associated with large SEP events. When the metric type II bursts were accompanied by DH type II bursts, the SEP association rate increased to 90%. Their DH type II bursts include all type II events with an ending frequency below 14 MHz. Because m-km type II bursts are associated with very fast CMEs, they must have the best association with SEP events.

### 5.5.6.2 On other stars

Although type II radio bursts hold promise as proxies for stellar CMEs, they have yet to be detected despite various attempts using ground-based observations (e.g., Crosley et al., 2016; Villadsen & Hallinan, 2019; Crosley & Osten, 2018a,b). Three major factors which may be responsible for this are: 1) the sensitivity and accumulated observing time required to unambiguously associate radio emission with a type II burst (e.g., via structure in dynamic spectra) at the relevant wavelengths has not yet been delivered by current instrumentation, and 2) the stars previously prioritized for targeted observations may be confining CMEs or inhibiting the shock formation necessary for type II bursts.

It is generally accepted that type II bursts require the presence of a shock. The strong magnetic fields of magnetically active stars could be an inhibitor in this respect—the Alfvén velocity associated with a kG field can be upwards of an order of magnitude higher than the highest CME speeds observed from the Sun (Alvarado-Gómez et al., 2020b). Although a shock may develop further in the stellar corona where the ratio of the plasma density to the magnetic energy density is conducive to shock formation, the distance in the corona—and the associated plasma density—may produce plasma emission that is not observable from ground-based radio observatories due to the ionospheric cutoff (Villadsen & Hallinan, 2019). Because of this, detecting type II bursts (and other types) from such systems would require a high-sensitivity, space-based or lunar array. Detecting stellar type II/III radio bursts is one of the key proposed science cases for the Farside Array for Radio Science Investigations of the Dark Ages and Exoplanets (FARSIDE) radio telescope, for example (Burns & Hallinan, 2020). To detect stellar type II and III bursts from the ground, active stars with moderate global field strengths (10s-100s G), may be ideal. Such stars may be more likely to accelerate material that is both fast enough to drive shocks and strong enough to escape magnetic fields overlying the acceleration region (Alvarado-Gómez et al., 2019).

A typical solar type II flux density ($\sim 10^5\,\mathrm{SFU} = 10^9\,\mathrm{Jy}$) corresponds to a required sensitivity of $\sigma_d \approx 4\,d_{\mathrm{pc}}^{-2}\,\mathrm{mJy}$ for a $5\sigma$ detection. The LOFAR low-band array (LBA) achieves a sensitivity of $\sim 4\,\mathrm{mJy}$ when integrated over $\sim 30\,\mathrm{min}$ and between 42-66 MHz (de Gasperin et al., 2023). Given that type II bursts last on the order of tens of minutes to an hour at these frequencies, this sensitivity should be sufficient to *detect* a type II burst, but not to provide the spectral and time resolution required to resolve the characteristic frequency drift of the burst. Long-term monitoring of hundreds to thousands of stars may facilitate the detection of rarer, brighter bursts with sufficient sensitivity to be characterized and classified. New generations of low-frequency (<100 MHz) radio telescopes with ultra-wide or all-sky field of view, such as the Owens Valley Radio Observatory Long Wavelength Array View (OVRO-LWA), New Extension in Nançay Upgrading LOFAR (NenuFAR), and Amsterdam-ASTRON Radio Transients Facility and Analysis Center (LOFAR/AARTFAAC), have the capability for broad searches, though data processing bottlenecks remain an obstacle.

Both coherent and incoherent stellar radio bursts can be orders of magnitude brighter than solar radio bursts (Güdel, 2002). While it is unclear how exactly plasma emission intensity might scale with stellar properties like, for example, ambient wind density and field strength, it is likely not unreasonable that the plasma emission from, for example, younger, more active stars would be more luminous than the emission we see from the Sun. Regardless, nearby targets with moderately strong field strengths should be prioritized for ground-based searches.

Although solar type II bursts can occur as high as 1 GHz, the most intense bursts occur at frequencies below 15 MHz (i.e., below the ionospheric cutoff). Such frequencies, therefore, should be well-suited for detecting stellar type II bursts and hence inferring stellar CMEs. Recently, Mohan et al. (2024) compiled the properties of CMEs and flares associated with interplanetary (IP) type II bursts and obtained the following relationship:

$$\sqrt{L_X \phi_{\mathrm{rec}}} = 10^{12.5 \pm 8.5} (\sqrt{L_R V_{\mathrm{CME}}^2})^{0.76 \pm 0.04} \tag{5.2}$$

where $L_X$ is the soft x-ray luminosity of the flare, $\phi_{\mathrm{rec}}$ is the total reconnection flux obtained from the magnetic flux underlying the post-eruption arcade, $L_R$ is the type II peak radio flux in the 3-7 MHz range, and $V_{\mathrm{CME}}$ is the "matured" speed of the CME averaged over the coronagraphic field of view. The essence of the equation is that flare power (left side of the equation) is proportional to the CME power (right side of the equation). The relationship between the flare power and CME power for

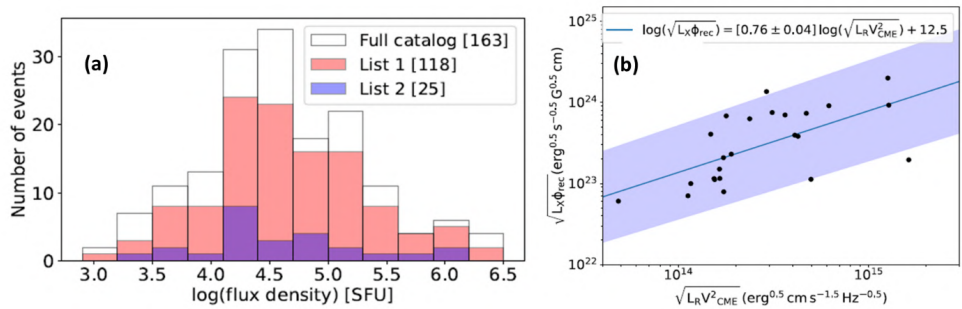

Figure 5.21: (a) Radio flux of 163 DH type II bursts observed by Wind/WAVES and/or STEREO. Of these, List1 is a subset of 118 events that have a good estimation of $L_X$ and $V_{CME}$. List 2 is further filtered based on the reliable measurement of all the parameters: $L_X$, $L_R$, $V_{CME}$, and $\phi_{rec}$. (b) The scaling law between flare power and CME power. The shaded region represents the fitting uncertainty interval. Adapted from Mohan et al. (2024).

163 IP type II bursts is shown in the right panel of Figure 5.21. This relationship is analogous to the Güdel-Benz relation that links the flare thermal power and the particle acceleration strength driven by flare reconnection in that a connection between the coronal particle acceleration (flare reconnection) and IP particle acceleration (CME-driven shock) is found. The relationship provides a robust way to link sun-as-a-star metrics from a CME event to spatially resolved parameters that are not observable in stars.

### 5.5.7 Type III Bursts

#### 5.5.7.1 On the Sun

Type III bursts originate from plasma that has been excited by mildly relativistic ($v \sim 0.2 - 0.3$c) electrons accelerated along magnetic field lines. Like type II bursts, type III bursts are a result of the plasma emission mechanism and, as such, trace the plasma density profile as the accelerated electrons propagate through the corona and IPM. Because the electrons driving the plasma instability to produce the emission move much faster than CMEs, the signature of type III bursts is much steeper and narrower in time-frequency space than type II bursts; Alvarez & Haddock (1973a) describe their drift rate in the range between 50 kHz to 550 MHz as $\dot{v} = -0.01 v_{MHz}^{1.84}$ MHz/s and Alvarez & Haddock (1973b) report the duration as $t_d = 10^{7.7} v^{-0.95}$. Additionally, the magnetic field's role in accelerating the electrons means the emission can be much more highly circularly polarized than type II bursts, upwards of 70% Dulk (1985).

Although type III bursts can occur without any clear particle enhancement being observed at Earth, a significant fraction ($\gtrsim 90\%$) of SEPs are associated with a type III burst (Cane et al., 2002; Duffin et al., 2015; Winter & Ledbetter, 2015). One of the distinguishing characteristics of the type III bursts associated with SEPs is that they are of long duration, typically more than 15 minutes. However, the reverse is not true—there exist long-duration type III bursts without SEP events. These events are often accompanied by CMEs, but those CMEs do not produce a shock. This suggests that the SEPs associated with long-duration type III bursts are produced from shock acceleration. Figure 5.22 shows the populations of SEP-associated and non-associated type III bursts, along with a potential correlation between SEP fluence and burst duration.

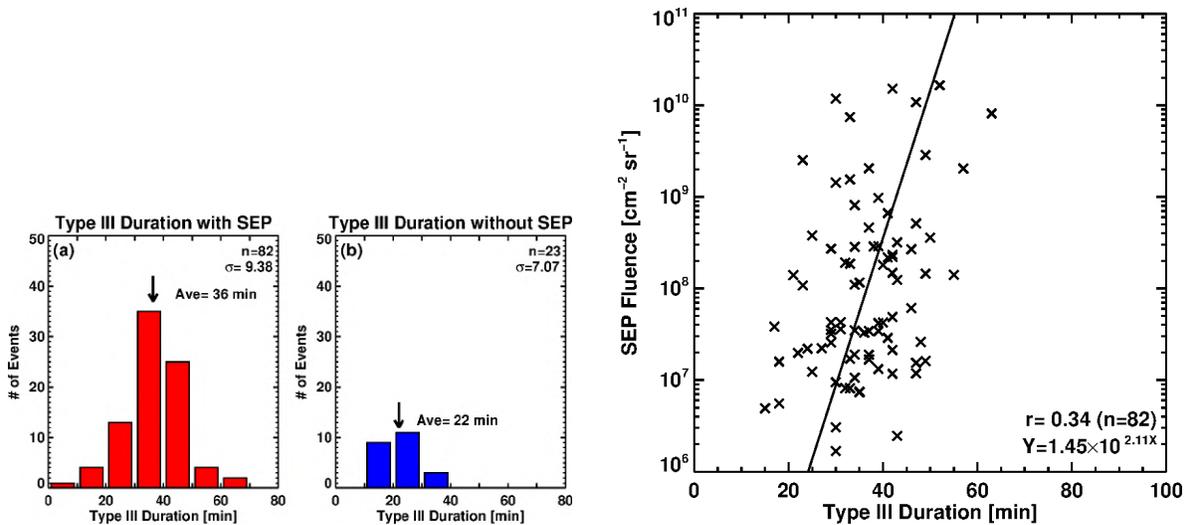

(a) Histograms of type III burst durations separated by their association with an SEP event. Type III bursts with SEPs have an average duration of ~36 min compared to 22 min for those without SEPs.

(b) Relationship between type III duration and SEP fluence for a large number of events. A weak but significant correlation is observed, which merits further investigation.

Figure 5.22: Statistics of SEPs and type III burst duration. Figures by Nat Gopalswamy.

It is quite common to see type III bursts without a detected SEP event due in no small part to geometry. To observe SEPs directly, there must be a magnetic connection between the observer and the SEP acceleration site. Because the accelerator for type III bursts is much narrower than for type II bursts, the probability of being in the right place to detect associated SEPs is lower.

Additionally, type III bursts can also be produced by SEPs traveling on closed field lines. Such bursts can have a distinctive reversal of frequency drift resulting in J and U bursts, named for their shape in the dynamic spectra (Maxwell & Swarup, 1958; Reid & Kontar, 2017). Because these SEPs are confined in the solar corona, their impacts on the larger Solar System environment are limited.

Overall, type III bursts have limited use as a tool for space weather forecasting. The observation of a type III burst does not indicate which magnetic field lines particles have been accelerated along and whether that line might steer those particles to impact a planet. That said, type III bursts serve as an unambiguous indicator of accelerated particles and, depending on the frequency of the emission, can be attributed to particles that have propagated into the IPM. Furthermore, they're important diagnostic tools for studying the coronal and IPM density structures.

### 5.5.7.2 On other stars

For stars, some type III, or type III-like, bursts have been observed. Osten & Bastian (2006) reported radio bursts from the nearby M dwarf AD Leo at 1.4 GHz, observing highly circularly polarized, positively and negatively drifting bursts over the 1120-1620 MHz range on timescales of tens of milliseconds. More recently, Mohan et al. (2024) report a possible detection of type III bursts from AD Leo between 550-850 MHz with the Giant Metrewave Radio Telescope (GMRT).

However, the time resolution required to resolve individual bursts and their drifts—especially at these frequencies—is expected to be relatively high. Indeed, the GMRT's resolution was insufficient to resolve the bursts reported by Mohan et al. (2024). Without this resolution, it becomes increasingly difficult to determine whether the emission is a type III burst. While a highly circular polarization fraction can be used to determine whether the emission is coherent, it is difficult to determine which coherent emission mechanism (either plasma emission or electron cyclotron maser emission) is responsible without first knowing, at a minimum, either the magnetic field structure and strength or the plasma density profile—the latter of which is especially poorly constrained for any star besides the Sun.

To assess the likelihood of being able to detect a type III burst on another star, we can start with the most extreme solar events, which reach flux densities as high as $10^{12}$ Jy at 1 MHz and would be closer to $\approx 10^{11}$ Jy at frequencies just above the Earth's ionospheric plasma cutoff (tens of MHz) (Dulk, 2000; Saint-Hilaire et al., 2013; Sasikumar Raja et al., 2022). A $5\sigma$ detection of this would require a sensitivity of $470 d_{pc}^{-2}$ mJy where $d_{pc}$ is the distance to the star in parsecs. Because type III bursts are so brief at frequencies above the Earth's ionospheric cutoff, this sensitivity needs to be achieved for integration times on the order of 1-10 s. The LOFAR low-band array (LBA) achieves $\sim 100$ mJy sensitivity on 1 s integration times when integrated between 42-66 MHz. As such, it should be capable of detecting an extreme solar type III burst out to a few parsecs, albeit with little to no information on the time-frequency structure of the burst with which to define it as a type III burst.

Without stellar examples to derive burst statistics, we can only derive anticipated burst rates by using solar statistics. Saint-Hilaire et al. (2013) provide burst statistics from observations made with the Nançay Radio Heliograph (NRH) between 150-400 MHz. Paired with the type III detection rate as a function of distance as provided by Vedantham (2020), this enables an estimate for the detection rate of type III bursts from a given star. Vedantham (2020) show detection rate as a function of distance $d_{pc}$ and flux density threshold $S_{mJy}$ can be expressed as:

$$N(S_{mJy}) = 0.17 d_{pc}^{-1.38} S_{mJy}^{-0.69} \text{ day}^{-1} \text{ (10 s integration)} \tag{5.3}$$

$$N(S_{mJy}) = 0.83 d_{pc}^{-1.38} S_{mJy}^{-0.69} \text{ day}^{-1} \text{ (1 s integration)} \tag{5.4}$$

The higher detection rate estimate for a 1 s integration compared to a 10 s integration arises from the $\sim 1\,s$ duration of solar type III bursts at these frequencies, causing longer integration times to dilute the observed flux density and reduce detectability. These expressions are not representative of burst rates at frequencies below about 100 MHz, even in the solar case. At lower frequencies, type III bursts tend to have both higher flux densities and last longer. Assuming a $10^{11}$ Jy flux density for type III bursts, Vedantham (2020) estimates that LOFAR's all-sky detection rate for the high-band array (HBA) would be $N \approx 0.06 \text{day}^{-1} (4\pi \text{sr})^{-1}$ out to 10 pc.

High time and spectral resolution at low frequencies would enable measurements of the rate of frequency drift during a burst. This may be true even if the drift cannot be observed in a dynamic spectrum. "De-dispersing" the signal before integrating in frequency—as is done for pulsar searches— can strengthen a signal if the de-dispersion parameters are tuned to the drift rate of the burst.

Overall, the advent of more sensitive instrumentation, the existence of candidate detections, and the possible employment of de-dispersion for type III burst searches make the future of studying their stellar counterpart promising.

### 5.5.8  Planetary Aurorae as Probes of Transient Events

The detection of auroral emission (introduced in Section 2.4) would be a strong indication of direct particle effects on extrasolar planetary atmospheres. In particular, time variability in this emission could indicate the presence of space weather such as CMEs. The pressure from a CME can compress planetary magnetospheres and drive reconnection in both the nose and the tail (Dungey, 1961; Cowley et al., 2004; Fuselier et al., 2014) of the planetary magnetopause, resulting in particle acceleration and enhanced aurorae. Particularly telling would be a stellar flare followed by an enhancement in planetary emission consistent with the expected travel time for a CME. With further theoretical development, these emissions may provide some constraints on the physical properties of the event, such as the energy of the precipitating particles in the case of $H_3^+$ emission (Pineda et al., 2024).

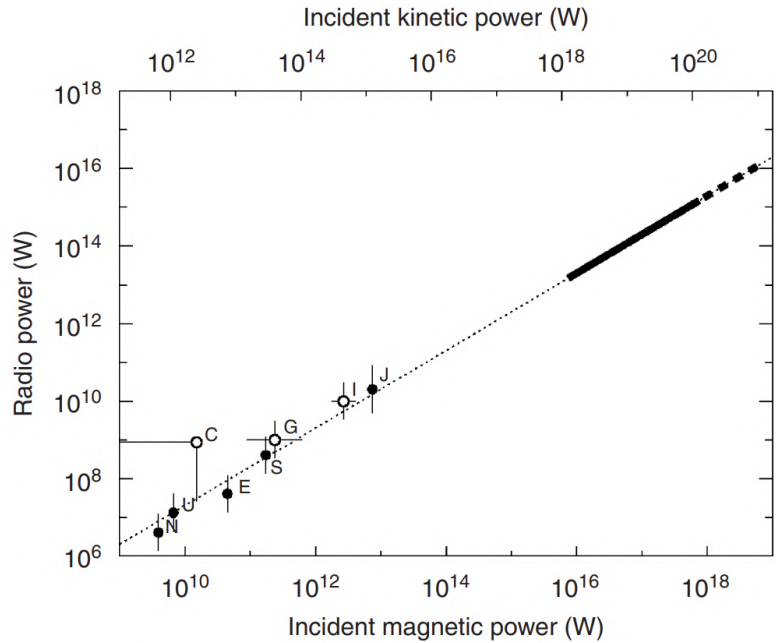

Figure 5.23: Graphical representation of the radiometric Bode's law that shows how emitted radio power scales with incident solar wind power. Figure 6 from Zarka (2007a).

For auroral radio emission via ECMI, however, some footing is provided by the radiometric Bode's law, shown in Figure 5.23 and described in detail in Lazio et al. (2004). Based on Solar System objects, this relates the auroral radio emission from a planet to the incident magnetic and kinetic power due to the solar wind impacting the planetary magnetosphere to order-of-magnitude precision. In this description, the kinetic power scales linearly with the number density of the impinging material (i.e., the solar wind or the solar wind and a CME), although the conversion efficiency from kinetic power to radio power is low ($\sim 10^{-5}$) (Lazio et al., 2004). The power of radio emission is also influenced by the density of planetary plasma, which can vary with the incident flux of ionizing photons, a potentially confounding factor.

It has also been proposed that optical signatures of enhanced aurorae associated with CME interaction may be detectable and offer unique opportunities to characterize the atmospheres of terrestrial exoplanets. Luger et al. (2017) estimate the power in the O I 5577Å auroral line from Proxima Cen b, assuming an Earth-like atmosphere, to reach a total power of $\sim$ 1 TW–10 TW during periods of strong magnetospheric disturbance, orders of magnitude higher than observed for Earth. The planet-star contrast ratio in this line may be as low as $10^{-4} - 10^{-5}$, with detection possible in < 1 day for space-based coronagraphic telescopes such as the proposed Habitable World Observatory. In addition to providing a potentially unique avenue for biosignature detection, such observations provide an indirect measure of the stellar wind conditions during a CME.

### 5.5.9  Magnetic Star-Planet Interactions

In the Solar System, the stellar wind at planets is nearly always super-Alfvénic, preventing direct magnetic interactions between the planets and the Sun. Many of the closely orbiting exoplanets, however, likely orbit within the sub-Alfvénic regime of their host star's wind. The resulting interaction, dubbed "star-planet-interaction" (SPI), is capable of generating radio emission via ECMI (see Sect. 2.4) that originates from the star itself, drawing parallels to the sub-Alfvénic interaction between Jupiter and its moon Io (Neubauer, 1980; Zarka, 1998, 2007b; Saur, 2004; Grießmeier, 2007; Zarka, 2018). Here, the host star and planet mimic Jupiter and Io. If the planet orbits within the sub-Alfvénic regime of its host star's wind, it can generate Alfvén waves that travel back towards the star (Ip et al., 2004; McIvor et al., 2006; Lanza, 2012; Turnpenney et al., 2018; Strugarek et al., 2019; Vedantham et al., 2020). Some of the wave energy produced in this interaction is expected to dissipate and generate radio emission via ECMI in the stellar corona.

Albeit not yet confidently observed in radio wavelengths, magnetic SPI has been observed in several systems at blue optical wavelengths (e.g., Shkolnik et al. 2005). For reviews on the observational and theoretical aspects of these, see Shkolnik & Llama (2018) and Strugarek & Shkolnik (2025), respectively.

As proposed by Alvarado-Gómez et al. (2022b), SPI and ECMI might provide a non-conventional way to detect flare-associated CME activity in close-in planet-hosting stars. The general idea is that the sub-Alfvénic conditions experienced by the planet that allow for the generation of radio waves via SPI-ECMI would be temporarily lifted by the large perturbations in plasma density, velocity, and magnetic field carried by an incoming CME. In this way, the space environment of the close-in planet would change from sub-Alfvénic to super-Alfvénic on a time scale comparable to the passage of the CME through the planetary magnetosphere (intrinsic or induced). This is illustrated in Figure 5.24, which shows results from 3D MHD simulations of a highly energetic CME from the flare star AU Mic as it travels and impacts planet b of this system (Cohen et al., 2022). The planet is assumed to be magnetized with a dipole magnetic field comparable to Earth (0.3 G).

Assuming that a pre-eruption SPI-ECMI emission baseline can be detected and isolated from other stellar sources, the CME-driven transition from sub-Alfvénic to super-Alfvénic would temporarily shut down the associated radio emission. This emission should reappear once the system returns to its steady, pre-CME state. Therefore, the SPI-ECMI emission would show a "dimming event" similar to the coronal dimming signatures discussed in Section 5.5.3, during the CME passage. Moreover, if such a CME event is linked to a stellar flare, the time difference between the flare onset and the beginning of the ECMI radio dimming could provide an estimate of the CME speed, assuming the exoplanet's orbit is known.

### 5.5.10  Exoplanet Outflow Variability

Highly irradiated planets with hydrogen-rich atmospheres can produce observable hydrodynamic outflows that interact with the stellar wind through ram pressure and charge exchange (see Sections 2.2 and 4.3.4). Temporal and spatial variability in the stellar wind, including CMEs and stream or corotating interaction regions, will influence the mass loss rate and structure of planetary outflows. This could be an avenue to observing evidence of transient space weather, and perhaps even to placing some physical constraints on the events observed. However, the viability of this may be limited by a confounding effect: variability in the stellar radiation, such as from the flares that often accompany solar CMEs. The outflow from the hot Jupiter HD 189733 b has been extensively studied in this regard due to variability in the transit produced by its outflow in the neutral hydrogen Lyman-$\alpha$ line

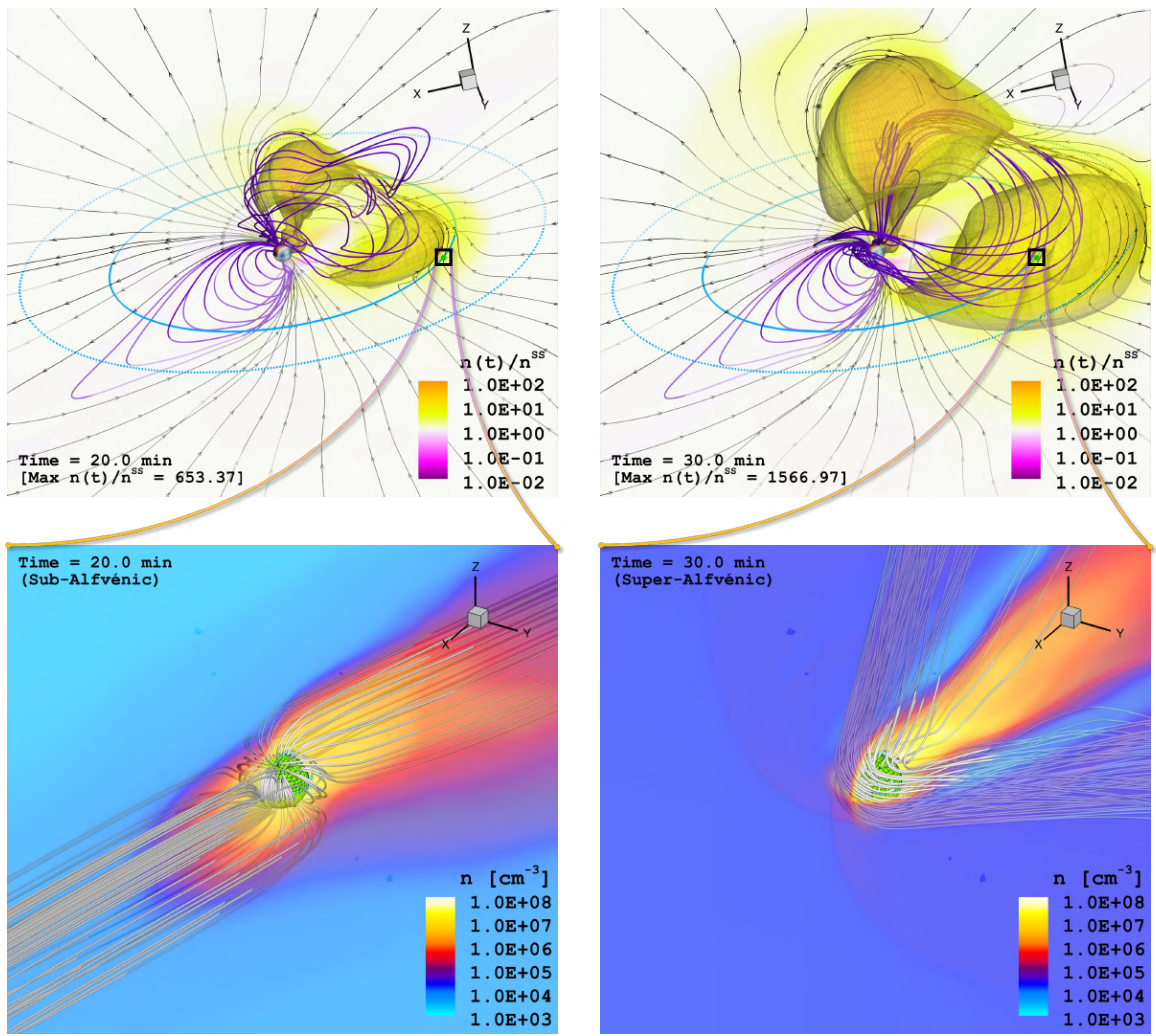

Figure 5.24: Snapshots of a CME simulation on the flare star AU Mic illustrating the CME-detection methodology based on SPI-ECMI. The planetary conditions before the CME arrival (left panels) are characterized by a sub-Alfvénic stellar wind. As such, the magnetospheric structure around the planet (assumed hosting an Earth-like magnetic field) displays a characteristic Alfvén wing geometry (lower-left panel). This SPI configuration could, in principle, induce radio emission via ECMI in the stellar corona. As the CME arrives (right panels), the large perturbations in plasma density, velocity, and magnetic field are sufficient to transform the local conditions to super-Alfvénic. The Alfvén wing configuration is then replaced by a closed magnetospheric structure, including a bow shock (lower-right panel). This change of regime should also shut down any associated SPI-ECMI emission during the CME passage until the pre-CME conditions are restored. The detection of such variability in the SPI-ECMI emission could provide an alternative method of detecting very energetic CME events on close-in exoplanetary systems. Adapted from: Alvarado-Gómez et al. (2022b) and Cohen et al. (2022).

(Lecavelier des Etangs et al., 2012). Changes in the mass loss rate and the ionization structure of the outflow in response to XUV radiation, as well as the stellar wind density and proton flux, have all been proposed as causes of the observed variability (Lecavelier des Etangs et al., 2012; Bourrier & Lecavelier des Etangs, 2013; Guo & Ben-Jaffel, 2016; Chadney et al., 2017).

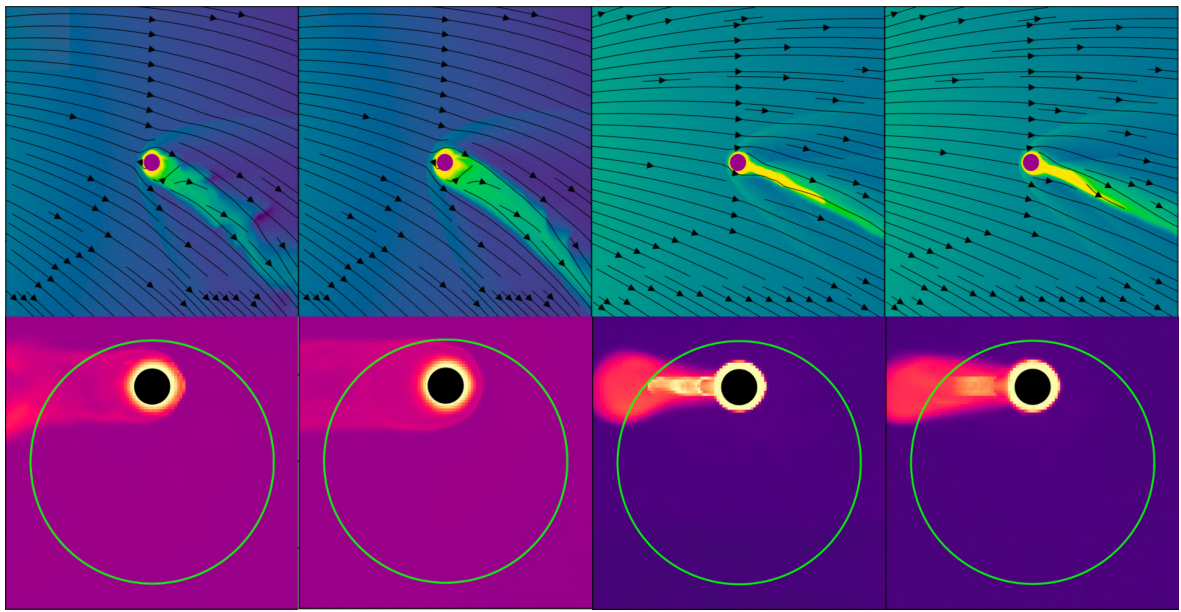

Figure 5.25: Simulated impact of a CME with the planetary outflow of HD 189733 b (Hazra et al., 2022). Left to right, the panels show the outflow during quiescent stellar wind and radiation, during a flare, during a CME, and during both combined. The top row shows the outflow density on a slice through the orbital plane of the planet, using a logarithmic color scale spanning three decades. The bottom row shows fractional absorption at -200 km s$^{-1}$ from the Lyman-$\alpha$ line center as seen in transit against the stellar disk (green circle), using a logarithmic scale spanning roughly five decades. Figure assembled from Hazra et al. (2022) Figures 6 & 7.

Just as hydrodynamic outflows can act as probes of the stellar wind, they can also act as probes of stellar wind variability, including the impact of CMEs. CMEs can sweep away material in the outflow that is beyond the planetary Roche lobe (Cherenkov et al., 2017). In 3D modeling of HD 189733 b during a CME impact without magnetic fields, the increased dynamic pressure produced a more confined and radially directed outflow tail, as well as a greater mass loss rate (Figure 5.25; Hazra et al., 2022). Radial acceleration of the tail and, to a lesser extent, increased mass loss, outcompeted the greater confinement to produce about a 50% increase in the transit depth in the blue wing of the Lyman-$\alpha$ line. Adding magnetic fields to a similar simulation produced magnetospheric oscillations, with effects on mass loss and Lyman-$\alpha$ transit depth that depended on the assumed orientation of the CME magnetic field (Hazra et al., 2024). Even without CMEs, variability in the stellar wind environment, including the magnetic field, experienced by HD 189733 b and other planets with hydrodynamic outflows will influence the rate and dynamics of their outflows (Cohen et al., 2011).

As observational capabilities grow, extended variability and flare monitoring of host stars ahead of transit could enable the association of transit variability with stellar flares and their potentially associated CMEs. If recent changes in a star's XUV emission could not account for changes in the signature of a planetary outflow, the difference might be attributed to CMEs or structures in the stellar wind. Contemporaneous ZDI monitoring could offer a means of testing whether real-time wind conditions based on models constructed using the ZDI observations agree with variations in transits between observations. Population studies offer another direction by examining the relationship between the variability of outflow transits and proxies for the activity of the host star, perhaps circumventing the need for detailed knowledge of each star's recent XUV emission ahead of transits.

### 5.5.11 Star-Disk Interactions

Just as stellar energetic particles can influence the chemistry of exoplanet atmospheres (Section 2.3), they may also leave observable signatures in the chemistry of star-forming disks. While the exact relationship between X-ray flaring and stellar particle fluxes remains uncertain, X-ray enhancements are often used as observational proxies for active stellar conditions likely to coincide with elevated particle emission. This starts with general expectations established by Feigelson et al. (2002) that young stars, with X-ray luminosities enhanced by factors of order $10^5$ compared with the present-day Sun, would have a similarly high enhancement of stellar energetic particles. An anomalously low abundance ratio of $HCO^+$ to $N_2H^+$ in a protostellar disk by Ceccarelli et al. (2014) appeared to confirm this, requiring high ionization rates that Padovani et al. (2015) interpreted as due to the action of energetic particles, although enhanced by orders of magnitude above the level hypothesized by Feigelson et al. (2002). Cleeves et al. (2017) found evidence for temporal variations in the molecular ion abundances in a protoplanetary disk around the time that a large X-ray flare was detected from the system. They noted that in one out of three observations of the T Tauri star IM Lup, the disk-integrated J=3-2 rotational transition of $H_{13}CO^+$ observed with the Atacama Large Millimeter/submillimeter Array (ALMA) doubled in flux compared to the continuum, which had the same level in all three epochs. The line emission in the other two epochs were similar to each other.

This was explored further in Waggoner & Cleeves (2019), who determined that strong X-ray flares with an increase in X-ray luminosity of a factor of 100 could temporarily increase the gas-phase $H_2O$ abundances relative to H by factors of 3–5 along the disk surface. This X-ray-driven chemistry could temporarily enhance $HCO^+$ abundances in the upper layers of the disk atmosphere. Particles accelerated in stellar magnetic reconnection flares also have the potential to produce chemical signatures in the planet-forming disk, which could be a signpost of their presence and characteristics.

Several studies have investigated the impact of stellar energetic particles on protoplanetary disks originating from a stellar eruption, with different treatments of particle transport, energy spectrum of accelerated particles, and inclusion of other effects like that of a disk wind (which have the effect of advecting energetic particles away from the disk) or impact of CME passage on the stellar and disk magnetic field structure. Fraschetti et al. (2018) found that the ionization of the protoplanetary disk is dominated by X-rays over much of its area, except within narrow regions where the energetic particles are channeled onto the disk by the magnetic field, potentially producing spatially patchy chemical signatures, or a "mottled" effect, across the disk. Rodgers-Lee et al. (2017) find that low-energy stellar energetic particles (of order 3 GeV) can significantly ionize the mid-plane of these disks out to roughly 1 AU. Rab et al. (2017) investigated the possible observational tracers of stellar energetic particles accelerated in flares by looking at the thermo-chemical changes in the gas component of the protoplanetary disk. They find that $HCO^+$ and $N_2H^+$ are particularly impacted by ionization produced from stellar energetic particles, with column densities increasing by factors of 3–10 out to disk radii up to 200 AU.

This is an area where modeling the impact of energetic particles, which may accompany X-ray flares from young planet-forming stars, combined with observations of optically thin line emission in protoplanetary disks, could help place observational limits on the presence and impact of flare-associated energetic particles in the youngest and most active stellar systems. Current studies disagree on how far from the star these effects may appear, with estimates ranging from highly localized or mottled regions, to within 1 AU, or even out to 200 AU. Further modeling is needed to clarify where such signatures are most likely to be observed. Given the large investment in ALMA time in the last several years to probe planet-forming disk morphologies and spectral properties, a first-pass observational approach could be a statistical analysis of disk properties correlated with stellar magnetic activity properties. This is especially promising for future observations with JWST mid-infrared observations that can probe the chemistry of the outer disk, and ALMA's wide-band sensitivity upgrade (Carpenter et al., 2022), which will increase sensitivity and bandwidth at mm and sub-mm wavelengths and enable study of the inner-disk regions. The proposed next-generation Very Large Array (ngVLA; Murphy et al., 2018) would have the resolution and sensitivity to probe the dynamics of inner planet-forming disks.

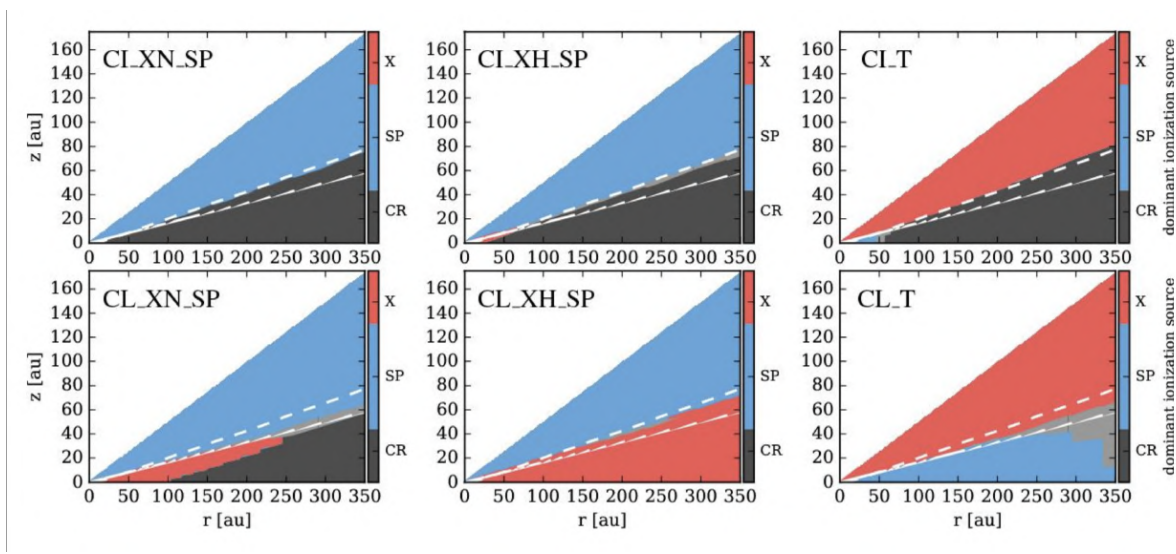

Figure 5.26: Figure 7 of Rab et al. (2017) indicating locations in a planet-forming disk where different sources of ionization (X: X-rays; SP: stellar energetic particles; CR: galactic cosmic rays) dominate. Top panels show combinations with an ISM cosmic ray ionization rate (CI), while the bottom row shows models with low cosmic ray ionization rates (CL). Models with an X-ray normal (XN; $L_X = 10^{30}$ erg s$^{-1}$) and X-ray high (XH; $L_X = 5 \times 10^{30}$ erg s$^{-1}$) spectrum are shown in the left and middle columns, respectively. The spectrum of stellar energetic particles (SP) here is taken to be the solar spectrum, but with flux enhanced by a factor of $10^5$. The right column compares results with an earlier study that made different assumptions about X-ray luminosity, galactic cosmic rays, and stellar energetic particles. The white solid line indicates a column density of $10^{25}$ cm$^{-2}$, while the white dashed line shows the CO ice line.

## 5.6   Potentially Relevant Solar Emission For Stellar Consideration

The techniques described in the previous section were those that have been used (either successfully or unsuccessfully) for identifying CME and SEP candidates, or at least had been evaluated for their potential use as such tools. However, there are additional emissions observed from the Sun that are used for CME and SEP detection and characterization, the utility of which has not been as rigorously explored for other stars. We review a few of these emissions here and encourage future evaluation of their potential for constraining stellar particle flux.

### 5.6.1 Nuclear Gamma-Ray Emission

Gamma-ray emission can indicate the presence of high-energy particle beams generated by magnetic reconnection on the Sun. This raises the question of whether gamma-ray emission could be diagnostic of SEPs produced by the same reconnection-driven particle acceleration. If so, they could theoretically be used to remotely infer SEP events for the Sun and, eventually, stars. Unfortunately, further consideration suggests this is unlikely.

Vilmer et al. (2011) provides a helpful review of solar gamma-ray emission. When downward-directed particles impact the Sun's dense photosphere, they generate bremsstrahlung emission (covered in Section 5.6.2) and excite nuclei that then emit gamma rays (among other processes). Above $\sim 1$ MeV, gamma-ray emission from the Sun is dominated by nuclear emission, and the discussion in this section is limited to this component. By fitting the various lines with known cross sections, the abundances of the elements can be extracted, providing a useful tool for determining solar or accelerated ion abundances.

Solar flare gamma-ray emissions are observed by space missions, such as Solar Maximum Mission (SMM) in the past (see SMM solar flare catalog: Vestrand et al. 1999), Reuven Ramaty High Energy Solar Spectroscopic Imager (RHESSI, Hurford et al. 2003), as well as the recent Fermi Gamma-ray Space Telescope (see Fermi-LAT Solar Flare Catalog: Ajello et al. 2021). The gamma-ray emission of solar flares is observed in nuclear lines with energies $\sim$1-10 MeV and a high-energy continuum with energies >10 MeV. Note that gamma-ray emission is observed only in flares that reach a certain threshold of peak X-ray flux (M and X flares in the GOES classification scheme), but above that threshold, gamma-ray and X-ray intensities are poorly correlated.

Long-duration gamma-ray events, in which the emission lasts much longer than that typically associated with solar flares, suggest that reconnection-based particle acceleration, expected to be short-lived, is not the sole cause of solar gamma rays (Ryan, 2000; Share et al., 2018). The mechanism to produce this extended emission has not been determined. Possibilities include ions accelerated back toward the Sun by a CME-driven shock, the extended acceleration of ions in flare-associated coronal loops, and/or trapping of ions in loops that gradually leak out to impact the surface. Each scenario has challenges, and thus, none has gained community-wide acceptance to explain long-duration gamma-ray events.

The weak correlations and uncertain mechanisms cast doubt on whether gamma rays could probe stellar SEP events. Further, detections of gamma rays from Sun-like or cool stars have been rare. A recent attempt to search for gamma-ray emission from stellar flares was inspired by the long-duration gamma-ray events on the Sun frequently observed by Fermi (Song et al., 2024). The search included 1505 flares from 200 flaring dwarf stars, yet yielded no detections despite a promising survey sensitivity $(1.9 \times 10^{-10} \text{ cm}^{-2} \text{ s}^{-1})$.

### 5.6.2 X-Ray Emission

Solar flares emit X-ray emission resulting from the impact of particle beams generated by flare-associated reconnection on the atmosphere. If these downward-directed beams were to be correlated well with "upward-directed" escaping beams, e.g., SEPs, stellar X-ray flares might provide a proxy for stellar SEPS. However, prospects here are not favorable.

Hard X-ray (HXR) emission and $< 1$ MeV gamma-ray emission from solar flares are primarily due to bremsstrahlung as high-energy electrons decelerate in the field of increasingly dense ions in the chromosphere. Most HXR emission is seen at the footpoints of reconnected loops in the chromosphere. It is also possible for electron-ion interactions to occur in the coronal loops themselves. Generally, the densities in coronal loops are too low to create HXRs, yet several nonthermal HXR sources in the corona have been observed, in particular when the footpoint emission was occulted by the solar limb (Krucker et al., 2008).

Impinging particle beams also generate soft X-ray (SXR) emission through an indirect process. The bulk of the energy of impinging particles dissipates into thermal energy of the chromospheric plasma they impact. This causes the plasma to explosively expand into the corona, a process termed "chromospheric evaporation," (Fisher et al., 1985) where they emit SXRs. Because SXR emission during flares results from cumulative heating, whereas HXR emission results from the instantaneous rate of particle deceleration, the SXR time profile typically resembles the integral of the HXR time profile, a phenomenon termed the "Neupert effect" (Dennis & Zarro, 1993; Veronig et al., 2005).

If the reconnection events generating X-ray emission were also primarily responsible for SEPs, that is, if they generated both downward- and upward-directed particle beams, one would expect X-ray emission to have some connection to SEPs. However, this does not seem to be the case. Comparisons of X-ray and gamma-ray emission with in-situ SEP measurements suggest a significant imbalance between particles that impact the solar atmosphere and those that escape to interplanetary space (e.g., de Nolfo et al., 2019; Krucker et al., 2007; Dresing et al., 2021). This implies it would be very difficult, if not impossible, to accurately infer the properties of SEP events caused by reconnection from flaring X-ray emission for the Sun and, by extension, stars. Unless this limitation can be overcome, X-ray emission is unlikely to prove useful for constraining stellar SEP behavior.

### 5.6.3 Nonthermal Incoherent Radio Emission

Solar CMEs can produce nonthermal[1], incoherent synchrotron and gyrosynchrotron emission that appears both in spatially integrated, Sun-as-a-star radio spectra, namely as type IV radio bursts, as well as in some spatially resolved observations as "radio CMEs." These could offer an avenue for detecting and characterizing stellar CMEs. Although low fluxes, relatively small spatial scales, and confounding signals (particularly with other coherent bursts) pose challenges, progress may be possible with sufficient research and development.

Synchrotron and gyrosynchrotron emission arise from relativistic and mildly relativistic electrons gyrating in the magnetic field of CMEs. This broadband emission has a positive spectral slope at low frequencies and a negative spectral slope at high frequencies (see Figure 5.27) and joins at a spectral peak (Figure 5.27). The peak brightness temperature $T_B^{\mathrm{pk}}$ reflects the effective temperature $T_{\mathrm{eff}}$ of the source electrons at an optical depth of $\tau \approx 1$. The peak frequency $\nu^{\mathrm{pk}}$ is highly sensitive to the magnetic field in the source. Together, these two metrics provide unique diagnostics of the energetic electrons and magnetic fields of the emitting plasma.

#### 5.6.3.1 Spatially unresolved emission, type IV bursts

On the Sun, incoherent (gyro)synchrotron emission can appear in total-power (spatially integrated) dynamic radio spectra as a moving type IV (type IVm) burst (Bain et al., 2014) or smooth type II-like burst (Bastian, 2007). However, it is often outshone by much brighter type II radio bursts (Section 5.5.6), making it difficult to distinguish.

---

[1]Thermal emissions from solar CMEs, including thermal bremsstrahlung and gyroresonance emission, are extremely weak and rarely reported.

Type IVm bursts are named based on their appearance in the total-power radio dynamic spectrum, exhibiting continuum emission that drifts slowly in frequency, indicative of a slowly moving source. They are often interpreted as CME-associated radio emission. In comparison to (stationary) type IV bursts, type IVm bursts are rare. They are also rare in comparison to type II bursts, being typically associated with events that exhibit rarer, stronger, and longer X-ray flares than those where type II bursts are seen. Only about 5% of CMEs have an associated type IVm burst (Gergely, 1986).

Their rarity, particularly those observed without imaging, has hindered studies of their emission mechanisms. Some type IVm emission has been identified as gyrosynchrotron (Bain et al., 2014; Carley et al., 2017). Yet there is evidence that a coherent process, such as plasma radiation (Gary et al., 1985) or electron cyclotron maser instability (ECMI) emission (Morosan et al., 2019a), may be involved in some events. Indeed, the relevant emission mechanism appears to evolve in time, with gyrosynchrotron being associated with an initial stationary type IV and a coherent process producing the type IVm (Morosan et al., 2019a). Even this picture is not perfectly simple, as some stationary type IV bursts have also been attributed to coherent emission, like ECMI emission (Liu et al., 2018).

The emission mechanism responsible for a type IVm is important because it determines what physical properties the emission probes. If gyrosynchrotron emission is observed near its turnover frequency with a sufficiently broad frequency bandwidth to sample both the optically thick and thin parts of the spectrum, the spectrum can be used to measure the source's magnetic field and effective temperature. Solar CMEs in the low to middle solar corona (up to a few solar radii) often have a magnetic field strength of a few to $\sim$10 G. As a result, the gyrosynchrotron spectrum usually peaks in the meter-decameter wavelength regime (or a few tens of MHz to $\sim$100 MHz). Flares, in contrast, usually occur at locations where the magnetic field is $\sim$100 G, resulting in peak frequencies approaching 10 GHz (falling in the microwave band).

Candidate stellar analogs of type IV bursts have been identified (Zic et al., 2020; Mohan et al., 2024). Both of the stars with a candidate type IV detection—Proxima Centauri and AD Leo—are nearby, active M dwarfs. The process attributed to the type IV emission from these stars was not gyrosynchrotron, but rather ECMI emission.

### 5.6.3.2 Spatially resolved emission, radio CMEs

On the Sun, faint radio emissions from CMEs that closely resemble their white light counterparts are dubbed "radio CMEs" (Figure 5.27). (Carley et al., 2020) posits that radio CMEs are the spatially resolved counterparts to type IVm bursts. Spectral imaging has shown that these are likely due to gyrosynchrotron radiation (see a recent review by Vourlidas et al. 2020). These emissions can be used to map the evolving CME magnetic field strength (and possibly direction, if polarimetry is available), nonthermal electron distribution, and the thermal electron number density (see, e.g., Bastian et al. 2001).

The spatially resolved nature of these observations can reveal CME radio emission where it would otherwise be difficult to distinguish from other bright emissions. Yet, even with imaging, distinguishing radio CME emission can be challenging in the face of bright coherent and incoherent gyrosynchrotron emission from energetic electrons accelerated by associated flares. Even on the Sun, reports of incoherent radio CMEs are rare. Next-generation interferometers are overcoming this problem by enabling high dynamic ranges through the use of a large number of antennas, which allows for the simultaneous detection of both faint and bright sources. More events have been reported recently (Mondal et al., 2020; Kansabanik et al., 2024), offering new insights into the magnetic structure of solar CMEs.

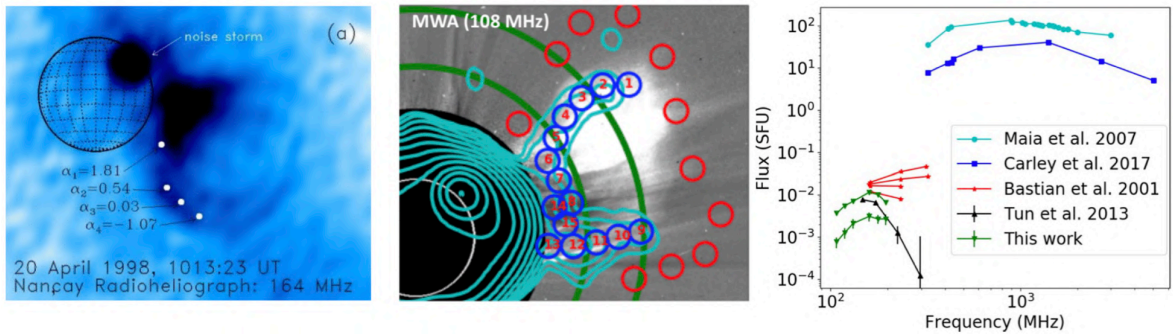

Figure 5.27: Observations of "radio CMEs" on the Sun. *(left)* A radio CME event observed by the Nançay Radioheliograph at 164 MHz. Figure from Bastian et al. (2001). *(center)* Another radio CME observed by the Murchison Widefield Array (MWA) at 108 MHz. The background is the white-light coronagraph image. Figure from Mondal et al. (2020). *(right)* (Gyro)synchrotron spectra of solar CMEs reported in the literature, compiled by Mondal et al. 2020. Figure from Mondal et al. (2020).

Although gyrosynchrotron and synchrotron emission have been observed from stars for decades (Dulk, 1985; Güdel, 2002), it has never been attributed to a stellar CME. However, with a sufficiently strong magnetic field, the peak frequency of the gyrosynchrotron emission of the CMEs could fall within the microwave regime. In this case, the peak flux density of the source, given by

$$S_{\mathrm{pk}} = 2k\Omega v_{\mathrm{pk}}^2 T_{\mathrm{eff}}/c^2 \propto v_{\mathrm{pk}}^2, \tag{5.5}$$

could be orders of magnitude larger than the solar case, increasing detectability. Nonetheless, distinguishing CME emission in the upper stellar corona from that produced by the often-associated stellar flares remains challenging if the emission is unresolved. Multiwavelength campaigns will be critical for distinguishing these emissions. Meanwhile, next-generation radio observatories with milliarcsecond resolutions, corresponding to $< 1\ R_\odot$ at 1 pc, have the potential to spatially resolve radio CMEs from nearby stars.

## 5.7  Paths Forward in Transient Exospace Weather

Many of the techniques discussed in this chapter hold promise for one day enabling the systematic study of stellar CMEs and SEPs, some of which have already yielded compelling results. Yet on their own, no single technique has proven capable of repeatedly and unambiguously identifying transient space weather events generated by stars. The primary challenge is also the key defining feature of these events: their transience. Because they are brief, they cannot be confirmed at a later date with independent observations—observers have only one shot at detecting and confirming a single event.

A key strategy to overcome this challenge is to simultaneously employ multiple techniques to detect and characterize events. Several considerations could help boost the power of this approach:

- Critically, for two techniques to complement each other, they must be sensitive to CMEs within an overlapping parameter space. A key parameter is viewing angle. Techniques favoring face-on detection (Doppler shifts, Section 5.5.2; obscuration dimming Section 5.5.4) could be used together or in combination with techniques that are comparatively insensitive to viewing angle, such as mass-loss dimming (Section 5.5.3) and radio bursts (Section 5.5.5).

- Pairing techniques that probe differing properties of CMEs would enhance the information yield, possibly even yielding more information than the sum of the parts. For instance, it might be possible to combine observations of type II bursts (Section 5.5.6) with speeds derived from Doppler spectroscopy (Section 5.5.2) to constrain both the CME direction and the density profile of the stellar corona through which it propagates.
- Pairing signals produced at the star with ones produced further out could alleviate the requirement that observations be simultaneous. For instance, the detection of a type II (Section 5.5.6) or III (Section 5.5.7) burst could trigger a coronagraph pointing (Section 5.5.1), a search for planetary auroral emission (Section 5.5.8), or monitoring for a dropout in SPI emission. (Section 5.5.9) in the ensuing hours or days.
- Coordinating targeted observations of single objects with multi-object, long-term monitoring surveys can mitigate scheduling logistics. The work of Zic et al. (2020) and Namekata et al. (2022c) has already demonstrated the power of this approach. Low-frequency ($\lesssim 200\,\mathrm{MHz}$) arrays are especially suited for this, as they can survey the entire sky (Prasad et al., 2016; Anderson et al., 2019).

There are many instruments currently in development that could revolutionize the observation of transient stellar space weather. The low-frequency band of the Square Kilometer Array (SKA-Low), currently undergoing science commissioning, will achieve unprecedented sensitivity in the frequency range 50–350 MHz (Braun et al., 2019). At higher frequencies, the 2000-element Deep Synoptic Array (DSA-2000) will lead surveys in the 0.7–2 GHz range (Hallinan et al., 2019). The potential development of radio arrays on the far side of the Moon will unlock a part of the electromagnetic spectrum that had previously only been accessible for Solar System science, offering the possibility of detecting stellar type II and III bursts associated with particles that have propagated beyond the stellar corona (Burns & Hallinan, 2020; Brinkerink & Ampe, 2024).

At shorter wavelengths, the exceptional field of view and observing cadence of Plato will offer a new optical-photometry survey that targeted observations with other instruments can leverage (Ragazzoni et al., 2016). Looking to the far future, HWO's UV spectrograph is likely to enable the detection of Doppler-shifted chromospheric emission or absorption (Section 5.5.2), mass loss dimmings (Section 5.5.3), and obscuration dimmings (Section 5.5.4). Meanwhile, the possible pairing of a starshade with HWO may enable direct imaging of stellar CMEs (Section 4.4.2 Allan et al., 2023; Damiano et al., 2024).

Progress in observing transient stellar space weather is, in many ways, influenced and aided by progress in other areas. Advances in observing stellar magnetic fields and their use in modeling space weather (Section 3.4) will aid in predicting or interpreting observations that require conditions such as CME breakout from overlying fields or shock development, or for which other processes can produce confounding signals. Observationally constraining stellar winds (Section 4.5) will aid in modeling transient particle propagation, and could be paired with observations of transient events to address the balance between transient events and the quasi-steady wind in cumulative stellar mass loss. Understanding planetary properties (Section 2.6) such as atmospheric composition, field strength, and position within the star's Alfvén surface will aid in predicting and interpreting signatures, such as auroral emission that rely on planets as probes.

This chapter closes with a key finding and recommendation for advancing observations of transient (and quasi-steady) exospace weather. Dedicated and strategic effort will eventually yield regular observations of stellar transient events. And with these observations, we will better understand the evolution of stars and the planets that orbit them, including our own.

**Zooming Out: Related Findings and Recommendations (FRs)**

The ephemeral nature of transient events makes them especially challenging to unambiguously detect and characterize. Dedicated efforts to establish the credibility of detection methods (FR4) and to mature new techniques through comparative and/or multiwavelength efforts (FR5) are foundational to establishing a robust body of bona-fide observations. Purpose-built observatories fueled by instrument development (FR6) would strongly facilitate observations of transient events (FR3), while next-generation models are critical for interpreting observations, particularly in non-solar contexts (FR2). To realize this progress, strong cross-disciplinary exchange is a must (FR7).

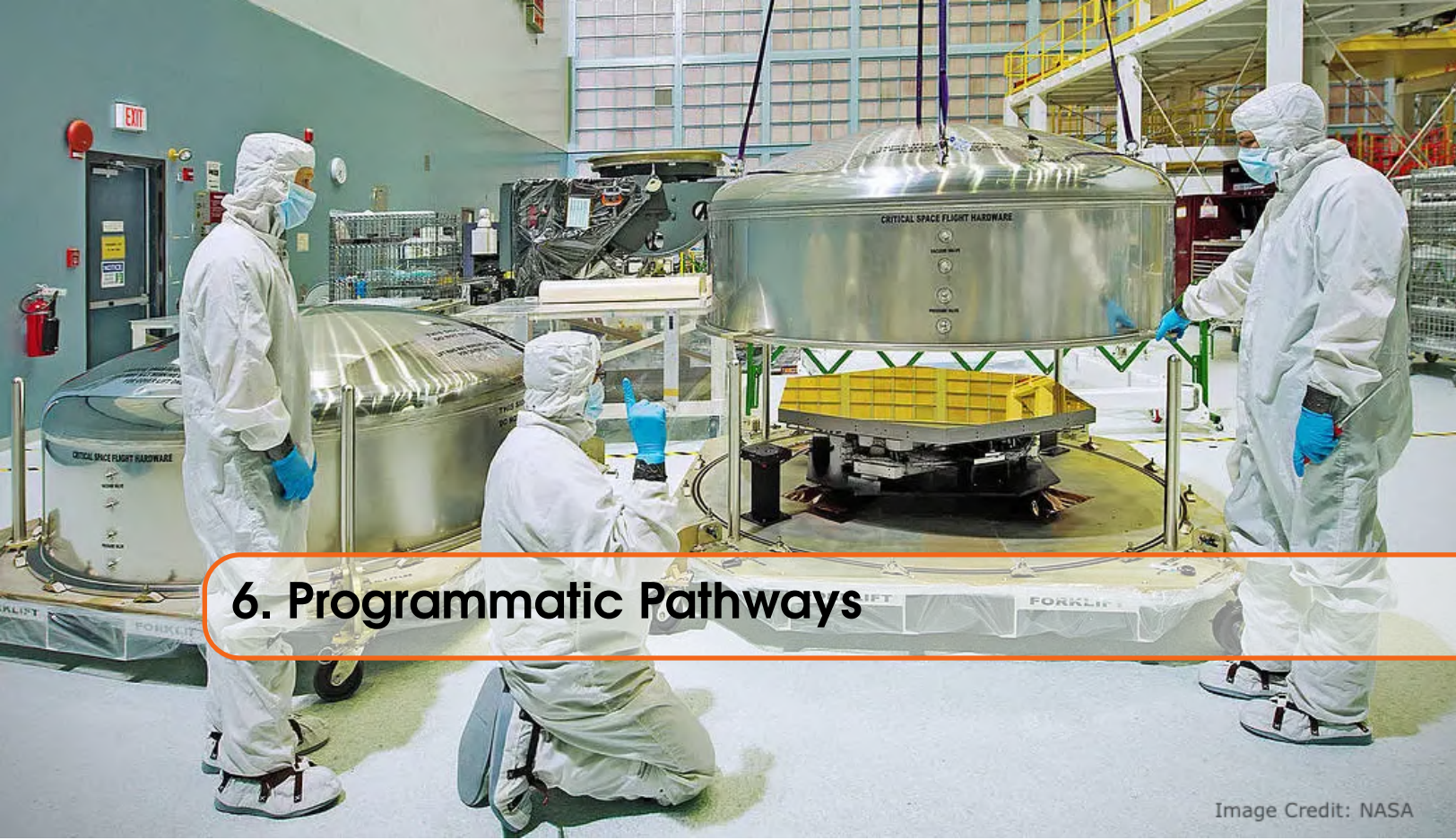
Image Credit: NASA

## 6. Programmatic Pathways

### 6.1 Aligning with NASEM's Decadal Visions

Even the most compelling science will not happen unless it is financially supported. If exospace weather is to progress as a field, it must be accessible, imperative, and exciting to funders. And as such, it should be influential within the strategic documents that shape the priorities of national agencies. The Decadal Surveys and reports published by the National Academies of Sciences, Engineering, and Medicine (NASEM) represent a powerful expression of community consensus and institutional direction. To unlock resources and credibility, exospace weather research must engage with their process and continue where they left off.

Space weather is now recognized as a fundamental factor in shaping planetary environments, particularly for the evolution and habitability of worlds beyond our Solar System. Yet, within the structures that guide long-term U.S. space science strategy, namely the three relevant Decadal Surveys (see below) plus the *Exoplanet Science Strategy Report*, the concept of exospace weather remains underdeveloped. While the latest reports from the heliophysics, astrophysics, and planetary science communities acknowledge the importance of stellar activity and its impacts on exoplanets, they stop short of outlining coordinated approaches to study particle environments beyond the Sun. This chapter examines how exospace weather intersects with these strategic documents and identifies where additional programmatic momentum is needed to fully realize its scientific potential.

### What the Decadals Say (and Don't Say)

We analyzed four recent reports, three of them Decadal Surveys: Solar & Space Physics (SSP; National Academies of Sciences, Engineering, and Medicine et al. 2024), Astronomy & Astrophysics (AA; National Academies of Sciences, Engineering, and Medicine 2021), Planetary Science & Astrobiology (PSA; National Academies of Sciences, Engineering, and Medicine 2022), and the Exoplanet Science Strategy Report (ESS; National Academies of Sciences, Engineering, and Medicine 2018).

[1]

This report helps bridge that gap by outlining concrete, interdisciplinary strategies that can connect stellar particle environments to planetary outcomes in Chapters 2–5, i.e., to translate high-level ambitions into actionable strategies grounded in specific observables, models, and collaborative structures.

## Key Gaps and Opportunities Across the Four Reports

- **The SSP (a.k.a. Heliophysics) Decadal Survey** provides the strongest technical and scientific foundation for studying stellar particle environments. It offers a deep treatment of solar wind, SEPs, flares, and the heliospheric system. However, its framing remains focused on the Sun-Earth system, treating other stars primarily only as solar analogs. It only briefly considers how to extrapolate solar-derived models or observational strategies to other stars or their planetary systems. Our report builds on the SSP's strong foundation by explicitly extending this knowledge to exoplanet systems, providing a framework for applying heliophysics tools to broader planetary science and astrophysics questions, and providing avenues to using observations of the stars to solidify our understanding of key aspects of the Sun's present and past magnetic activity.

- **The AA Decadal Survey (a.k.a. Astro2020)** emphasizes the importance of stellar activity, magnetic fields, and variability in shaping planetary environments, particularly in connection with exoplanet characterization and the need for UV observing capabilities. However, it provides limited detail on how to measure or model particle phenomena like stellar winds, SEPs, or CMEs, or how to integrate them into exoplanet atmospheric modeling. Our report addresses this by outlining specific observational techniques (e.g., radio, Ly-$\alpha$, X-ray dimming) and modeling approaches to close this gap.

- **The PSA Decadal Survey** recognizes atmospheric loss and space weather as important to the evolution of planetary atmospheres and habitability of Solar System bodies. It notes the importance of heliospheric context and space weather impacts, but it does not provide a clear path for modeling the impacts of stellar magnetic fields or SEPs on exoplanets or early Solar System planets, particularly through empirical constraints on space weather inputs through time. Our report picks up here by offering a modeling framework to assess the consequences of particle weather, with direct observational tie-ins.

- **The ESS Report** explicitly calls for stronger interdisciplinary connections between exoplanet and stellar science, and identifies the need to understand planetary habitability in its stellar context. However, it does not elaborate on how to observe or model stellar particle environments in a way that can feed into habitability assessments. Our report provides those next steps, laying out how exospace weather can be observed, modeled, and connected to atmospheric chemistry and biosignature interpretation.

## Cross-Cutting Opportunities and Leverage Points

There are clear places where exospace weather can advance the goals of multiple Decadal communities simultaneously:

- **Atmospheric escape modeling** connects heliophysics, planetary science, and exoplanet detection. (See Section 2.2, 4.3.4, & 5.5.10.)

---

[1]Although not technically a "Decadal Survey," NASEM's ESS report is a consensus study commissioned by NASA as input to both the AA and PSA Surveys. It reflects broad community priorities and represents the first formal, community-wide effort to define a comprehensive strategy for exoplanet science. It followed the National Academies' rigorous consensus process and is widely treated as a companion reference to the Decadal reports. Each highlights questions and goals that intersect with exospace weather, but none provides a clear, detailed roadmap for how to tackle them.

- **Auroral emissions and magnetic field detection** are frontier observables in both planetary and exoplanetary science. (See Section 2.4, & 5.5.8).
- **Prebiotic chemistry and biosignature evolution** require inputs from particle radiation models and space weather variability, while chemical disequilibrium probes particle radiation inputs. (See Section 2.5 for particle-driven prebiotic chemistry and Chapter 5 for observational strategies to constrain flare-SEP relationships.)
- **Solar and stellar** observations provide complementary depth and breadth, anchoring physical frameworks with deep in-situ study, testing them in non-solar contexts, and probing space weather generated by the Sun across time. (See Chapters 4 & 5)
- **MHD models of space weather** transform sparse observations into spatially and temporally complete descriptions of space weather, both providing benefit to stellar and planetary science while benefiting from the observational constraints they can provide. (See Chapter 3 and techniques explored in Chapters 4 & 5)

Programs like NASA's Research Opportunities in Space and Earth Science (ROSES) interdisciplinary calls and the Diversify, Realize, Integrate, Venture, Educate (DRIVE) initiative in heliophysics represent key leverage points. The ROSES program offers opportunities to propose cross-divisional research that brings together heliophysics, astrophysics, and planetary science, and could be strengthened further by emphasizing exospace weather themes in targeted solicitations. The DRIVE initiative, while designed to enhance heliophysics discovery, explicitly encourages integration of theory, modeling, and observations——core components of exospace weather science.

In addition, existing programs such as NASA's Habitable Worlds program (within astrobiology), the Planetary Data Ecosystem Independent Review Board (PDE IRB), the Exoplanet Research Program (XRP), and the Interdisciplinary Consortia for Astrobiology Research (ICAR) all provide channels for exospace weather research to flourish, particularly when encouraged by programmatic language that prioritizes interdisciplinary integration. Joint review panels, co-sponsored workshops, and programmatic coordination between divisions (e.g., through inter-divisional ROSES calls or mission science working groups) could significantly accelerate progress in this domain.

The National Science Foundation (NSF) offers interdisciplinary opportunities in both an ad hoc and directed fashion. The Astronomy and Astrophysics Research Grants (AAG) program within the Division of Astronomical Sciences (AST) is the core funding opportunity for a broad range of observational, theoretical, and experimental astronomy. Simultaneously, solar studies of space weather are often directed to programs such as the Solar, Heliospheric, and Interplanetary Environment (SHINE) within the Division of Atmospheric and Geospace Sciences (GEO/AGS). Cross-cutting programs can be submitted to a proposal call in one division but can undergo co-review and joint funding when recognized as clearly interdisciplinary.

Examples of directed interdisciplinary opportunities include Dear Colleague Letters (DCLs). The NSF often issues DCLs to encourage the submission of proposals in specific emerging or high-priority interdisciplinary areas. Geoscience Lessons for and from Other Worlds (GLOW) is a recent example of a DCL with some relevance to the material of this report, for example.

Finally, it is worth highlighting the opportunities for developing new ground-based instrumentation that can address some of the scientific opportunities highlighted in this report. The Major Research Instrumentation Program (MRI) and the Mid-scale Research Infrastructure Program (MSRI-I; MSRI-2) together facilitate instrumentation and facility development ranging in cost from \$1 M–\$100M and spanning NSF-wide across science and engineering research disciplines. Cross-cutting science applications are viewed as a strong positive.

**Integrating Exospace Weather into Future Strategy**

Exospace weather touches nearly every priority outlined across the Decadal Surveys, from understanding planetary atmospheres and climate evolution to detecting biosignatures and identifying habitable environments. While it is not yet fully woven into the fabric of these strategic visions, the opportunity is clear. Our report provides a roadmap for how to fill this gap: connecting Decadal goals to new methods, partnerships, and observables that can bring the space weather component of planetary systems into a clearer scientific and programmatic view. Strengthening these connections will help ensure that exospace weather becomes a more natural part of future planning, collaboration, and discovery. We hope that future Decadal Surveys include explicit charges to evaluate the state of exospace weather science and identify coordinated strategies for its advancement.

---

**Zooming Out: Related Findings and Recommendations (FRs)**

Exospace weather cannot grow into a fully fledged field unless it receives programmatic support. Two areas where programmatic support would especially enable progress are 1) the development of observatories or instruments purpose-built for exospace weather research (FR3), and 2) efforts that enhance cross-pollination between the wide range of disciplines—heliophysics, astrophysics, planetary science, and astrobiology—necessary to understand exospace weather and its impacts (FR7).

# Bibliography

Aarnio, A. N., Stassun, K. G., Hughes, W. J., & McGregor, S. L. 2011, SoPh, 268, 195, doi: 10.1007/s11207-010-9672-7

Abraham, R. G., & van Dokkum, P. G. 2014, PASP, 126, 55, doi: 10.1086/674875

Aime, C., Theys, C., Prunet, S., & Ferrari, A. 2024, A comparison of solar and stellar coronagraphs that make use of external occulters. https://ui.adsabs.harvard.edu/abs/2024arXiv241103144A

Airapetian, V., Glocer, A., Gronoff, G., Hebrard, E., & Danchi, W. 2016, Nature Geoscience, 9, 452

Airapetian, V., Barnes, R., Cohen, O., et al. 2020, International Journal of Astrobiology, 19, 136

Airapetian, V. S., Jin, M., Lüftinger, T., et al. 2021, ApJ, 916, 96, doi: 10.3847/1538-4357/ac081e

Ajello, M., Baldini, L., Bastieri, D., et al. 2021, ApJS, 252, 13, doi: 10.3847/1538-4365/abd32e

Allan, G., Ruane, G., Walter, A. B., et al. 2023, in Society of Photo-Optical Instrumentation Engineers (SPIE) Conference Series, Vol. 12680, Society of Photo-Optical Instrumentation Engineers (SPIE) Conference Series, 1268019, doi: 10.1117/12.2676876

Alvarado-Gómez, J. D., Drake, J. J., Cohen, O., et al. 2022a, Astronomische Nachrichten, 343, e10100, doi: 10.1002/asna.20210100

Alvarado-Gómez, J. D., Drake, J. J., Cohen, O., Moschou, S. P., & Garraffo, C. 2018, ApJ, 862, 93, doi: 10.3847/1538-4357/aacb7f

Alvarado-Gómez, J. D., Drake, J. J., Moschou, S. P., et al. 2019, ApJL, 884, L13, doi: 10.3847/2041-8213/ab44d0

Alvarado-Gómez, J. D., Hussain, G. A. J., Cohen, O., et al. 2016a, A&A, 588, A28, doi: 10.1051/0004-6361/201527832

—. 2016b, A&A, 594, A95, doi: 10.1051/0004-6361/201628988

Alvarado-Gómez, J. D., Hussain, G. A. J., Grunhut, J., et al. 2015, A&A, 582, A38, doi: 10.1051/0004-6361/201525771

Alvarado-Gómez, J. D., Drake, J. J., Fraschetti, F., et al. 2020a, ApJ, 895, 47, doi: 10.3847/1538-4357/ab88a3

—. 2020b, ApJ, 895, 47, doi: 10.3847/1538-4357/ab88a3

Alvarado-Gómez, J. D., Cohen, O., Drake, J. J., et al. 2022b, ApJ, 928, 147, doi: 10.3847/1538-4357/ac54b8

Alvarez, H., & Haddock, F. T. 1973a, SoPh, 29, 197, doi: 10.1007/BF00153449

—. 1973b, SoPh, 30, 175, doi: 10.1007/BF00156186

Anche, R. M., Sen, A. K., Anupama, G. C., Sankarasubramanian, K., & Skidmore, W. 2018, Journal of Astronomical Telescopes, Instruments, and Systems, 4, 018003, doi: 10.1117/1.JATIS.4.1.018003

Anche, R. M., Williams, G., Tailor, H., et al. 2023a, in Society of Photo-Optical Instrumentation Engineers (SPIE) Conference Series, Vol. 12690, Polarization Science and Remote Sensing XI, ed. M. K. Kupinski, J. A. Shaw, & F. Snik, 126900F, doi: 10.1117/12.2676777

Anche, R. M., Ashcraft, J. N., Haffert, S. Y., et al. 2023b, A&A, 672, A121, doi: 10.1051/0004-6361/202245651

Anderson, M. M., Hallinan, G., Eastwood, M. W., et al. 2019, ApJ, 886, 123, doi: 10.3847/1538-4357/ab4f87

Anglada-Escudé, G., Amado, P. J., Barnes, J., et al. 2016, Nature, 536, 437, doi: 10.1038/nature19106

Antiochos, S. K., MacNeice, P. J., Spicer, D. S., & Klimchuk, J. A. 1999, ApJ, 512, 985, doi: 10.1086/306804

Antonucci, E., Kohl, J. L., Noci, G., et al. 1997, ApJL, 490, L183, doi: 10.1086/311028

Aplin, K. L. 2013, Venus (Dordrecht: Springer Netherlands), 13–20, doi: 10.1007/978-94-007-6633-4_3

Ardila, D. R., Shkolnik, E., Ziemer, J., et al. 2022, in Society of Photo-Optical Instrumentation Engineers (SPIE) Conference Series, Vol. 12181, Space Telescopes and Instrumentation 2022: Ultraviolet to Gamma Ray, ed. J.-W. A. den Herder, S. Nikzad, & K. Nakazawa, 1218104, doi: 10.1117/12.2629000

Argiroffi, C., Reale, F., Drake, J. J., et al. 2019, Nature Astronomy, 3, 742, doi: 10.1038/s41550-019-0781-4

Arney, G., Domagal-Goldman, S. D., Meadows, V. S., et al. 2016, Astrobiology, 16, 873, doi: `10.1089/ast.2015.1422`

Attrill, G. D. R., van Driel-Gesztelyi, L., Démoulin, P., et al. 2008, SoPh, 252, 349, doi: `10.1007/s11207-008-9255-z`

Atwood, J., Skidmore, W., Anupama, G. C., et al. 2014, in Society of Photo-Optical Instrumentation Engineers (SPIE) Conference Series, Vol. 9150, Modeling, Systems Engineering, and Project Management for Astronomy VI, ed. G. Z. Angeli & P. Dierickx, 915013, doi: `10.1117/12.2056965`

Avice, G., Parai, R., Jacobson, S., et al. 2022, Space Science Reviews, 218, 60

Axford, W. I., Leer, E., & Skadron, G. 1977, in International Cosmic Ray Conference, Vol. 11, International Cosmic Ray Conference, 132

Ayres, T. R., Brown, A., Osten, R. A., et al. 2001, ApJ, 549, 554, doi: `10.1086/319051`

Babcock, H. W. 1953, ApJ, 118, 387, doi: `10.1086/145767`

Bain, H. M., Krucker, S., Glesener, L., & Lin, R. P. 2012, ApJ, 750, 44, doi: `10.1088/0004-637X/750/1/44`

Bain, H. M., Krucker, S., Saint-Hilaire, P., & Raftery, C. L. 2014, ApJ, 782, 43, doi: `10.1088/0004-637X/782/1/43`

Bastian, T. S. 2007, ApJ, 665, 805, doi: `10.1086/519246`

Bastian, T. S., Pick, M., Kerdraon, A., Maia, D., & Vourlidas, A. 2001, ApJL, 558, L65, doi: `10.1086/323421`

Beaudoin, E., Lilensten, J., Gronoff, G., et al. 2025, Journal of Space Weather and Space Climate, doi: `10.1051/swsc/2025012`

Belikov, R., Bendek, E., Sirbu, D., & Quarles, B. 2019, in NASA Technical Reports, Washington, DC, United States. `https://ntrs.nasa.gov/citations/20190028237`

Bellotti, S., Fares, R., Vidotto, A. A., et al. 2023, A&A, 676, A139, doi: `10.1051/0004-6361/202346675`

Bellotti, S., Petit, P., Jeffers, S. V., et al. 2024, arXiv e-prints, arXiv:2412.09365, doi: `10.48550/arXiv.2412.09365`

Bertaux, J.-L., Leblanc, F., Witasse, O., et al. 2005, Nature, 435, 790

Bignall, H. E., Macquart, J.-P., Jauncey, D. L., et al. 2006, ApJ, 652, 1050, doi: `10.1086/508406`

Birkeland, K. 1896, Electrical Review, 38, 968

Bisikalo, D., Kaygorodov, P., Ionov, D., et al. 2013, The Astrophysical Journal, 764, 19, doi: `10.1088/0004-637X/764/1/19`

Bobra, M. G., van Ballegooijen, A. A., & DeLuca, E. E. 2008, ApJ, 672, 1209, doi: `10.1086/523927`

Bond, A., & Martin, A. R., eds. 1978, Project Daedalus: Final Report (London: British Interplanetary Society)

Boro Saikia, S., Lueftinger, T., Airapetian, V. S., et al. 2023, ApJ, 950, 124, doi: 10.3847/1538-4357/acca14

Borovikov, D., Sokolov, I. V., Roussev, I. I., Taktakishvili, A., & Gombosi, T. I. 2018, ApJ, 864, 88, doi: 10.3847/1538-4357/aad68d

Bourrier, V., & Lecavelier des Etangs, A. 2013, Astronomy and Astrophysics, 557, A124, doi: 10.1051/0004-6361/201321551

Bourrier, V., Lecavelier des Etangs, A., Ehrenreich, D., Tanaka, Y., & Vidotto, A. 2016, Astronomy and Astrophysics, 591, A121

Brain, D., Lillis, R., Ma, Y., & Ramstad, R. 2021, 53, 263, doi: 10.3847/25c2cfeb.15a827c3

Brain, D., Barabash, S., Boesswetter, A., et al. 2010, Icarus, 206, 139, doi: 10.1016/j.icarus.2009.06.030

Braun, R., Bonaldi, A., Bourke, T., Keane, E., & Wagg, J. 2019, arXiv e-prints, arXiv:1912.12699, doi: 10.48550/arXiv.1912.12699

Brinkerink, C., & Ampe, A. 2024, in Society of Photo-Optical Instrumentation Engineers (SPIE) Conference Series, Vol. 13092, Space Telescopes and Instrumentation 2024: Optical, Infrared, and Millimeter Wave, ed. L. E. Coyle, S. Matsuura, & M. D. Perrin, 130922M, doi: 10.1117/12.3018919

Broomhall, A. M., Chaplin, W. J., Davies, G. R., et al. 2009, MNRAS, 396, L100, doi: 10.1111/j.1745-3933.2009.00672.x

Brown, S. A., Fletcher, L., & Labrosse, N. 2016, A&A, 596, A51, doi: 10.1051/0004-6361/201628390

Brueckner, G. E., Howard, R. A., Koomen, M. J., et al. 1995, SoPh, 162, 357, doi: 10.1007/BF00733434

Brun, A. S., & Browning, M. K. 2017, Living Reviews in Solar Physics, 14, 4, doi: 10.1007/s41116-017-0007-8

Burlaga, L. F., & Ness, N. F. 2014, The Astrophysical Journal, 784, 146, doi: 10.1088/0004-637X/784/2/146

Burns, J. O., & Hallinan, G. 2020, in American Astronomical Society Meeting Abstracts, Vol. 235, American Astronomical Society Meeting Abstracts #235, 130.01

Calisto, M., Usoskin, I., & Rozanov, E. 2013, Environmental Research Letters, 8, 045010

Cane, H. V., Erickson, W. C., & Prestage, N. P. 2002, Journal of Geophysical Research (Space Physics), 107, 1315, doi: 10.1029/2001JA000320

Canou, A., Amari, T., Bommier, V., et al. 2009, ApJL, 693, L27, doi: 10.1088/0004-637X/693/1/L27

Caramazza, M., Flaccomio, E., Micela, G., et al. 2007, A&A, 471, 645, doi: 10.1051/0004-6361:20077195

Carley, E. P., Vilmer, N., Simões, P. J. A., & Ó Fearraigh, B. 2017, A&A, 608, A137, doi: 10.1051/0004-6361/201731368

Carley, E. P., Vilmer, N., & Vourlidas, A. 2020, Frontiers in Astronomy and Space Sciences, 7, 79, doi: 10.3389/fspas.2020.551558

Carolan, S., Vidotto, A. A., Villarreal D'Angelo, C., & Hazra, G. 2021, Monthly Notices of the Royal Astronomical Society, 500, 3382, doi: 10.1093/mnras/staa3431

Carpenter, J., Brogan, C., Iono, D., & Mroczkowski, T. 2022, arXiv e-prints, arXiv:2211.00195, doi: 10.48550/arXiv.2211.00195

Carrington, R. C. 1859, MNRAS, 20, 13, doi: 10.1093/mnras/20.1.13

Cash, W. 2006, Nature, 442, 51, doi: 10.1038/nature04930

Cauley, P. W., Shkolnik, E. L., Llama, J., & Lanza, A. F. 2019, Nature Astronomy, 3, 1128, doi: 10.1038/s41550-019-0840-x

Ceccarelli, C., Dominik, C., López-Sepulcre, A., et al. 2014, ApJL, 790, L1, doi: 10.1088/2041-8205/790/1/L1

Chadney, J. M., Koskinen, T. T., Galand, M., Unruh, Y. C., & Sanz-Forcada, J. 2017, Astronomy and Astrophysics, 608, A75, doi: 10.1051/0004-6361/201731129

Chamberlin, P. C. 2016, SoPh, 291, 1665, doi: 10.1007/s11207-016-0931-0

Chebly, J. J., Alvarado-Gómez, J. D., Poppenhäger, K., & Garraffo, C. 2023, MNRAS, 524, 5060, doi: 10.1093/mnras/stad2100

Chen, H., Tian, H., Li, H., et al. 2022, ApJ, 933, 92, doi: 10.3847/1538-4357/ac739b

Chen, H., Zhan, Z., Youngblood, A., et al. 2021, Nature Astronomy, 5, 298

Chen, P. F. 2011, Living Reviews in Solar Physics, 8, 1, doi: 10.12942/lrsp-2011-1

Cheng, L., Zhang, M., Kwon, R. Y., & Lario, D. 2024, Simulation of solar energetic particle events originated from coronal mass ejection shocks with a data-driven physics-based transport model. https://ui.adsabs.harvard.edu/abs/2024arXiv241104095C

Cheng, Z., Wang, Y., Liu, R., Zhou, Z., & Liu, K. 2019, ApJ, 875, 93, doi: 10.3847/1538-4357/ab0f2d

Cherenkov, A., Bisikalo, D., Fossati, L., & Möstl, C. 2017, The Astrophysical Journal, 846, 31, doi: 10.3847/1538-4357/aa82b2

Cheung, M. C. M., & DeRosa, M. L. 2012, ApJ, 757, 147, doi: 10.1088/0004-637X/757/2/147

Chikunova, G., Dissauer, K., Podladchikova, T., & Veronig, A. M. 2020, ApJ, 896, 17, doi: 10.3847/1538-4357/ab9105

Chintzoglou, G., Zhang, J., Cheung, M. C. M., & Kazachenko, M. 2019, ApJ, 871, 67, doi: 10.3847/1538-4357/aaef30

Chyba, C. F. 2000, Nature, 403, 381

Ciaravella, A., Raymond, J. C., Fineschi, S., et al. 1997, ApJL, 491, L59, doi: 10.1086/311048

Ciaravella, A., Raymond, J. C., Thompson, B. J., et al. 2000, ApJ, 529, 575, doi: 10.1086/308260

Cleeves, L. I., Bergin, E. A., Öberg, K. I., et al. 2017, ApJL, 843, L3, doi: 10.3847/2041-8213/aa76e2

Cliver, E. W., Kahler, S. W., & Reames, D. V. 2004, ApJ, 605, 902, doi: 10.1086/382651

Coffaro, M., Stelzer, B., Orlando, S., et al. 2020, Astronomy and Astrophysics, 636, A49, doi: 10.1051/0004-6361/201936479

Cohen, C. M. S., Li, G., Mason, G. M., Shih, A. Y., & Wang, L. 2021, in Solar Physics and Solar Wind, ed. N. E. Raouafi & A. Vourlidas, Vol. 1, 133, doi: 10.1002/9781119815600.ch4

Cohen, C. M. S., Mewaldt, R. A., Leske, R. A., et al. 1999, Geophysical Research Letters, 26, 2697, doi: 10.1029/1999GL900560

Cohen, O. 2011, MNRAS, 417, 2592, doi: 10.1111/j.1365-2966.2011.19428.x

Cohen, O., Alvarado-Gómez, J. D., Drake, J. J., et al. 2022, ApJ, 934, 189, doi: 10.3847/1538-4357/ac78e4

Cohen, O., Kashyap, V. L., Drake, J. J., et al. 2011, The Astrophysical Journal, 733, 67, doi: 10.1088/0004-637X/733/1/67

Cohen, O., Yadav, R., Garraffo, C., et al. 2017, ApJ, 834, 14, doi: 10.3847/1538-4357/834/1/14

Cowley, S. W. H., Bunce, E. J., & Prangé, R. 2004, Annales Geophysicae, 22, 1379, doi: 10.5194/angeo-22-1379-2004

Cravens, T. E. 1997, Geophys. Res. Lett., 24, 105, doi: 10.1029/96GL03780

Cremades, H., & Bothmer, V. 2004, A&A, 422, 307, doi: 10.1051/0004-6361:20035776

Crider, D. H. 2004, Advances in Space Research, 33, 152, doi: 10.1016/j.asr.2003.04.013

Crosley, M. K., & Osten, R. A. 2018a, ApJ, 856, 39, doi: 10.3847/1538-4357/aaaec2

—. 2018b, ApJ, 862, 113, doi: 10.3847/1538-4357/aacf02

Crosley, M. K., Osten, R. A., Broderick, J. W., et al. 2016, ApJ, 830, 24, doi: 10.3847/0004-637X/830/1/24

Cruise, M., Guainazzi, M., Aird, J., et al. 2024, Nature Astronomy, 9, 36, doi: 10.1038/s41550-024-02416-3

Cruise, M., Guainazzi, M., Aird, J., et al. 2025, Nature Astronomy, 9, 36, doi: 10.1038/s41550-024-02416-3

Currie, M. H., Stark, C. C., Kammerer, J., Juanola-Parramon, R., & Meadows, V. S. 2023, The Astronomical Journal, 166, 197, doi: 10.3847/1538-3881/acfda7

Curry, S. M., Luhmann, J., Ma, Y., et al. 2015a, Planetary and Space Science, 115, 35

Curry, S. M., Luhmann, J. G., Ma, Y., et al. 2015b, Geophysical Research Letters, 42, 9095, doi: https://doi.org/10.1002/2015GL065304

Curry, S. M., Hara, T., Luhmann, J. G., et al. 2025, S c i e n c e A d v a n c e s

Czechowski, A., & Mann, I. 2003, Journal of Geophysical Research (Space Physics), 108, 8038, doi: 10.1029/2003JA009917

Daley-Yates, S., & Stevens, I. R. 2018, MNRAS, 479, 1194, doi: 10.1093/mnras/sty1652

Damiano, M., Shaklan, S., Hu, R., et al. 2024, Starshade Exoplanet Data Challenge: What We Learned. https://ui.adsabs.harvard.edu/abs/2024arXiv241009183D

de Gasperin, F., Edler, H. W., Williams, W. L., et al. 2023, A&A, 673, A165, doi: 10.1051/0004-6361/202245389

de Juan Ovelar, M., Snik, F., Keller, C. U., & Venema, L. 2014, A&A, 562, A8, doi: 10.1051/0004-6361/201321717

de Nolfo, G. A., Bruno, A., Ryan, J. M., et al. 2019, ApJ, 879, 90, doi: 10.3847/1538-4357/ab258f

DeForest, C. E., Howard, T. A., & McComas, D. J. 2013, The Astrophysical Journal, 769, 43, doi: 10.1088/0004-637X/769/1/43

Deighan, J., Jain, S., Chaffin, M., et al. 2018, Nature Astronomy, 2, 802

Denevi, B. W., Noble, S. K., Christoffersen, R., et al. 2023, Reviews in Mineralogy and Geochemistry, 89, 611, doi: 10.2138/rmg.2023.89.14

Dennis, B. R., & Zarro, D. M. 1993, SoPh, 146, 177, doi: 10.1007/BF00662178

Desai, M., & Giacalone, J. 2016, Living Reviews in Solar Physics, 13, 3, doi: 10.1007/s41116-016-0002-5

Dialynas, K., Krimigis, S. M., Decker, R. B., & Mitchell, D. G. 2019, Geophysical Research Letters, 46, 7911, doi: 10.1029/2019GL083924

Dissauer, K., Veronig, A. M., Temmer, M., & Podladchikova, T. 2019, ApJ, 874, 123, doi: 10.3847/1538-4357/ab0962

Dissauer, K., Veronig, A. M., Temmer, M., Podladchikova, T., & Vanninathan, K. 2018, ApJ, 863, 169, doi: 10.3847/1538-4357/aad3c6

Dobrijevic, M., Loison, J. C., Hickson, K. M., & Gronoff, G. 2016, Icarus, 268, 313, doi: 10.1016/j.icarus.2015.12.045

Donati, J. F. 2011, in Astrophysical Dynamics: From Stars to Galaxies, ed. N. H. Brummell, A. S. Brun, M. S. Miesch, & Y. Ponty, Vol. 271, 23–31, doi: 10.1017/S1743921311017431

Donati, J. F., & Landstreet, J. D. 2009, ARA&A, 47, 333, doi: 10.1146/annurev-astro-082708-101833

Dong, C., Jin, M., Lingam, M., et al. 2018, Proceedings of the National Academy of Sciences, 115, 260

Dong, C., Lingam, M., Ma, Y., & Cohen, O. 2017a, The Astrophysical Journal Letters, 837, L26, doi: 10.3847/2041-8213/aa6438

Dong, Y., Fang, X., Brain, D., et al. 2017b, Journal of Geophysical Research: Space Physics, 122, 4009

Dos Santos, L. A. 2022, Observations of planetary winds and outflows, Tech. rep., doi: 10.48550/arXiv.2211.16243

Drake, J. J., Cohen, O., Yashiro, S., & Gopalswamy, N. 2013, The Astrophysical Journal, 764, 170, doi: 10.1088/0004-637X/764/2/170

Drake, J. J., & Stelzer, B. 2023, in Handbook of X-ray and Gamma-ray Astrophysics (Springer), 132, doi: 10.1007/978-981-16-4544-0_78-1

Drake, S. A., Simon, T., & Brown, A. 1993, The Astrophysical Journal, 406, 247, doi: 10.1086/172436

Dravins, D., Lindegren, L., & Nordlund, A. 1981, A&A, 96, 345

Dresing, N., Warmuth, A., Effenberger, F., et al. 2021, A&A, 654, A92, doi: 10.1051/0004-6361/202141365

Drossart, P., Maillard, J. P., Caldwell, J., et al. 1989, Nature, 340, 539, doi: 10.1038/340539a0

Duffin, R. T., White, S. M., Ray, P. S., & Kaiser, M. L. 2015, in Journal of Physics Conference Series, Vol. 642, Journal of Physics Conference Series (IOP), 012006, doi: 10.1088/1742-6596/642/1/012006

Dulk, G. A. 1985, ARA&A, 23, 169, doi: 10.1146/annurev.aa.23.090185.001125

—. 2000, Geophysical Monograph Series, 119, 115, doi: 10.1029/GM119p0115

Dumusque, X., Udry, S., Lovis, C., Santos, N. C., & Monteiro, M. J. P. F. G. 2011, A&A, 525, A140, doi: 10.1051/0004-6361/201014097

Dungey, J. W. 1961, PhRvL, 6, 47, doi: 10.1103/PhysRevLett.6.47

Echim, M. M., Lemaire, J., & Lie-Svendsen, Ø. 2011, Surveys in Geophysics, 32, 1, doi: 10.1007/s10712-010-9106-y

Edberg, N. J. T., Nilsson, H., Futaana, Y., et al. 2011, Journal of Geophysical Research (Space Physics), 116, A09308, doi: 10.1029/2011JA016749

Eiroa, C., Marshall, J. P., Mora, A., et al. 2013, Astronomy and Astrophysics, 555, A11, doi: 10.1051/0004-6361/201321050

Emslie, A. G., Kucharek, H., Dennis, B. R., et al. 2004, Journal of Geophysical Research (Space Physics), 109, A10104, doi: 10.1029/2004JA010571

Engelbrecht, N. E., Effenberger, F., Florinski, V., et al. 2022, SSRv, 218, 33, doi: 10.1007/s11214-022-00896-1

Evensberget, D., & Vidotto, A. A. 2024, MNRAS, 529, L140, doi: 10.1093/mnrasl/slae010

Evensberget, D., Marsden, S. C., Carter, B. D., et al. 2023, MNRAS, 524, 2042, doi: 10.1093/mnras/stad1650

Faherty, J. K., Burningham, B., Gagné, J., et al. 2024, Nature, 628, 511, doi: 10.1038/s41586-024-07190-w

Fanson, J., Bernstein, R., Ashby, D., et al. 2022, in Society of Photo-Optical Instrumentation Engineers (SPIE) Conference Series, Vol. 12182, Ground-based and Airborne Telescopes IX, ed. H. K. Marshall, J. Spyromilio, & T. Usuda, 121821C, doi: 10.1117/12.2631694

Farrell, W. M., Desch, M. D., & Zarka, P. 1999, Journal of Geophysical Research, 104, 14025, doi: 10.1029/1998JE900050

Favata, F., & Schmitt, J. H. M. M. 1999, A&A, 350, 900, doi: 10.48550/arXiv.astro-ph/9909041

Feigelson, E. D., Garmire, G. P., & Pravdo, S. H. 2002, ApJ, 572, 335, doi: 10.1086/340340

Feinstein, A. D., France, K., Youngblood, A., et al. 2022, The Astronomical Journal, 164, 110, doi: 10.3847/1538-3881/ac8107

Fichtinger, B., Güdel, M., Mutel, R. L., et al. 2017, A&A, 599, A127, doi: 10.1051/0004-6361/201629886

Fisher, G. H., Canfield, R. C., & McClymont, A. N. 1985, ApJ, 289, 425, doi: 10.1086/162902

Fisher, G. H., Abbett, W. P., Bercik, D. J., et al. 2015, Space Weather, 13, 369, doi: 10.1002/2015SW001191

Forbush, S. E. 1946, Physical Review, 70, 771, doi: 10.1103/PhysRev.70.771

France, K., Linsky, J. L., & Parke Loyd, R. O. 2014, Ap&SS, 354, 3, doi: 10.1007/s10509-014-1947-2

France, K., Linsky, J. L., Yang, H., Stocke, J. T., & Froning, C. S. 2011, Ap&SS, 335, 25, doi: 10.1007/s10509-011-0611-3

France, K., Stocke, J. T., Yang, H., et al. 2010, ApJ, 712, 1277, doi: 10.1088/0004-637X/712/2/1277

France, K., Loyd, R. O. P., Youngblood, A., et al. 2016, The Astrophysical Journal, 820, 89, doi: 10.3847/0004-637X/820/2/89

France, K., Fleming, B., Youngblood, A., et al. 2022, Journal of Astronomical Telescopes, Instruments, and Systems, 8, 014006, doi: 10.1117/1.JATIS.8.1.014006

Franciosini, E., Pallavicini, R., & Tagliaferri, G. 2001, A&A, 375, 196, doi: 10.1051/0004-6361:20010830

Fraschetti, F., Alvarado-Gómez, J. D., Drake, J. J., Cohen, O., & Garraffo, C. 2022, ApJ, 937, 126, doi: 10.3847/1538-4357/ac86d7

Fraschetti, F., Drake, J. J., Alvarado-Gómez, J. D., et al. 2019, ApJ, 874, 21, doi: 10.3847/1538-4357/ab05e4

Fraschetti, F., Drake, J. J., Cohen, O., & Garraffo, C. 2018, ApJ, 853, 112, doi: 10.3847/1538-4357/aaa48b

Frisch, P. 2000, American Scientist, 88, 52. https://www.americanscientist.org/article/the-galactic-environment-of-the-sun

Funke, B., Dudok de Wit, T., Ermolli, I., et al. 2024, Geoscientific Model Development, 17, 1217, doi: 10.5194/gmd-17-1217-2024

Fuselier, S. A., Frahm, R., Lewis, W. S., et al. 2014, Journal of Geophysical Research (Space Physics), 119, 2563, doi: 10.1002/2013JA019684

Futaana, Y., Stenberg Wieser, G., Barabash, S., & Luhmann, J. G. 2017, SSRv, 212, 1453, doi: 10.1007/s11214-017-0362-8

Gabici, S., Evoli, C., Gaggero, D., et al. 2019, International Journal of Modern Physics D, 28, 1930022, doi: 10.1142/S0218271819300222

Garraffo, C., Drake, J. J., & Cohen, O. 2016, A&A, 595, A110, doi: 10.1051/0004-6361/201628367

Garraffo, C., Drake, J. J., Dotter, A., et al. 2018, ApJ, 862, 90, doi: 10.3847/1538-4357/aace5d

Gary, D. E., Dulk, G. A., House, L. L., et al. 1985, A&A, 152, 42

Gastine, T., Morin, J., Duarte, L., et al. 2013, A&A, 549, L5, doi: 10.1051/0004-6361/201220317

Geballe, T. R., Jagod, M. F., & Oka, T. 1993, ApJL, 408, L109, doi: 10.1086/186843

Gergely, T. E. 1986, SoPh, 104, 175, doi: 10.1007/BF00159959

Giacalone, J., Fahr, H., Fichtner, H., et al. 2022, Space Science Reviews, 218, 22, doi: 10.1007/s11214-022-00890-7

Gibbs, A., & Fitzgerald, M. P. 2022, AJ, 164, 63, doi: 10.3847/1538-3881/ac7718

Gibson, S. E. 2018, Living Reviews in Solar Physics, 15, 7, doi: 10.1007/s41116-018-0016-2

Gibson, S. E., & Low, B. C. 1998, ApJ, 493, 460, doi: 10.1086/305107

Giegenback, R. 2015, Proceedings of the American Philosophical Society, 159, 421. http://www.jstor.org/stable/26159195

Gilbert, H. R., Holzer, T. E., Burkepile, J. T., & Hundhausen, A. J. 2000, ApJ, 537, 503, doi: 10.1086/309030

Gilster, P. 2004, Centauri Dreams: Imagining and Planning Interstellar Exploration (New York: Copernicus Books / Springer)

Gombosi, T. I., van der Holst, B., Manchester, W. B., & Sokolov, I. V. 2018, Living Reviews in Solar Physics, 15, 4, doi: `10.1007/s41116-018-0014-4`

Gopalswamy, N. 2006, SSRv, 124, 145, doi: `10.1007/s11214-006-9102-1`

Gopalswamy, N. 2016, Geoscience Letters, 3, 8, doi: `10.1186/s40562-016-0039-2`

Gopalswamy, N. 2022, Atmosphere, 13, 1781, doi: `10.3390/atmos13111781`

Gopalswamy, N., Aguilar-Rodriguez, E., Yashiro, S., et al. 2005a, Journal of Geophysical Research (Space Physics), 110, A12S07, doi: `10.1029/2005JA011158`

Gopalswamy, N., Lara, A., Lepping, R. P., et al. 2000, Geophys. Res. Lett., 27, 145, doi: `10.1029/1999GL003639`

Gopalswamy, N., Mäkelä, P., & Yashiro, S. 2019, Sun and Geosphere, 14, 111, doi: `10.31401/SunGeo.2019.02.03`

Gopalswamy, N., Michałek, G., Yashiro, S., et al. 2024, The SOHO LASCO CME Catalog – Version 2, arXiv, doi: `10.48550/arXiv.2407.04165`

Gopalswamy, N., Shimojo, M., Lu, W., et al. 2003, ApJ, 586, 562, doi: `10.1086/367614`

Gopalswamy, N., Xie, H., Yashiro, S., et al. 2012, SSRv, 171, 23, doi: `10.1007/s11214-012-9890-4`

Gopalswamy, N., & Yashiro, S. 2013, PASJ, 65, S11, doi: `10.1093/pasj/65.sp1.S11`

Gopalswamy, N., Yashiro, S., Kaiser, M. L., Howard, R. A., & Bougeret, J. L. 2001, J. Geophys. Res., 106, 29219, doi: `10.1029/2001JA000234`

Gopalswamy, N., Yashiro, S., Liu, Y., et al. 2005b, Journal of Geophysical Research (Space Physics), 110, A09S15, doi: `10.1029/2004JA010958`

Gopalswamy, N., Yashiro, S., Thakur, N., et al. 2016, ApJ, 833, 216, doi: `10.3847/1538-4357/833/2/216`

Gosling, J. T. 1993, J. Geophys. Res., 98, 18937, doi: `10.1029/93JA01896`

Gosling, J. T., Hildner, E., MacQueen, R. M., et al. 1974, J. Geophys. Res., 79, 4581, doi: `10.1029/JA079i031p04581`

Grechnev, V. V., Uralov, A. M., Slemzin, V. A., et al. 2008, SoPh, 253, 263, doi: `10.1007/s11207-008-9178-8`

Greene, T. P., Bell, T. J., Ducrot, E., et al. 2023, Nature, 618, 39, doi: `10.1038/s41586-023-05951-7`

Grenfell, J. L., Grießmeier, J.-M., Patzer, B., et al. 2007, Astrobiology, 7, 208, doi: `10.1089/ast.2006.0129`

Grießmeier, J. M. 2007, Planet. Space Sci., 55, 530, doi: `10.1016/j.pss.2006.11.003`

Grießmeier, J. M., Stadelmann, A., Motschmann, U., et al. 2005, Astrobiology, 5, 587, doi: 10.1089/ast.2005.5.587

Grießmeier, J. M., Tabataba-Vakili, F., Stadelmann, A., Grenfell, J. L., & Atri, D. 2016, A&A, 587, A159, doi: 10.1051/0004-6361/201425452

Gronoff, G., Lilensten, J., Desorgher, L., & Flückiger, E. 2009a, A&A, 506, 955, doi: 10.1051/0004-6361/200912371

Gronoff, G., Lilensten, J., & Modolo, R. 2009b, A&A, 506, 965, doi: 10.1051/0004-6361/200912125

Gronoff, G., Simon Wedlund, C., Hegyi, B., et al. 2025, Advances in Space Research, doi: https://doi.org/10.1016/j.asr.2025.03.061

Gronoff, G., Simon Wedlund, C., Mertens, C. J., & Lillis, R. J. 2012, Journal of Geophysical Research (Space Physics), 117, A04306, doi: 10.1029/2011JA016930

Gronoff, G., Arras, P., Baraka, S., et al. 2020, Journal of Geophysical Research (Space Physics), 125, e27639, doi: 10.1029/2019JA02763910.1002/essoar.10502458.1

Güdel, M. 2002, ARA&A, 40, 217, doi: 10.1146/annurev.astro.40.060401.093806

——. 2007, Living Reviews in Solar Physics, 4, 3, doi: 10.12942/lrsp-2007-3

Gunell, H., Maggiolo, R., Nilsson, H., et al. 2018, Astronomy and Astrophysics, 614, L3, doi: 10.1051/0004-6361/201832934

Guo, F., Li, X., French, O., et al. 2023, PhRvL, 130, 189501, doi: 10.1103/PhysRevLett.130.189501

Guo, J. H., & Ben-Jaffel, L. 2016, The Astrophysical Journal, 818, 107, doi: 10.3847/0004-637X/818/2/107

Guo, Y., Ding, M. D., Schmieder, B., et al. 2010, ApJL, 725, L38, doi: 10.1088/2041-8205/725/1/L38

Gustafson, B. A. S. 1994, Annual Review of Earth and Planetary Sciences, 22, 553, doi: 10.1146/annurev.ea.22.050194.003005

Guzmán-Marmolejo, A., Segura, A., & Escobar-Briones, E. 2013, Astrobiology, 13, 550

Güdel, M. 2002, Annual Review of Astronomy and Astrophysics, 40, 217, doi: 10.1146/annurev.astro.40.060401.093806

Habbal, S. R., Morgan, H., & Druckmüller, M. 2014, ApJ, 793, 119, doi: 10.1088/0004-637X/793/2/119

Habbal, S. R., Woo, R., Fineschi, S., et al. 1997, ApJL, 489, L103, doi: 10.1086/310970

Habbal, S. R., Druckmüller, M., Alzate, N., et al. 2021, ApJL, 911, L4, doi: 10.3847/2041-8213/abe775

Haisch, B. M., Linsky, J. L., Bornmann, P. L., et al. 1983, ApJ, 267, 280, doi: 10.1086/160866

Hale, G. E., Ellerman, F., Nicholson, S. B., & Joy, A. H. 1919, ApJ, 49, 153, doi: 10.1086/142452

Hallinan, G., Ravi, V., Weinreb, S., et al. 2019, in Bulletin of the American Astronomical Society, Vol. 51, 255, doi: 10.48550/arXiv.1907.07648

Hand, K., Sotin, C., Hayes, A., & Coustenis, A. 2020, Space science reviews, 216, 95

Hand, K. P., Chyba, C. F., Priscu, J. C., Carlson, R. W., & Nealson, K. H. 2009, Europa, 589

Hansen, R. T., Garcia, C. J., Grognard, R. J. M., & Sheridan, K. V. 1971, PASA, 2, 57

Hapke, B. 2001, Journal of Geophysical Research: Planets, 106, 10039, doi: 10.1029/2000JE001338

Hara, H., Watanabe, T., Harra, L. K., Culhane, J. L., & Young, P. R. 2011, ApJ, 741, 107, doi: 10.1088/0004-637X/741/2/107

Harbach, L. M., Moschou, S. P., Garraffo, C., et al. 2021, ApJ, 913, 130, doi: 10.3847/1538-4357/abf63a

Harra, L. K., Schrijver, C. J., Janvier, M., et al. 2016, SoPh, 291, 1761, doi: 10.1007/s11207-016-0923-0

Hathaway, D. H. 2015, Living Reviews in Solar Physics, 12, 4, doi: 10.1007/lrsp-2015-4

Hawley, S. L., Walkowicz, L. M., Allred, J. C., & Valenti, J. A. 2007, PASP, 119, 67, doi: 10.1086/510561

Hayakawa, H., Ribeiro, J. R., Ebihara, Y., Correia, A. P., & Sôma, M. 2020, Earth, Planets and Space, 72, 122, doi: 10.1186/s40623-020-01249-4

Haywood, R. D., Collier Cameron, A., Unruh, Y. C., et al. 2016, MNRAS, 457, 3637, doi: 10.1093/mnras/stw187

Hazra, G., Vidotto, A. A., Carolan, S., D'Angelo, C. V., & Ó Fionnagáin, D. 2024, Monthly Notices of the Royal Astronomical Society, stae2559, doi: 10.1093/mnras/stae2559

Hazra, G., Vidotto, A. A., Carolan, S., Villarreal D'Angelo, C., & Manchester, W. 2022, MNRAS, 509, 5858, doi: 10.1093/mnras/stab3271

Hazra, G., Vidotto, A. A., Carolan, S., Villarreal D'Angelo, C., & Manchester, W. 2022, Monthly Notices of the Royal Astronomical Society, 509, 5858, doi: 10.1093/mnras/stab3271

Heller, R., & Armstrong, J. 2014, Astrobiology, 14, 50, doi: 10.1089/ast.2013.1088

Heptonstall, A., Andersen, D., Terada, H., et al. 2024, in Ground-based and Airborne Instrumentation for Astronomy X, ed. J. J. Bryant, K. Motohara, & J. R. D. Vernet, Vol. 13096, International Society for Optics and Photonics (SPIE), 130960V, doi: 10.1117/12.3020687

Herbst, K., Papaioannou, A., Banjac, S., & Heber, B. 2019, A&A, 621, A67, doi: 10.1051/0004-6361/201832789

Herbst, K., Grenfell, J. L., Sinnhuber, M., et al. 2019, Astronomy & Astrophysics, 631, A101

Herbst, K., Scherer, K., Ferreira, S. E. S., et al. 2020, ApJL, 897, L27, doi: 10.3847/2041-8213/ab9df3

Herbst, K., Bartenschlager, A., Grenfell, J. L., et al. 2024, ApJ, 961, 164, doi: 10.3847/1538-4357/ad0895

Hewish, A., Scott, P. F., & Wills, D. 1964, Nature, 203, 1214, doi: 10.1038/2031214a0

Hinton, P., Brain, D., Schnepf, N., Jarvinen, R., & Bagenal, F. 2024, 14406, doi: 10.5194/egusphere-egu24-14406

Hodgson, R. 1859, MNRAS, 20, 15, doi: 10.1093/mnras/20.1.15a

Holmström, M., Ekenbäck, A., Selsis, F., et al. 2008, Nature, 451, 970, doi: 10.1038/nature06600

Hou, Z., Tian, H., Yan, J., et al. 2025, Astronomy & Astrophysics, 695, A12, doi: 10.1051/0004-6361/202453282

Houdebine, E. R., Foing, B. H., & Rodono, M. 1990, A&A, 238, 249

House, L. L., Wagner, W. J., Hildner, E., Sawyer, C., & Schmidt, H. U. 1981, ApJL, 244, L117, doi: 10.1086/183494

Howard, R. 1974, SoPh, 38, 283, doi: 10.1007/BF00155067

Hu, J., Airapetian, V. S., Li, G., Zank, G., & Jin, M. 2022, Science Advances, 8, eabi9743, doi: 10.1126/sciadv.abi9743

Hu, J., Li, G., Ao, X., Zank, G. P., & Verkhoglyadova, O. 2017, Journal of Geophysical Research (Space Physics), 122, 10,938, doi: 10.1002/2017JA024077

Hu, R., Peterson, L., & Wolf, E. T. 2020, The Astrophysical Journal, 888, 122

Hudson, H. S., Woods, T. N., Chamberlin, P. C., et al. 2011, SoPh, 273, 69, doi: 10.1007/s11207-011-9862-y

Huenemoerder, D. P., Schulz, N. S., Testa, P., et al. 2010, ApJ, 723, 1558, doi: 10.1088/0004-637X/723/2/1558

Huff, R. L., Calvert, W., Craven, J. D., Frank, L. A., & Gurnett, D. A. 1988, J. Geophys. Res., 93, 11445, doi: 10.1029/JA093iA10p11445

Hughes, A. M., Duchêne, G., & Matthews, B. C. 2018, Annual Review of Astronomy and Astrophysics, 56, 541, doi: 10.1146/annurev-astro-081817-052035

Hurford, G. J., Schwartz, R. A., Krucker, S., et al. 2003, ApJL, 595, L77, doi: 10.1086/378179

Illing, R. M. E., & Hundhausen, A. J. 1985, J. Geophys. Res., 90, 275, doi: 10.1029/JA090iA01p00275

Inoue, S., Maehara, H., Notsu, Y., et al. 2023, ApJ, 948, 9, doi: 10.3847/1538-4357/acb7e8

Inoue, S., Iwakiri, W. B., Enoto, T., et al. 2024, ApJL, 969, L12, doi: `10.3847/2041-8213/ad5667`

Ip, W.-H., Kopp, A., & Hu, J.-H. 2004, ApJL, 602, L53, doi: `10.1086/382274`

Jakosky, B. M., Brain, D., Chaffin, M., et al. 2018, Icarus, 315, 146, doi: `10.1016/j.icarus.2018.05.030`

Jardine, M. 2024, in AAS/Division for Extreme Solar Systems Abstracts, Vol. 56, AAS/Division for Extreme Solar Systems Abstracts, 203.03

Jasinski, J. M., Regoli, L. H., Cassidy, T. A., et al. 2020, Nature Communications, 11, 4350, doi: `10.1038/s41467-020-18220-2`

Jeffers, S. V., Kiefer, R., & Metcalfe, T. S. 2023, SSRv, 219, 54, doi: `10.1007/s11214-023-01000-x`

Jewitt, D. 2024, Non-Gravitational Forces in Planetary Systems. `https://ui.adsabs.harvard.edu/abs/2024arXiv241110923J`

Jian, L. K., Russell, C. T., Luhmann, J. G., Galvin, A. B., & Simunac, K. D. C. 2013, AIP Conference Proceedings, 1539, 191, doi: `10.1063/1.4811020`

Jiang, C., Wu, S. T., Feng, X., & Hu, Q. 2016, Nature Communications, 7, 11522, doi: `10.1038/ncomms11522`

Jin, M., Manchester, W. B., van der Holst, B., et al. 2017a, ApJ, 834, 173, doi: `10.3847/1538-4357/834/2/173`

—. 2017b, ApJ, 834, 173, doi: `10.3847/1538-4357/834/2/173`

Jin, M., Cheung, M. C. M., DeRosa, M. L., et al. 2019, Proceedings of the International Astronomical Union, 15, 426–432, doi: `10.1017/S1743921320000575`

Jivani, A., Sachdeva, N., Huang, Z., et al. 2023, Space Weather, 21, e2022SW003262, doi: `10.1029/2022SW003262`

Johnstone, C. P., Bartel, M., & Güdel, M. 2021, Astronomy and Astrophysics, 649, A96, doi: `10.1051/0004-6361/202038407`

Kahler, S. W. 2001, J. Geophys. Res., 106, 20947, doi: `10.1029/2000JA002231`

Kahler, S. W., & Ling, A. G. 2023, The Astrophysical Journal, 956, 24, doi: `10.3847/1538-4357/acf1ff`

Kahler, S. W., & Reames, D. V. 2003, The Astrophysical Journal, 584, 1063, doi: `10.1086/345780`

Kajdič, P., Sánchez-Cano, B., Neves-Ribeiro, L., et al. 2021, Journal of Geophysical Research (Space Physics), 126, e28442

Kansabanik, D., Mondal, S., & Oberoi, D. 2024, ApJ, 968, 55, doi: `10.3847/1538-4357/ad43e9`

Kao, M. M., Hallinan, G., Pineda, J. S., et al. 2016, The Astrophysical Journal, 818, 24, doi: `10.3847/0004-637X/818/1/24`

Kao, M. M., & Pineda, J. S. 2024, Monthly Notices of the Royal Astronomical Society, doi: `10.1093/mnras/stae905`

Käpylä, P. J., Browning, M. K., Brun, A. S., Guerrero, G., & Warnecke, J. 2023, SSRv, 219, 58, doi: `10.1007/s11214-023-01005-6`

Kasper, J., Lazio, T. J. W., Romero-Wolf, A., Lux, J. P., & Neilsen, T. 2022, in 2022 IEEE Aerospace Conference (AERO), 1–8, doi: `10.1109/AERO53065.2022.9843607`

Kavanagh, R. D., & Vidotto, A. A. 2020, Monthly Notices of the Royal Astronomical Society, 493, 1492, doi: `10.1093/mnras/staa422`

Kavanagh, R. D., Vidotto, A. A., Klein, B., et al. 2021, MNRAS, 504, 1511, doi: `10.1093/mnras/stab929`

Kavanagh, R. D., Vidotto, A. A., Ó. Fionnagáin, D., et al. 2019, Monthly Notices of the Royal Astronomical Society, 485, 4529, doi: `10.1093/mnras/stz655`

Kawaguchi, Y. 2019, in Astrobiology, ed. A. Yamagishi, T. Kakegawa, & T. Usui, 419, doi: `10.1007/978-981-13-3639-3_27`

Kay, C., Opher, M., & Kornbleuth, M. 2016, The Astrophysical Journal, 826, 195, doi: `10.3847/0004-637X/826/2/195`

Khurana, K. K., Pappalardo, R. T., Murphy, N., & Denk, T. 2007, Icarus, 191, 193, doi: `10.1016/j.icarus.2007.04.022`

Kislyakova, K. G., Güdel, M., Koutroumpa, D., et al. 2024, Nature Astronomy, 8, 596, doi: `10.1038/s41550-024-02222-x`

Kislyakova, K. G., Holmström, M., Lammer, H., Odert, P., & Khodachenko, M. L. 2014, Science, 346, 981, doi: `10.1126/science.1257829`

Klein, B., Donati, J.-F., Moutou, C., et al. 2021, MNRAS, 502, 188, doi: `10.1093/mnras/staa3702`

Kobayashi, K., Kaneko, T., Saito, T., & Oshima, T. 1998, Origins of Life and Evolution of the Biosphere, 28, 155

Kobayashi, K., Ise, J.-i., Aoki, R., et al. 2023, Life, 13, 1103

Kochukhov, O. 2016, in Lecture Notes in Physics, Berlin Springer Verlag, ed. J.-P. Rozelot & C. Neiner, Vol. 914 (Springer), 177, doi: `10.1007/978-3-319-24151-7_9`

Kochukhov, O. 2018, Stellar Magnetic Fields, ed. J. Sánchez Almeida & M. J. Martínez González, Canary Islands Winter School of Astrophysics (Cambridge University Press), 47–86

Kochukhov, O. 2021, A&A Rv, 29, 1, doi: `10.1007/s00159-020-00130-3`

Kochukhov, O., Hackman, T., & Lehtinen, J. J. 2023, A&A, 680, L17, doi: `10.1051/0004-6361/202347930`

Kochukhov, O., Hackman, T., Lehtinen, J. J., & Wehrhahn, A. 2020a, A&A, 635, A142, doi: `10.1051/0004-6361/201937185`

—. 2020b, A&A, 635, A142, doi: 10.1051/0004-6361/201937185

Kohl, J. L., Noci, G., Antonucci, E., et al. 1997, SoPh, 175, 613, doi: 10.1023/A:1004903206467

Kopparapu, R. K., Ramirez, R., Kasting, J. F., et al. 2013, ApJ, 765, 131, doi: 10.1088/0004-637X/765/2/131

Koutchmy, S. 1988, SSRv, 47, 95, doi: 10.1007/BF00223238

Krissansen-Totton, J. 2023, The Astrophysical Journal Letters, 951, L39

Krucker, S., Kontar, E. P., Christe, S., & Lin, R. P. 2007, ApJL, 663, L109, doi: 10.1086/519373

Krucker, S., Battaglia, M., Cargill, P. J., et al. 2008, A&A Rv, 16, 155, doi: 10.1007/s00159-008-0014-9

Kumar, S., Srivastava, N., Gopalswamy, N., & Dash, A. 2024, On The Influence Of The Solar Wind On The Propagation Of Earth-impacting Coronal Mass Ejections, doi: 10.48550/arXiv.2411.01165

Kwon, R.-Y., Zhang, J., & Vourlidas, A. 2015, The Astrophysical Journal, 799, L29, doi: 10.1088/2041-8205/799/2/L29

Lammer, H., Kasting, J. F., Chassefière, E., et al. 2008, Space Science Reviews, 139, 399, doi: 10.1007/s11214-008-9413-5

Lammer, H., & Kasting, J.F., Chassefière, E. 2008, Space Science Reviews, 139, 399–436, doi: 10.1007/s11214-017-0362-8

Landi, E., Raymond, J. C., Miralles, M. P., & Hara, H. 2010, ApJ, 711, 75, doi: 10.1088/0004-637X/711/1/75

Lanza, A. F. 2012, A&A, 544, A23, doi: 10.1051/0004-6361/201219002

Lazio, J., Farrell, W. M., Dietrick, J., et al. 2004, arXiv e-prints, astro, doi: 10.48550/arXiv.astro-ph/0405343

Lazio, T. J. W. 2024, arXiv e-prints, arXiv:2404.12348, doi: 10.48550/arXiv.2404.12348

Lecavelier des Etangs, A., Bourrier, V., Wheatley, P. J., et al. 2012, Astronomy and Astrophysics, 543, L4, doi: 10.1051/0004-6361/201219363

Lehmann, L. T., Donati, J. F., Fouqué, P., et al. 2024, MNRAS, 527, 4330, doi: 10.1093/mnras/stad3472

Leitzinger, M., & Odert, P. 2022, Serbian Astronomical Journal, 205, 1, doi: 10.2298/SAJ2205001L

Leitzinger, M., Odert, P., Greimel, R., et al. 2014, MNRAS, 443, 898, doi: 10.1093/mnras/stu1161

Leitzinger, M., Odert, P., & Heinzel, P. 2022, MNRAS, 513, 6058, doi: 10.1093/mnras/stac1284

Leitzinger, M., Odert, P., Ribas, I., et al. 2011, A&A, 536, A62, doi: 10.1051/0004-6361/201015985

Leitzinger, M., Odert, P., Greimel, R., et al. 2020, MNRAS, 493, 4570, doi: 10.1093/mnras/staa504

Lenz, L. F., Reiners, A., Seifahrt, A., & Käufl, H. U. 2016, A&A, 589, A99, doi: 10.1051/0004-6361/201525675

Lewis, B. L., Fitzgerald, M. P., Esposito, T. M., et al. 2024, arXiv e-prints, arXiv:2407.15986, doi: 10.48550/arXiv.2407.15986

Li, G., Jin, M., Ding, Z., et al. 2021, ApJ, 919, 146, doi: 10.3847/1538-4357/ac0db9

Li, H., Wang, C., & Richardson, J. D. 2008, Geophysical Research Letters, 35, 2008GL034869, doi: 10.1029/2008GL034869

Li, J., Raymond, J. C., Acton, L. W., et al. 1998, ApJ, 506, 431, doi: 10.1086/306244

Lin, H., Penn, M. J., & Tomczyk, S. 2000, ApJL, 541, L83, doi: 10.1086/312900

Lin, Y.-C. 2024, Unveiling habitable planets: Toy coronagraph tackles the exozodiacal dust challenge, doi: 10.48550/arXiv.2409.05797

Linsky, J. 2019, Host Stars and their Effects on Exoplanet Atmospheres, Vol. 955, doi: 10.1007/978-3-030-11452-7

Linsky, J. L. 2025, Astrophysics and Space Science Library, Vol. 473, Host Stars and Their Effects on Exoplanet Atmospheres: An Introductory Overview (Cham: Springer International Publishing), doi: 10.1007/978-3-031-75208-7

Linsky, J. L., Rickett, B. J., & Redfield, S. 2008, ApJ, 675, 413, doi: 10.1086/526420

Lionello, R., Downs, C., Mason, E. I., et al. 2023, ApJ, 959, 77, doi: 10.3847/1538-4357/ad00be

Lionello, R., Velli, M., Downs, C., et al. 2014, ApJ, 784, 120, doi: 10.1088/0004-637X/784/2/120

Liu, H., Chen, Y., Cho, K., et al. 2018, SoPh, 293, 58, doi: 10.1007/s11207-018-1280-y

Livingston, W. C., Harvey, J., Slaughter, C., & Trumbo, D. 1976, ApOpt, 15, 40, doi: 10.1364/AO.15.000040

Loeb, A., & Lingam, M. 2018, Journal of the British Interplanetary Society, 71, 2

Loison, J. C., Hébrard, E., Dobrijevic, M., et al. 2015, Icarus, 247, 218, doi: 10.1016/j.icarus.2014.09.039

Loyd, R. O. P., Shkolnik, E. L., Schneider, A. C., et al. 2021, The Astrophysical Journal, 907, 91, doi: 10.3847/1538-4357/abd0f0

Loyd, R. O. P., Mason, J. P., Jin, M., et al. 2022, The Astrophysical Journal, 936, 170, doi: 10.3847/1538-4357/AC80C1

Loyd, R. O. P., Vissapragada, S., Barman, T. S., et al. 2024, STELa: Survey of Transiting Exoplanets in Lyman-alpha, HST Proposal. Cycle 32, ID. #17804

Lu, H.-p., Tian, H., Chen, H.-c., et al. 2023, ApJ, 953, 68, doi: 10.3847/1538-4357/acd6a1

Lubin, P. 2016, Journal of the British Interplanetary Society, 69, 40

Lugaz, N., Zhuang, B., Scolini, C., et al. 2024, The Astrophysical Journal, 962, 193, doi: 10.3847/1538-4357/ad17b9

Luger, R., Lustig-Yaeger, J., Fleming, D. P., et al. 2017, ApJ, 837, 63, doi: 10.3847/1538-4357/aa6040

Luhmann, J. 1986, Space science reviews, 44, 241

Luhmann, J. G. 1990, The Solar Wind Interaction with Unmagnetized Planets: A Tutorial (American Geophysical Union (AGU)), 401–411, doi: https://doi.org/10.1029/GM058p0401

Luhmann, J. G., Kasprzak, W. T., & Russell, C. T. 2007, Journal of Geophysical Research, 112, E04S10, doi: 10.1029/2006JE002820

Luhmann, J. G., Dong, C. F., Ma, Y. J., et al. 2017, Journal of Geophysical Research: Space Physics, 122, 6185, doi: https://doi.org/10.1002/2016JA023513

Lynch, B. J., Wood, B. E., Jin, M., et al. 2023, in Bulletin of the American Astronomical Society, Vol. 55, 254, doi: 10.3847/25c2cfeb.2dd884d5

Lyot, B. 1939, MNRAS, 99, 580, doi: 10.1093/mnras/99.8.580

Lyu, X., Koll, D. D. B., Cowan, N. B., et al. 2024, ApJ, 964, 152, doi: 10.3847/1538-4357/ad2077

Macintosh, B., Graham, J. R., Ingraham, P., et al. 2014, Proceedings of the National Academy of Science, 111, 12661, doi: 10.1073/pnas.1304215111

Macquart, J. P., & de Bruyn, A. G. 2007, MNRAS, 380, L20, doi: 10.1111/j.1745-3933.2007.00341.x

Maehara, H., Notsu, Y., Namekata, K., et al. 2021, PASJ, 73, 44, doi: 10.1093/pasj/psaa098

Maggiolo, R., Maes, L., Cessateur, G., et al. 2022, Journal of Geophysical Research (Space Physics), 127, e2022JA030899, doi: 10.1029/2022JA030899

Maillard, J., & Miller, S. 2011, 450, 19. https://ui.adsabs.harvard.edu/abs/2011ASPC..450..19M

Malanushenko, A., Schrijver, C. J., DeRosa, M. L., & Wheatland, M. S. 2014, ApJ, 783, 102, doi: 10.1088/0004-637X/783/2/102

Mandt, K., Luspay-Kuti, A., Lustig-Yaeger, J., Felton, R., & Domagal-Goldman, S. 2022, The Astrophysical Journal, 930, 73

Mann, I., Köhler, M., Kimura, H., Cechowski, A., & Minato, T. 2006, The Astronomy and Astrophysics Review, 13, 159, doi: 10.1007/s00159-006-0028-0

Marsch, E. 2006, Living Reviews in Solar Physics, 3, 1, doi: 10.12942/lrsp-2006-1

Marsden, S. C., Evensberget, D., Brown, E. L., et al. 2023, MNRAS, 522, 792, doi: 10.1093/mnras/stad925

Martinez, P., Boccaletti, A., Kasper, M., et al. 2008, Astronomy & Astrophysics, 492, 289, doi: `10.1051/0004-6361:200810650`

Mason, E. I., Lionello, R., Downs, C., et al. 2023, ApJL, 959, L4, doi: `10.3847/2041-8213/ad00bd`

Mason, G. M., Li, G., Cohen, C. M. S., et al. 2012, ApJ, 761, 104, doi: `10.1088/0004-637X/761/2/104`

Mason, J. P., Attie, R., Arge, C. N., Thompson, B., & Woods, T. N. 2019, The Astrophysical Journal Supplement Series, 244, 13, doi: `10.3847/1538-4365/ab380e`

Mason, J. P., Woods, T. N., Caspi, A., Thompson, B. J., & Hock, R. A. 2014, ApJ, 789, 61, doi: `10.1088/0004-637X/789/1/61`

Mason, J. P., Woods, T. N., Webb, D. F., et al. 2016, ApJ, 830, 20, doi: `10.3847/0004-637X/830/1/20`

Masunaga, K., Futaana, Y., Persson, M., et al. 2019, Icarus, 321, 379

Matloff, G. L., Mallove, E. F., & Andrews, D. G. 1988, Project Longshot: An Unmanned Probe to Alpha Centauri, Tech. rep., NASA/US Naval Academy

Matsakos, T., Uribe, A., & Königl, A. 2015, Astronomy and Astrophysics, 578, A6, doi: `10.1051/0004-6361/201425593`

Mauk, B., Clarke, J., Grodent, D., et al. 2002, Nature, 415, 1003

Mawet, D., Wizinowich, P., Dekany, R., et al. 2016, in Society of Photo-Optical Instrumentation Engineers (SPIE) Conference Series, Vol. 9909, Adaptive Optics Systems V, ed. E. Marchetti, L. M. Close, & J.-P. Véran, 99090D, doi: `10.1117/12.2233658`

Maxwell, A., & Swarup, G. 1958, Nature, 181, 36, doi: `10.1038/181036a0`

McCann, J., Murray-Clay, R. A., Kratter, K., & Krumholz, M. R. 2019, The Astrophysical Journal, 873, 89, doi: `10.3847/1538-4357/ab05b8`

McComas, D. J., Ebert, R. W., Elliott, H. A., et al. 2008a, Geophys. Res. Lett., 35, L18103, doi: `10.1029/2008GL034896`

—. 2008b, Geophys. Res. Lett., 35, L18103, doi: `10.1029/2008GL034896`

McComas, D. J., Rankin, J. S., Schwadron, N. A., & Swaczyna, P. 2019, The Astrophysical Journal, 884, 145, doi: `10.3847/1538-4357/ab441a`

McComas, D. J., Alexashov, D., Bzowski, M., et al. 2012, Science, 336, 1291, doi: `10.1126/science.1221054`

McComas, D. J., Swaczyna, P., Szalay, J. R., et al. 2021, ApJS, 254, 19, doi: `10.3847/1538-4365/abee76`

McIvor, T., Jardine, M., & Holzwarth, V. 2006, MNRAS, 367, L1, doi: `10.1111/j.1745-3933.2005.00098.x`

McLean, D. J., & Labrum, N. R. 1985, Solar radiophysics : studies of emission from the sun at metre wavelengths

Mcnichol, J. C., & Gordon, R. 2012, in Genesis - In The Beginning: Precursors of Life, Chemical Models and Early Biological Evolution, ed. J. Seckbach (Dordrecht: Springer Netherlands), 591–619, doi: `10.1007/978-94-007-2941-4_30`

Meadows, V. S., Reinhard, C. T., Arney, G. N., et al. 2018, Astrobiology, 18, 630

Mennesson, B., Belikov, R., Por, E., et al. 2024, Current laboratory performance of starlight suppression systems, and potential pathways to desired Habitable Worlds Observatory exoplanet science capabilities, doi: `10.48550/arXiv.2404.18036`

Mertens, C. J., Gronoff, G. P., Zheng, Y., et al. 2024, NASA Scientific and Technical Information Program

Mesquita, A. L., Rodgers-Lee, D., Vidotto, A. A., Atri, D., & Wood, B. E. 2022a, MNRAS, 509, 2091, doi: `10.1093/mnras/stab3131`

Mesquita, A. L., Rodgers-Lee, D., Vidotto, A. A., & Kavanagh, R. D. 2022b, MNRAS, 515, 1218, doi: `10.1093/mnras/stac1624`

Meunier, N., Desort, M., & Lagrange, A. M. 2010, A&A, 512, A39, doi: `10.1051/0004-6361/200913551`

Mewaldt, R. A., Mason, G. M., Gloeckler, G., et al. 2001, in American Institute of Physics Conference Series, Vol. 598, Joint SOHO/ACE workshop "Solar and Galactic Composition", ed. R. F. Wimmer-Schweingruber (AIP), 165–170, doi: `10.1063/1.1433995`

Meyer, P., Parker, E. N., & Simpson, J. A. 1956, Physical Review, 104, 768, doi: `10.1103/PhysRev.104.768`

Mierla, M., Chifu, I., Inhester, B., Rodriguez, L., & Zhukov, A. 2011, A&A, 530, L1, doi: `10.1051/0004-6361/201016295`

Milligan, R. O., & Dennis, B. R. 2009, ApJ, 699, 968, doi: `10.1088/0004-637X/699/2/968`

Mironova, I. A., Aplin, K. L., Arnold, F., et al. 2015, Space science reviews, 194, 1

Mohan, A., Gopalswamy, N., Raju, H., & Akiyama, S. 2024, Novel scaling laws to derive spatially resolved flare and CME parameters from sun-as-a-star observables. `https://ui.adsabs.harvard.edu/abs/2024arXiv240919145M`

Mohan, A., Mondal, S., Wedemeyer, S., & Gopalswamy, N. 2024, arXiv e-prints, arXiv:2402.00185, doi: `10.48550/arXiv.2402.00185`

Mondal, S., Oberoi, D., & Vourlidas, A. 2020, ApJ, 893, 28, doi: `10.3847/1538-4357/ab7fab`

Moore, R. L., Sterling, A. C., Hudson, H. S., & Lemen, J. R. 2001, ApJ, 552, 833, doi: `10.1086/320559`

Morosan, D. E., Kilpua, E. K. J., Carley, E. P., & Monstein, C. 2019a, A&A, 623, A63, doi: 10.1051/0004-6361/201834510

Morosan, D. E., Carley, E. P., Hayes, L. A., et al. 2019b, Nature Astronomy, 3, 452, doi: 10.1038/s41550-019-0689-z

Moschou, S.-P., Drake, J. J., Cohen, O., Alvarado-Gomez, J. D., & Garraffo, C. 2017, ApJ, 850, 191, doi: 10.3847/1538-4357/aa9520

Moschou, S.-P., Drake, J. J., Cohen, O., et al. 2019a, ApJ, 877, 105, doi: 10.3847/1538-4357/ab1b37

—. 2019b, ApJ, 877, 105, doi: 10.3847/1538-4357/ab1b37

Muheki, P., Guenther, E. W., Mutabazi, T., & Jurua, E. 2020, A&A, 637, A13, doi: 10.1051/0004-6361/201936904

Murphy, E. J., Bolatto, A., Chatterjee, S., et al. 2018, in Astronomical Society of the Pacific Conference Series, Vol. 517, Science with a Next Generation Very Large Array, ed. E. Murphy, 3, doi: 10.48550/arXiv.1810.07524

Murphy, N. A., Raymond, J. C., & Korreck, K. E. 2011, ApJ, 735, 17, doi: 10.1088/0004-637X/735/1/17

Murray-Clay, R. A., Chiang, E. I., & Murray, N. 2009, The Astrophysical Journal, 693, 23

Muslimov, E., Neiner, C., & Bouret, J.-C. 2024, arXiv e-prints, arXiv:2410.01491, doi: 10.48550/arXiv.2410.01491

Nail, F., MacLeod, M., Oklopčić, A., et al. 2024, arXiv e-prints, arXiv:2410.19381, doi: 10.48550/arXiv.2410.19381

Nakamura, Y., Leblanc, F., Terada, N., et al. 2023, Journal of Geophysical Research: Space Physics, 128, e2022JA031250

Namekata, K., Ichimoto, K., Ishii, T. T., & Shibata, K. 2022a, ApJ, 933, 209, doi: 10.3847/1538-4357/ac75cd

Namekata, K., Maehara, H., Honda, S., et al. 2022b, arXiv e-prints, arXiv:2211.05506, doi: 10.48550/arXiv.2211.05506

—. 2022c, Nature Astronomy, 6, 241, doi: 10.1038/s41550-021-01532-8

Namekata, K., Airapetian, V. S., Petit, P., et al. 2024, ApJ, 961, 23, doi: 10.3847/1538-4357/ad0b7c

National Academies of Sciences, Engineering, and Medicine. 2018, Exoplanet Science Strategy, National Academies of Sciences, Engineering, and Medicine. 2018. Exoplanet Science Strategy. Washington, DC: The National Academies Press. https://doi.org/10.17226/25187., doi: 10.17226/25187

—. 2021, Pathways to Discovery in Astronomy and Astrophysics for the 2020s (Washington, DC: The National Academies Press)

—. 2022, Origins, Worlds, and Life: A Decadal Strategy for Planetary Science and Astrobiology 2023–2032 (Washington, DC: The National Academies Press)

—. 2024, The Next Decade of Discovery in Solar and Space Physics: Exploring and Safeguarding Humanity's Home in Space, doi: 10.17226/27938

National Academies of Sciences, Engineering, and Medicine, et al. 2024

Nava-Sedeño, J. M., Ortiz-Cervantes, A., Segura, A., & Domagal-Goldman, S. D. 2016, Astrobiology, 16, 744, doi: 10.1089/ast.2015.1435

Neiner, C., Girardot, A., & Reess, J.-M. 2025, arXiv e-prints, arXiv:2503.05556, doi: 10.48550/arXiv.2503.05556

Neubauer, F. M. 1980, J. Geophys. Res., 85, 1171, doi: 10.1029/JA085iA03p01171

Newton, E. R., Irwin, J., Charbonneau, D., et al. 2017, ApJ, 834, 85, doi: 10.3847/1538-4357/834/1/85

Nilsson, H., Barghouthi, I., Slapak, R., Eriksson, A. I., & André, M. 2012, Journal of Geophysical Research: Space Physics, 117

Notsu, Y., Kowalski, A. F., Maehara, H., et al. 2024, ApJ, 961, 189, doi: 10.3847/1538-4357/ad062f

Ó Fionnagáin, D., Vidotto, A. A., Petit, P., et al. 2019, MNRAS, 483, 873, doi: 10.1093/mnras/sty3132

Odert, P., Leitzinger, M., Guenther, E. W., & Heinzel, P. 2020, MNRAS, 494, 3766, doi: 10.1093/mnras/staa1021

Odert, P., Leitzinger, M., Hanslmeier, A., & Lammer, H. 2017, Monthly Notices of the Royal Astronomical Society, 472, 876, doi: 10.1093/mnras/stx1969

Olnon, F. M. 1975, Astronomy and Astrophysics, 39, 217. https://ui.adsabs.harvard.edu/abs/1975A&A....39..217O

Olson, S. L., Schwieterman, E. W., Reinhard, C. T., & Lyons, T. W. 2018, Earth: Atmospheric Evolution of a Habitable Planet, ed. H. J. Deeg & J. A. Belmonte (Cham: Springer International Publishing), 2817–2853, doi: 10.1007/978-3-319-55333-7_189

Opher, M., Loeb, A., & Peek, J. E. G. 2024a, Nature Astronomy, 8, 983, doi: 10.1038/s41550-024-02279-8

Opher, M., Loeb, A., Zucker, C., et al. 2024b, ApJ, 972, 201, doi: 10.3847/1538-4357/ad596e

Oran, R., van der Holst, B., Landi, E., et al. 2013, ApJ, 778, 176, doi: 10.1088/0004-637X/778/2/176

Orell-Miquel, J., Murgas, F., Pallé, E., et al. 2024, The MOPYS project: A survey of 70 planets in search of extended He I and H atmospheres. No evidence of enhanced evaporation in young planets, doi: 10.48550/arXiv.2404.16732

Osten, R. A. 2023, in IAU Symposium, Vol. 370, Winds of Stars and Exoplanets, ed. A. A. Vidotto, L. Fossati, & J. S. Vink, 25–36, doi: 10.1017/S1743921322003714

Osten, R. A., & Bastian, T. S. 2006, ApJ, 637, 1016, doi: 10.1086/498410

Osten, R. A., & Crosley, M. K. 2017, arXiv e-prints, arXiv:1711.05113, doi: 10.48550/arXiv.1711.05113

Osten, R. A., & Wolk, S. J. 2016, Proceedings of the International Astronomical Union, 12, 243–251, doi: 10.1017/S1743921317004252

Otsu, T., Asai, A., Ichimoto, K., Ishii, T. T., & Namekata, K. 2022, ApJ, 939, 98, doi: 10.3847/1538-4357/ac9730

Ottmann, R., & Schmitt, J. H. M. M. 1996, A&A, 307, 813

Owen, J. E. 2019, Annual Review of Earth and Planetary Sciences, 47, 67

Owen, J. E., & Wu, Y. 2017, The Astrophysical Journal, 847, 29, doi: 10.3847/1538-4357/aa890a

Owen, J. E., Murray-Clay, R. A., Schreyer, E., et al. 2023, MNRAS, 518, 4357, doi: 10.1093/mnras/stac3414

Padovani, M., Hennebelle, P., Marcowith, A., & Ferrière, K. 2015, A&A, 582, L13, doi: 10.1051/0004-6361/201526874

Padovani, P., & Cirasuolo, M. 2023, Contemporary Physics, 64, 47, doi: 10.1080/00107514.2023.2266921

Panagia, N., & Felli, M. 1975, A&A, 39, 1

Pandey, J. C., & Singh, K. P. 2012, MNRAS, 419, 1219, doi: 10.1111/j.1365-2966.2011.19776.x

Papaioannou, A., Herbst, K., Ramm, T., et al. 2023, A&A, 671, A66, doi: 10.1051/0004-6361/202243407

Parenti, S. 2014, Living Reviews in Solar Physics, 11, 1, doi: 10.12942/lrsp-2014-1

Park, N., Shkolnik, E. L., & Llama, J. 2025, Constraining the Mass Loss and the Kinetic Energy of Solar Coronal Mass Ejections with Far-Ultraviolet Flares, arXiv, doi: 10.48550/arXiv.2509.16477

Pavlov, A. A., McLain, H., Glavin, D. P., et al. 2024, Astrobiology, 24, 698, doi: 10.1089/ast.2023.0120

Pesses, M. E., Eichler, D., & Jokipii, J. R. 1981, The Astrophysical Journal, 246, L85, doi: 10.1086/183559

Petrie, G. J. D. 2015, Living Reviews in Solar Physics, 12, 5, doi: 10.1007/lrsp-2015-5

Pevtsov, A. A., Balasubramaniam, K. S., & Hock, R. A. 2007, Advances in Space Research, 39, 1781, doi: 10.1016/j.asr.2007.02.058

Pevtsov, A. A., Bertello, L., Nagovitsyn, Y. A., Tlatov, A. G., & Pipin, V. V. 2021, Journal of Space Weather and Space Climate, 11, 4, doi: 10.1051/swsc/2020069

Pietrow, A. G. M., Cretignier, M., Druett, M. K., et al. 2024, A&A, 682, A46, doi: 10.1051/0004-6361/202347895

Pineda, J. S., Hallinan, G., Desert, J.-M., & Harding, L. K. 2024, The Astrophysical Journal, 966, 58, doi: 10.3847/1538-4357/ad2f9e

Pineda, J. S., Youngblood, A., & France, K. 2021, The Astrophysical Journal, 911, 111, doi: 10.3847/1538-4357/abe8d7

Plavchan, P., Jura, M., & Lipscy, S. J. 2006, 357, 127. https://ui.adsabs.harvard.edu/abs/2006ASPC..357..127P

Pontin, D. I., & Priest, E. R. 2022, Living Reviews in Solar Physics, 19, 1, doi: 10.1007/s41116-022-00032-9

Prasad, P., Huizinga, F., Kooistra, E., et al. 2016, Journal of Astronomical Instrumentation, 5, 1641008, doi: 10.1142/S2251171716410087

Presa, A., Driessen, F. A., & Vidotto, A. A. 2024, MNRAS, 534, 3622, doi: 10.1093/mnras/stae2325

Rab, C., Güdel, M., Padovani, M., et al. 2017, A&A, 603, A96, doi: 10.1051/0004-6361/201630241

Ragazzoni, R., Magrin, D., Rauer, H., et al. 2016, in Society of Photo-Optical Instrumentation Engineers (SPIE) Conference Series, Vol. 9904, Space Telescopes and Instrumentation 2016: Optical, Infrared, and Millimeter Wave, ed. H. A. MacEwen, G. G. Fazio, M. Lystrup, N. Batalha, N. Siegler, & E. C. Tong, 990428, doi: 10.1117/12.2236094

Ramsay, G., & Doyle, J. G. 2015, Monthly Notices of the Royal Astronomical Society, 449, 3015

Ramstad, R., & Barabash, S. 2021, Space Science Reviews, 217, 36

Ramstad, R., Barabash, S., Futaana, Y., Nilsson, H., & Holmström, M. 2018, Journal of Geophysical Research: Planets, 123, 3051

Rankin, J. S., Bindi, V., Bykov, A. M., et al. 2022, SSRv, 218, 42, doi: 10.1007/s11214-022-00912-4

Raouafi, N. E., Stenborg, G., Seaton, D. B., et al. 2023, ApJ, 945, 28, doi: 10.3847/1538-4357/acaf6c

Raychaudhuri, P. 2005, in IAU Symposium, Vol. 226, Coronal and Stellar Mass Ejections, ed. K. Dere, J. Wang, & Y. Yan, 211–212, doi: 10.1017/S1743921305000529

Raymond, J. C., & Ciaravella, A. 2004, ApJL, 606, L159, doi: 10.1086/421391

Raymond, J. C., Ciaravella, A., Dobrzycka, D., et al. 2003, ApJ, 597, 1106, doi: 10.1086/378663

Reames, D. V. 2013, Space Science Reviews, 175, 53, doi: 10.1007/s11214-013-9958-9

Reid, H. A. S., & Kontar, E. P. 2017, A&A, 606, A141, doi: 10.1051/0004-6361/201730701

Reinard, A. A., & Biesecker, D. A. 2008, ApJ, 674, 576, doi: 10.1086/525269

Reiners, A. 2012, Living Reviews in Solar Physics, 9, 1, doi: 10.12942/lrsp-2012-1

Reiners, A., Shulyak, D., Käpylä, P. J., et al. 2022a, A&A, 662, A41, doi: 10.1051/0004-6361/202243251

—. 2022b, A&A, 662, A41, doi: 10.1051/0004-6361/202243251

Réville, V., Velli, M., Panasenco, O., et al. 2020, ApJS, 246, 24, doi: 10.3847/1538-4365/ab4fef

Ribas, I., Guinan, E. F., Güdel, M., & Audard, M. 2005, The Astrophysical Journal, 622, 680

Richardson, I. G. 2018, Living Reviews in Solar Physics, 15, 1, doi: 10.1007/s41116-017-0011-z

Richardson, J. D., Burlaga, L. F., Elliott, H., et al. 2022, SSRv, 218, 35, doi: 10.1007/s11214-022-00899-y

Richey-Yowell, T., Shkolnik, E. L., Llama, J., Sikora, J., & Smith, P. 2025, arXiv e-prints, arXiv:2504.15393, doi: 10.48550/arXiv.2504.15393

Ridgway, R. J., Zamyatina, M., Mayne, N. J., et al. 2023, MNRAS, 518, 2472, doi: 10.1093/mnras/stac3105

Robbrecht, E., & Berghmans, D. 2004, Astronomy and Astrophysics, 425, 1097, doi: 10.1051/0004-6361:20041302

Robbrecht, E., Berghmans, D., & Van der Linden, R. A. M. 2009, The Astrophysical Journal, 691, 1222, doi: 10.1088/0004-637X/691/2/1222

Rockcliffe, K. E., Newton, E. R., Youngblood, A., et al. 2023, AJ, 166, 77, doi: 10.3847/1538-3881/ace536

Rodgers-Lee, D., Taylor, A. M., Ray, T. P., & Downes, T. P. 2017, MNRAS, 472, 26, doi: 10.1093/mnras/stx1889

Rodgers-Lee, D., Taylor, A. M., Vidotto, A. A., & Downes, T. P. 2021a, MNRAS, 504, 1519, doi: 10.1093/mnras/stab935

Rodgers-Lee, D., Vidotto, A. A., & Mesquita, A. L. 2021b, MNRAS, 508, 4696, doi: 10.1093/mnras/stab2788

Rodgers-Lee, D., Vidotto, A. A., Taylor, A. M., Rimmer, P. B., & Downes, T. P. 2020, MNRAS, 499, 2124, doi: 10.1093/mnras/staa2737

Ryan, J. M. 2000, SSRv, 93, 581, doi: 10.1023/A:1026547513730

Saar, S. H., & Donahue, R. A. 1997, ApJ, 485, 319, doi: 10.1086/304392

Sabine, E. 1852, Philosophical Transactions of the Royal Society of London Series I, 142, 103

Saint-Hilaire, P., Vilmer, N., & Kerdraon, A. 2013, ApJ, 762, 60, doi: 10.1088/0004-637X/762/1/60

Sandel, B., & Broadfoot, A. 1981, Nature, 292, 679

Sasikumar Raja, K., Maksimovic, M., Kontar, E. P., et al. 2022, ApJ, 924, 58, doi: `10.3847/1538-4357/ac34ed`

Saur, J. 2004, Journal of Geophysical Research (Space Physics), 109, A01210, doi: `10.1029/2002JA009354`

Savcheva, A., & van Ballegooijen, A. 2009a, ApJ, 703, 1766, doi: `10.1088/0004-637X/703/2/1766`

—. 2009b, ApJ, 703, 1766, doi: `10.1088/0004-637X/703/2/1766`

Scherer, K., Herbst, K., Engelbrecht, N. E., et al. 2025, A&A, 694, A106, doi: `10.1051/0004-6361/202450324`

Scherrer, P. H., Wilcox, J. M., Svalgaard, L., et al. 1977, SoPh, 54, 353, doi: `10.1007/BF00159925`

Scheucher, M., Grenfell, J. L., Wunderlich, F., et al. 2018, The Astrophysical Journal, 863, 6

Scheucher, M., Herbst, K., Schmidt, V., et al. 2020, ApJ, 893, 12, doi: `10.3847/1538-4357/ab7b74`

Schillings, A., Slapak, R., Nilsson, H., et al. 2019, Earth, Planets and Space, 71, 70

Schreyer, E., Owen, J. E., Loyd, R. O. P., & Murray-Clay, R. 2024, Monthly Notices of the Royal Astronomical Society, 533, 3296, doi: `10.1093/mnras/stae1976`

Schrijver, C. J., & Zwaan, C. 2000, Solar and Stellar Magnetic Activity

Schwieterman, E. W., Kiang, N. Y., Parenteau, M. N., et al. 2018, Astrobiology, 18, 663

See, V., Yuxi Lu, Amard, L., & Roquette, J. 2024, The impact of stellar metallicity on rotation and activity evolution in the Kepler field using gyro-kinematic ages. `https://ui.adsabs.harvard.edu/abs/2024arXiv240500779S`

See, V., Jardine, M., Vidotto, A. A., et al. 2016, MNRAS, 462, 4442, doi: `10.1093/mnras/stw2010`

Segura, A., Walkowicz, L. M., Meadows, V., Kasting, J., & Hawley, S. 2010, Astrobiology, 10, 751

Share, G. H., Murphy, R. J., White, S. M., et al. 2018, ApJ, 869, 182, doi: `10.3847/1538-4357/aaebf7`

Sheoran, J., Pant, V., Patel, R., & Banerjee, D. 2023, Frontiers in Astronomy and Space Sciences, 10, 27, doi: `10.3389/fspas.2023.1092881`

Shkolnik, E., Gaidos, E., & Moskovitz, N. 2006, AJ, 132, 1267, doi: `10.1086/506476`

Shkolnik, E., Walker, G. A. H., Bohlender, D. A., Gu, P. G., & Kürster, M. 2005, ApJ, 622, 1075, doi: `10.1086/428037`

Shkolnik, E. L., & Barman, T. S. 2014, AJ, 148, 64, doi: `10.1088/0004-6256/148/4/64`

Shkolnik, E. L., & Llama, J. 2018, in Handbook of Exoplanets, ed. H. J. Deeg & J. A. Belmonte, 20, doi: `10.1007/978-3-319-55333-7_20`

Shulyak, D., Reiners, A., Engeln, A., et al. 2017, Nature Astronomy, 1, 0184, doi: `10.1038/s41550-017-0184`

Simnett, G. M. 2000, Journal of Atmospheric and Solar-Terrestrial Physics, 62, 1479, doi: `10.1016/S1364-6826(00)00088-2`

Sinnhuber, M., Nieder, H., & Wieters, N. 2012, Surveys in Geophysics, 33, 1281

Skumanich, A. 1972, The Astrophysical Journal, 171, 565, doi: `10.1086/151310`

Slapak, R., Schillings, A., Nilsson, H., et al. 2017, Annales Geophysicae, 35, 721, doi: `10.5194/angeo-35-721-2017`

Slavin, J. D., Frisch, P. C., Müller, H.-R., et al. 2012, ApJ, 760, 46, doi: `10.1088/0004-637X/760/1/46`

Smith, R. K., Brickhouse, N. S., Liedahl, D. A., & Raymond, J. C. 2001, ApJL, 556, L91, doi: `10.1086/322992`

Sokolov, I. V., & Gombosi, T. I. 2023, ApJ, 955, 126, doi: `10.3847/1538-4357/aceef5`

Sokolov, I. V., van der Holst, B., Oran, R., et al. 2013, ApJ, 764, 23, doi: `10.1088/0004-637X/764/1/23`

Song, Y., Paglione, T. A. D., & Ilin, E. 2024, MNRAS, 531, 3215, doi: `10.1093/mnras/stae1347`

Sterken, V. J., Altobelli, N., Kempf, S., et al. 2013, A&A, 552, A130, doi: `10.1051/0004-6361/201219609`

Sterling, A. C., & Hudson, H. S. 1997, ApJL, 491, L55, doi: `10.1086/311043`

Stief, F., Löhner-Böttcher, J., Schmidt, W., Steinmetz, T., & Holzwarth, R. 2019, A&A, 622, A34, doi: `10.1051/0004-6361/201834538`

Stone, E. C., Alkalai, L., & Freedman, L. 2015, doi: `10.26206/3PKN-N663`

Stone, E. C., Cummings, A. C., Heikkila, B. C., & Lal, N. 2019, Nature Astronomy, 3, 1013, doi: `10.1038/s41550-019-0928-3`

Stone, E. C., Cummings, A. C., McDonald, F. B., et al. 2005, Science, 309, 2017, doi: `10.1126/science.1117684`

Strugarek, A., Brun, A. S., Donati, J. F., Moutou, C., & Réville, V. 2019, ApJ, 881, 136, doi: `10.3847/1538-4357/ab2ed5`

Strugarek, A., & Shkolnik, E. 2025, arXiv e-prints, arXiv:2502.13262, doi: `10.48550/arXiv.2502.13262`

Su, Y., van Ballegooijen, A., Lites, B. W., et al. 2009, ApJ, 691, 105, doi: `10.1088/0004-637X/691/1/105`

Temmer, M. 2021, Living Reviews in Solar Physics, 18, 4, doi: `10.1007/s41116-021-00030-3`

Temmer, M., Holzknecht, L., Dumbović, M., et al. 2021, Journal of Geophysical Research (Space Physics), 126, e28380, doi: 10.1029/2020JA028380

Thomas, T. B., Hu, R., & Lo, D. Y. 2023, The Planetary Science Journal, 4, 41

Thompson, B. J., Plunkett, S. P., Gurman, J. B., et al. 1998, Geophys. Res. Lett., 25, 2465, doi: 10.1029/98GL50429

Thompson, M. A., Krissansen-Totton, J., Wogan, N., Telus, M., & Fortney, J. J. 2022, Proceedings of the National Academy of Sciences, 119, e2117933119

Tian, H., Tomczyk, S., McIntosh, S. W., et al. 2013, SoPh, 288, 637, doi: 10.1007/s11207-013-0317-5

Tilley, M. A., Segura, A., Meadows, V., Hawley, S., & Davenport, J. 2019, Astrobiology, 19, 64

Titov, V. S., & Démoulin, P. 1999, A&A, 351, 707

Titov, V. S., Török, T., Mikic, Z., & Linker, J. A. 2014, ApJ, 790, 163, doi: 10.1088/0004-637X/790/2/163

Tokumaru, M., Kojima, M., Fujiki, K., Yamashita, M., & Yokobe, A. 2003, Journal of Geophysical Research (Space Physics), 108, 1220, doi: 10.1029/2002JA009574

Tomczyk, S., McIntosh, S. W., Keil, S. L., et al. 2007, Science, 317, 1192, doi: 10.1126/science.1143304

Török, T., Lugaz, N., Lee, C. O., et al. 2023, in Bulletin of the American Astronomical Society, Vol. 55, 393, doi: 10.3847/25c2cfeb.395f28b4

Tousey, R., Bartoe, J. D. F., Bohlin, J. D., et al. 1973, SoPh, 33, 265, doi: 10.1007/BF00152418

Trafton, L. M., Geballe, T. R., Miller, S., Tennyson, J., & Ballester, G. E. 1993, ApJ, 405, 761, doi: 10.1086/172404

Trainer, G. M. 2013, Current Organic Chemistry, 17, 1710

Tsuboi, Y., Koyama, K., Murakami, H., et al. 1998, ApJ, 503, 894, doi: 10.1086/306024

Tu, L., Johnstone, C. P., Güdel, M., & Lammer, H. 2015, Astronomy & Astrophysics, 577, L3, doi: 10.1051/0004-6361/201526146

Turnpenney, S., Nichols, J. D., Wynn, G. A., & Burleigh, M. R. 2018, ApJ, 854, 72, doi: 10.3847/1538-4357/aaa59c

van Ballegooijen, A. A. 2004, ApJ, 612, 519, doi: 10.1086/422512

van der Holst, B., Sokolov, I. V., Meng, X., et al. 2014a, ApJ, 782, 81, doi: 10.1088/0004-637X/782/2/81

—. 2014b, ApJ, 782, 81, doi: 10.1088/0004-637X/782/2/81

van der Holst, B., Huang, J., Sachdeva, N., et al. 2022, ApJ, 925, 146, doi: 10.3847/1538-4357/ac3d34

Vedantham, H. K. 2020, A&A, 639, L7, doi: 10.1051/0004-6361/202038576

Vedantham, H. K., Callingham, J. R., Shimwell, T. W., et al. 2020, Nature Astronomy, 4, 577, doi: 10.1038/s41550-020-1011-9

Veronig, A. M., Brown, J. C., Dennis, B. R., et al. 2005, ApJ, 621, 482, doi: 10.1086/427274

Veronig, A. M., Gömöry, P., Dissauer, K., Temmer, M., & Vanninathan, K. 2019, The Astrophysical Journal, 879, 85, doi: 10.3847/1538-4357/ab2712

Veronig, A. M., Odert, P., Leitzinger, M., et al. 2021, Nature Astronomy, 5, 697, doi: 10.1038/s41550-021-01345-9

Veronig, A. M., Dissauer, K., Kliem, B., et al. 2025, Coronal dimmings and what they tell us about solar and stellar coronal mass ejections, arXiv. https://ui.adsabs.harvard.edu/abs/2025arXiv250519228V

Vestrand, W. T., Share, G. H., J. Murphy, R., et al. 1999, ApJS, 120, 409, doi: 10.1086/313180

Vial, J.-C., & Engvold, O., eds. 2015, Solar Prominences, Astrophysics and Space Science Library (Cham: Springer International Publishing), doi: 10.1007/978-3-319-10416-4

Vial, J.-C., & Engvold, O., eds. 2015, Astrophysics and Space Science Library, Vol. 415, Solar Prominences, doi: 10.1007/978-3-319-10416-4

Vida, K., Kriskovics, L., Oláh, K., et al. 2016a, A&A, 590, A11, doi: 10.1051/0004-6361/201527925

—. 2016b, A&A, 590, A11, doi: 10.1051/0004-6361/201527925

Vida, K., Kővári, Z., Leitzinger, M., et al. 2024, Universe, 10, 313, doi: 10.3390/universe10080313

Vidotto, A. A. 2021, Living Reviews in Solar Physics, 18, 3, doi: 10.1007/s41116-021-00029-w

Vidotto, A. A., Bourrier, V., Fares, R., et al. 2023, A&A, 678, A152, doi: 10.1051/0004-6361/202347237

Vidotto, A. A., & Donati, J. F. 2017, Astronomy and Astrophysics, 602, A39, doi: 10.1051/0004-6361/201629700

Vidotto, A. A., Fares, R., Jardine, M., Moutou, C., & Donati, J. F. 2015, MNRAS, 449, 4117, doi: 10.1093/mnras/stv618

Villadsen, J., & Hallinan, G. 2019, ApJ, 871, 214, doi: 10.3847/1538-4357/aaf88e

Villadsen, J., Hallinan, G., Bourke, S., Güdel, M., & Rupen, M. 2014, ApJ, 788, 112, doi: 10.1088/0004-637X/788/2/112

Villadsen, J., Hallinan, G., Bourke, S., Güdel, M., & Rupen, M. 2014, The Astrophysical Journal, 788, 112, doi: 10.1088/0004-637X/788/2/112

Villarreal D'Angelo, C., Esquivel, A., Schneiter, M., & Sgró, M. A. 2018, MNRAS, 479, 3115, doi: 10.1093/mnras/sty1544

Vilmer, N., MacKinnon, A. L., & Hurford, G. J. 2011, SSRv, 159, 167, doi: 10.1007/s11214-010-9728-x

Vitkevich, V. V., & Vlasov, V. I. 1970, Soviet Ast., 13, 669

Vourlidas, A., Carley, E. P., & Vilmer, N. 2020, Frontiers in Astronomy and Space Sciences, 7, 43, doi: 10.3389/fspas.2020.00043

Waggoner, A. R., & Cleeves, L. I. 2019, ApJ, 883, 197, doi: 10.3847/1538-4357/ab3d38

Wagner, K., Boehle, A., Pathak, P., et al. 2021, Nature Communications, 12, 922, doi: 10.1038/s41467-021-21176-6

Waite, J. H., Young, D. T., Cravens, T. E., et al. 2007, Science, 316, 870, doi: 10.1126/science.1139727

Wang, H. P., Poedts, S., Lani, A., et al. 2025, A&A, 694, A234, doi: 10.1051/0004-6361/202452279

Wang, Y., Tuntsov, A., Murphy, T., et al. 2021, MNRAS, 502, 3294, doi: 10.1093/mnras/stab139

Wang, Y. M., & Sheeley, Jr., N. R. 2002, ApJ, 575, 542, doi: 10.1086/341145

Wargelin, B. J., & Drake, J. J. 2001, ApJL, 546, L57, doi: 10.1086/318066

—. 2002, ApJ, 578, 503, doi: 10.1086/342270

Webb, D. F., & Howard, T. A. 2012, Living Reviews in Solar Physics, 9, 3, doi: 10.12942/lrsp-2012-3

Webb, D. F., Kahler, S. W., McIntosh, P. S., & Klimchuck, J. A. 1997, J. Geophys. Res., 102, 24161, doi: 10.1029/97JA01867

Whitman, K., Egeland, R., Richardson, I. G., et al. 2023, Advances in Space Research, 72, 5161, doi: 10.1016/j.asr.2022.08.006

Wilson, D. J., Froning, C. S., Duvvuri, G. M., et al. 2021, The Astrophysical Journal, 911, 18, doi: 10.3847/1538-4357/abe771

Wilson, M. L., & Raymond, J. C. 2022, AJ, 164, 108, doi: 10.3847/1538-3881/ac80c4

Wilson, M. L., Raymond, J. C., Lepri, S. T., et al. 2022, ApJ, 927, 27, doi: 10.3847/1538-4357/ac4d35

Winter, L. M., & Ledbetter, K. 2015, ApJ, 809, 105, doi: 10.1088/0004-637X/809/1/105

Wood, B. E. 2004, Living Reviews in Solar Physics, 1, 2, doi: 10.12942/lrsp-2004-2

Wood, B. E., Linsky, J. L., Müller, H.-R., & Zank, G. P. 2001, ApJL, 547, L49, doi: 10.1086/318888

Wood, B. E., Müller, H.-R., Zank, G. P., & Linsky, J. L. 2002, ApJ, 574, 412, doi: 10.1086/340797

Wood, B. E., Müller, H. R., Zank, G. P., Linsky, J. L., & Redfield, S. 2005, ApJL, 628, L143, doi: 10.1086/432716

Wood, B. E., Müller, H.-R., Redfield, S., et al. 2021, ApJ, 915, 37, doi: 10.3847/1538-4357/abfda5

Wood, B. E., Müller, H.-R., Redfield, S., et al. 2021, The Astrophysical Journal, 915, 37, doi: 10.3847/1538-4357/abfda5

Wordsworth, R., & Kreidberg, L. 2022, Annual Review of Astronomy and Astrophysics, 60, 159, doi: https://doi.org/10.1146/annurev-astro-052920-125632

Wright, A. E., & Barlow, M. J. 1975, MNRAS, 170, 41, doi: 10.1093/mnras/170.1.41

Xu, Y., Alvarado-Gómez, J. D., Tian, H., et al. 2024, arXiv e-prints, arXiv:2406.08194, doi: 10.48550/arXiv.2406.08194

Xu, Y., Tian, H., Alvarado-Gómez, J. D., Drake, J. J., & Guerrero, G. 2025, Simulations of coronal mass ejections on a young solar-type star and their detectability through coronal spectral observations, arXiv, doi: 10.48550/arXiv.2504.04144

Xu, Y., Tian, H., Hou, Z., et al. 2022, ApJ, 931, 76, doi: 10.3847/1538-4357/ac69d5

Xu, Y., Tian, H., Veronig, A. M., & Dissauer, K. 2024, The Astrophysical Journal, 970, 60, doi: 10.3847/1538-4357/ad500b

Yamashiki, Y. A., Maehara, H., Airapetian, V., et al. 2019, ApJ, 881, 114, doi: 10.3847/1538-4357/ab2a71

Yang, Z., Bethge, C., Tian, H., et al. 2020, Science, 369, 694, doi: 10.1126/science.abb4462

Yashiro, S., & Gopalswamy, N. 2009, in IAU Symposium, Vol. 257, Universal Heliophysical Processes, ed. N. Gopalswamy & D. F. Webb, 233–243, doi: 10.1017/S1743921309029342

Yashiro, S., Gopalswamy, N., Michalek, G., et al. 2004, Journal of Geophysical Research (Space Physics), 109, A07105, doi: 10.1029/2003JA010282

Yau, A., Shelley, E., Peterson, W., & Lenchyshyn, L. 1985, Journal of Geophysical Research: Space Physics, 90, 8417

Yeates, A. R., Nandy, D., & Mackay, D. H. 2008, ApJ, 673, 544, doi: 10.1086/524352

Youngblood, A., France, K., Loyd, R. O. P., et al. 2017, The Astrophysical Journal, 843, 31, doi: 10.3847/1538-4357/aa76dd

Yung, Y. L., & DeMore, W. B. 1999, Photochemistry of planetary atmospheres (Oxford University Press, USA)

Zahnle, K. J., & Catling, D. C. 2017, The Astrophysical Journal, 843, 122

Zarka, P. 1998, J. Geophys. Res., 103, 20159, doi: 10.1029/98JE01323

—. 2007a, Planet. Space Sci., 55, 598, doi: 10.1016/j.pss.2006.05.045

—. 2007b, Planet. Space Sci., 55, 598, doi: `10.1016/j.pss.2006.05.045`

—. 2018, in Handbook of Exoplanets, ed. H. J. Deeg & J. A. Belmonte, 22, doi: `10.1007/978-3-319-55333-7_2210.1007/978-3-319-30648-3_22-1`

Zhang, J., Temmer, M., Gopalswamy, N., et al. 2021, Progress in Earth and Planetary Science, 8, 56, doi: `10.1186/s40645-021-00426-7`

Zic, A., Murphy, T., Lynch, C., et al. 2020, ApJ, 905, 23, doi: `10.3847/1538-4357/abca90`

Zieba, S., Kreidberg, L., Ducrot, E., et al. 2023, No thick carbon dioxide atmosphere on the rocky exoplanet TRAPPIST-1 c, doi: `10.48550/arXiv.2306.10150`

Ó Fionnagáin, D., Vidotto, A. A., Petit, P., et al. 2019, Monthly Notices of the Royal Astronomical Society, 483, 873, doi: `10.1093/mnras/sty3132`

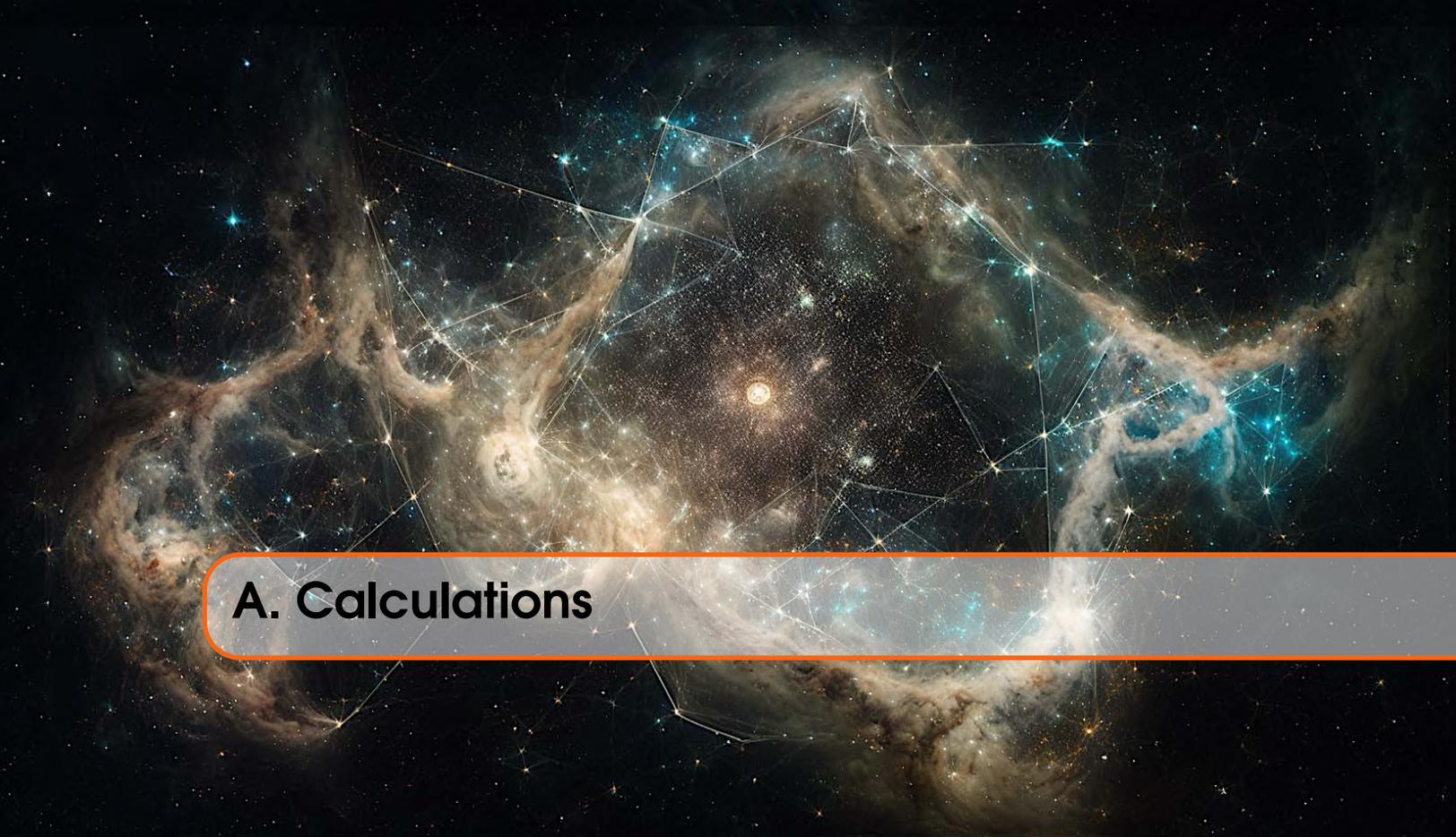

# A. Calculations

## A.1 Quasi-Steady: Direct Imaging of Astrospheres

Some simple back-of-the-envelope calculations establish potential feasibility for the direct imaging of astrospheres. As a limiting case, we consider optically thin H$\alpha$ emission, and estimate the integrated line flux using standard equations for the flux expected from a plasma in collisional ionization equilibrium:

$$f_{\text{line}} = \frac{A E_{\text{line}} \varepsilon(T) n_e^2 V}{4\pi d^2} \quad , \tag{A.1}$$

where $A$ is the elemental abundance relative to solar, $E_{\text{line}}$ is the energy of the transition in erg, $\varepsilon$ is the emissivity of the transition (units of photons cm$^3$ s$^{-1}$), generally a function of temperature, $n_e$ is the electron density, $V$ the physical volume of the emitting region, and $d$ is the distance to the object.

Wood et al. (2002) used their astrospheric models to infer what the variation of hydrogen temperature and density would be from the best match of the model to the excess blue-ward absorption in the Lyman$\alpha$ emission line. Examining these trends, most of $\varepsilon$ Eri's astrosphere is expected to have a temperature of roughly $4\times10^4$ K, but the temperature where the electron density is at its peak (n$_e$ $\sim$0.4 cm$^{-3}$ is about $2\times10^4$ K. This temperature is near the peak temperature of H$\alpha$ emissivity.

We assume $A = A_\odot = 1$ for the hydrogen abundance. The plasma emissivity is taken from the calculations of ATOMDB (Smith et al., 2001) and summed over the individual transitions of H at this central wavelength. For volume, we assume a rectangular prism with a square face of size one arcsecond on the sky and length $l$. To estimate $l$, we assume that the length-to-width ratio of the astrosphere is on the flat side, and take a value of $l/w \approx 0.3$ with the full width as stated above of $\approx$8000 AU. For $\varepsilon$ Eri's astrosphere at a distance of 3.2 parsecs, this leads to an estimation of the integrated line flux of $1.4\times10^{-20}$ erg cm$^{-2}$ s$^{-1}$. To estimate a flux density, we use the full width at half maximum (FWHM) of the R band filter, which contains the H$\alpha$ emission line, 1380 Å. This leads to a flux density of $10^{-23}$ erg cm$^{-2}$ s$^{-1}$ Å$^{-1}$ or an R magnitude of 35.6.

## A.2    Hα Emission from Astrosphere Hydrogen Wall

## A.3    Radio Scintillation

The idea described in Section 4.3.5 is to use a constellation of background quasars to determine the extent of a star's astrosphere, by identifying those quasars that exhibit scintillations as containing a line of sight through the astrosphere. Assuming the CME is undergoing spherical expansion, the density would decrease as $n_{e,CME}(r) = n_{e,CME}(r_0)(r_0/r)^3$. There would be a critical distance from the star $d_{crit}$ where this enhanced CME wind would exceed the background interstellar medium (ISM) density,

$$n_{\text{ISM}} = n_{e,CME}(d_{\text{crit}}) = n_{e,CME}(r_0)(r_0/d_{\text{crit}})^3 \; , \qquad (A.2)$$

and where any density irregularities could plausibly induce scintillation. Taking numbers from the solar system, the number density of one fast CME was measured at $3R_\odot$ to be in the range $5 \times 10^5$-$10^7$ cm$^{-3}$ (Raymond & Ciaravella, 2004). For the local interstellar medium electron density of 0.1 cm$^{-3}$, this leads to $d_{crit}$ values of 2.4-6.4 AU. As this number was the density associated with one CME, a CME-dominated wind could be several times more dense and thus push this distance further out.

## A.4    CME Thomson-scattering Contrast Estimation

For this estimate, we do an order-of-magnitude calculation for the ideal case that a CME is launched off the limb of the star from our viewing perspective. The luminosity of a single electron at a distance $r$ from the stellar surface due to reflected stellar flux is:

$$L_{e^-} = F_\star \sigma_{\text{T}} = \frac{L_\star \sigma_{\text{T}}}{4\pi r^2} \qquad (A.3)$$

and the total luminosity from electrons in a CME due to Thomson scattering is:

$$L_{\text{CME}} = L_{\text{e}^-} \times N_{\text{e}^-} = L_{\text{e}^-} \times n_{\text{e}^-} \cdot V, \qquad (A.4)$$

where $\sigma_{\text{T}}$ is the Thomson cross section of an electron, $L_\star$ is the luminosity of the star, $N_{\text{e}^-}$ is the total number of scattering electrons, $n_{\text{e}^-}$ is the number density of electrons, and $V$ is the volume of the CME. If we assume the mass of the CME is dominated by ions of various species $X$ and that the number density of ions scales with the number density of electrons, then:

$$n_{\text{e}^-} \cdot V \approx \frac{M_{\text{CME}}}{m_X} \qquad (A.5)$$

so that contrast $C$ is:

$$C = \frac{L_{\text{CME}}}{L_\star} = \frac{L_{\text{e}^-}}{L_\star} \times n_{\text{e}^-} \cdot V = \frac{\sigma_{\text{T}}}{4\pi r^2} \frac{M_{\text{CME}}}{m_X} \qquad (A.6)$$

For simplicity, we assume a purely hydrogen CME so that $m_X$ is just the proton mass. Contrast, then, is simply a function of CME mass and radius.

There are many additional implicit assumptions made in this approximation—for instance, we neglect chromatic effects (e.g., due to Rayleigh scattering) and assume the CME is optically thin and contained within a single pixel. The optically thin assumption is satisfied for column densities such that $n_{\text{e}^-} l << 1/\sigma_{\text{T}} = 1.5 \times 10^{24}$ cm$^{-2}$, where $l$ is the line-of-sight depth of the CME. From this expression, the optically thin assumption should be a reasonable expectation for even truly extreme values of $n_{e^-}$ and CME line-of-sight depth (e.g., $n_{e^-} > 10^9 \text{cm}^{-3}$ and depths of several AU). However, for the high angular resolution of direct-imaging instrumentation ($\theta \lesssim 0.5$"), we should consider the possibility of stellar CMEs spanning multiple pixels.

To address this, we now consider the per-pixel contrast to evaluate the detectability of a CME in any given pixel it spans. For this consideration, let us instead represent the CME volume as a function of its cross-sectional area $A_{CME}$ and its line-of-sight depth $l$. The number of pixels the CME spans will depend on the distance-scaled pixel area, $A_{pix} = (\theta d)^2 [\text{AU}^2]$ where $d$ is the distance to the star in parsecs (pc) and $\theta$ is the plate scale (units arcsec/pix). The measured intensity of the CME then depends on 1) the number of Thomson-scattering electrons measured by each pixel and 2) the number of pixels that contain the CME $N_{pix} = \lceil A_{CME}/A_{pix} \rceil$, where $\lceil \; \rceil$ indicate taking the ceiling value since the number of pixels is a discrete integer value. The contrast per pixel can now be represented as:

$$C = \frac{\sigma_T}{4\pi r^2} n_{e^-, pix} N_{pix} = \frac{\sigma_T}{4\pi r^2} (\theta \, d)^2 \, l \, n_{e^-} \, N_{pix} = \frac{\sigma_T}{4\pi r^2} \frac{A_{pix}}{A_{CME}} \frac{M_{CME}}{m_X} \qquad (A.7)$$

where $n_{e^-, pix}$ is the number of Thomson-scattering electrons per pixel and $A_{pix}/A_{CME} < 1$ for this case that the CME spans multiple pixels. If we know the pixels that the CME spans, then the contrast scales by a factor of $\lceil \frac{A_{CME}}{A_{pix}} \rceil$; in the case that the area of the CME is exactly the area of the pixels that it spans, this equation simplifies to equation A.6, assuming the correct pixels are summed. Now, the contrast *explicitly* depends on the resolution of the instrument and distance of the star (while still also implicitly depending on these values via the minimum value of $r$ that can be resolved).

## A.5    Stellar Wind Thomson-scattering Contrast Estimation

A first-order estimate of the contrast of Thomson-scattered light from the stellar wind is possible under the simplifying assumptions that the wind is isotropic, of constant velocity, made entirely of ionized hydrogen, and scatters isotropically. The volume number density of such a wind is $n(r) = \dot{M}/4\pi r^2 v m_p$, where $\dot{M}$ is the mass loss rate of the stellar wind, $v$ is the wind speed, and $m_p$ is the proton mass, and it is irradiated with a flux of $L/4\pi r^2$, neglecting losses from scattering and considering only points far enough from the star that treating it as a point source is a reasonable approximation. Under these assumptions, the volume scattering rate, $\varepsilon$, of the wind is

$$\varepsilon = \frac{nL\sigma_T}{4\pi r^2} = \frac{\dot{M}L\sigma_T}{16\pi^2 r^4 v m_p}, \qquad (A.8)$$

where $\sigma_T$ is the Thomson-scattering cross section, although this expression and those following could be generalized to any scattering mechanism that is approximately isotropic by substituting the appropriate cross section. Integrating this along a line of sight with a closest approach of $r_\perp$ to the star gives a flux from the wind (per unit area on the plane of the sky passing through the star). This integral is

$$F_{wind}(r_\perp) = \int_{-\infty}^{\infty} \varepsilon \; ds \qquad (A.9)$$

where $s$ is an integration coordinate along the line of sight defined as $s^2 = r^2 - r_\perp^2$. The solution is

$$F_{wind}(r_\perp) = \frac{\dot{M}L\sigma_T}{32\pi r_\perp^3 v m_p}. \qquad (A.10)$$

In starlight suppression systems, a theoretical contrast value as a function of radial distance from the central star is generally defined as the flux of the diffracted and scattered starlight at that distance relative to the flux of the star at the center of the Airy disk for an image of the star without any suppression (e.g., Martinez et al., 2008). To enable comparisons with the contrast of scattered light from the wind, we compute an analogous contrast by comparing the diffuse flux of the wind to the diffuse flux of the star at the center of the Airy disk. We find it convenient to do this by treating the Airy disk of the star as imaged by the instrument projected onto the sky. We derive this "apparent" flux from the expression for the flux of a point source imaged through a circular aperture at the center of the Airy disk, $F = \pi P D^2 / 4\lambda^2 l^2$, where $P$ is the power entering the aperture, $D$ is the diameter of the circular aperture, $\lambda$ is the wavelength, and $l$ is the distance from aperture to image plane. Projecting onto the sky plane, power within the Airy disk will be spread out by a factor of $d^2/l^2$ relative to the image plane. And just as integrating the flux across the image plane of the telescope would yield the full power entering the aperture, $P$, we wish to integrate the Airy disk of the starlight projected on the sky to yield the power emitted by the star, $L$, justifying the substitution $P \to L$. Thus, the starlight will appear to be coming from a diffuse source with a surface flux of

$$F_\star = \frac{L\pi D^2}{4\lambda^2 d^2} \tag{A.11}$$

at its center on the plane of the sky, where $D$ is the diameter of the instrument aperture (larger apertures produce smaller Airy disks, increasing the apparent diffuse flux), $\lambda$ is the wavelength of light, and $d$ is the distance to the object (for more distant objects, the Airy disk projects onto a larger physical area at the star, reducing the apparent diffuse flux). The contrast, $C$, of the wind relative to the star is then

$$C = \frac{F_{\text{wind}}}{F_\star} = \frac{\dot{M}\lambda^2 d^2 \sigma_T}{8\pi^2 D^2 r_\perp^3 v m_p}. \tag{A.12}$$

This formulation neglects the smearing of light from the wind due to the point spread function of the instrument, though this is not a major concern beyond $\sim 2\lambda/D$ from the star.

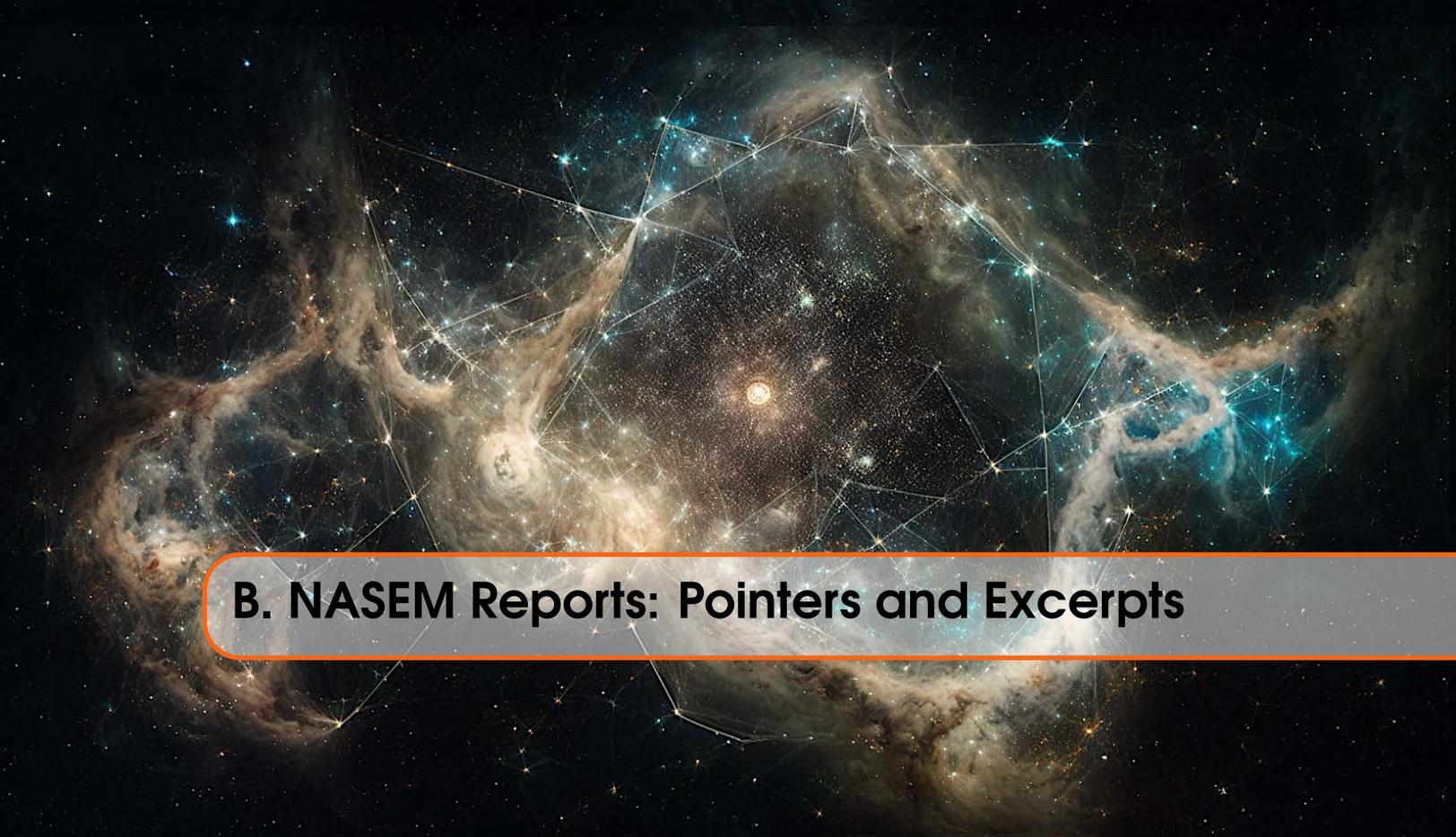

# B. NASEM Reports: Pointers and Excerpts

This appendix provides pointers to specific text in four relevant NASEM science survey or strategy reports that may prove helpful for developing or motivating exospace weather research programs. These are not meant to be complete or comprehensive, but rather provide starting points for further exploration of these visionary reports.

**Solar & Space Physics (aka, Heliophysics) Decadal**

(National Academies of Sciences, Engineering, and Medicine et al., 2024)

- Key guiding questions
    - "What can we learn from comparative studies of planetary systems?"
    - "Why does the Sun and its environment differ from other similar stars?"
    - "What internal and external characteristics have played a role in creating a space environment conducive to life?"
- Call for cross-disciplinary research: "transformative research is often found at the boundaries between disciplines"
- comparison of the Sun-Earth-heliosphere to other stellar systems identified as a "particularly promising research direction" and necessary to determine whether life exists in other systems.
- "heliospheric models can be used to generate synthetic observations to quantify detectability thresholds for other systems"
- recommendation of increasing the levels of investment to expand theory and modeling efforts (p. 30)

**Astronomy & Astrophysics Decadal**

(National Academies of Sciences, Engineering, and Medicine, 2021)

- Panel on Exoplanets, Astrobiology, and the Solar System
    - Science frontier questions

          * "What are the properties of individual planets, and which processes lead to planetary diversity?"
          * "How do habitable environments arise and evolve within the context of their planetary systems?"
          * "How can signs of life be identified and interpreted in the context of their planetary environments?"
    – Research avenues:
          * measuring or constraining the magnetic fields of extrasolar planets, likely via radio wavelength observations
          * improving our understanding of atmospheric escape processes, particularly those due to star-planet interactions
          * identifying how stellar processes might affect determinations of biosignatures.

- Panel on Stars, the Sun, and Stellar Populations
    – Science frontier questions
          * "What are the most extreme stars and stellar populations?"
          * "How does multiplicity affect the way a star lives and dies?"
          * "What would stars look like if we could view them like we do the Sun?"
          * "How do the Sun and other stars create space weather?"

- Section (1.1.1 and (2.1): "Worlds and Suns in Context, builds on revolutionary advances in our observations of exoplanets and stars and aims to understand their formation, evolution, and interconnected nature and to characterize other Solar Systems, including potentially habitable analogs to our own."
- "A high spatial and spectral resolution X-ray space observatory to probe stellar activity across the entire range of stellar types, including host stars of potentially life-sustaining exoplanets." p.15
- Section 2.2: "Many aspects of how stars live and die are currently uncertain enough to limit the ability to model and interpret the effects of stellar feedback, be it in the form of radiation, stellar winds, or supernovae. Even in the local universe, for example, there are significant gaps in the understanding of stellar winds." p. 69

## Planetary Science & Astrobiology Decadal

(National Academies of Sciences, Engineering, and Medicine, 2022)

- Priority Question 1 "What were the initial conditions in the Solar System? What processes led to the production of planetary building blocks, and what was the nature and evolution of these materials?"
    – Link to exospace weather: "for example, stellar and galactic cosmic rays can cause spallation—fragmentation of larger nuclides (e.g., carbon, oxygen, magnesium, and iron, or C, O, Mg, and Fe, respectively) into smaller ones (e.g., lithium, beryllium, and boron, or Li, Be, B, respectively)—which can result in detectable changes in the composition of dust."
    – Supporting activity: "telescopic observations that support cross-disciplinary studies relevant to early Solar System processes, particularly protoplanetary disks"
- Priority Question 6 "What establishes the properties and dynamics of solid body atmospheres and exospheres, and what governs material loss to and gain from space and exchange between the atmosphere and the surface and interior? Why did planetary climates evolve to their current varied states?"
    – Link to exospace weather: Figures 9-3, 9-5, and 9-6 illustrate how the solar wind incident on a planet is one of the important processes to understand.

- – Supporting activity: improved modeling to "explore the connections between the solar wind, magnetic fields, and the neutral atmosphere/exosphere"
- – Q6.2 "what processes govern the evolution of planetary atmospheres and climates over geologic timescales?"
- • Priority Question 9 "What conditions and processes led to the emergence and evolution of life on Earth; what is the range of possible metabolisms in the surface, subsurface, and/or atmosphere; and how can this inform our understanding of the likelihood of life elsewhere?"
  - – Q9.1a "How was the emergence and evolution of life on Earth influenced by volatiles, impacts, and planetary evolution in early Solar System environments?" encompasses a need to understand "the principal components of the early Solar System environment."
- • Priority Question 10 "What other potentially habitable environments exist in the Solar System, how did they form, and how do planetary and habitable environments coevolve?"
  - – "organic synthesis requires carbon and energy. In planetary surface environments, simple organic molecules like methane ($CH_4$) or formaldehyde ($CH_2O$) initially seed organic chemistry driven by ultraviolet light from the Sun, charged particles, and cosmic rays,"
  - – Q10.4b "What processes have enabled abiotic in-situ organic synthesis and cycling on planetary bodies?"
  - – Q10.7a "What exogenous factors control the continuity of habitability?"
- • Priority Question 12 "What does our planetary system and its circumplanetary systems of satellites and rings reveal about exoplanetary systems, and what can circumstellar disks and exoplanetary systems teach us about the Solar System?"
  - – Q12.6d "How does the evolution of the host star affect planetary atmospheres, including photochemistry and escape processes?"
  - – Q12.6e "What processes impact the evolution of atmospheric chemistry, cloud, and haze formation in diverse planetary atmospheres?"
  - – Q12.10c "What external factors influence the loss or maintenance of surface habitability over time on rocky-type exoplanets?", noting "a priority in this area is stellar emission: characterizing the high-energy radiation emitted from host stars (particularly extreme ultraviolet radiation and stellar winds) is vital to determining how rapidly exoplanet atmospheres are lost to space."

### NASEM Exoplanet Science Strategy

(National Academies of Sciences, Engineering, and Medicine, 2018)

- • Goal 2 tasks the exoplanet research community with creating methods to study potentially habitable planets orbiting stars similar to our Sun that go beyond the habitable zone and include the full context of the stellar and planetary environment (p. 2)
- • The section on Star-Planet Radiative Interactions and Evolution notes, especially for planets orbiting M dwarfs, that assessing the potential for habitability requires interdisciplinary studies that explore how the levels and evolution of stellar luminosity, activity, and winds interact with a planet's atmosphere and magnetic field (p. 54-59)
- • Comparisons of exoplanets with Solar System planets can lend insight into a planet's ability to maintain an atmosphere through a balance of composition, magnetic field strength, and instellation, a concept known as the "cosmic shoreline" (p. 54-55)
- • Producing a comprehensive framework for biosignature identification and interpretation is a fundamental research area dependent upon exospace weather (p. 57)

- Understanding the context of stellar and planetary environments is central to understanding biosignatures and their potential false positives and negatives (p. 59)
- A significant challenge facing the goals of understanding planets and the search for life is the limited laboratory and ab initio data available for exoplanets, especially when it comes to understanding biosignatures in light of nonthermal particle inputs (p. 57,59)
- This endeavor is too vast for any single team or method to handle alone, and the report calls for collaborative efforts that integrate the expertise of astrophysicists, planetary scientists, Earth scientists, and heliophysicists to pool their insights and resources to make meaningful progress (p. 51,113)

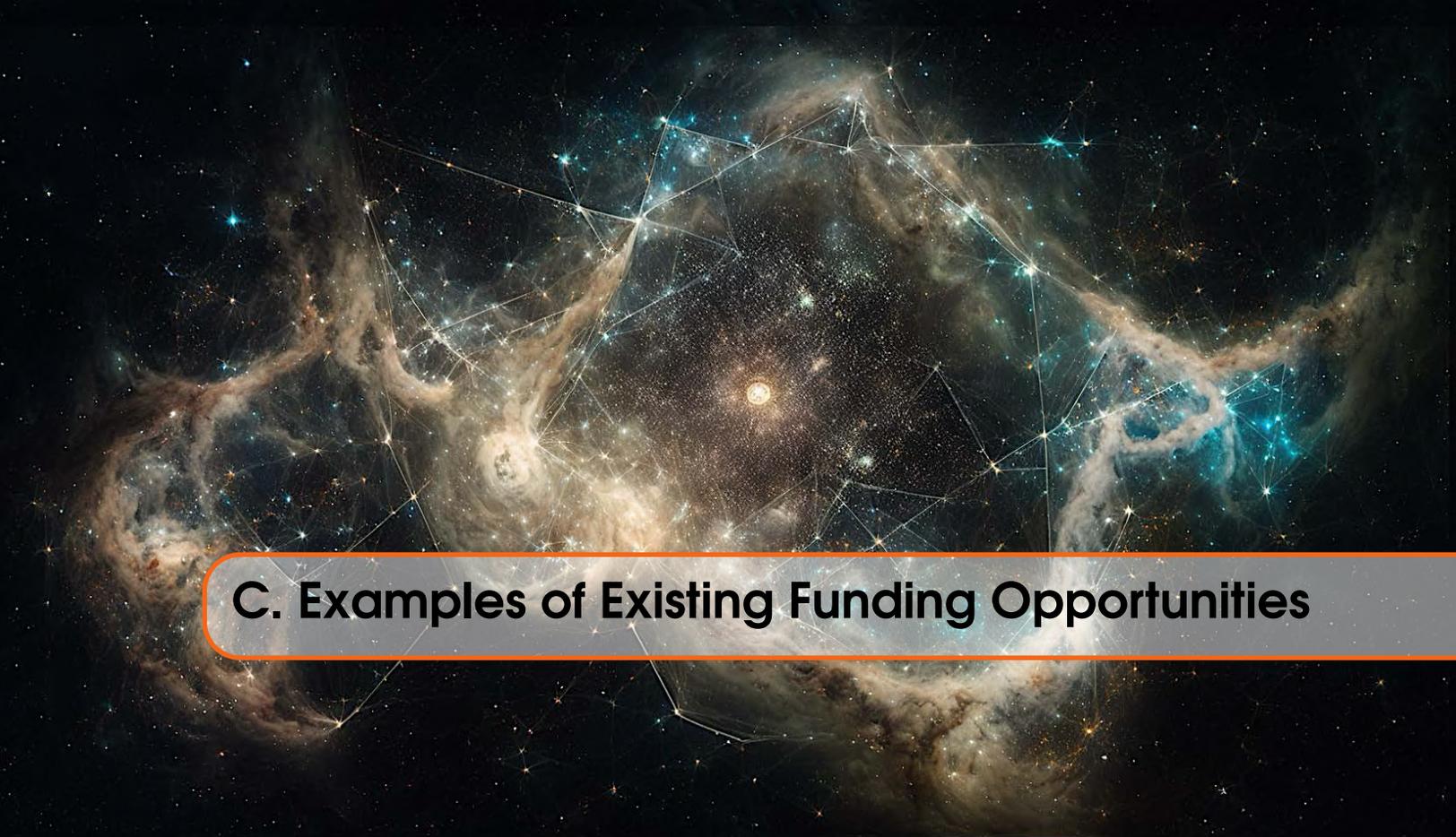

## C. Examples of Existing Funding Opportunities

This incomplete list of programs offering support for cross-disciplinary research, distilled primarily from the 2024 NASA Research Opportunities in Space and Earth Science (ROSES) omnibus solicitation, may provide a starting point for funding new exospace weather research programs.

- **B.2: Heliophysics-supporting Research**. Proposals submitted to this element should include data analysis and/or interpretation of NASA spacecraft data, as well as modeling, theory, or simulations. Proposals may aim to "Advance our understanding of the Sun's activity, and the connections between solar variability and Earth and planetary space environments, the outer reaches of our Solar System, and the interstellar medium."

- **B.5: Living with a Star**. In the 2024 ROSES call, proposal objectives were required to include one of the following: 1) Connecting auroral phenomena with magnetospheric phenomena; 2) Understanding SEP transport through the inner heliosphere; and 3) Understanding atmospheric loss and habitability in the presence of a star, including implications for exoplanets.

- **C.2: Emerging Worlds**. Proposals to this element can include studies of planetary formation, as long as the specific research case will result in an increased understanding of our own Solar System.

- **C.5: Exobiology**. This element includes soliciting research associated with the study of the macro-evolution of life on Earth, including an evaluation of environmental factors, like the influence of latitudinal variations, or extraterrestrial (including solar variation, and gamma-ray bursts) and planetary processes (e.g., climate change, etc.) on the large-scale evolution of life on Earth. Mass-radiation events are also of interest. These topics, as they relate to non-Earth bodies, are also allowed.

- **C.20: Interdisciplinary Consortia for Astrobiology Research (ICAR)**. This element includes solicitations for habitability and the detection of life on ocean worlds, including the identification of other factors beyond liquid water that affect habitability (e.g., free energy sources, physical and chemical environmental factors, and the presence of bio-essential elements). It also solicits studies in prebiotic chemistry in early Earth environments: planetary and molecular processes that set the physical and chemical conditions within which living systems may have arisen.

- **F.4: Exoplanets Research Program (XRP).** This element solicits basic research proposals about our knowledge of the formation, detection, and characterization of planets outside of our Solar System, including interactions with their host stars.

# Acronyms

**ACE**  Advanced Composition Explorer.
**ACR**  Anomalous Cosmic Ray.
**AFRL**  Air Force Research Laboratory.
**AIA**  Atmospheric Imaging Assembly.
**ALMA**  Atacama Large Millimeter Array.
**AR**  Active Region.
**AWSoM**  Alfvén Wave Solar Model.
**AXIS**  Advanced X-ray Imaging Satellite.

**CDAW**  Coordinated Data Analysis Workshop.
**CFHT**  Canada-France-Hawaii Telescope.
**CGEM**  Coronal Global Evolutionary Model.
**CIR**  Co-rotating Interaction Region.
**CME**  Coronal Mass Ejection.
**CR**  Carrington Rotation.
**CRIRES**  CRyogenic high-resolution InfraRed Echelle Spectrograph.

**DEM**  Differential Emission Measure.
**DH**  decameter-hectometric.
**DRIVE**  Diversify, Realize, Integrate, Venture, Educate.
**DSA**  Deep Synoptic Array.

**ECMI**  Electron Cyclotron Maser Instability.
**EIT**  Extreme ultraviolet Imaging Telescope.
**ELT**  Extremely Large Telescope.
**EMD**  Emission Measure Distribution.

**ESA** European Space Agency.
**ESCAPE** Extreme-ultraviolet Stellar Characterization for Atmospheric Physics and Evolution.
**ESO** European Southern Observatory.
**ESP** Energetic Storm Particle.
**ESPaDOnS** Echelle SpectroPolarimetric Device for the Observation of Stars.
**ESS** Exoplanet Science Strategy Report.
**EUV** Extreme Ultraviolet.
**EUVE** Extreme Ultraviolet Explorer.
**EVE** Extreme ultraviolet Variability Experiment.

**FIR** Far Infrared.
**FUMES** Far-Ultraviolet M-dwarf Evolution Survey.
**FUV** Far Ultraviolet.

**GALEX** Galaxy Evolution Explorer.
**GCR** Galactic Cosmic Ray.
**GL** Gibson & Low.
**GLE** Ground Level Enhancement.
**GMRT** Giant Metrewave Radio Telescope.
**GMT** Giant Magellan Telescope.
**GOES** Geostationary Operational Environmental Satellite.
**GPI** Gemini Planet Imager.

**HAO** High Altitude Observatory.
**HARPS** High Accuracy Radial Velocity Planet Searcher.
**HAZMAT** HAbitable Zones and M dwarf Activity across Time.
**HBA** High Band Array.
**HMI** Helioseismic and Magnetic Imager.
**HST** Hubble Space Telescope.
**HWO** Habitable Worlds Observatory.
**HXR** Hard X-ray.

**ICAR** Interdisciplinary Consortia for Astrobiology Research.
**ICME** Interplanetary Coronal Mass Ejection.
**IMF** Interplanetary Magnetic Field.
**IP** Interplanetary.
**iPATH** improved Particle Acceleration and Transport.
**IPM** Interplanetary Medium.
**IR** Infrared.
**ISM** Interstellar Medium.

**JVLA** Jansky Very Large Array.
**JWST** James Webb Space Telescope.

**KISS** Keck Institute for Space Studies.

**LASCO** Large Angle and Spectrometric Coronagraph.
**LBA** Low-Band Array.

**LBT** Large Binocular Telescope.

**LISM** Local Interstellar Medium.

**LOFAR** LOw Frequency ARray.

**LOS** Line of Sight.

**LWA** Long Wavelength Array.

**LWS** Living With a Star.

**M-FLAMPA** Multiple Field Line Advection Model for Particle Acceleration.

**MDI** Michelson Doppler Imager.

**MFR** Magnetic Flux Rope.

**MHD** Magnetohydrodynamics.

**MUSCLES** Measurements of the Ultraviolet Spectral Characteristics of Low-mass Exoplanetary Systems.

**MWA** Murchison Widefield Array.

**NASA** National Aeronautics and Space Administration.

**NASEM** National Academies of Sciences, Engineering, and Medicine.

**NCAR** National Center for Atmospheric Research.

**NeXSS** Nexus for Exoplanet System Science.

**ngVLA** next-generation Very Large Array.

**NICER** Neutron star Interior Composition Explorer.

**NIR** Near Infrared.

**NOAA** National Oceanic and Atmospheric Administration.

**NRH** Nançay Radioheliograph.

**NSF** National Science Foundation.

**NUV** Near Ultraviolet.

**OSO** Orbiting Solar Observatory.

**OVRO** Owens Valley Radio Observatory.

**PAG** Program Analysis Group.

**PAMELA** Payload for Antimatter Matter Exploration and Light-nuclei Astrophysics.

**PDE IRB** Planetary Data Ecosystem Independent Review Board.

**PEPSI** Potsdam Echelle Polarimetric and Spectroscopic Instrument.

**PI** Principal Investigator.

**PIC** Particle-in-Cell.

**POS** Plane of the Sky.

**PPS** Proton Prediction System.

**PROTONS** Solar Energetic Particle (SEP) proton prediction model.

**PSA** Planetary Science & Astrobiology.

**PSP** Parker Solar Probe.

**RHESSI** Reuven Ramaty High Energy Solar Spectroscopic Imager.

**ROSES** Research Opportunities in Space and Earth Sciences.

**SDO** Solar Dynamics Observatory.

**SEP** Solar/Stellar Energetic Particle.

**SFU** Solar Flux Unit.

**SIR** Stream Interaction Region.

**SKA** Square Kilometer Array.

**SMM** Solar Maximum Mission.

**SOHO** Solar and Heliospheric Observatory.

**SPARCS** Star-Planet Activity Research CubeSat.

**SPI** Star-Planet Interaction.

**SPIRou** SPectropolarimètre InfraROUge.

**SSN** Sunspot Number.

**SSP** Solar & Space Physics.

**STELa** Survey of Transiting Exoplanets in Lyman-$\alpha$.

**STEREO** Solar TErrestrial RElations Observatory.

**STIS** Space Telescope Imaging Spectrograph.

**SunRISE** Sun Radio Interferometer Space Experiment.

**SWMF** Space Weather Modeling Framework.

**SWPC** Space Weather Prediction Center.

**SXR** Soft X-ray.

**TBL** Télescope Bernard Lyot.

**TD** Titov Demoulin.

**TMT** Thirty Meter Telescope.

**TRAPPIST** Transiting Planets and Planetesimals Small Telescope.

**ULTRASAT** Ultraviolet Transient Astronomy Satellite.

**UV** Ultraviolet.

**UV-SCOPE** Ultraviolet Spectroscopic Characterization Of Planets and their Environments.

**UVCS** Ultraviolet Coronagraph Spectrometer.

**UVEX** Ultraviolet Explorer.

**VLA** Very Large Array.

**VLT** Very Large Telescope.

**WAVES** Radio and Plasma Wave Investigation.

**XMM** X-Ray Multi-Mirror Mission.

**XRP** Exoplanets Research Program.

**XUV** X-ray and Extreme Ultraviolet.

**ZAMS** Zero-Age Main Sequence.

**ZB** Zeeman Broadening.

**ZDI** Zeeman-Doppler Imaging.

**ZI** Zeeman Intensification.

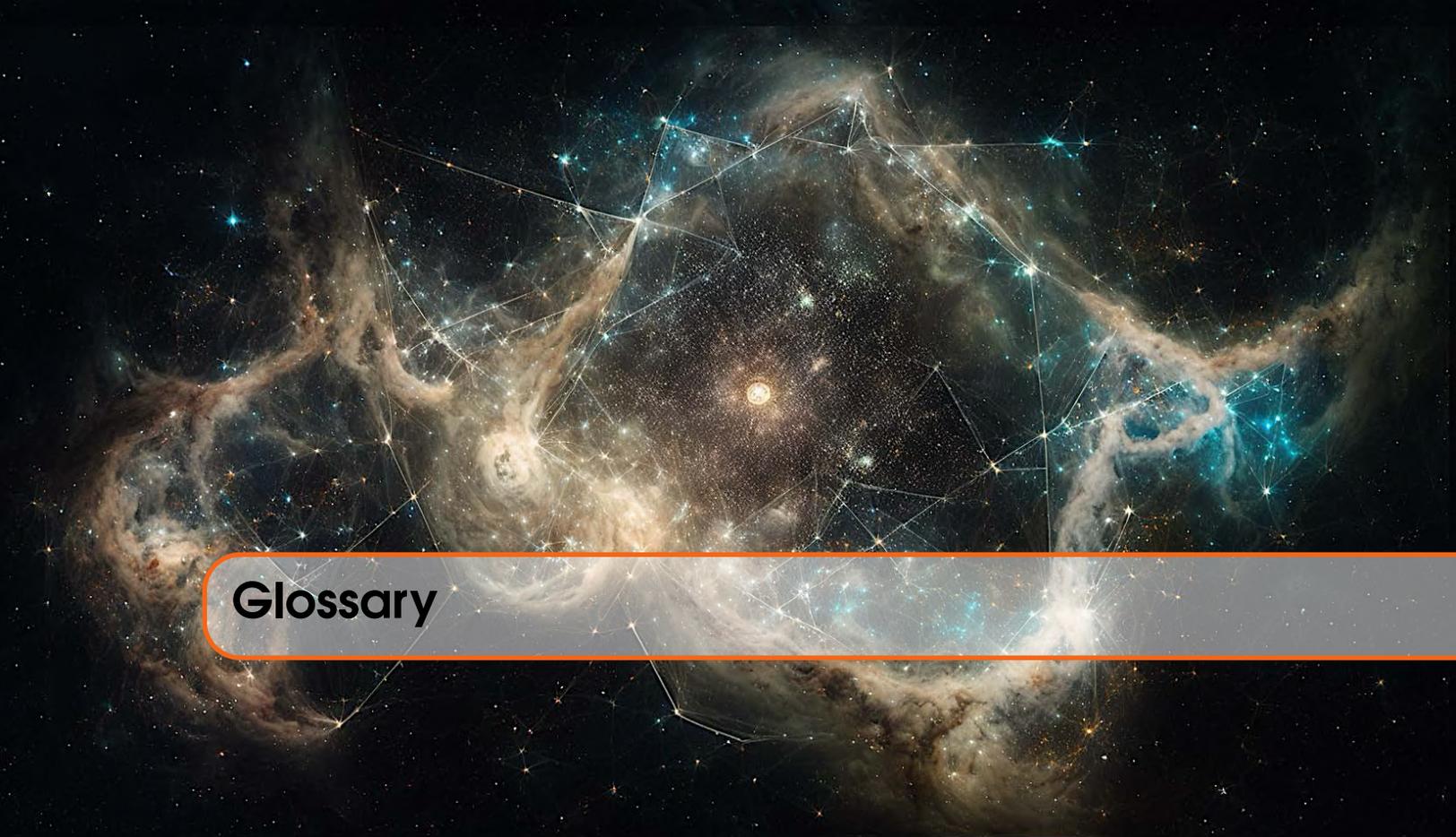

**Glossary**

**airglow**  At Earth, airglow is the emission of light from the atmosphere after absorbing UV and X-ray light. Airglow is a source of red and green line emissions in Earth's atmosphere and nitrogen emission at Mars and Venus. Airglow can also be caused by particles precipitating from the perturbed radiation belt in Earth's South Atlantic Anomaly.

**astro- / heliosphere**  The cavity carved out by a star's wind in the surrounding interstellar medium. The boundary region includes the termination shock, the astropause (or heliopause for the Sun), and a bow shock/wave. See also Section 4.2.

**atmospheric escape/outflow**  The process by which atmospheric gas is lost to space from a celestial body. Escape can be driven by thermal processes (e.g., hydrodynamic escape) or nonthermal processes involving stellar winds and CMEs (e.g., ion pickup).

**aurora**  Light emitted by atmospheric atoms and molecules that have been excited by the precipitation of energetic charged particles (electrons and protons) funneled along magnetic field lines.

**Carrington Event**  The most powerful geomagnetic storm in recorded history, occurring in 1859, which was preceded by the first-ever observation of a solar flare by Richard Carrington and R. Hodgson.

**CME, halo**  A CME that appears as an expanding ring, or "halo," in coronagraph images, which occurs when the opening angle of the CME intersects the observer's line of sight. These are often CMEs with wide opening angles.

**coronagraphy**  An instrumental technique used to block the light from a central bright source (like a star) to observe faint nearby phenomena, such as a corona, stellar wind, or exoplanets.

**dimming, coronal / mass-loss**  A temporary reduction in EUV and Soft X-ray (SXR) emission from a stellar corona, caused by the density reduction from an evacuating CME. It is a strong proxy for CME occurrence on the Sun.

**dimming, obscuration** A temporary dimming caused by cool, optically thick material (like an erupting filament) absorbing background emission along the line of sight.

**dynamo** The process by which the motion of an electrically conductive fluid within a celestial body generates a large-scale magnetic field.

**energetic neutral atom (ENA)** Neutral particles (particles with no charge) with energies that can range from 10s of eV to more than 1 MeV. They originate from charge exchange between high-energy ions and neutral atoms (typically hydrogen) at the boundary and just outside of the heliosphere.

**flare** An intense, impulsive brightening on a star's surface, caused by the release of magnetic energy through magnetic reconnection.

**flux rope** A bundle of magnetic field lines twisted helically around a common axis, thought to be the core structure of a CME.

**GOES flare class** A scheme that divides solar flares into five categories—A, B, C, M, and X—based on peak soft X-ray flux. Each class represents a 10-fold increase in peak flux. Classes A, B, C, and M can be subdivided into levels ranging from 1—9. X-class flares can go beyond 9 and have no upper limit, e.g., an X28-class flare.

**heliosphere** See "Astro- / Heliosphere".

**magnetic confinement** The prevention of a plasma eruption by a strong overlying "confining" magnetic field.

**magnetosphere** The region of space around a celestial body where the magnetic field of that body dominates over the interplanetary magnetic field. It can be intrinsic (generated by a dynamo), remnant (frozen into the mineral crust), induced (generated by the interaction of the stellar wind with an atmosphere/ionosphere or solid body), or a combination of the above. Some fields consider an internal magnetic field to be required for the region around the planet to be called a magnetosphere.

**particle** In the context of space weather, an atomic or subatomic mass like an electron, proton, or ion. In the context of this Report, the term "particle" generally excludes photons, dust, or anything other than the definition given here.

**prominence / filament** A large, cool, dense structure of plasma suspended in the hot solar corona by magnetic fields. When seen against the bright solar disk, it is called a filament; when seen above the limb, it is called a prominence. Erupting prominences are often the dense core of CME.

**radiation belt** Regions of enhanced populations of energetic electrons and protons surrounding the Earth in space. These belts are highly dynamic, increasing and decreasing on time scales of minutes to years. The high levels of radiation caused by the energetic electrons and protons makes this a very harsh region for satellites.

**radio bursts (solar/stellar)** Brief, intense appearances of radio emission often associated with flares and CMEs.

**Rossby number (Ro)** A dimensionless number produced from the ratio of inertial to Coriolis forces. In order for a dynamo to exist, the Ro must be $<< 1$, so the Coriolis force is dynamically important.

**slingshot prominence**  A very large stellar prominence typically found on fast rotating stars that co-rotate with the star at distances up to a few stellar radii.

**solar-like stars**  There are three categories of stars that can be compared to the Sun. The broadest group are called "solar-like", which are stars with spectral types ranging from F to K. This group of stars has a large range of ages and metallicities. The next sub-category is "solar-analog" stars, which are stars that have spectral types between G0 and G5 with metallicities within a factor of 2–3 times that of the Sun. The most strict category is "solar twin", which are stars that are as identical as possible to the Sun, including in temperature, gravity, age, and composition.

**solar/stellar wind**  A continuous outflow of charged particles (plasma) from the Sun's or a star's corona that permeates the space around the star. See also Section 4.1.

**space weather**  Refers to the physical and phenomenological state of natural space environments, particularly concerning the Sun or stars and their interactions with interplanetary and planetary environments. Space weather encompasses various phenomena including the density, velocity, temperature, and magnetic field intensity of stellar wind, CMEs, changes in light flux, particle acceleration and propagation, and the interaction of stellar radiation and particles with surrounding planetary bodies.

**sputtering**  A process of collisional atmospheric escape where something (e.g. a small body or a particle) hits the surface or atmosphere of a planet and, through the collision, ejects material beyond the planet's effective gravitational boundary.

**switchback, magnetic field**  Traveling disturbances in the solar wind that caused the magnetic field to bend back on itself, an as-yet unexplained phenomenon that might help scientists uncover more information about how the solar wind is accelerated from the Sun.

**type II radio burst**  Slow-drifting radio bursts produced by electrons accelerated at CME-driven shock fronts.

**type III radio burst**  Fast-drifting radio bursts from mildly relativistic electrons traveling along open magnetic field lines.

**type IV radio burst**  Broadband continuum radio bursts from nonthermal electrons trapped in magnetic structures, sometimes associated with moving CMEs.